\documentclass[iop]{emulateapj}
\usepackage{psfig}
\usepackage{graphicx}

\font\aipsfont = cmsy9 scaled\magstep1

\newcommand\aips {{\aipsfont AIPS}}

\def\arcmper{\ifmmode \rlap.{^{\prime}}\else
    $\rlap.{^{\prime}}$\fi}

\def\lya{\ifmmode {\rm Ly\alpha}\else{\rm Ly$\alpha$}\fi}
\def\Lya{\ifmmode {\rm Ly\alpha}\else{\rm Ly$\alpha$}\fi}
\def\LFIR{\ifmmode {\rm \,L_{FIR}}\else ${\rm \,L_{FIR}}$\fi}
\newcommand{\Lsun}{\hbox{L$_{\odot}$}}
\newcommand{\Msun}{\hbox{M$_{\odot}$}}
\def\Msunpyr{\ifmmode {\rm\,M_\odot\,yr^{-1}} \else {${\rm\,M_\odot\,yr^{-1}}$}\fi}
\def\pyr{\ifmmode {\rm\,yr^{-1}} \else {${\rm\,yr^{-1}}$}\fi}
\def\kms{\ifmmode {\rm\,km~s^{-1}} \else ${\rm\,km\,s^{-1}}$\fi}
\def\kmps{\ifmmode {\rm\,km~s^{-1}} \else ${\rm\,km\,s^{-1}}$\fi}
 
\def\ergps{\ifmmode {\rm\,erg\,s^{-1}} \else {${\rm\,erg\,s^{-1}}$}\fi}
\def\ergpspcm{\ifmmode {\rm\,erg\,s^{-1}\,cm^{-2}} \else {${\rm\,erg\,s^{-1}\,cm^{-2}}$}\fi}
\def\surfbr{\ifmmode {\rm\,erg\,s^{-1}\,cm^{-2}\,arcsec^{-2}} \else {${\rm\,erg\,s^{-1}\,cm^{-2}\,arcsec^{-2}}$}\fi}
\newcommand{\simlt}{\la}
\newcommand{\simgt}{\ga}

\def\eg{{e.g.,}}
\def\ie{{i.e.,}}
\def\etal{{et al.}}

\def\erg{{\rm\thinspace erg}}

\def\uJy{{\rm\thinspace \mu Jy}}

\def\km{{\rm\thinspace km}}

\def\Mpc{{\rm\thinspace Mpc}}

\def\s{{\rm\thinspace s}}

\def\ergps{\mbox{$\erg\s^{-1}$}}

\def\kmps{\hbox{$\km\s^{-1}\,$}}

\def\kmpspMpc{\hbox{$\km\s^{-1}\Mpc^{-1}$}}

\def\um{\hbox{$\mu {\rm m}$}}

\slugcomment{Accepted for publication on 2010 September 25}

\shorttitle{The Spitzer High Redshift Radio Galaxy Survey}
\shortauthors{De Breuck et al.}

\begin{document}


\title{The Spitzer High Redshift Radio Galaxy Survey}


\author{Carlos De Breuck\altaffilmark{1}, 
Nick Seymour\altaffilmark{2},
Daniel Stern\altaffilmark{3},
S.~P.~Willner\altaffilmark{4}, 
P.~R.~M.~Eisenhardt\altaffilmark{3},
G.~G.~Fazio\altaffilmark{4},
Audrey Galametz\altaffilmark{1,3},
Mark Lacy\altaffilmark{5},
Alessandro Rettura\altaffilmark{6},
Brigitte Rocca-Volmerange\altaffilmark{7,8} \& 
Jo\"el Vernet\altaffilmark{1}}

\altaffiltext{1}{European Southern Observatory, Karl Schwarzschild Stra\ss e 2, 85748 Garching bei M\"unchen, Germany; e-mail: {\tt cdebreuc@eso.org}}
\altaffiltext{2}{Mullard Space Science Laboratory, University College London, Holmbury St Mary, Dorking, Surrey, RH5 6NT, United Kingdom}
\altaffiltext{3}{Jet Propulsion Laboratory, California Institute of Technology, 4800 Oak Grove Drive, Pasadena, CA 91109, USA}
\altaffiltext{4}{Harvard-Smithsonian Center for Astrophysics, 60 Garden Street, Cambridge, MA 02138, USA}
\altaffiltext{5}{National Radio Astronomy Observatory, 520 Edgemont Road, Charlottesville, VA 22903, USA}
\altaffiltext{6}{Department of Physics and Astronomy, University of California, Riverside, CA 92521, USA}
\altaffiltext{7}{Institut d'Astrophysique de Paris, UMR7095 CNRS, Universit\'e Pierre et Marie Curie - Paris 6, 98bis boulevard Arago, 75014 Paris, France}
\altaffiltext{8}{Universit\'e Paris Sud, B\^at 121, 91405 Orsay Cedex, France}

\begin{abstract}

We present results from a comprehensive imaging survey of 70 radio
galaxies at redshifts $1<z<5.2$ using all three cameras onboard the
{\it Spitzer Space Telescope}.\/ The resulting spectral energy
distributions unambiguously show a stellar
population in 46 sources and hot dust emission associated with the
active nucleus in 59.  Using a new {\em restframe} $S_{3\,
  \micron}/S_{1.6\,\micron}$ versus $S_{5\,\micron}/S_{3\,\micron}$
criterion, we identify 42 sources where the restframe 1.6\,$\mu$m
emission from the stellar population can be measured.  For these radio
galaxies, the median stellar mass is high, $2 \times
10^{11}$\,\Msun, and remarkably constant within the range
$1<z<3$. At $z>3$, there is tentative evidence for a factor of two
decrease in stellar mass. This suggests that radio galaxies have
assembled the bulk of their stellar mass by $z \sim 3$, but
confirmation by more detailed decomposition of stellar and AGN
emission is needed.

The restframe 500\,MHz radio luminosities are only marginally
correlated with stellar mass but are strongly correlated with the
restframe 5\,$\mu$m hot dust luminosity.  This suggests that the radio
galaxies have a large range of Eddington ratios. We also present new
Very Large Array 4.86 and 8.46\,GHz imaging of 14 radio galaxies and
find that radio core dominance --- an indicator of jet orientation ---
is strongly correlated with hot dust luminosity. While all of our
targets were selected as narrow-lined, type~2 AGNs, this result can
be understood in the context of orientation-dependent models if there
is a continuous distribution of orientations from obscured type~2 to
unobscured type~1 AGNs rather than a clear dichotomy. Finally, four
radio galaxies have nearby ($<$6\arcsec) companions whose mid-IR
colors are suggestive of their being AGNs.  This may indicate an
association between radio galaxy activity and major mergers.

\end{abstract}


\section{Introduction}

Across cosmic time, powerful radio sources are robust beacons of the
most massive galaxies in the Universe.  At low redshift, this has been
known since the first visible counterparts of extragalactic radio
sources were shown to be giant elliptical (gE and cD) galaxies
\citep{matthews64}.  In the more distant Universe, indirect evidence
for this correlation initially came from a variety of observations,
including the near-IR $r^{1/4}$ light profiles of high redshift radio
galaxies (HzRGs) at $1\lesssim z \lesssim 2$ in {\it Hubble Space
  Telescope} images \citep[\eg][]{pen00,zirm03} and the tendency for
HzRGs to reside in moderately rich (proto-)cluster environments
\citep[\eg][]{ven07,galametz10}. The most direct evidence comes from
the remarkably tight scatter of the observed $K$-band magnitudes of
HzRGs in the Hubble $K$--$z$ diagram
\citep[\eg][]{lilly84,eales97,best98,wvb98,jarvis01,deb02,willott03,rocca04,brookes06,bryant09}.
Assuming that the observed-frame $K$-band light is dominated by
emission from old stellar populations, these studies put HzRGs at the
top end of the stellar mass function out to $z=5.2$. However, this
assumption is uncertain due to large band-shifting effects and
remaining contributions from the AGN and young stellar populations,
which undoubtedly contribute to the observed $K$-band magnitudes
particularly at the highest redshifts where the observed $K$-band measures
rest-frame UV light. As Seymour
\etal\ (2007, hereafter S07)\nocite{sey07} argued in the precursor to this paper, the only solution to
minimize these uncertainties is to observe at longer wavelengths with
the {\it Spitzer Space Telescope}. Such observations avoid
k-correction effects by consistently observing the same restframe
wavelengths where the old stellar population peaks.  S07 argued that
the restframe $H$-band luminosity (corrected for non-stellar emission)
is the most efficacious band for deriving stellar masses because it
corresponds to the minimum in the opacity of the H$^-$ ion
\citep[e.g.,][]{john88} and is associated with a bump in the stellar
population SED for almost all stellar populations \citep[e.g.,][see
  S07 for details]{simpson99}.

The initial observations for our program, the {\it Spitzer}
High-Redshift Radio Galaxy or SHzRG project, were obtained during {\it
  Spitzer} Cycle~1 (Program ID 3329) and involved mid-IR ($\lambda>3.5$\,\micron)
imaging of 70 HzRGs at $1<z<5.2$. S07 reported the sample selection,
data processing, and initial results from the full survey. Using
imaging from all three instruments on {\it Spitzer}, S07 decomposed
the restframe visible to infrared spectral energy distributions (SEDs)
of HzRGs into stellar, AGN, and dust components and determined the
contribution of host galaxy stellar emission at restframe $H$-band to
derive stellar masses.  S07 found that $>$60\% of restframe
$H$-band light is from stars for the majority of well-studied HzRGs.
As expected from unified models of AGNs, the fraction of restframe
$H$-band luminosity due to stars is not correlated with redshift,
radio luminosity, or restframe mid-IR (5\,\micron) luminosity.  In
addition, while the stellar $H$-band luminosity was not found to vary
with stellar fraction, the total $H$-band luminosity was found to
anti-correlate with stellar fraction, as expected if the underlying
HzRG hosts comprise a homogeneous population.  From a comparison with
predicted luminosities from stellar evolution models, S07 found that
the HzRG restframe $H$-band stellar luminosities imply host stellar
masses of $10^{11}$--$10^{11.5}$\,\Msun\ for HzRGs, even at the highest
redshifts, implying an early formation epoch for these massive
galaxies. However, with a large fraction of the sources poorly sampled
longward of observed 8\,$\mu$m, many of the results of S07 were based
on upper limits or on the subset of galaxies with well-sampled SEDs.

In addition to the comprehensive results presented by S07, our SHzRG
observations have been used to complement samples of lower redshift
radio galaxies \citep[e.g.,][]{haas08} and to explore HzRG
environments \citep[e.g.,][]{galametz09, doh10, galametz10, kuiper10,
  tanaka10}.  The legacy value of the SHzRG program is also
seen in the numerous studies of individual HzRGs which have drawn upon
the SHzRG data set --- e.g., Stern \etal\ (2006;
LBDS~53W091)\nocite{stern06}, Villar-Mart{\'\i}n \etal\ (2006;
MRC~2104$-$242)\nocite{vil06}, Broderick \etal\ (2007;
PKS~0529$-$549)\nocite{bro07}, Greve \etal\ (2007;
4C~41.17)\nocite{gre07}, Ivison \etal\ (2008; 4C~60.07)\nocite{ivi08},
Nesvadba \etal\ (2009; USS~0828+193), Smith \etal\ (2010;
TXS~J1908+7220), and De~Breuck \etal\ (in prep.; 4C~23.56).  Follow-up
mid-IR spectroscopy of two SHzRG targets (4C~23.56 and 4C~72.26) is
also reported by \citet{sey08}.

This paper reports the results of a  Cycle~4 GTO program (Program ID 40093)
to complete the {\it Spitzer} observations of the SHzRG sample.  The
observations presented here entail complete six-band 3.6 to 24\,$\mu$m
mid-IR imaging photometry for the entire SHzRG sample using all three
instruments on-board {\it Spitzer}. In addition, new radio observations complete
the radio morphological parameters for our sample.
Throughout we assume a concordance model of Universe expansion with 
$\Omega_M = 1 - \Omega_{\Lambda} = 0.3$, $\Omega_0 = 1$, and $H_0 =
70$\,\kmpspMpc.  Infrared luminosities presented in this paper are of
the form $\nu L_\nu/L_\odot$, where
$L_\odot=3.9\times10^{26}$\,W. Following common practice, radio
luminosities at 500\,MHz and 3\,GHz are expressed as monochromatic
luminosity densities.

\section{Data}

\subsection{Sample}

The SHzRG sample was described in detail by S07 and is listed in
Table~\ref{table.shzrg}.  In short, we selected 69 powerful radio
galaxies\footnote{We define radio galaxies as galaxies (i.e., not type~1  unobscured quasars) with restframe 3\,GHz luminosities greater than
  $10^{26}$\,W\,Hz$^{-1}$.} from the literature covering a uniform
range in both redshift (requiring $z > 1$) and 3\,GHz restframe
luminosity. In practice, all $z>3.5$ radio galaxies known in 2004 February
were included in our sample. As no low-power radio sources are
known as yet at the highest redshifts, we unfortunately are unable to
study radio power dependence at $z\simgt3$.  At lower redshifts, we
preferentially selected those sources with the most supporting data.
During Cycle~1, we inadvertently observed MRC~0211$-$256 ($z=1.300$)
instead of TXS~0211$-$122 ($z=2.340$).  As the latter is a
well-studied radio galaxy with supporting {\it Hubble Space Telescope}
data \citep{ojik97, pen01, ver01}, we have added this source to
our sample in Cycle~4, which thus now contains 70 radio galaxies at
$1.0<z<5.2$.
\begin{figure}[ht]
\psfig{file=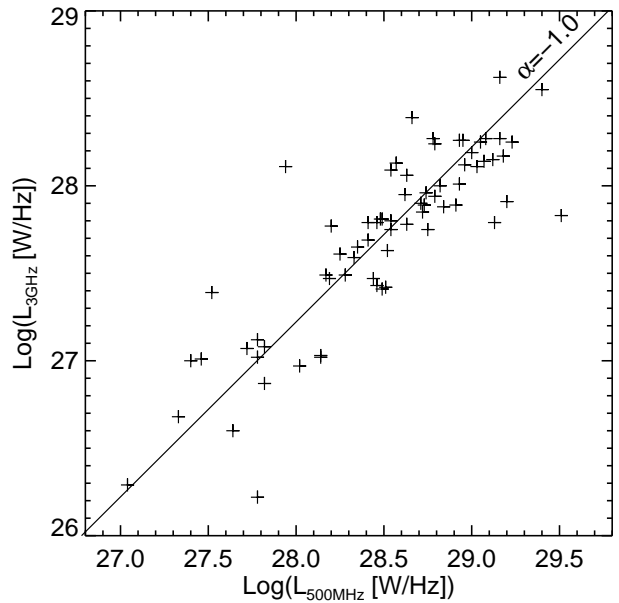,width=12cm}
\caption{Restframe 3\,GHz luminosity plotted against restframe
500\,MHz luminosity. The low scatter shows that the contribution
by Doppler-boosted radio emission in our sample is nearly always negligible. The
solid line shows spectral
index $\alpha= -1.0$, typical of the radio galaxies in our sample.}
\label{L500L3}
\end{figure}

We initially calculated the radio luminosities at a restframe
frequency of 3~GHz in order to be able to interpolate between the
all-sky, low-frequency (327/352, 365 and 1400\,MHz) radio surveys
available at the time of our Cycle~1 proposal \citep{ren97, dou96,
  con98}.  Since then, the 74\,MHz VLA Low Frequency Survey
\citep[VLSS;][]{coh07} has been completed, covering the entire sky at
$\delta>-30\degr$.  Following \citet{mil08}, we therefore now also
calculate restframe 500\,MHz luminosities for our sample using the
VLSS and the NRAO VLA Sky Survey \citep[NVSS;][]{con98}\footnote{For
  the few sources not detected in the VLSS, we used the 325\,MHz
  Westerbork Northern Sky Survey \citep[WENSS;][]{ren97} or the
  365\,MHz Texas survey \citep{dou96}.}.  Table~\ref{table.radiodata}
lists both the 500\,MHz and 3\,GHz restframe radio luminosities.  The
median flux ratio between the two frequencies is 6.3 for our sample
(see Fig.~\ref{L500L3}), corresponding to a spectral index $\alpha =
-1.0$ ($S_{\nu} \propto \nu^{\alpha}$). This steeper-than-average
spectral index reflects the steep spectrum selection criteria adopted in
most of the parent samples from which our targets were drawn.

\begin{figure}[ht]
\psfig{file=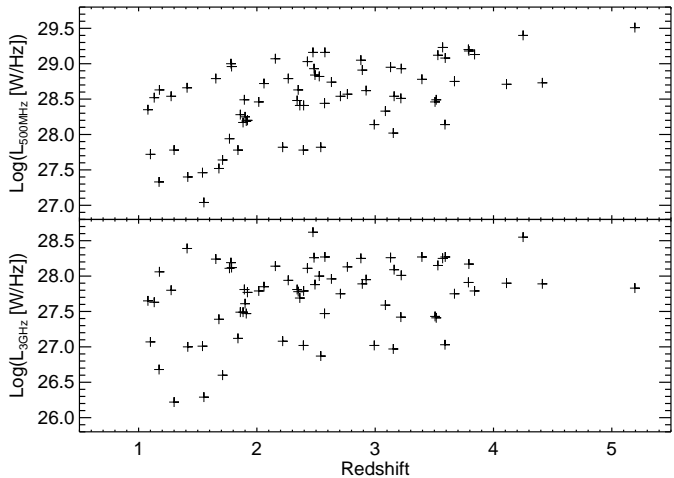,width=9cm}
\caption{Restframe 500\,MHz (top) and 3\,GHz (bottom) luminosity
plotted against redshift.  Redshift and radio power are more degenerate at 500\,MHz.}
\label{zradio}
\end{figure}

Figure~\ref{zradio} revisits the radio power--redshift plane used in
the definition of our sample.  Compared to the 3\,GHz luminosity
versus redshift plot (bottom panel), there is a slightly larger
degeneracy with $L_{\rm 500\,MHz}$ (top panel), particularly at $z>3.5$ and
$z<2$. Our sample still contains sources with $\log(L_{\rm 500\,MHz})
\sim 28.5$ throughout the redshift range $1<z<4.5$ but covers only
$\sim$1~dex at a given redshift.  As argued by S07, we do not expect
Doppler-boosted emission to significantly contribute in our sample,
even at restframe 3\,GHz.  However, in this paper we will use the
500\,MHz rather than the 3\,GHz luminosities because emission at the
lower frequency is expected to be  more nearly isotropic \citep[\eg][]{blu98}.

\subsection{Spitzer observations}

Our Cycle~1 SHzRG observations used a combination of the imaging modes
for all three instruments onboard {\it Spitzer}.  We observed (i) all
69 sources from our (initial) sample in all four bands\footnote{With
  the exception of 3C~65, which was only observed in the 3.6 and
  5.8\,$\mu$m bands (see S07). As the source is dominated by hot dust
  emission at $\lambda_{\rm obs}=5.8$\,$\mu$m, we did not re-observe
  it at 4.5 and 8.0\,$\mu$m during Cycle~4} --- 3.6, 4.5, 5.8 and
8.0\,$\mu$m --- of the Infrared Array Camera \citep[IRAC;][]{faz04},
(ii) all 46 sources at $z>2$ using the 16\,$\mu$m peak-up imaging
camera of the InfraRed Spectrograph \citep[IRS;][]{hou04}, and (iii)
all 26 sources with ``low'' predicted mid-IR background ($S_{\rm 24\,\micron
  }<\rm 20$\,MJy\,sr$^{-1}$) in all three bands --- 24, 70, and
160\,$\mu$m --- of the Multiband Imaging Photometer for {\it Spitzer}
\citep[MIPS;][]{rie04}. The Cycle~4 program 
completes the full six-band 3.6 to 24\,$\mu$m photometry.
Table~\ref{table.shzrg} summarizes all {\it Spitzer} imaging
observations of our sample, including those previously published by
S07.  A handful of sources have significantly deeper {\it
  Spitzer} exposures in one or more bands (e.g., deep IRAC
observations of 4C~41.17; IRAC and MIPS observations of B2~0902+34).
These were obtained by other programs (GO and GTO) and are included
here as part of the SHzRG sample.  We now
describe the Cycle~4 SHzRG program in more detail.
Table~\ref{table.photometry} lists the full six-band photometry for
our sample, and Appendix~A shows the individual SEDs.

\begin{figure*}[ht]
\psfig{file=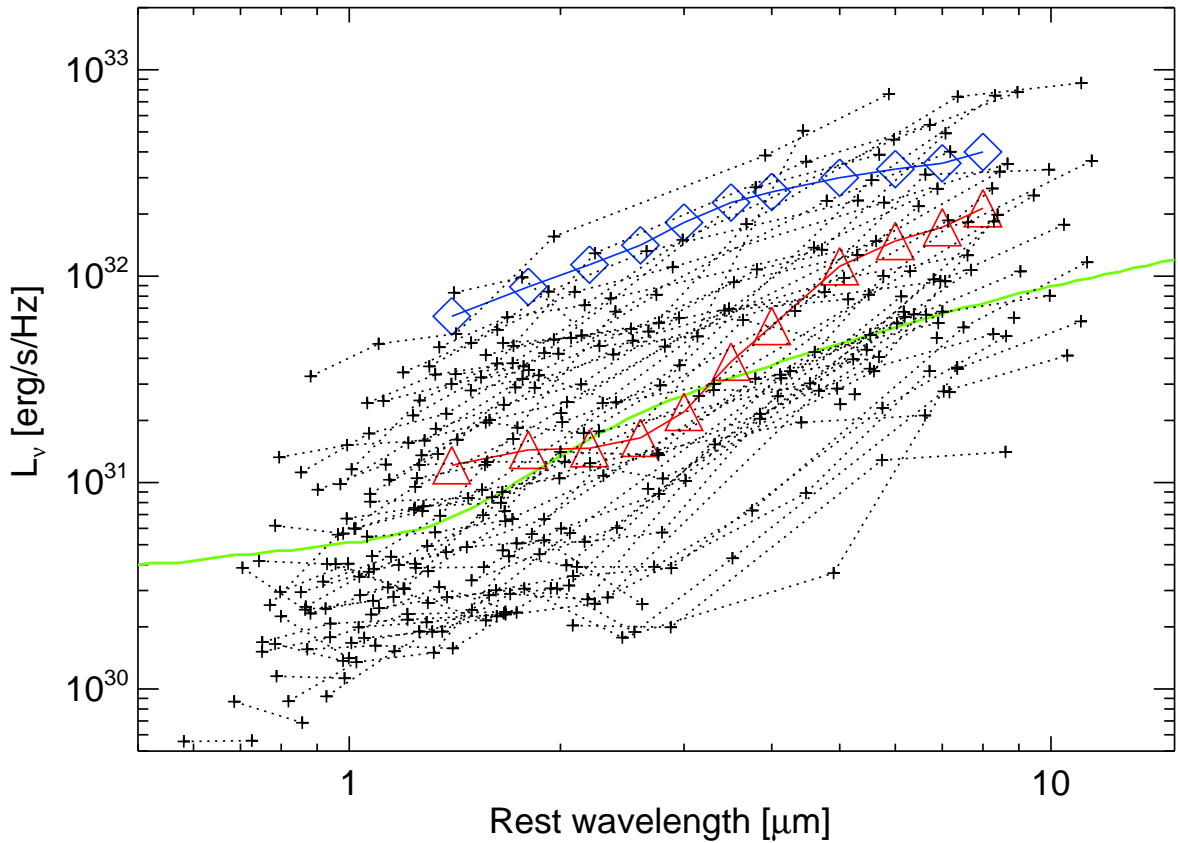,width=17cm}
\caption{Restframe luminosity density of the {\it Spitzer} data in
our sample. The red triangles show the average 3CRR radio galaxy SED
and the blue diamonds the average quasar SED from \citet{leipski10},
both based on objects with $1<z<1.4$ and normalized to radio flux
density.  The green solid line is the mean SED of all SDSS (visible
color selected, radio quiet) type-1 AGNs from \citet{ric06} normalized
to 0.08\,mJy at 2500~\AA.}
\label{datacomposite}
\end{figure*}

{\em IRAC:} We observed only TXS~0211$-$122 (see \S2.1) with IRAC
during Cycle~4.  The observations were identical to our Cycle~1
program, consisting of four dithered exposures in each of the
channels.  Data reduction followed the methods described by S07
except that we used IRAC pipeline version 16.1 instead of 13.2. As
there are no calibration differences between the
pipelines, we did not re-reduce the other IRAC data.

{\em IRS:} We observed the 24 $z < 2$ objects using the 16\,$\mu$m
peak-up imaging camera of IRS.  During Cycle~1, this observing mode
was not officially supported, and we used an experimental double
nodding sequence.  For the new Cycle~4 observations, we switched to
the more efficient peak-up astronomical observing template consisting
of five random positions of 30\,s ramp duration each. To ensure a fair
comparison between our Cycle~1 and Cycle~4 photometry, we re-obtained
all our data from the {\it Spitzer} archive (pipeline version S16.0.0)
and re-mosaiced the data with {\tt MOPEX}. Since the Cycle~1 data
consisted of just two exposures, we mosaiced the difference images
after subtracting a median to remove the background (including
residual images in a few cases). The final images were resampled on a
grid with 1\farcs8$\times$1\farcs8 pixels. We extracted the fluxes in
circular apertures with a 6-pixel radius and a background subtraction
annulus between 6 and 10.7~pixels. We applied an aperture correction
of 1.154 to our photometry to obtain total fluxes\footnote{See
  http://ssc.spitzer.caltech.edu/irs/irsinstrumenthandbook/45/}. The
new fluxes listed in Table~\ref{table.photometry} are on average
50\,\% higher than the one published by S07 due to the improved
calibration.

{\em MIPS:} Our Cycle~1 observations gave a low detection rate in
the MIPS 70\,$\mu$m imaging (five out of 24: 21\%) and no detections
at 160\,$\mu$m.  The 44 HzRGs not observed with MIPS during Cycle~1
were omitted because of the larger predicted background
emission ($S_{24\,\mu m} > 20$\,MJy\,sr$^{-1}$, see S07) at their positions. Because of the
low detection rate at 70 and 160\,$\mu$m and the higher expected
background emission due to zodiacal and Galactic cirrus emission, an
observing request aimed at detecting our complete sample at these
wavelength was considered impractical.  We therefore decided to
concentrate on 24\,$\mu$m imaging during Cycle~4.  To compensate
for the higher expected sky background, the new 24\,$\mu$m
observations consisted of two cycles of 30\,s exposures each
(totaling 28 individual images compared to a single cycle, i.e., 14
images, in our Cycle~1 observations).  One radio galaxy,
MRC~0156$-$252, was not observed before the end of the {\it Spitzer}
cryogenic mission. To ensure a fair comparison of flux densities
between Cycles~1 and 4, we re-obtained the data from the archive
(pipeline version S16.1.0) and mosaiced and derived photometry in the
same fashion as S07. Sources with predicted sky background levels
$S_{24\,\mu m} > 40$\,MJy,sr$^{-1}$ are affected by strong remaining
gradients in the combined images.  While such fields are not
appropriate to study the environment of the HzRGs (Mayo et al., in
prep.), the images can still be used to obtain reliable photometry for
the central radio galaxies.  The one exception is 5C~7.269, the source
with the highest observed background level in our sample ($S_{24\,\mu
  m} = 65$\,MJy,sr$^{-1}$), which saturated the MIPS detector.

\subsection{New radio data}

Of the 70 sources in our SHzRG sample, 49 already had high resolution
(0\farcs5) radio maps available from the Very Large Array
\citep[VLA;][]{napier83} or the Australia Telescope Compact Array
\citep[ATCA;][]{frater92}. Most of these observations are contained in
the compendia published by \citet{car97} and \citet{pen00}. To
complete the radio observations for our sample, we observed the
remaining 21 sources with the VLA in the A configuration between 2006
February 20 and 2006 March 16 (observing project ID AD520). For
consistency, we used the same observational setup as \citet{car97} and
\citet{pen00}; \ie\ 10 to 40\,min of snapshot observations in C-band
(4.86\,GHz) and 20 to 80\,min in X-band (8.46\,GHz). We followed the
standard calibration and imaging reduction steps in the Astronomical
Imaging Processing System (\aips) and the Common Astronomy Software
Applications (CASA).

Table~\ref{table.radiodata} summarizes the archival and new radio data
for our sample. We classified the radio structures into three classes: S
= single component (i.e., spatially unresolved sources); D = double
component sources without a core detection; and T = core-detected,
resolved sources, which are generally triple component sources with a
core and two radio lobes.  For all sources, we measured the largest
angular size $\theta$ \citep{car97} of the radio structure.
For sources with a core detection, we list the core fraction at a
restframe frequency of 20\,GHz, $CF_{20}$ (following the definition of
Carilli et al. 1997\nocite{car97}).  Appendix~B shows contour plots of
the new radio data.

\section{Mid-IR Spectral Energy Distributions}

\subsection{Mean HzRG SED}

As our {\it Spitzer} survey presents the largest collection of mid-IR photometry
of targeted, high-redshift type~2 AGNs, we first examine the general
trends in their SEDs.  Figure~\ref{datacomposite} shows the restframe
infrared luminosity densities from our {\it Spitzer} photometry.
There is a scatter of $\sim$2~dex at any given rest wavelength with a
notable exception around $\lambda_{\rm rest} = 1.6$\,\micron, where the
scatter decreases to 1.5~dex.  This wavelength corresponds to the peak in old
stellar populations \citep[\eg][]{saw02}, and the decreased scatter suggests that the
stellar population dominates the SEDs at this wavelength.  The 1.6\,\micron\ peak is
most pronounced in the lowest luminosity objects and tends to become
less obvious in sources with greater mid-IR luminosity.
Almost all the SEDs show a sharp rise at $\lambda_{\rm rest} > 2$\,\micron\
due to the onset of hot dust emission at longer wavelengths.

The average radio galaxy SED as derived by \citet{leipski10} is a good match to our
sample, though the most luminous sources have SEDs compatible with
either of the two quasar SEDs shown in Figure~\ref{datacomposite}.  The radio galaxy points have a steeper
rise between restframe $\sim$2 and $\sim$8\,$\mu$m than the quasar
composites.  This indicates that the hottest dust emission is highly
extincted or absent in radio galaxies, as suggested by orientation
unification models \citep[\eg][]{ogle06,cle07,haas08,leipski10}. At
$\lambda_{\rm rest} < 2$\,\micron, the radio galaxy SEDs are on average
about half as luminous as the quasar composite for a given $\lambda_{\rm
  rest}=8$\,\micron\ luminosity density.  
The radio galaxy SEDs at $\lambda_{\rm rest} <
2$\,\micron\ are at least as blue as those of
the quasars and are consistent with a stellar population rather
than hot dust emission.  In fact, for many sources a clear dip is seen
between the stellar peak at restframe 1.6\,\micron\ and the rising dust
emission at longer wavelengths.  It therefore seems feasible to derive
robust stellar masses from our {\it Spitzer} photometry, at least for
a subset of our sample, though given the large spread in SED shapes,
great care needs to be taken to separate stellar and dust emission in
each individual galaxy.  The next two subsections describe our SED
modeling and how well we can separate these components.

\subsection{SED modeling}

We used a slightly updated version of the model described by S07 to
fit the {\it Spitzer} radio galaxy SEDs. For the host galaxy
component, we used elliptical galaxy templates calculated from the
P\'EGASE\footnote{http://www.iap.fr/pegase} spectrophotometric model
\citep{fioc97} with an assumed formation redshift $z_{\rm form}$=10
and a \citet{kroupa01} initial mass function. (S07 give further
details about the template and the influences of changing the IMF, the
formation redshift, and the treatment of extinction and thermally
pulsing asymptotic giant branch stars.)  Templates for individual
galaxies differ only in overall stellar mass and in the amount of
passive evolution from $z=10$ to the individual galaxy redshift.  To
describe the dust emission, we used a simple model consisting of three
pure blackbody components: (i) a $T=60$\,K component, which was used
only for the five sources with 70\,$\mu$m detections; (ii) a
$T=250$\,K component\footnote{For 6C~0032+412, we had to modify this
  to a $T=650$\,K component, see \S4.3.2.}; and (iii) a component with
$500<T< 1500$\,K, where the temperature was a free parameter.  This
model with four free parameters (five for those sources with
70\,$\mu$m detections) was fit to our six (or seven) broad-band data
points.  For sources undetected in the longer wavelength data, we fit
a maximum dust emission model through the 3$\sigma$ upper limit at
$\lambda_{\rm obs}=16$\,\micron\ or 24\,$\mu$m, whichever is more
sensitive, in order to obtain reliable upper limits to the hot dust
emission.  When the two limits were within 10\,\% of each other, the
maximum dust emission was fitted through both. The contributions from
stellar populations at $\lambda_{\rm obs}=24$\,$\mu$m are expected to
be more than an order of magnitude lower than the depth of our MIPS
photometry and can therefore be ignored.  Physical dust models with a
proper treatment of a clumpy torus, viewing angle geometry, and
radiative transfer exist in the literature \citep[\eg][]{pier92,
  gra97, nen02, elitzur06, fri06, sie07}.  However, a detailed
analysis using these models would overinterpret our data as the models
require mid-IR spectroscopy for a detailed fitting of spectral
features.  \citet{sey08} presented IRS spectroscopy of two sources
from our sample, and further IRS spectroscopy of five sources in our
sample will be presented in a future manuscript (J. Rawlings
\etal\ 2011, in prep.). As these data cover only 10\% of our sample,
here we use only broad-band photometry and restrict our analysis to
separating the stellar and hot dust emission in order to determine
interpolated restframe luminosities.

Other components besides dust and stars can contribute to the mid-IR emission.
Non-thermal synchrotron emission has been detected in {\it
  Spitzer}/IRS spectroscopy of several 3CR sources
\citep[\eg][]{cle07, lei09}.  However, this component is only seen in
type~1 AGNs dominated by flat radio spectra. Our radio galaxies have
much steeper radio spectral indices (see Fig.~\ref{L500L3}) and have
low radio core dominance (see Table~\ref{table.radiodata}).
\citet{mil08} showed the SED of 4C~23.56, which is one of the most
core-dominated sources in our sample ($CF_{20}=14.6$\%), and even for this
galaxy, the synchrotron contribution near the peak of the far-IR SED
at $\lambda_{\rm rest}=50$\,$\mu$m is at least five orders of
magnitude lower than the dust emission. We therefore do not consider a
synchrotron component in our models.

More important contributions can be expected from mid-IR spectral
features such as PAH emission at 7.7\,$\mu$m and silicate absorption
at 9.7\,$\mu$m.  Both of these features have been detected in
individual galaxies in our sample \citep{sey08,leipski10}. However, our
photometric bands are wide, and the equivalent width of PAH emission
is low compared to the underlying hot dust continuum emission.  We
therefore do not expect these lines to have a significant effect on
our photometry.  The silicate feature may have some discernible
effect, but as we cover restframe 9.7\,$\mu$m only for $z<1.5$, at most a
small subset of our sample can be affected.  Furthermore, the
composite radio galaxy template of \citet{haas08}
implies that neither PAH emission nor silicate absorption is expected
to have strong effects on the mid-IR SEDs, and thus neither of them is
considered in our modeling.

\begin{figure*}[ht]
\psfig{file=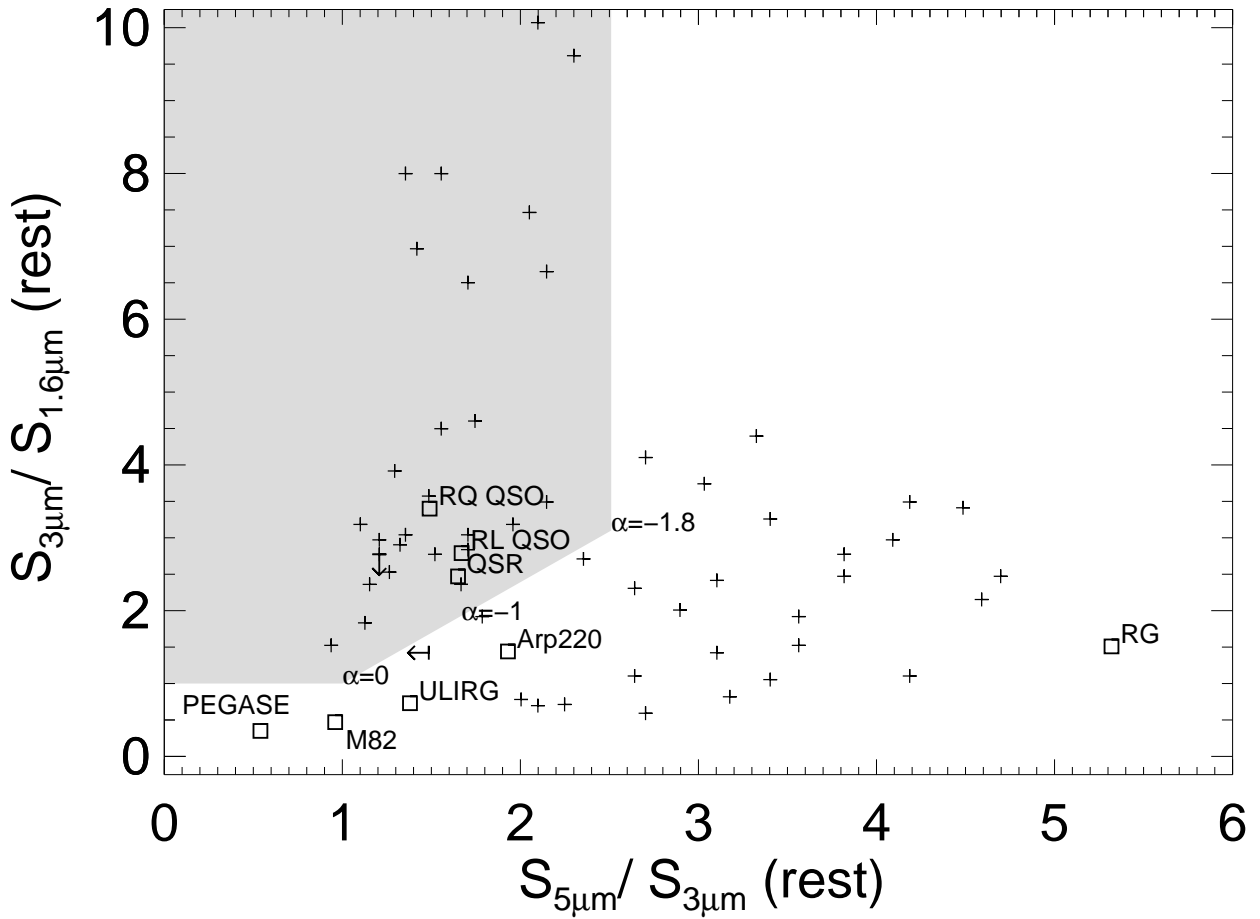,width=19cm}
\caption{Restframe mid-IR color-color diagram. The shaded area marks
the region where $\lambda_{\rm rest} \approx 1.6\,\mu$m is dominated
by hot dust rather than stellar emission. The slanted boundary marks
the locus of straight power-law SEDs with $-1.8<\alpha<0$ as
labeled. Open squares mark the locations of two well known objects
\citep[M82, Arp~220;][]{devriendt99}, the average quasar (QSR) and
radio galaxy (RG) SED of \citet{leipski10}, the radio-loud (RL) and
radio-quiet (RQ) QSO composite of \citet{elvis94}, the
ultraluminous infrared galaxy (ULIRG) SED of \citet{rieke09}, and the
P\'EGASE galaxy spectrophotometric model described in \S3.2.}
\label{restcolorcolor}
\end{figure*}

Finally, the H$\alpha$ emission line shifts into the
IRAC 3.6\,$\mu$m band at $z>3.7$ \citep[\eg][]{assef10}.  In at least one of the
seven SHzRG sources at $z > 3.7$, TN~J1338$-$1942 at $z=4.1$,
we indeed notice an unusual factor of $\sim$2 increase in the
3.6\,$\mu$m photometry compared to the $K$-band and 4.5\,$\mu$m points
(see Appendix~A).  We therefore use IRAC 3.6\,$\mu$m photometry at
$z>3.7$ only to verify the consistency of our stellar population but
not to fit the SED.

Appendix~A shows the SEDs including model fitting of the 70 sources in
our sample, ordered by increasing redshift. As already apparent from
Fig.~\ref{datacomposite}, the HzRG SEDs are quite diverse.  Some
galaxies such as 7C~1751+6809 ($z = 1.540$) show a clear stellar
contribution recognized by a decline in flux density from restframe
1.6\,$\mu$m to longer wavelengths and a  distinct hot dust
component steeply rising beyond restframe 3\,$\mu$m.  Other galaxies
such as 3C~257 ($z = 2.474$) or PKS~1138$-$262 ($z = 2.156$) have
mid-IR SEDs consistent with a power-law.  For such sources, we can
only derive upper limits to the mass of the host galaxy stellar
population. The SED of TXS~J1908+7220 ($z=3.530$) is dominated by a
transmitted quasar continuum (see \S4.3.2) and also allows only an
upper limit to be derived on the stellar population.

\subsection{Separation of stellar and dust emission}

{\it Spitzer} multi-band photometry is a powerful tool to trace the
mid-IR emission from AGNs. Various IRAC and MIPS color selection
criteria have been designed to identify AGNs \citep[e.g.,][]{ivi04,
  lacy04, ste05, pope08}.  S07 showed that 71\% and 88\% of the
SHzRG sources fall within the \citet{ste05} and \citet{lacy04} IRAC
color-color AGN selections, respectively. Because 16 sources in our
sample have only upper limits in the pivotal 8\,$\mu$m band, the
$S_{24\,\micron}/S_{8\,\micron}$ versus $S_{8\,\micron}/S_{4.5\,\micron}$ diagrams of
\citet{ivi04} and \citet{pope08} are not very useful for our sample.
These {\it observed-frame} color-color selection techniques have two
major limitations for our purposes: (1) our sample consists, by
definition, of type~2 AGNs, so we are not interested in {\it finding}
AGNs but rather in determining the AGN contribution relative to the
host galaxy; and (2) our sample covers $1<z<5.2$, while these
techniques are not uniformly sensitive over this full redshift range
\citep[e.g.,][]{assef10}.

As we have full redshift information for our sample, we set
up a new {\it restframe} color selection criterion to identify those
sources with strong dust contributions at restframe 1.6\,$\mu$m, where
the stellar population peaks.  To interpolate the data between our six
photometric data points, we used our fitted model (see \S3.2). As our
bands are wide and often contiguous, the use of this model should not
introduce significant interpolation errors as can be seen from the
individual SEDs in Appendix~A. \citet{haas08} introduced a restframe
$S_{3\,\micron}/S_{1.6\,\micron}$ versus $S_{5\,\micron}/S_{8\,\micron}$
diagram to separate the quasar and radio galaxy SEDs in their 3CR
sample at $1<z<2.5$.  We modified this criterion for use at $z>2.5$,
where MIPS 24\,$\mu$m photometry only covers out to
$\lambda_{\rm rest} \simlt 5$\,$\mu$m.  On the blue end, the data cover the
peak of the stellar emission at $\lambda_{\rm rest} \le 1.6$\,$\mu$m for
our entire sample.  Considering an even bluer $\lambda_{\rm rest}$
would be possible observationally but would increase the
contributions from younger stellar populations (see S07) as well as
non-stellar contributions such as scattered quasar emission and
nebular emission \citep[see][]{mil08}.  We also wanted our criterion to
pick up a possible minimum in the SEDs longward of the Rayleigh-Jeans
fall-off of the stellar emission at $\lambda_{\rm rest} \simgt 2$\,$\mu$m
and the onset of the steeply rising, hot dust continuum.  While the
stellar peak has a unique shape, the dust continuum slope depends on the
temperature and obscuration of the hot dust emission.  Moreover, the
relative offset between the stellar and hot dust components can vary
substantially as discussed in \S3.1.  We therefore identified
$\lambda_{\rm rest}=3$\,$\mu$m as the pivot wavelength for
our new criterion. This wavelength is also located centrally between
1.6 and 5\,$\mu$m, which minimizes the effect of measurement errors
on our photometry.  Table~\ref{table.parameters} lists the
monochromatic luminosities at 1.6, 3, and 5\,$\mu$m as derived from
our SED fitting.  Table~\ref{table.parameters}  also lists the fraction of stellar light at
$\lambda_{\rm rest}=1.6$\,$\mu$m and the stellar masses derived 
following the procedures of S07.

Figure~\ref{restcolorcolor} shows the restframe $S_{3\,\micron}/S_{1.6\,\micron}$
versus $S_{5\,\micron}/S_{3\,\micron}$ color-color diagram
for our sample. All AGN-dominated SEDs cluster around $(S_{5\,\micron}
  /S_{3\,\micron},S_{3\,\micron}/S_{1.6\,\micron}) \sim (1.5,3)$.  This
corresponds to an SED which is still steepening at $\lambda_{\rm
  rest}<3$\,$\mu$m, indicative of the warm temperature limit of the
dust emission with a possible contribution from extinction
\citep[\eg][]{haas08,leipski10}. Sources with very red SEDs
at $\lambda_{\rm rest}<3$\,$\mu$m (upper left corner
in Fig.~\ref{restcolorcolor}) are therefore likely to be dominated by
AGN dust emission. \citet{park10} showed that {\it Spitzer} power law
galaxies have 3--24\,$\mu$m spectral indices $-1.8<\alpha<0$
($S_{\nu}\propto \nu^{\alpha}$). We therefore defined the straight
power law locus as a boundary in Fig.~\ref{restcolorcolor} to separate
AGN dust emission and host galaxy (stellar) dominated HzRGs. Sources
with $\alpha < -1.8$ are indicative of strong extinction already in
the $\lambda_{\rm rest}=3$--5\,$\mu$m range and are not expected to
have significant remaining dust emission at 1.6\,$\mu$m. Sources with
$S_{3\,\micron}/S_{1.6\,\micron} < 1$ are consistent with stellar
populations as confirmed by the positions of the P\'EGASE and M82 SED
in Fig.~\ref{restcolorcolor}. Summarized, we defined sources dominated by an AGN at 1.6\,$\mu$m by the criterion:
$$S_{3\,\micron}/S_{1.6 \,\micron} > 1$$ $$\cap$$  $$S_{3\,\micron}/S_{1.6 \,\micron} > 1.39 \times S_{5\,\micron}/S_{3 \,\micron} -0.39$$ $$\cap$$ $$S_{5\,\micron}/S_{3 \,\micron} < 2.5$$

Even with our criterion,
classification of individual objects is uncertain, especially near the
boundaries: (1) measurement errors may shift objects across the
boundaries, (2) variations in the interpolation between the bands
using the SED model described in \S3.2 cause some uncertainties, and
(3) contributions from other non-stellar components such as a
transmitted or scattered quasar continuum might require a more complex
model for some sources (e.g., TXS~J1908+7220, see
\S4.3.2).  For individual sources, a more accurate separation would be
possible when using additional supporting data such as IRS spectra or
multi-band near-IR and visible imaging. This is beyond the scope of
the current paper and would also introduce redshift dependence in the results.
We are mainly interested in looking for statistical correlations, and
our rest-frame color criterion allows a uniform
treatment for all sources within our sample.

For 15 sources in our sample (mostly at $z>2$), we could obtain only
upper limits on the hot dust emission; these were determined from the
MIPS 24\,$\mu$m limits only, as our MIPS images are significantly more
sensitive than the IRS 16\,$\mu$m imaging.  However, the IRS and MIPS
photometry is sufficiently sensitive to assure that the IRAC
detections are dominated by stellar emission (e.g., TN~J1338$-$1942 at
$z = 4.1$). In total, we derived reliable stellar masses for 42 HzRGs
(see also \S 4.1). For the remaining 28 sources, we used the same
procedure to correct the $\lambda_{\rm rest}=1.6$\,$\mu$m flux from
the extrapolated hot dust model and fit the stellar template to the
remaining photometry. This procedure does become less reliable: the
median 1.6\,$\mu$m stellar fraction of these 28 sources is 0.71
compared to 0.96 for the sources outside the AGN-dominated area (see
table~\ref{table.parameters}). While the stellar masses may be close to
the real ones (especially for sources near the boundary in
Fig.~\ref{restcolorcolor}), we conservatively treat them as upper
limits.

\section{Results}
\label{s:results}

Having identified that both stellar mass and AGN-heated dust emission
contribute significantly to the near- and mid-IR SEDs of HzRGs, we now
examine the correlation between the physical parameters in our
sample. In order to properly consider the upper limits in the
statistical tests of the significance of correlations, we make use of
survival analysis and in particular the generalized Spearman's
$\rho$ coefficient \citep[\eg][]{isobe86}.  Results are expressed in
terms of the probability $P$ of seeing the observed or a higher value
of $\rho$ if the null hypothesis that the data are uncorrelated
holds. Thus a small value of $P$ indicates a high likelihood that a
correlation exists.

\begin{figure*}[ht]
\psfig{file=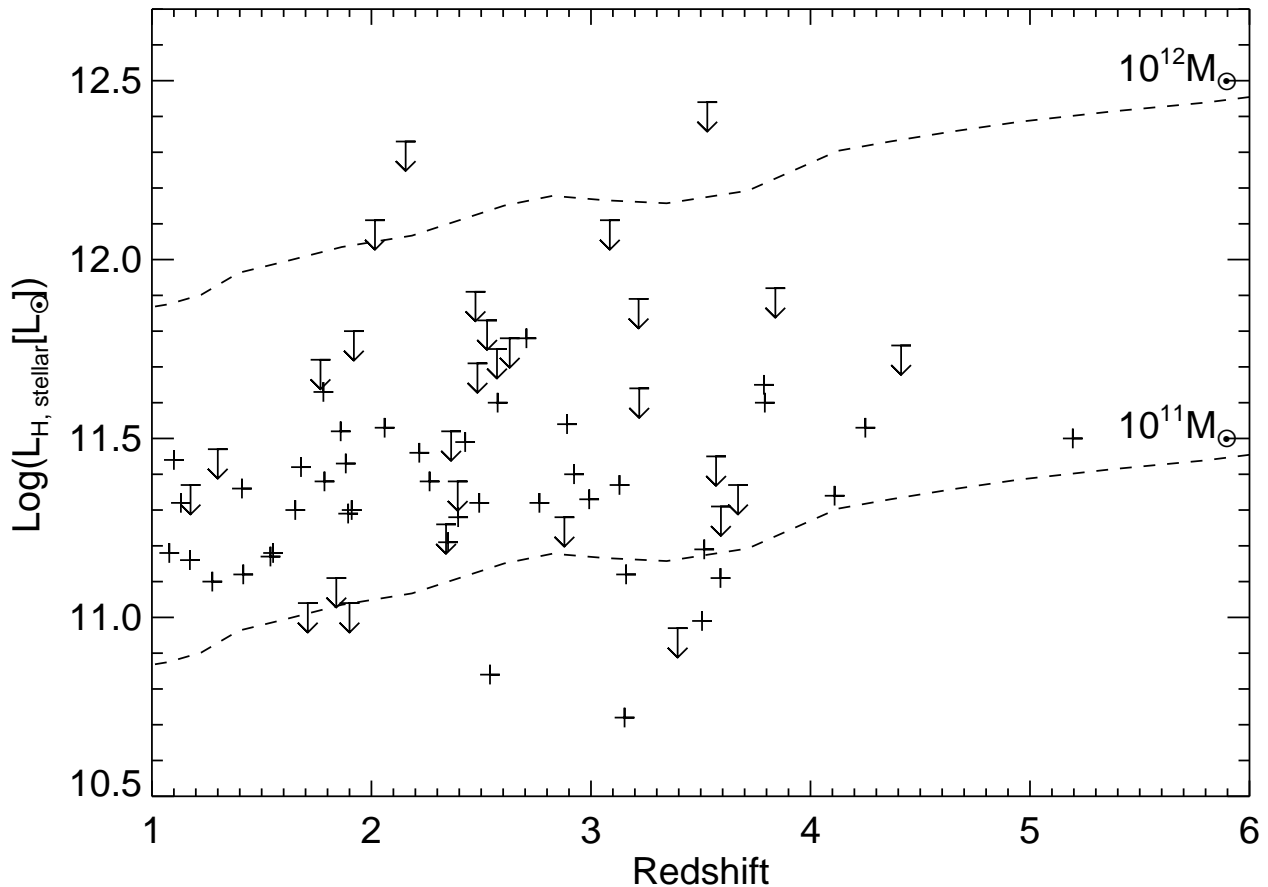,width=18cm}
\caption{Restframe $H$-band stellar luminosity versus redshift for
the {\it Spitzer} HzRG sample, derived from the best-fit models to the
multi-band photometry.  Upper limits indicate radio galaxies where
only a maximum fit to the stellar SED was possible.  The dashed lines
show the luminosities of  {\tt P\'EGASE.2} elliptical galaxy models \citep{rocca04}
with $z_{\rm form}=10$ and masses $10^{11}$ and $10^{12}$\,\Msun.}
\label{z_LHstellar}
\end{figure*}

\subsection{Stellar masses}

The original motivation of our {\it Spitzer} survey was to derive
accurate stellar masses by consistently observing the same restframe
wavelength where the old stellar population peaks, which
S07 argued is at restframe $H$-band. Figure~\ref{z_LHstellar} updates the
HzRG restframe $H$-band stellar luminosity density--redshift
relation.  The only difference from the S07 version is that
with full IRS and MIPS photometry, we now have a better handle on the
hot dust emission, particularly for sources where previously we obtained
only upper limits on the stellar masses (see \S3.3). The stellar
rest-frame $H$-band luminosity shows no correlation with redshift ($P=0.42$).

\begin{figure}[ht]
\vspace{0.8cm}
\psfig{file=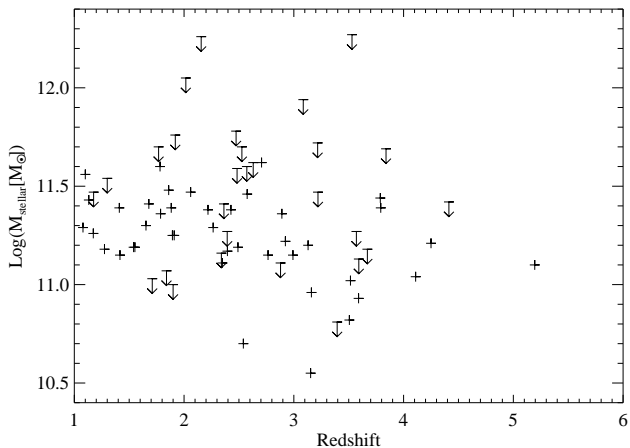,width=9cm}
\vspace{0.8cm}
\caption{Stellar mass versus redshift for the {\it Spitzer} HzRG
sample.  Masses are based on the SED decomposition described in
\S\ref{s:results} and the  {\tt P\'EGASE.2} models \citep{rocca04}.}
\label{z_Mstellar}
\end{figure}

Figure~\ref{z_Mstellar} shows the redshift evolution of the derived
stellar masses.  We confirm the result of S07 that HzRGs have high
stellar masses at every redshift.  However, we now find a trend for
the $z>3$ radio galaxies to be slightly less massive ($P=0.0044$).
The drop in stellar mass becomes apparent only at $z\simgt 3$.
Considering only the subset of sources at $z<3$, the correlation
becomes insignificant ($P=0.19$).  We caution that the putative mass decrease is based
on only 11 reliable stellar mass determinations at $z>3$, and  the
decrease in stellar mass is very small: the median stellar mass of
$z<3$ HzRGs is $2.3 \times 10^{11}$\,\Msun\ versus 
$1.6 \times 10^{11}$\,\Msun\ for  $z>3$.  If
confirmed, the difference would indicate that radio galaxies are still forming
at $z>3$ but have already assembled the bulk of their stellar mass by
$z\sim 3$. This pivotal redshift is consistent with the increase in
the submillimeter detection rate of radio galaxies at $z>3$,  where
submm flux
densities above a few mJy imply extreme star formation rates exceeding
1000\,\Msun\,yr$^{-1}$ \citep{archibald01, reuland04}.  A more
elaborate spectral decomposition would be required to better quantify
the reality of this small drop in stellar mass. In particular,
rest-frame visible and UV photometry and polarimetry are needed to
properly extrapolate the scattered and transmitted AGN
contributions. In addition, dust obscuration may affect even
restframe near-IR emission in the host galaxy
\citep[\eg][]{dey08}. While the bright submm emission in some $z>3$
radio galaxies suggests they may indeed contain large amounts of dust
(though the submm brightness is most likely dominated by higher star
formation rates), we do not expect this dust to cause large
obscurations because HzRG extended emission lines, in particular
Ly$\alpha$, are less obscured at $z>3$ than at $z<3$ \citep{villar07}.

\subsubsection{Dependence of stellar mass on radio luminosity and black hole activity}

\begin{figure}[ht]
\psfig{file=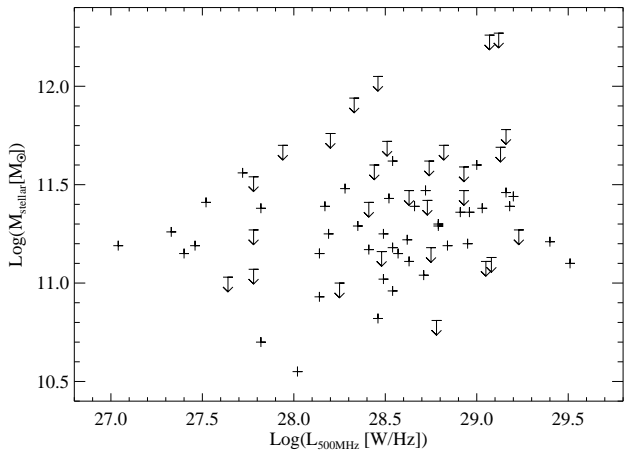,width=9cm}
\caption{Stellar mass (\S\ref{s:results}) versus restframe 500\,MHz
  radio luminosity for the {\it Spitzer} HzRG sample.}
\label{L500_Mstellar}
\end{figure}

Figure~\ref{L500_Mstellar} revisits the marginal correlation between
stellar mass and radio power discussed by S07. Our new plot is more
reliable in two ways: (i) it uses the more nearly isotropic 500\,MHz
rather than the 3\,GHz radio luminosity, and (ii) it has more reliable
determinations of stellar mass upper limits (see \S 3.3).  Our full
data set confirms the result of S07 that radio power and stellar mass
are not significantly correlated ($P=0.26$).  The absence of a
stronger $M_{\rm stellar}$--$L_{\rm 500\,MHz}$ correlation is slightly
surprising given previous claims of a correlation between radio power
and {\it observed} $K$-band magnitude \citep{eales97, best98,
  jarvis01, deb02, willott03}.  Apart from the difficulty in
disentangling redshift and radio luminosity effects in  samples
affected by a strong Malmquist bias, the correlations with $K$-band
magnitude apparently stem from other components that correlate with
radio power such as transmitted and/or scattered quasar light,
nebular emission (continuum and/or lines), or (in the low-redshift
objects) the tail of the hot dust emission.

The limited radio luminosity range of our sample and the  small
spread in stellar masses make it difficult to detect any but a
strong correlation. However, we can also turn this argument around
and suggest that the low stellar mass spread is proof that radio power, stellar
mass, or both are not strongly correlated with the accretion
properties of the central supermassive black hole (SMBH). The
correlation between the mass of the nuclear black hole and the stellar
bulge is now well established \citep{magorrian98,haering04}.  If this
relation also holds for HzRGs, the small range in stellar masses would
also imply a small range in black hole masses. Based on spectroscopy
of six broad H$\alpha$ lines in HzRGs\footnote{Five of which are also
  in our SHzRG sample.}, \citet{nesvadba10} do indeed find all black
hole masses in the range $M_{\rm BH} = 3\times 10^9$ to
$2\times10^{10}$\,M$_{\odot}$, which puts them only slightly offset from
the local $M_{\rm BH}$--$M_{\rm bulge}$ relation.  While further $M_{\rm BH}$
determinations would be needed to check this trend, it seems likely
that HzRGs contain SMBHs with masses similar to those expected.

On the other hand, low frequency radio luminosity is only one of the
estimators of the jet kinetic power \citep{willott99,cattaneo09} and
hence black hole activity.  Comparisons with other measurements, such
as the amount of kinetic power needed to create X-ray cavities
\citep{birzan08, birzan10,cavagnolo10}, show that there can be an
uncertainty of more than an order of magnitude in this
determination. Intrinsic spread in the radio luminosity--jet power
relation, environmental effects (e.g., density of the medium in which
the radio source is expanding), and variations in the Eddington ratios
of SMBHs of similar mass may therefore imply that the 500\,MHz
luminosity does not provide a very reliable measure of the black hole
properties, at least not in the limited 2~dex range of the most
luminous HzRGs studied by our sample.

\subsection{Mid-IR luminosities}

Before the launch of {\it Spitzer},
AGN-heated dust emission remained a virtually unexplored component in
$z>1$ radio galaxies, apart from mostly upper limits obtained by the
{\it Infrared Space Observatory} \citep[\eg][]{meisenheimer01, and02,
  haas04}.  S07 presented {\it Spitzer} results regarding the
mid-IR luminosities of HzRGs, and Section~3  advances the argument that the
$\lambda_{\rm rest} > 3$\,$\mu$m emission of 
HzRGs is dominated by hot dust.
The complete IRS 16\,$\mu$m and MIPS 24\,$\mu$m
photometry presented herein permit further analysis.  Hot dust emission was detected in 57 out
of 70 HzRGs; the non-detections are mainly at $z>3$ and likely are a result
of using a fixed integration time per object. This
section examines the characteristics of the hot dust emission and
what it can reveal about AGN activity. We selected a restframe
wavelength of 5\,$\mu$m in order to use a consistent luminosity
throughout the redshift range in our sample. 
Contributions from stellar emission are  negligible at this wavelength (see
Appendix~A).

\subsubsection{Orientation effects}

Low-frequency radio emission is considered to be one of the most nearly
isotropic components seen in distant AGNs. In contrast, mid-IR emission
originates from the innermost regions of the torus and is expected to be much
less isotropic due to obscuration effects.  The mid-IR luminosity in
type~2 AGNs does indeed appear to be fainter (= more obscured) than in
type~1 AGNs from the same parent radio-selected samples, providing
strong support for the orientation unification model
\citep{shi05,haas08,leipski10}.

\begin{figure}[ht]
\psfig{file=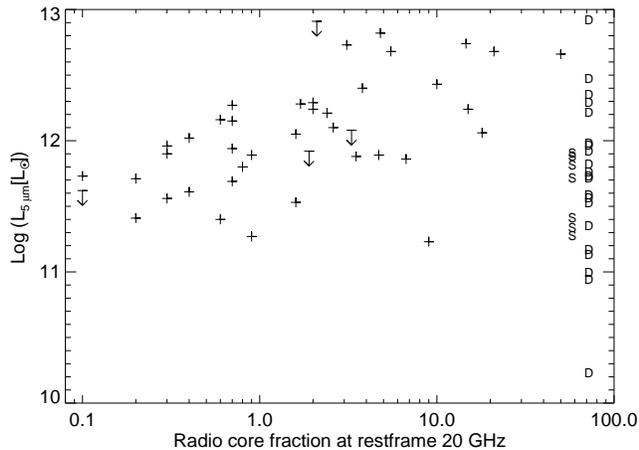,width=9cm}
\caption{5\,$\mu$m luminosity versus radio core fraction at
restframe frequency 20\,GHz. The 5\,\micron\ luminosities of the
double- and single-component radio sources that lack measurements of
their core fractions are indicated on the right hand side of the plot
by D and S respectively. The evident correlation indicates an
orientation-dependence in our sample.}
\label{CF20_L5um}
\end{figure}

While our sample is also selected at low frequency, it contains only
type~2 AGNs as determined from their restframe ultraviolet (observed
visible) spectra.  However, our mid-IR SEDs show a large variety,
ranging from clearly host-galaxy-dominated sources to SEDs that are
almost indistinguishable from those of type~1 AGNs
(Fig.~\ref{datacomposite}). This suggests a gradual transition from
highly obscured type~2 AGNs all the way to unobscured type~1 AGNs.  To
verify this hypothesis, we have checked whether radio morphological parameters
can be used as independent indicators for orientation. Both radio size
\citep[\eg][]{barthel89} and radio core dominance \citep[\eg][]{orr82}
have been suggested as measures of orientation.  The largest radio
size, $\theta$, and the hot-dust-dominated $L_{5\,\micron}$ show a likely
correlation ($P=0.014$). This suggests that $\theta$ is probably an
indicator of orientation but not a very reliable one.  On the other
hand, Figure~\ref{CF20_L5um} shows that the core fraction at
restframe frequency 20\,GHz, $CF_{20}$, is strongly correlated with
$L_{5\,\micron}$ ($P<0.0001$).  As we could determine core fractions
of only 34 of our 70 sources, we caution that this analysis is not
complete. Most of the 25 double-component sources have radio cores too
faint to be detected in the current radio maps and are expected to
have core fractions $CF_{20} << 1$\%. These 22 sources (marked by a
'D' on the righthand side of Fig~\ref{CF20_L5um}) have a median
$\langle \log (L_{5\,\micron}/L_\odot) \rangle = 11.7$ and will fall
close to the existing correlation. The remaining 8 single-component
sources have a median $\langle \log (L_{5\,\micron} / L_\odot) \rangle =
11.8$, but without higher spatial resolution observations we cannot
estimate their $CF_{20}$. However, they represent only 11\% of the
sample and are unlikely to be able to remove the
correlation completely.  We now examine the physical origin of this correlation
between the synchrotron-dominated radio emission and the hot dust
dominated $\lambda_{\rm rest}=5$\,$\mu$m emission. (As argued in \S3.2,
a direct contribution from synchrotron emission at $\lambda_{\rm
  rest}=5$\,$\mu$m can be neglected.)

Higher core fraction sources are more strongly beamed, which can be
understood if the radio jet is oriented closer along the line of sight
of the observer \citep[e.g.,][]{orr82}. These sources are therefore
more similar to type~1 AGNs. The higher hot dust luminosity in them (Fig.~\ref{CF20_L5um})
also indicates less obscuration, consistent with a more direct view of
the inner parts of an obscuring torus perpendicular to the radio
jet. This interpretation was recently confirmed by \citet{leipski10},
who used {\it Spitzer}/IRS spectroscopy to show that the silicate
absorption depth $\tau_{9.7\,\micron}$ increases with decreasing radio
core fraction (but see Landt \etal\ 2010\nocite{landt10} for a
different view).  Based on a sample of 31 3C radio galaxies,
\citet{ogle06} claimed that radio morphology is {\it not} a reliable
predictor of nuclear mid-IR luminosity. However, considering only
their 13 ``mid-IR luminous'' radio galaxies, there does seem to be a
correlation, consistent with our results which are based on a sample
three times larger. If the radio jets in the more core-dominated and
more luminous hot dust sources are indeed oriented closer to the line
of sight, we would expect other anisotropic parameters to
increase also.  Examples include the equivalent widths of the
[\ion{O}{2}]~$\lambda$3727\,\AA\ and [\ion{O}{3}]~$\lambda$5007\,\AA\ lines, both of
which are known to correlate with radio core dominance in type~1 AGNs
\citep{baker95}.  Deep restframe visible spectroscopy of our sample is
currently underway and should be able to test this hypothesis
(Nesvadba \etal\ 2011, in prep.).

We conclude that the radio core fraction is a powerful indicator of
the expected mid-IR luminosity because it provides a measure of the
obscuration. As a result, the spread in $L_{5\,\micron}$ in our sample is
in large part driven by obscuration/orientation effects with the
most core-dominated sources being the least obscured ones.

\begin{figure}[ht]
\psfig{file=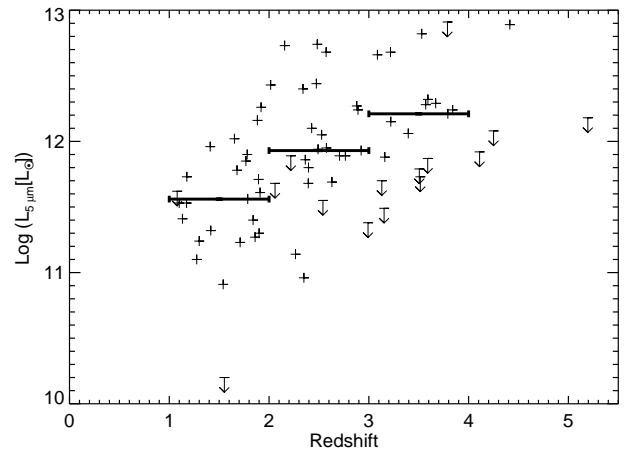,width=9cm}
\caption{Monochromatic luminosity at 5\,$\mu$m versus redshift. 
Horizontal bars indicate median luminosities in each of three redshift bins. 
Upper limits, indicated by arrows, cluster at $z>3$ owing to the use of a fixed integration
time for all sources.}
\label{z_L5}
\end{figure}

\subsubsection{Evolutionary effects}

Figure~\ref{z_L5} suggests a trend of rising restframe $5\,\mu$m
luminosity with redshift, which is confirmed by our generalized
Spearman rank analysis ($P=0.0021$).  Before interpreting this
correlation in terms of evolution of the dust content or emissivity,
it is important to consider three selection effects. First, the
5\,$\mu$m luminosity is even more strongly correlated with radio power
(see below), and there is a small remaining Malmquist bias in our
radio-selected sample (see \S2.1). This combined effect can amplify
the redshift evolution of $L_{5\,\,\micron}$, though it is unlikely to
imprint a correlation as strong as observed. Second, as shown in
\S4.3.1, orientation effects can strongly influence $L_{5\,\mu
  m}$. However, the core fractions are distributed quite evenly in
redshift, suggesting that there is no redshift-dependent obscuration
effect. Third, 9 out of 11 upper limits on $L_{5\,\,\micron}$ are at
$z>3$ compared to 13 detections at $z>3$. As shown by the survival
analysis, the data still indicate a real correlation, though deeper
mid-IR photometry at $z>3$ will be needed to quantify the actual
strength of the correlation. If confirmed, this would indicate that
higher-redshift AGNs are more efficient in heating dust.

\subsubsection{Correlation with radio power}

\begin{figure}
\psfig{file=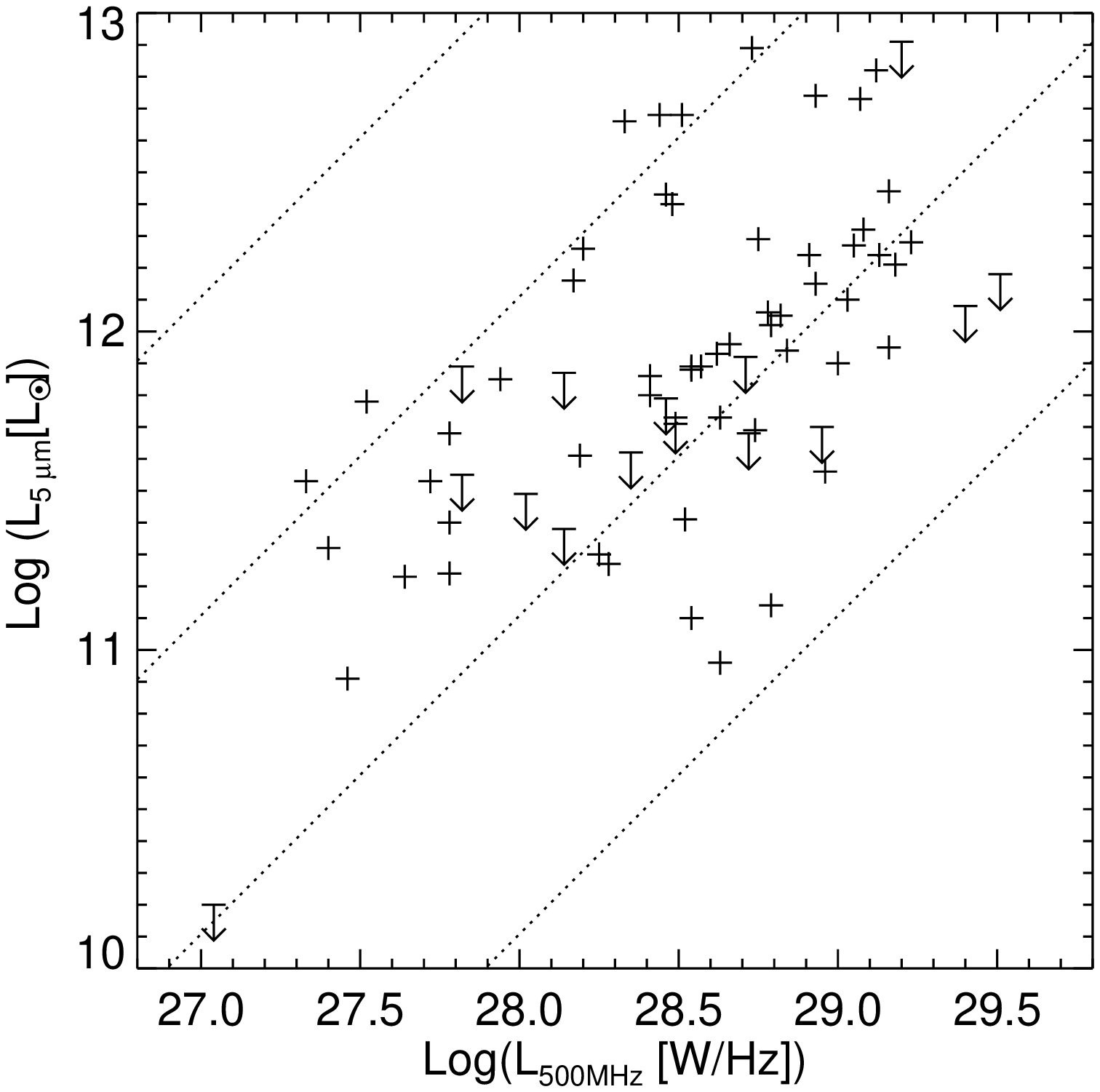,width=11.8cm}
\caption{Restframe 5\,$\mu$m luminosity against monochromatic 500\,MHz luminosity. Dotted lines indicate mid-IR to radio ratios of 1, 10, 100 and 1000 (bottom to top).} 
\label{Lradio_L5um}
\end{figure}

Several authors have compared the hot dust emission from distant
radio-loud AGNs at $8<\lambda_{\rm rest}<30$\,$\mu$m with the 178\,MHz
radio luminosity \citep[\eg][]{shi05, ogle06, cle07, haas08}. They
found that type~1 AGNs show a clear positive correlation with a scatter
of roughly an order of magnitude.
Type~2 AGNs show more scatter and generally 
fainter hot dust emission because of obscuration, though the upper
end of the type~2 range reaches the locus of
the type~1 AGNs.  Figure~\ref{Lradio_L5um} plots 5\,$\mu$m luminosity
against 500\,MHz radio luminosity density for our sample and shows a
strong correlation ($P<0.0001$). The correlation is equally
significant when using the 3\,GHz luminosity instead.
Obscuration will scatter
sources downwards in Fig.~\ref{Lradio_L5um}, and
it is remarkable that the observed correlation is so strong despite the
orientation effects discussed in \S4.3.1. We can get an idea of the
intrinsic scatter by considering only the sources with $\log L_{\rm
  5\,\micron}/L_{\odot} > 12.5$, which are likely to experience only
minimal extinction. These 8 sources cover a range of radio powers
$28.3< \log L_{\rm 500\,MHz} <29.3$. If all sources within this radio
power range have the same intrinsic $\log L_{5\,\micron}$, they
exhibit up to 2 orders of magnitude of extinction. The fact that none
of the 23 sources with $\log L_{\rm 500\,MHz} <28.3$ has $\log L_{5\mu
  m}/L_{\odot} > 12.5$ indicates that the ratio $L_{5\,\micron} /
L_{\rm 500\,MHz}$ has an upper limit around a few hundred (see dotted
lines in Fig.~\ref{Lradio_L5um}).  However, it is impossible to determine the intrinsic spread in
$L_{5\,\micron} / L_{\rm 500\,MHz}$ because of
the range of extinction in $L_{5\,\micron}$.  Orientation effects will
increase the spread in the $L_{5\,\micron}$--$L_{\rm 500\,MHz}$
plane, and the intrinsic correlation may be even stronger than
observed.

The $L_{5\,\micron}$--$L_{\rm 500\,MHz}$ correlation implies that both
are related to the energy output of the SMBH despite the vastly
different spatial scales of the emission. The hot dust emission
originates from the inner regions of the host galaxy near the SMBH on
scales which may be as small as a few pc \citep[\eg][]{jaffe04}, while
the radio luminosity is dominated by the extended radio lobes often
extending beyond the host galaxy and reaching scales up to hundreds of
kpc (Table~\ref{table.radiodata}).  It is therefore remarkable that
the scatter in the relation is limited to only two orders of
magnitude.  Naively, one might have expected a much larger spread, as
it may be influenced by several effects such as (i) orientation
effects as discussed above, (ii) different Eddington accretion rates
onto the SMBH, (iii) time delays related to the very different size
scales of the emitting regions, (iv) different dust distributions or
compositions (\eg\ in 6C~0032+412, see \S4.1.2), and (v) environmental
effects related to the medium in which the extended radio sources are
expanding.  The fact that the radio luminosity is more strongly
correlated with hot dust emission than with the stellar mass of the
host galaxy (\S 4.2.1) suggests that variations in the Eddington ratio
of the AGN may be the dominant factor in driving both the radio power
and hot dust emission. \citet{hickox09} and \citet{griffith10} also
suggest that radio-loud AGNs show a larger range in Eddington
luminosity than X-ray and mid-IR selected AGNs.

\subsubsection{Absence of a correlation with stellar mass}

\begin{figure}[ht]
\psfig{file=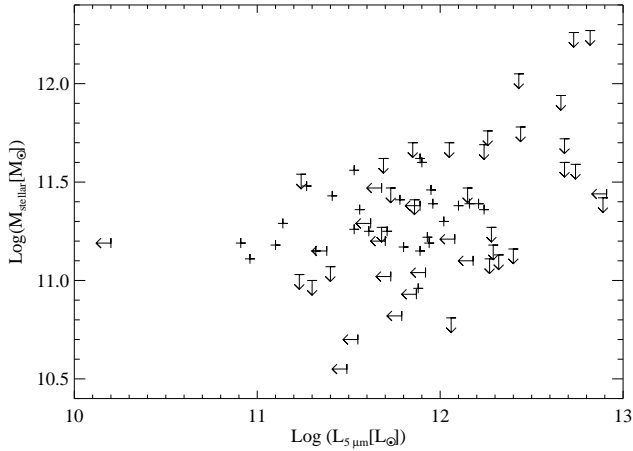,width=9cm}
\caption{Stellar mass versus 5\,$\mu$m luminosity.  The lack of
  correlation implies that
  our criteria to isolate hot-dust dominated SEDs are successful and
  that stellar mass is uncorrelated with 5\,$\mu$m luminosity.} 
\label{L5um_logM}
\end{figure}

Does hot dust emission depend on the mass of the host galaxy?
Figure~\ref{L5um_logM} plots the hot dust dominated 5\,$\mu$m
luminosity against stellar mass, showing that the two are uncorrelated
($P=0.73$). The upper limits in the top right corner illustrate the
success of our color-color selection criterion to identify hot dust
dominated SEDs (\S 3.3, Fig.~\ref{restcolorcolor}). The absence of a
correlation also suggests that AGN luminosity is independent of
stellar mass, though we warn that orientation effects (\S 4.2.1) make
$L_{5\,\micron}$ not a reliable measure of AGN luminosity.

\begin{figure*}[ht]
\psfig{file=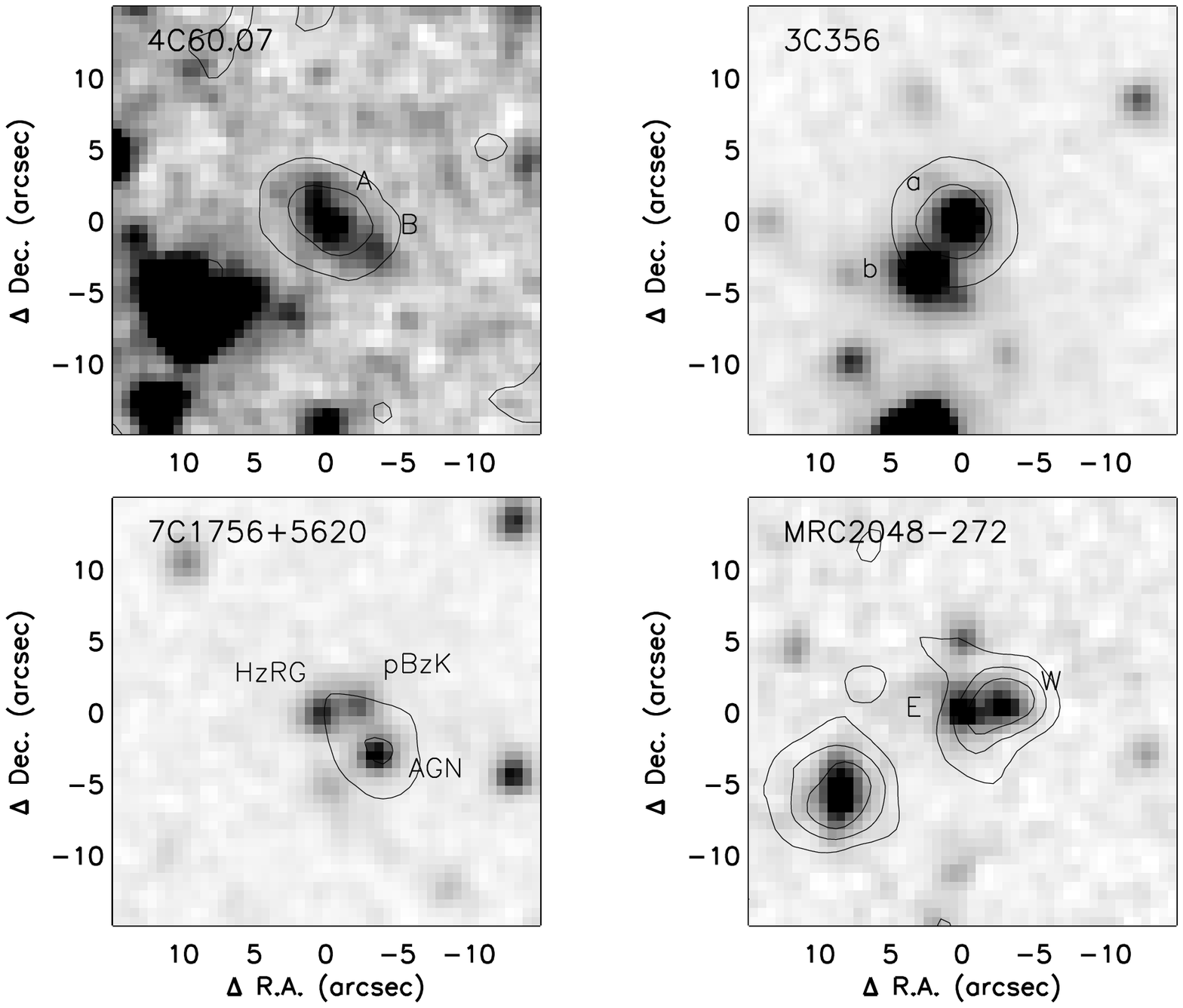,width=18cm}
\caption{\small {\it Spitzer}/IRAC 3.6\,$\mu$m images of the four
binary AGNs in our sample. Contours show the MIPS 24\,$\mu$m
images. Photometry for the HzRG and companions are listed in
Tables~\ref{table.photometry} and \ref{table.binaryAGN},
respectively.}
\label{doubleAGN}
\end{figure*}

\subsection{Notes on individual sources}

Radio galaxies are often complex systems consisting of several
components.  Therefore, our {\it Spitzer} photometry is occasionally
affected by companion and/or foreground objects.  We have visually
inspected the six-band {\it Spitzer} images for our entire sample,
using multi-wavelength data in the literature to determine the
correct astrometric identifications as listed in Table~\ref{table.shzrg}.
In most cases, we found an unambiguous radio source counterpart. 
The next two subsections describe all special cases, first
sources with multiple AGNs (\S4.3.1) followed by detailed
discussions of a few additional interesting sources (\S 4.3.2).

\subsubsection{Binary AGNs}

As shown in \S\ref{s:results}, HzRGs are massive galaxies hosting a
SMBH.  If such massive galaxies are formed in a series of merger
events as predicted by the standard $\Lambda$ Cold Dark Matter
paradigm, we expect that merging massive galaxies will coincide with
merging central SMBHs \citep[\eg][]{comerford09}.  Observationally,
this may manifest itself as a second AGN near the HzRG, or --- at
later stages in the merger event --- by perturbed morphologies as
observed in low redshift radio galaxies \citep[\eg][]{ramos10}. The
low spatial resolution of {\it Spitzer}, particularly at the longest
wavelengths, limits such studies to companion AGNs at many-kpc
distances rather than the sub-kpc separations in later stages of
merging.  The mid-IR images show evidence for a second AGN in four
HzRGs in our sample (Fig.~\ref{doubleAGN}; discussed in RA order):

{\it 4C~60.07 ($z=3.788$):} \citet{ivi08} showed that this system
consists of two AGNs separated by 4\arcsec\ (30\,kpc).  The radio
galaxy (A), as identified by the radio core, remains undetected in the
hot dust emission beyond 4.5\,$\mu$m, while the companion object~B,
which currently lacks a spectroscopic redshift, has very red hot dust
emission characteristic of an obscured AGN. An initial claim of a
CO(1-0) redshift for this second AGN \citep{greve04} could not be
confirmed in the re-analysis of those data \citep{ivi08}, but the
(sub)mm dust continuum morphology looks consistent with both targets
being at the same redshift.

{\it 3C~356 ($z=1.079$):} There are two potential identifications for
the location of the radio-loud AGN.  Based on the detection of
scattered broad line regions with a polarization angle perpendicular
to the ``dumbell'' structure of the galaxy, \citet{sim02} argued that
the northwestern object\,{\em a} is the most likely location, but
object\,{\em b}, offset by 4\farcs5 (36\,kpc), may also contain an
AGN. Both targets were spectroscopically confirmed to be at $z=1.079$.
\citet{cim97} showed that object\,{\em a} has a high restframe UV
polarization, clearly indicating the presence of a hidden AGN.  We
detected both objects in all four IRAC bands, though only object\,{\em
  a} has strong mid-IR dust emission seen with IRS and MIPS. This
provides further support for the conclusion that object\,{\em a} is
the location of the radio galaxy host, while object\,{\em b} is a
companion object. Tables~\ref{table.photometry} and
\ref{table.binaryAGN} list the photometry for both objects.

{\it 7C~1756+6520 ($z=1.416$,):} \citet{galametz09} found three objects
within 6\arcsec\ (48\,kpc) of the radio galaxy, two of which were
spectroscopically confirmed by \citet{galametz10} to be at the same
redshift as the radio galaxy.  One of them is a mid-IR selected AGN.
Both this AGN and a passively evolving $BzK$-selected (p$BzK$) galaxy
were individually detected with IRAC but only marginally resolved with
IRS and MIPS.  Tables~\ref{table.photometry} and \ref{table.binaryAGN}
provide photometry of both objects; as the HzRG and AGN are not fully
spatially resolved in the IRS 16\,$\mu$m and MIPS 24\,$\mu$m images,
we estimated the relative fluxes between the HzRG and AGN by summing
the pixel values in a fixed circular 6 pixel radius aperture using the
positions from the IRAC 3.6\,$\mu$m image (Fig.~\ref{doubleAGN}) as
centroids. The derived ratios of 50\%/50\% at 16\,$\mu$m and and
20\%/80\% at 24\,$\mu$m should be considered approximative.  This
paper uses the revised redshift of the HzRG from \citet{galametz10}.

{\it MRC~2048$-$272 ($z=2.060$):} No radio or X-ray core was detected
in this radio galaxy \citep{car97,ove05}. \citet{pen01} detected two
sources, separated by 2\farcs5 (21\,kpc), in NICMOS $H$-band imaging
and considered the easternmost one as the most likely AGN host. This
is consistent with the Ly$\alpha$ halo, which extends for $\sim$5\arcsec\
along the radio axis \citep{ven07}. We detected both components in
the IRAC 3.6 and 4.5\,$\mu$m images. However, at longer wavelengths
only the western source is detected, suggesting it also contains
an AGN. This configuration is reminiscent of 4C~60.07.

\subsubsection{Other Sources}

{\it 6C~0032+412 ($z=3.670$):} This HzRG is the only one where our
standard 3-component dust model (using $T=60$, 250, and 500--1500\,K,
see \S3.2) could not fit the SED. First, we had to increase the
temperature of the $T=250$ component to $T=650$\,K. In addition, this
is the only source where we required  $T=1500$\,K for the hottest dust component
compared to a median $\langle T \rangle=800$\,K .
A near-IR bump caused by a $T\sim1500$\,K dust
component has been seen in several quasars
\citep[\eg][]{gallagher07,mor09,leipski10} and is generally
associated with graphite dust near the sublimation temperature
\citep[\eg][]{barvainis87}. The very high temperature necessarily
locates the dust very close to the AGN. \citet{leipski10} argued that
this very hot dust emission is either more highly extincted or
completely absent in radio galaxies.  While this object is the only example
in our sample, it does represent the first counter-example to the
radio galaxy trend. 6C~0032+412 is a bona-fide type~2 AGN as derived
from its narrow-line UV and visible spectroscopy \citep{jarvis01}. The
modest 2\% radio core fraction (Table~\ref{table.radiodata})
also does not indicate a line of sight close to that of quasars (see
\S4.3.1). A possible explanation would be that 6C~0032+412 presents a
rare unobscured line of sight through the entire dusty torus all the
way to the central regions near the AGN.

{\it MRC~0037$-$258 ($z=1.10$):} The new radio imaging clearly detected
the radio core at 4.86 and 8.46\,GHz.

{\it MRC~0114$-$211 ($z=1.41$):} We identify the faint central
component, which is spatially resolved only in the 8.46\,GHz map, as the
core.

{\it 6C*~0132+330 ($z=1.71$):} We identify the central component seen
at both frequencies as the radio core.

{\it 6C~0140+326 ($z=4.41$):} \citet{rawlings96} argued that this
radio galaxy is gravitationally lensed by a foreground ($z=0.927$)
galaxy 1\farcs6 to the southeast. This foreground $L_*$ galaxy has an
ellipsoidal shape in both the Keck/NIRC $K$-band image \citep{wvb98}
and the {\it HST}/NICMOS F160W image \citep{lacy99}. Our {\it Spitzer}
photometry at 3.6 and 4.5\,$\mu$m is dominated by strong a strong
foreground star, while the 16 and 24\,$\mu$m photometry are dominated
by mid-IR emission from the lensing galaxy. The radio galaxy itself
appears undetected in all but the 5.8 or 8\,$\mu$m emission, where it
is only marginally detected and contaminated by the foreground galaxy.
We therefore provide only relatively shallow upper limits based on the
detection of the foreground galaxy.

{\it MRC~0251$-$273 ($z=3.16$):} Our radio map (Appendix~B) shows
five bright components at 8.46\,GHz as well as diffuse emission to
the north-northwest (mainly at 4.85\,GHz). The lower-resolution radio
map of \cite{kap98} confirms that this diffuse emission is
real. Within the radio structure, there is only a single IRAC
identification, coinciding within 0\farcs3 with the northernmost of
the bright components in the 8.46\,GHz map. We therefore identify this
component as the radio core. This radio morphology resembles that of
B2~0902+34 \citep{car95}. The radio jet is probably aligned relatively
closely to the line of sight, with the complex south-southeastern
radio structure being the approaching radio jet.

{\it MRC~0324$-$228 ($z=1.894$):} The NICMOS image of this radio
galaxy \citep{pen01} reveals two potential host galaxies separated by
1\farcs5 (12\,kpc).  Our {\it Spitzer} imaging does not resolve them
as individual sources, though it does show an extended structure along
the same north--south orientation. Our new radio data tentatively
detected a faint radio core at 4.86\,GHz, coinciding with the
southernmost of these two identifications, which we have adopted as the
host galaxy.  There are also two nearby objects that may contaminate
the MIPS photometry.

{\it MRC~0350$-$279 ($z=1.90$):} No radio core is detected in our new
radio maps.

{\it 5C~7.269 ($z=2.218$):} No radio core is detected in our new
radio maps.

{\it 6C~0820+3642 ($z=1.860$):} The new radio imaging detects a faint
radio core, which is most obvious in the 8.46\,GHz map.

{\it 6C~0901+3551 ($z=1.91$):} No radio core is detected in our new
radio maps.

{\it USS~1243+036 ($z=3.570$):} Our {\it Spitzer} photometry is not
sufficiently sensitive to exclude a possible contribution from hot
dust at rest-frame 1.6\,$\mu$m. The derived stellar masses are
therefore formally considered as upper limits, although the 3.5 and
4.5\,$\mu$m photometry suggests they are most likely uncontaminated by
hot dust.

{\it USS~1707+105 ($z=2.349$):} The new radio imaging detects a faint
radio core, which is most obvious in the 4.86\,GHz map.

{\it LBDS~53W002 ($z=2.293$):} The IRS and MIPS photometry are possibly
contaminated by a galaxy $\sim$4\arcsec\ to the northwest
\citep{keel02}. This galaxy is most likely at a different redshift
than the AGN host as it does not appear in the [\ion{O}{3}] and H$\alpha$
narrow-band imaging of \citet{keel02}.

{\it 7C~1751+6809 ($z=1.54$):} No radio core is detected in our new
radio maps.

{\it 7C~1805+6332 ($z=1.84$):} The new radio imaging detects a faint
radio core, which is only seen in the 8.46\,GHz map.

{\it TXS~J1908+7220 ($z=3.530$):} Using integral field spectroscopy,
\citet{smi10} argued that this broad line radio galaxy is a system of
two vigorously star-forming galaxies superimposed along the line of
sight and separated by $\sim$1300\,km\,s$^{-1}$ in velocity.  The detailed
SED decomposition of \citet{smi10} also showed that the near-IR
emission of this HzRG is dominated by transmitted quasar
continuum. While the hot dust continuum does not contribute
substantially at $\lambda_{\rm rest}=1.6$\,$\mu$m, this transmitted
quasar continuum prevents us from measuring the old stellar
population, and we provide only an upper limit to the stellar mass.

{\it TN~J2007-1316 ($z=3.84$):} We identify the second component from
the north as the radio core. This component is spatially resolved only
in the 8.46\,GHz map.

{\it MG~2144+1928 ($z=3.592$):} Our {\it Spitzer} photometry is not
sufficiently sensitive to exclude a possible contribution from hot
dust at rest-frame 1.6\,$\mu$m. The derived stellar masses are
therefore formally considered as upper limits, although the 3.5 and
4.5\,$\mu$m photometry suggests they are most likely uncontaminated by
hot dust.

\section{Summary and concluding remarks}

The main results from our full six-band {\it Spitzer} imaging survey
of 70 radio galaxies at $1<z<5.2$ can be summarized as follows:

\begin{enumerate}

\item Using a newly designed criterion to isolate sources dominated
by hot dust emission at $\lambda_{\rm rest}$=1.6\,$\mu$m, we
unambiguously detect a stellar population in 46 radio galaxies and
hot dust emission in 59.

\item Four of the 70 HzRGs show a second AGN within 6\arcsec, as revealed 
by their mid-IR colors. This may be an indication of an ongoing or
imminent major merger event.

\item The stellar masses have a remarkably low scatter around ${\sim}
2 \times 10^{11}$\,\Msun.  There is tentative evidence for a small
drop in stellar mass at $z>3$. If confirmed by more detailed spectral
decompositions, this result suggests that radio galaxies have built the
bulk of their stellar populations by $z\sim 3$. This is consistent
with the observed increase in their submillimeter luminosities,
suggesting higher star formation rates, at $z>3$. 

\item The radio core fraction is strongly correlated with restframe 
5\,$\mu$m luminosity. The most core-dominated radio galaxies are also
the most luminous (i.e., least obscured) hot dust emitters. This
suggests a gradual increase of the hot dust obscuration from type~2
to type~1 AGNs, consistent with the orientation paradigm of
AGN unification. It also complicates the use of  $\lambda_{\rm
  rest}=5$\,$\mu$m luminosity as an indicator of AGN power.

\item The 500\,MHz radio luminosity is only marginally correlated
with stellar mass but strongly correlated with the hot dust
luminosity at $\lambda_{\rm rest}=5$\,$\mu$m. Variations in the
Eddington ratio of the SMBH could explain these trends.

\end{enumerate}

Our {\it Spitzer} sample provides an unique database to study the
massive host galaxies of powerful radio galaxies, their AGN-heated
dust emission, as well as their environments.  In future papers we
plan to study the stellar populations in greater detail by combining
the {\it Spitzer} data with visible and near-infrared data to better
constrain the HzRG star formation histories and thus derive more
accurate stellar masses and ages.  We also plan to observe the cold
dust component with the Atacama Large Millimeter and submillimeter
Array (ALMA) and the peak of the dust emission with {\it Herschel
  Space Observatory}. This full SED will allow to separate the hot
AGN from the cool, starburst-heated dust emission, providing more
accurate measures of the instantaneous star formation rates.  This
combined dataset will provide a unique view of the star formation
history in these very massive galaxies at every epoch.

\acknowledgments

CDB and NS would like to thank the Jet Propulsion Laboratory and
the California Institute of Technology for their hospitality during
an extended visit in Spring 2010.  We thank Patrick Ogle for
stimulating discussions.  This work is based on observations made
with the {\it Spitzer Space Telescope} and made use of the NASA/IPAC
Extragalactic Database (NED). Both NED and {\it Spitzer} are operated
by the Jet Propulsion Laboratory, Caltech under contracts with
NASA.  The work of DS and PRME was carried out at Jet Propulsion
Laboratory, California Institute of Technology, under a contract
with NASA.  Support for this work was provided by NASA through an
award issued by the Jet Propulsion Laboratory/Caltech.       The
National Radio Astronomy Observatory is a facility of the National
Science Foundation operated under cooperative agreement by Associated
Universities, Inc.

{\it Facilities:} \facility{Spitzer,VLA}



\appendix

\section{SED Fits to Radio Galaxies and Notes on Individual Sources}

\begin{figure}
\vspace{-10pt}
\begin{tabular}{r@{}c@{}l}
\includegraphics[angle=-90,width=160pt,trim= 0 -7 70 0,clip=true]{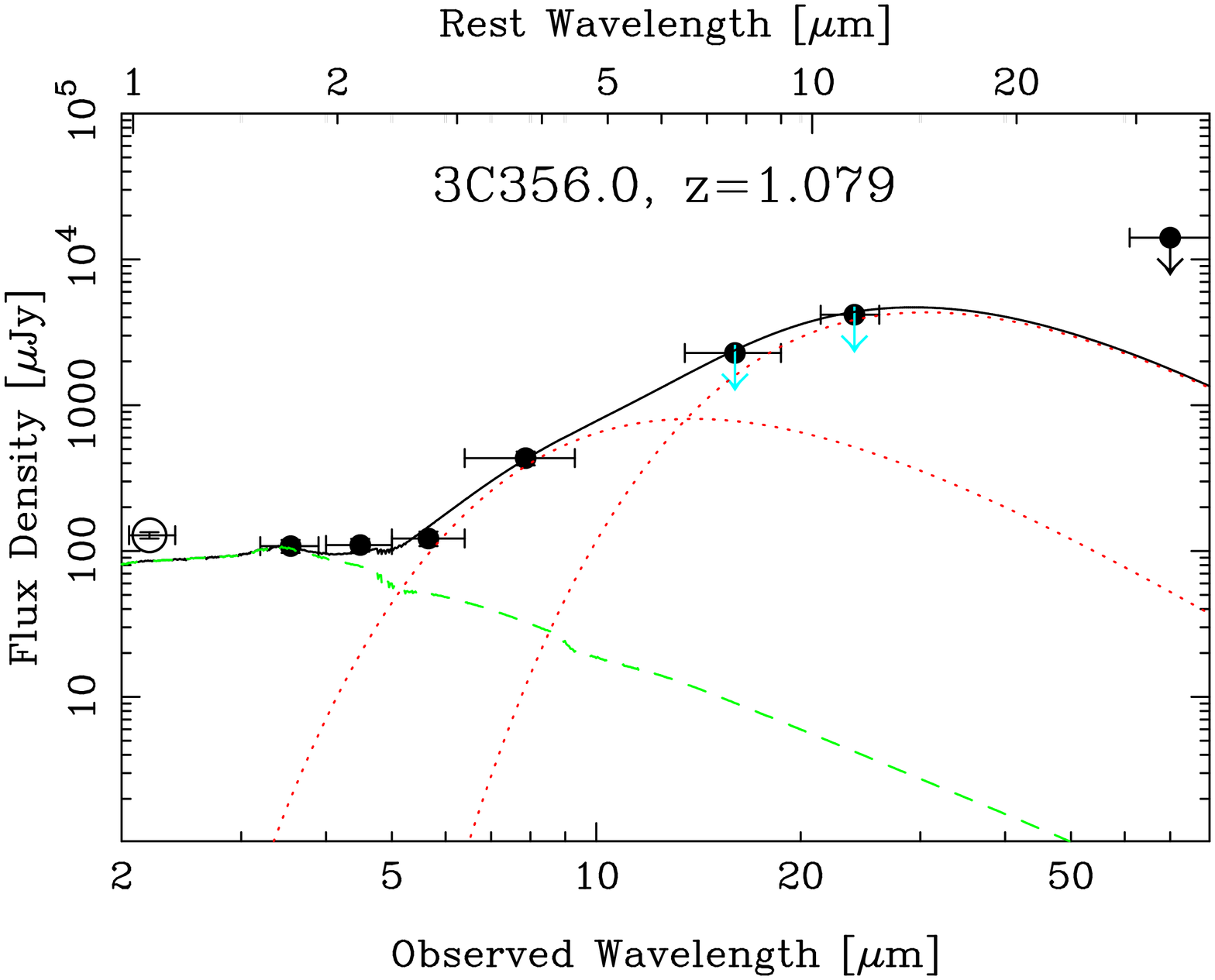} &
\includegraphics[angle=-90,width=144pt,trim= 0 59 70 0,clip=true]{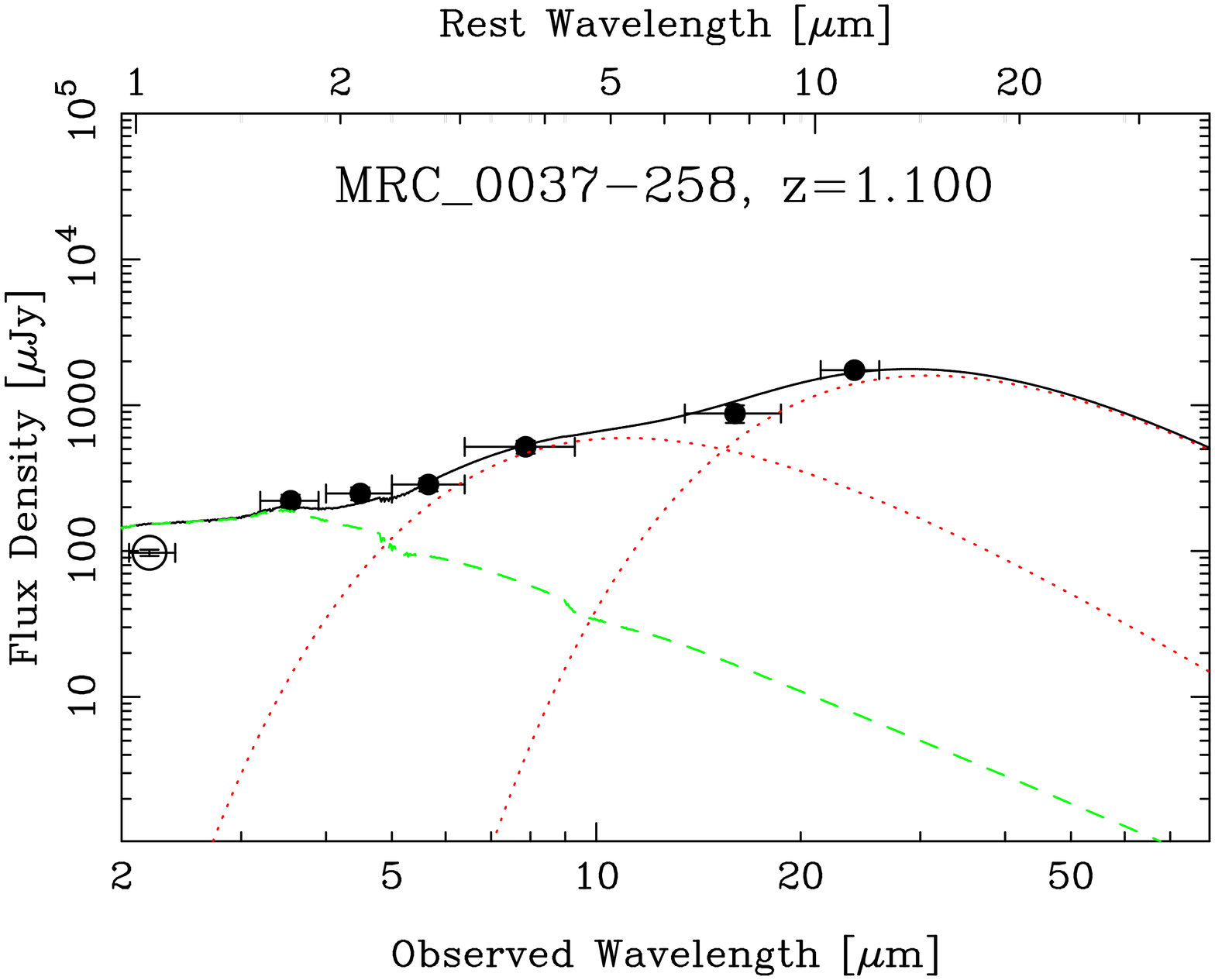} &
\includegraphics[angle=-90,width=144pt,trim= 0 59 70 0,clip=true]{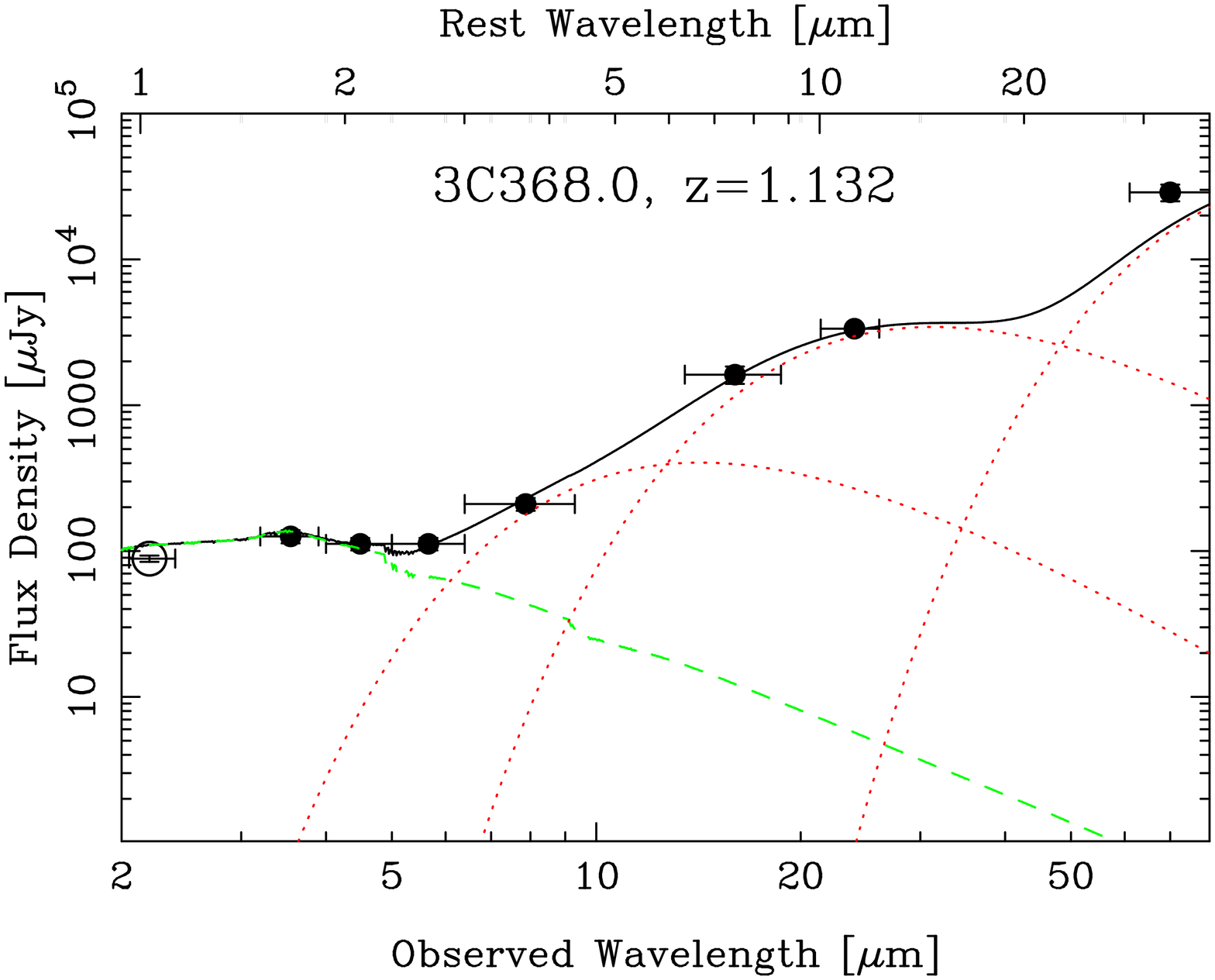} \\[-5pt]
\includegraphics[angle=-90,width=160pt,trim=28 -7 70 0,clip=true]{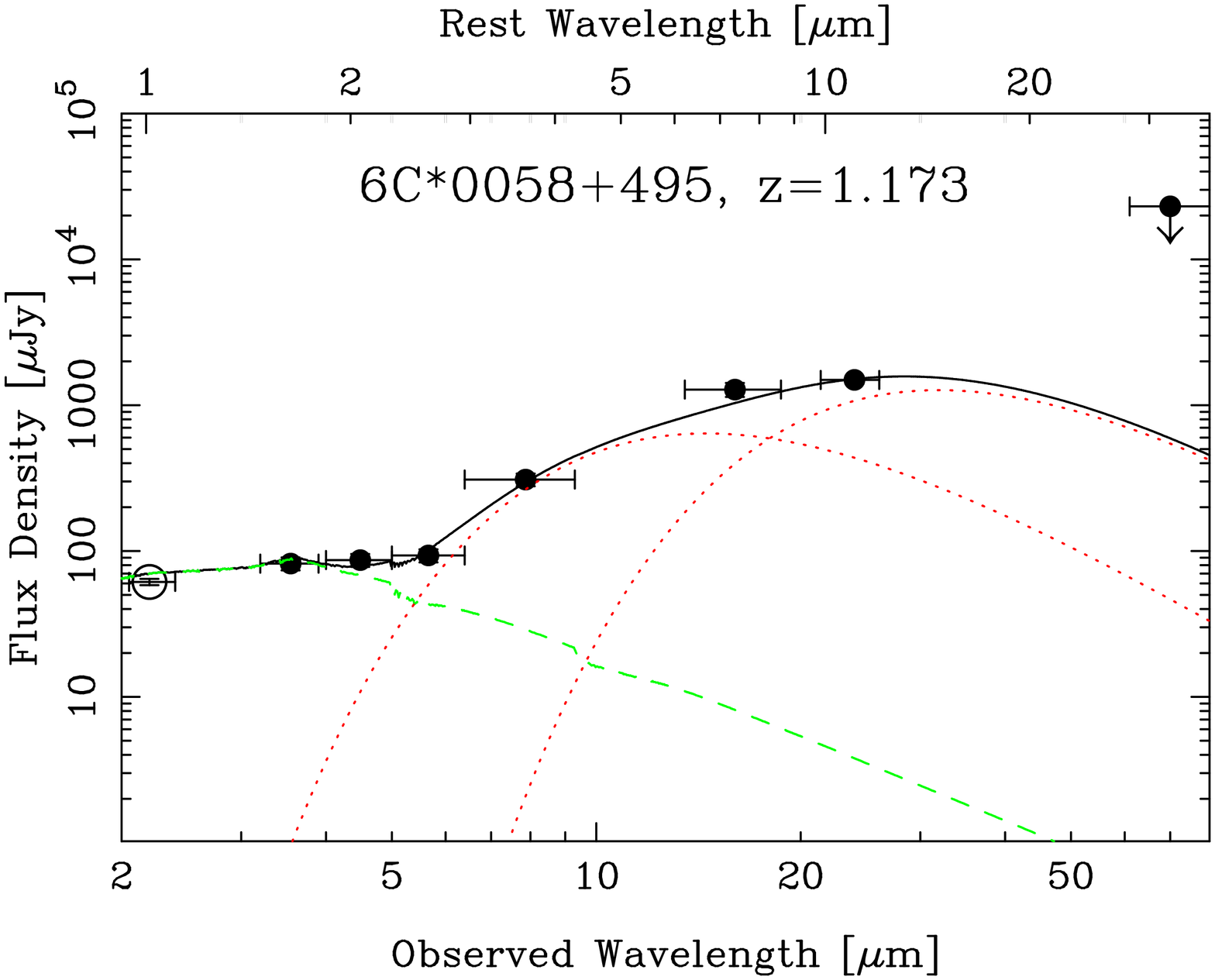} &
\includegraphics[angle=-90,width=144pt,trim=28 59 70 0,clip=true]{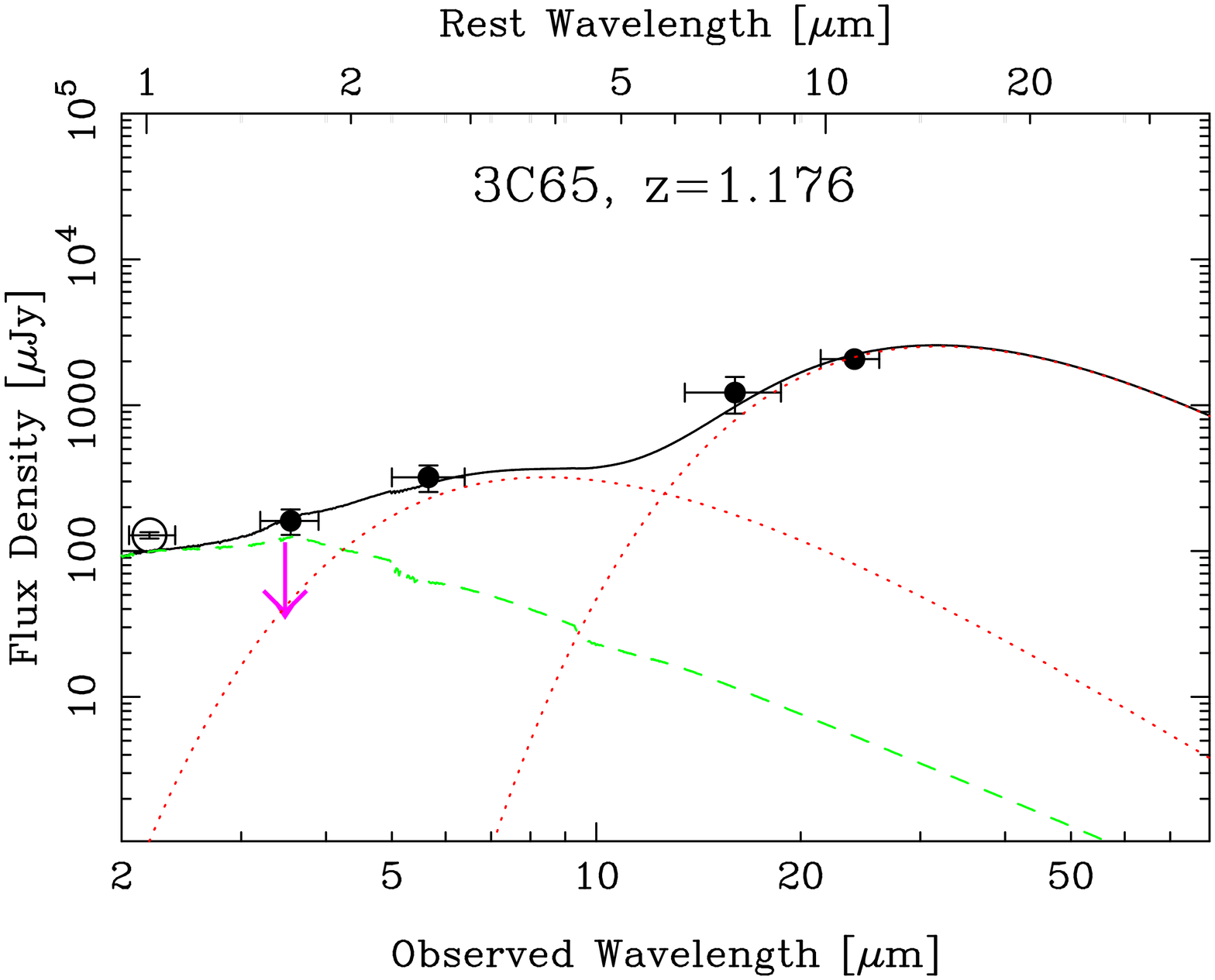} &
\includegraphics[angle=-90,width=144pt,trim=28 59 70 0,clip=true]{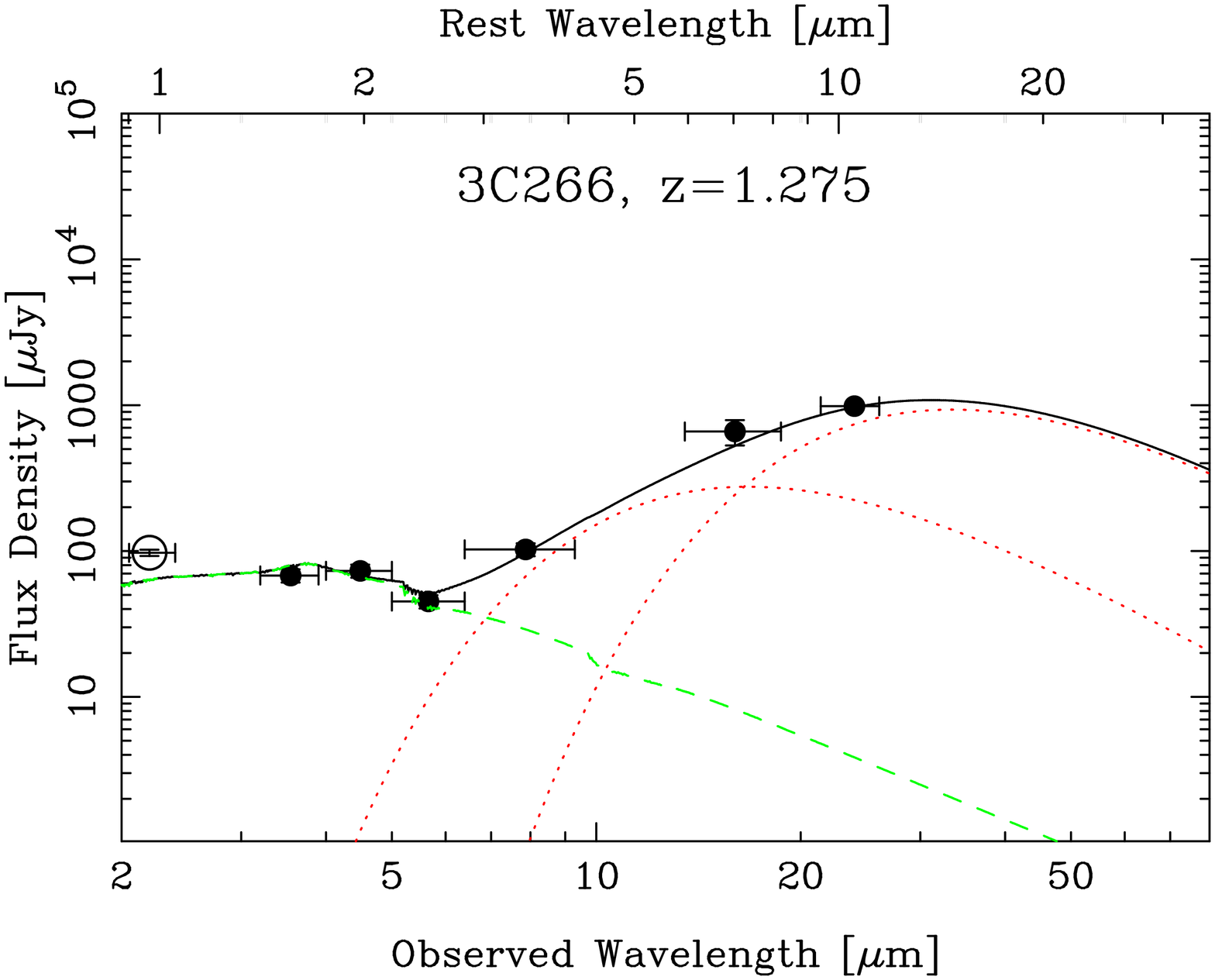} \\[-5pt]
\includegraphics[angle=-90,width=160pt,trim=28 -7 70 0,clip=true]{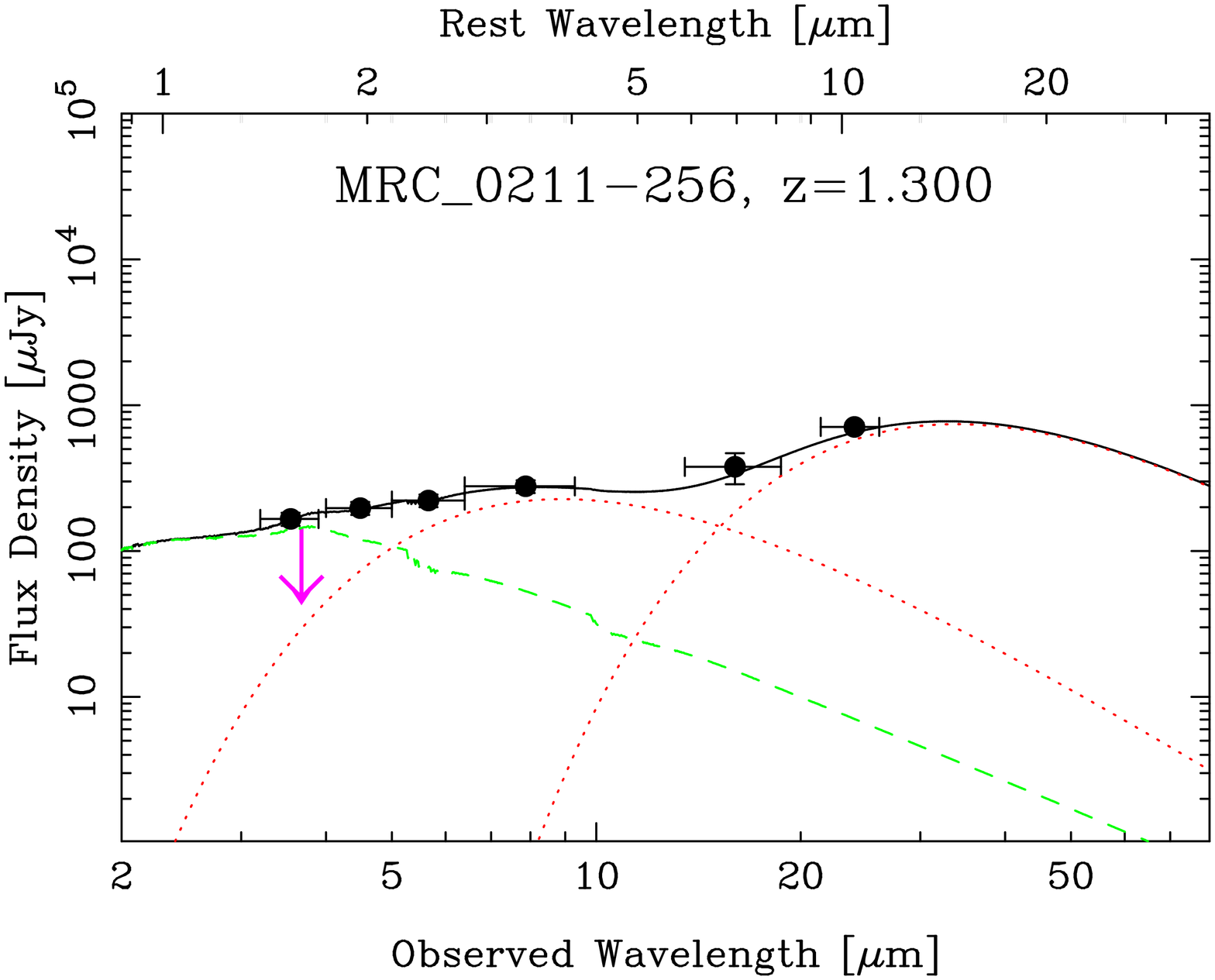} &
\includegraphics[angle=-90,width=144pt,trim=28 59 70 0,clip=true]{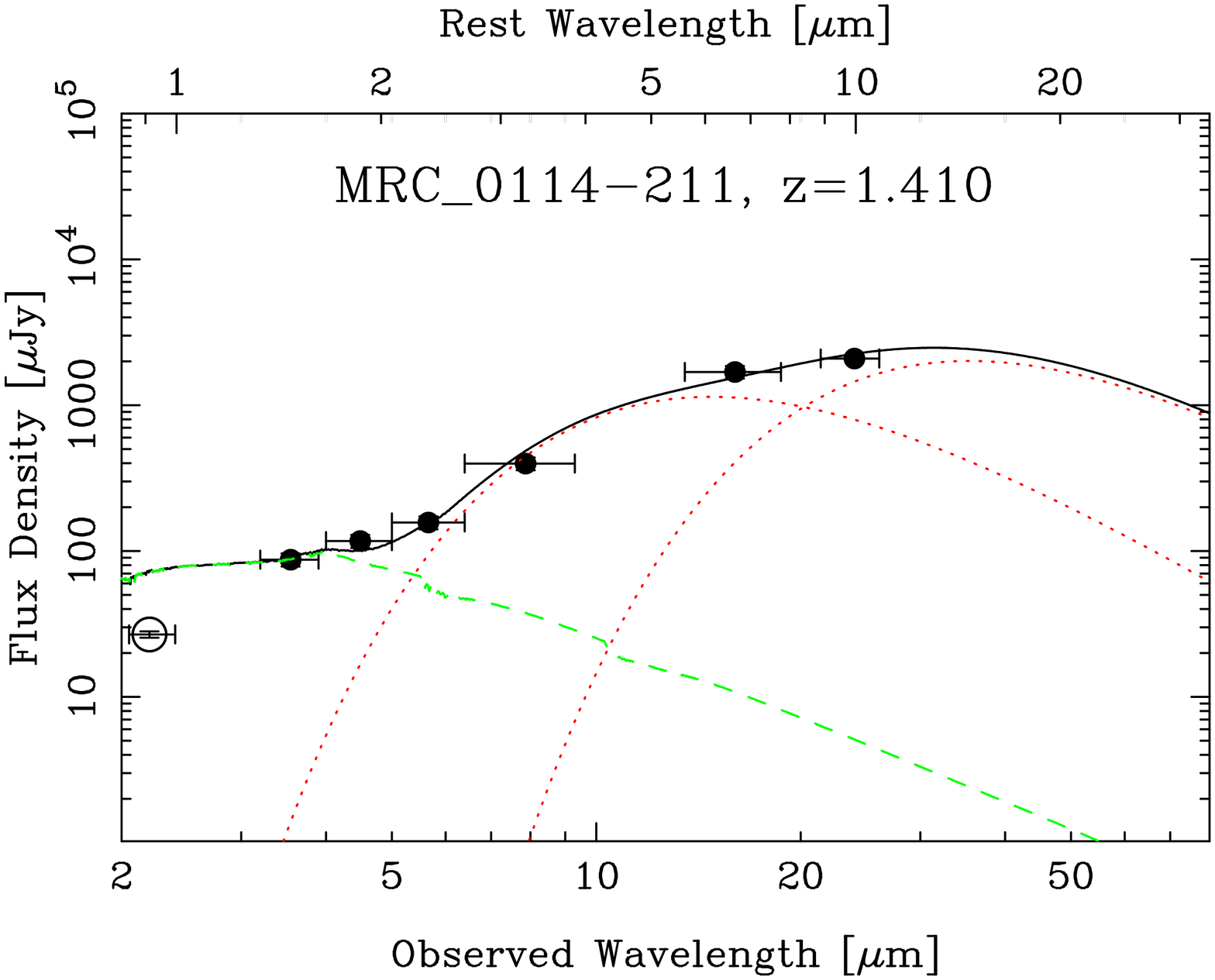} &
\includegraphics[angle=-90,width=144pt,trim=28 59 70 0,clip=true]{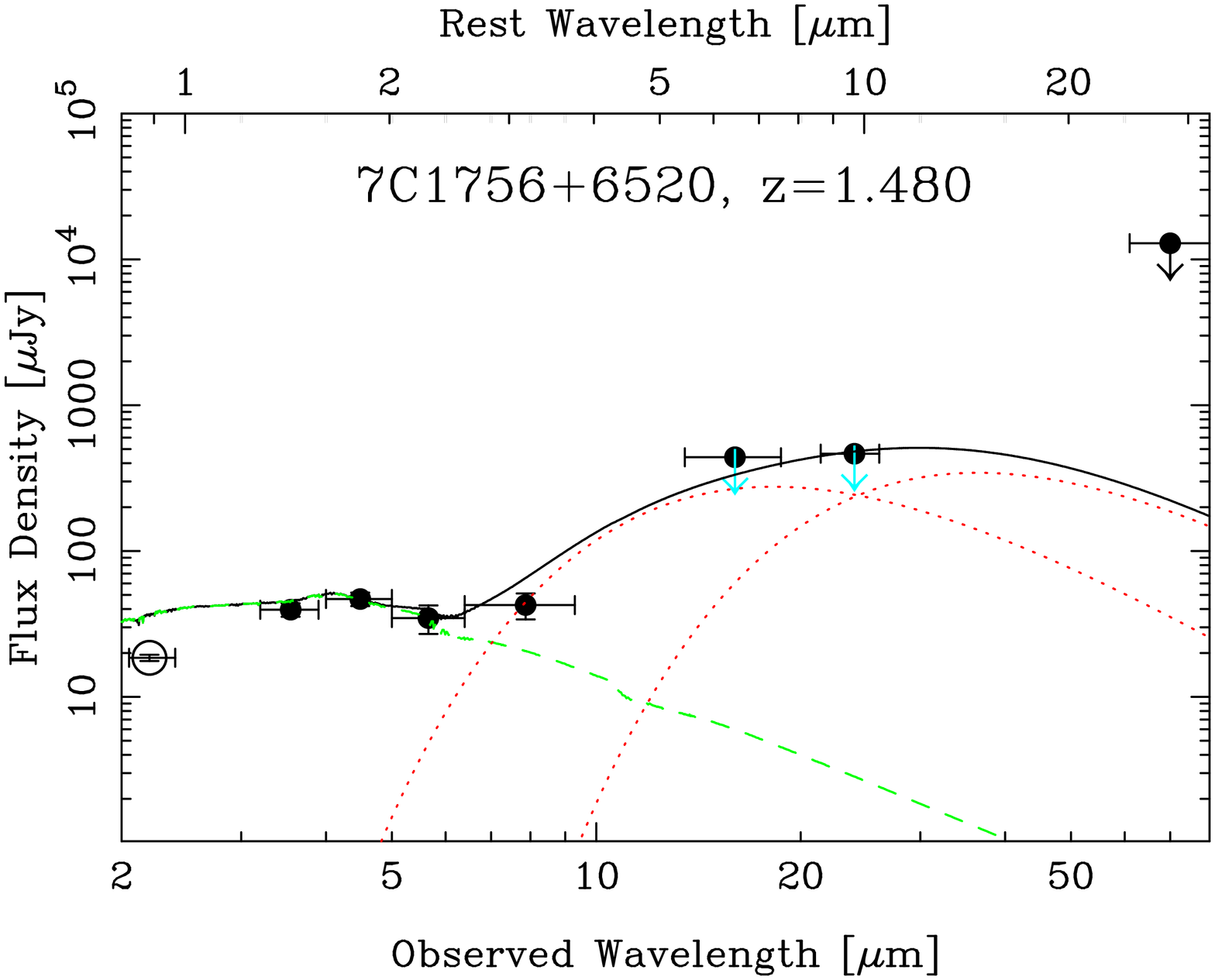} \\[-5pt]
\includegraphics[angle=-90,width=160pt,trim=28 -7 70 0,clip=true]{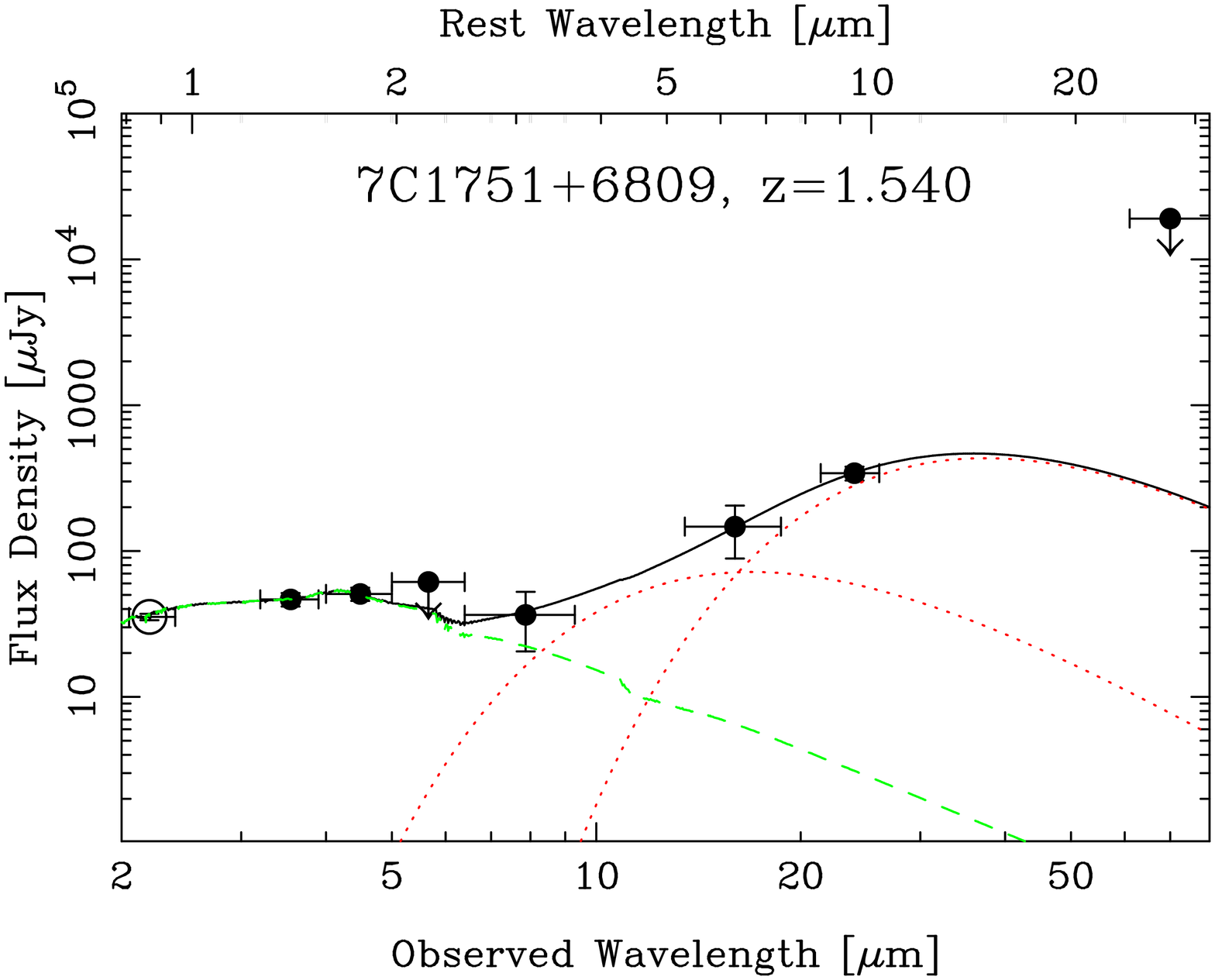} &
\includegraphics[angle=-90,width=144pt,trim=28 59 70 0,clip=true]{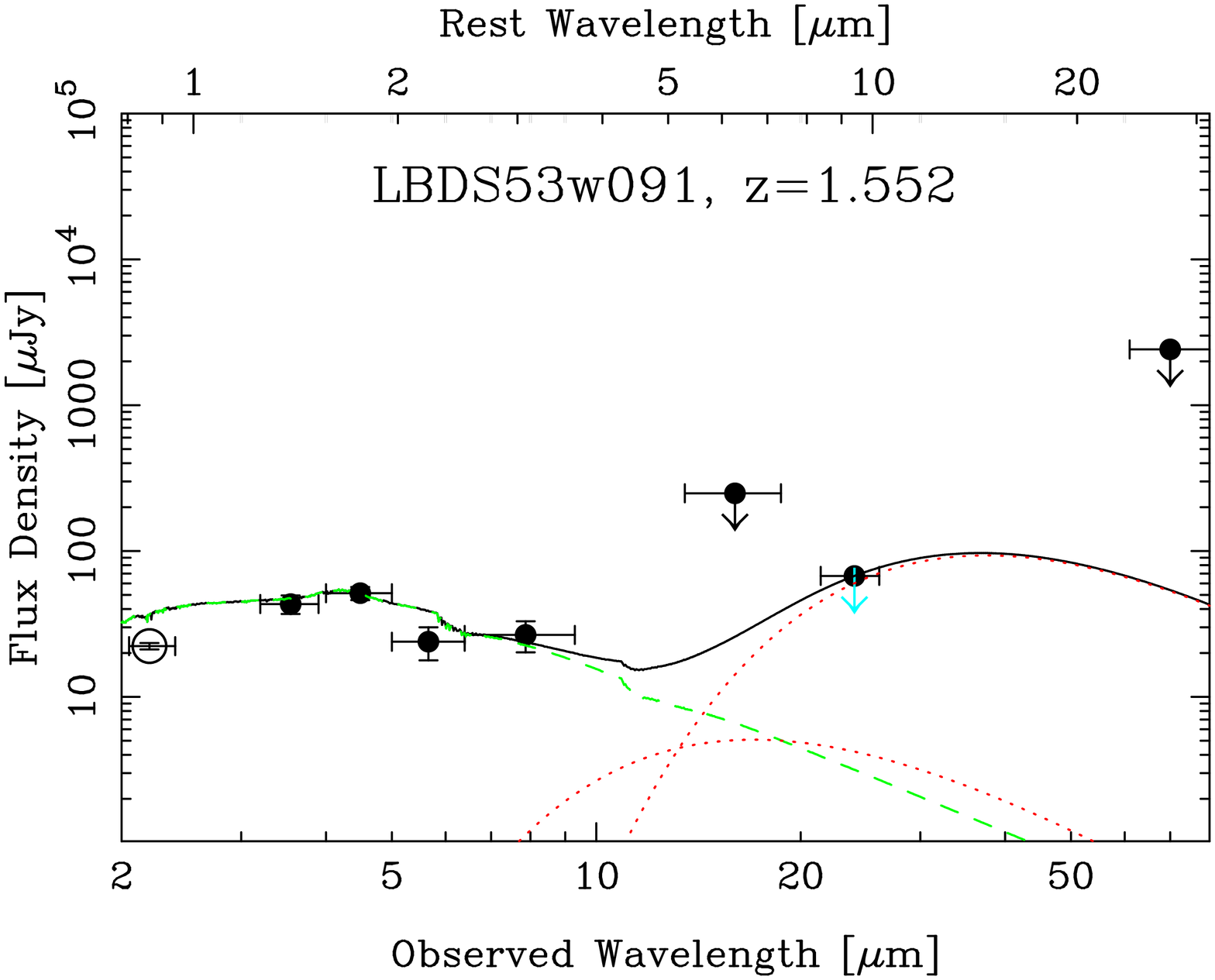} &
\includegraphics[angle=-90,width=144pt,trim=28 59 70 0,clip=true]{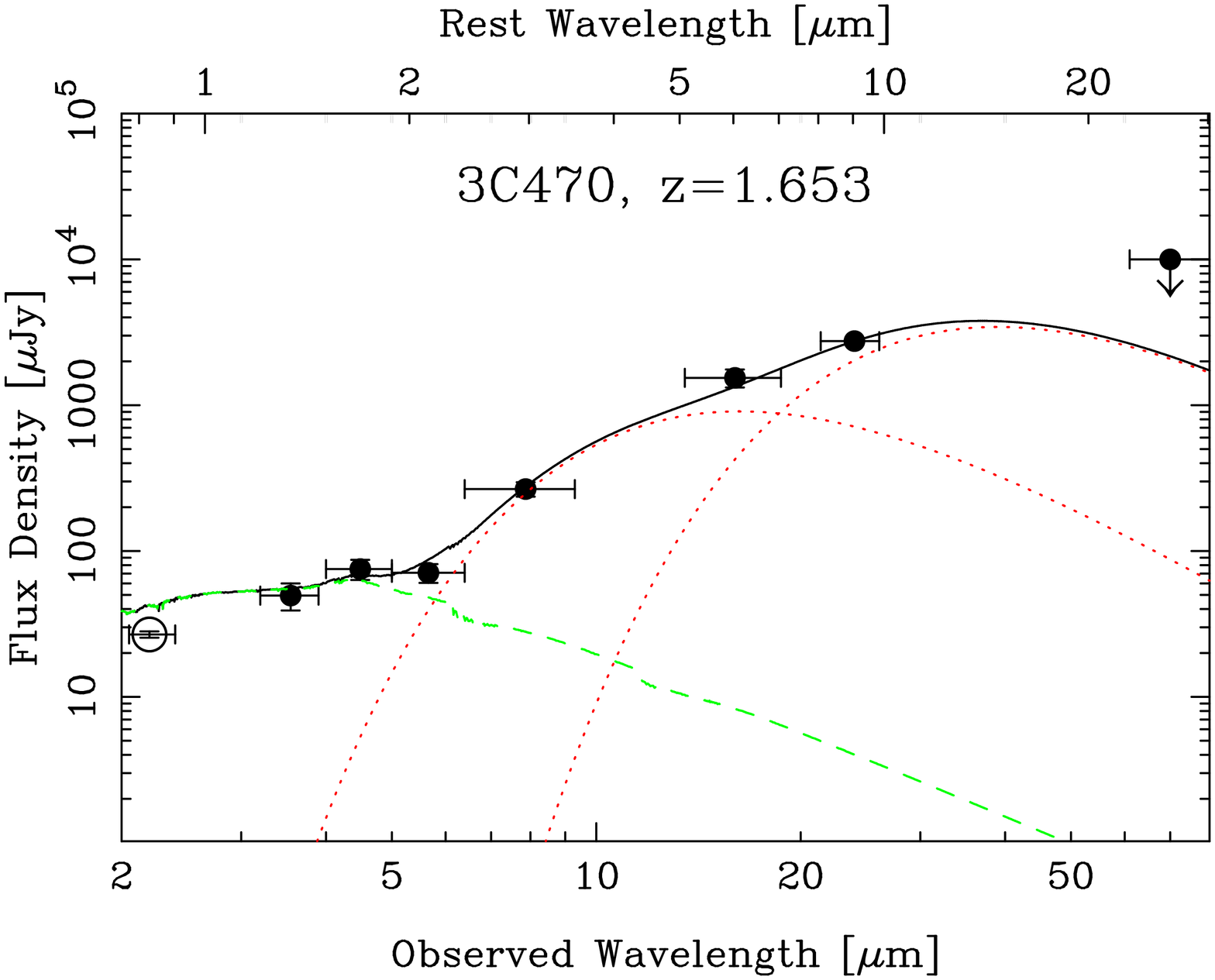} \\[-5pt]
\includegraphics[angle=-90,width=160pt,trim=28 -7 70 0,clip=true]{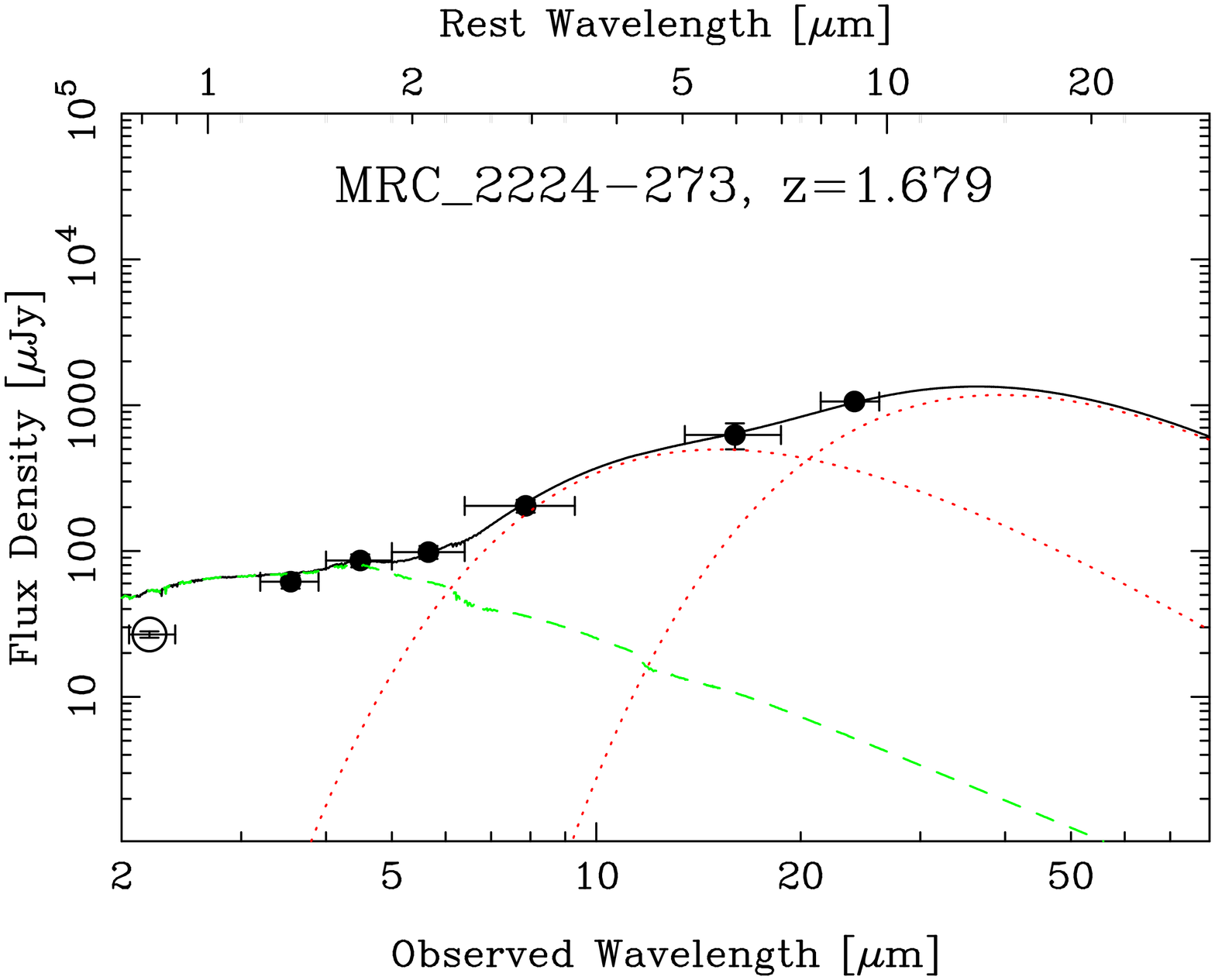} &
\includegraphics[angle=-90,width=144pt,trim=28 59 70 0,clip=true]{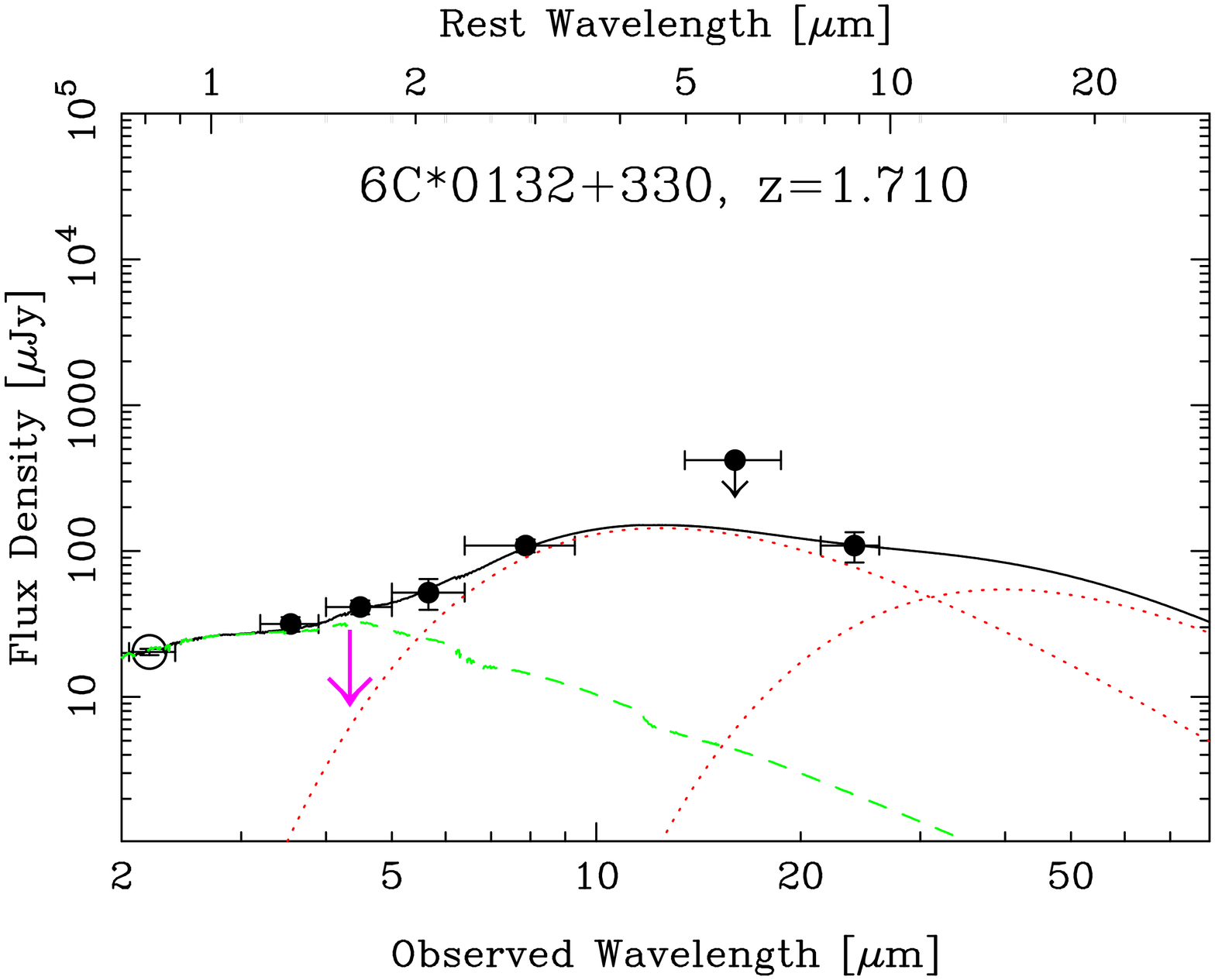} &
\includegraphics[angle=-90,width=144pt,trim=28 59 70 0,clip=true]{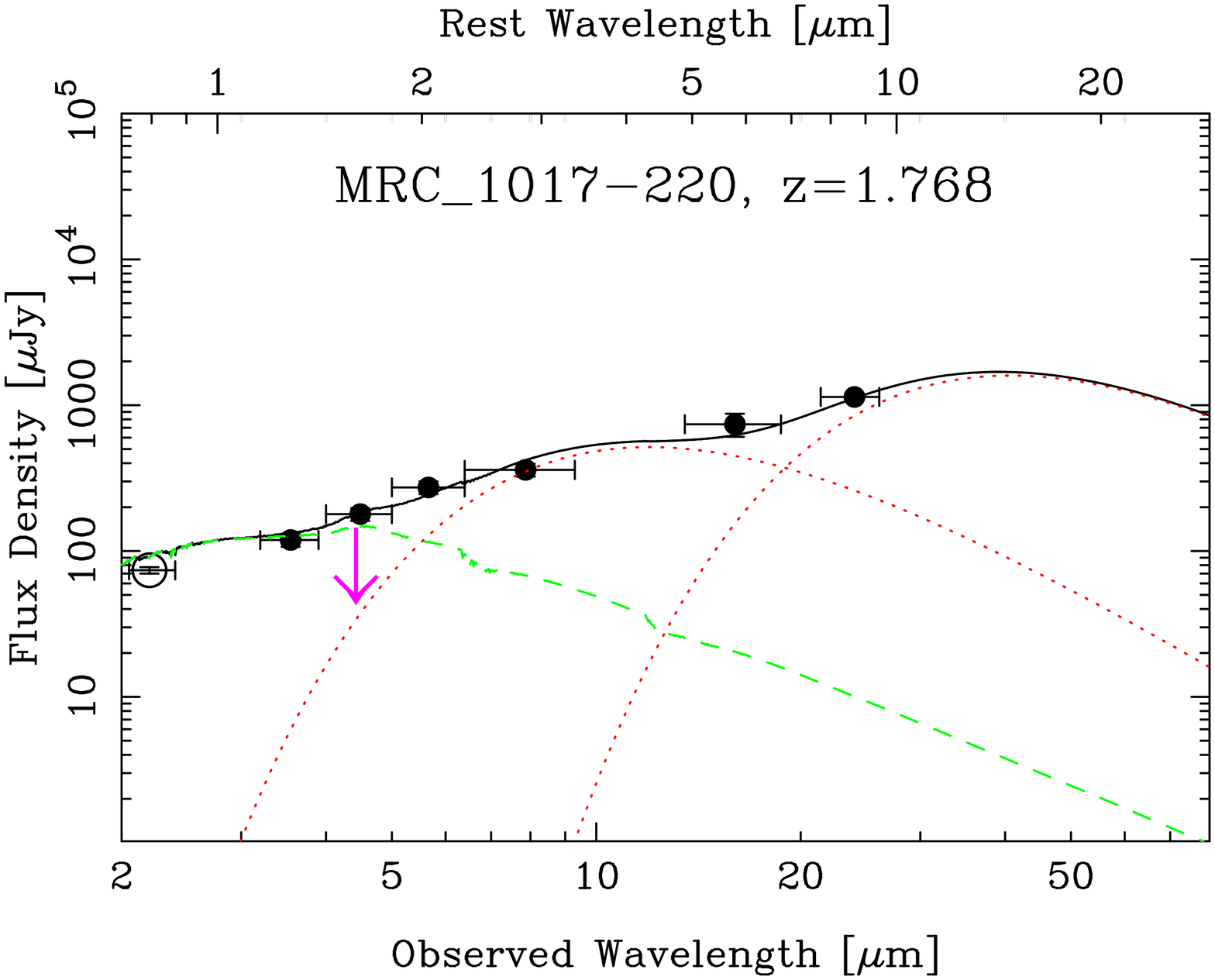} \\[-5pt]
\includegraphics[angle=-90,width=160pt,trim=28 -7  0 0,clip=true]{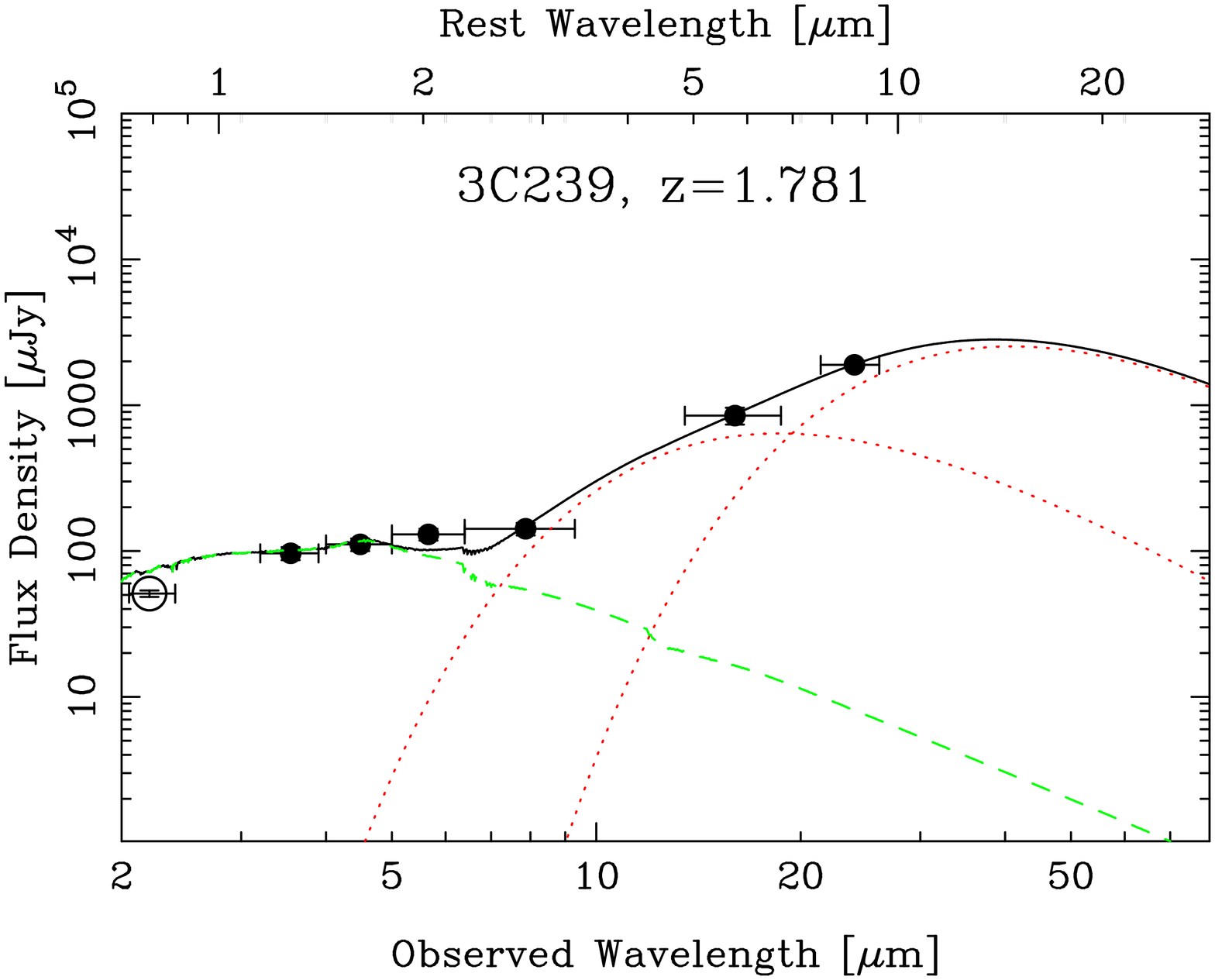} &
\includegraphics[angle=-90,width=144pt,trim=28 59  0 0,clip=true]{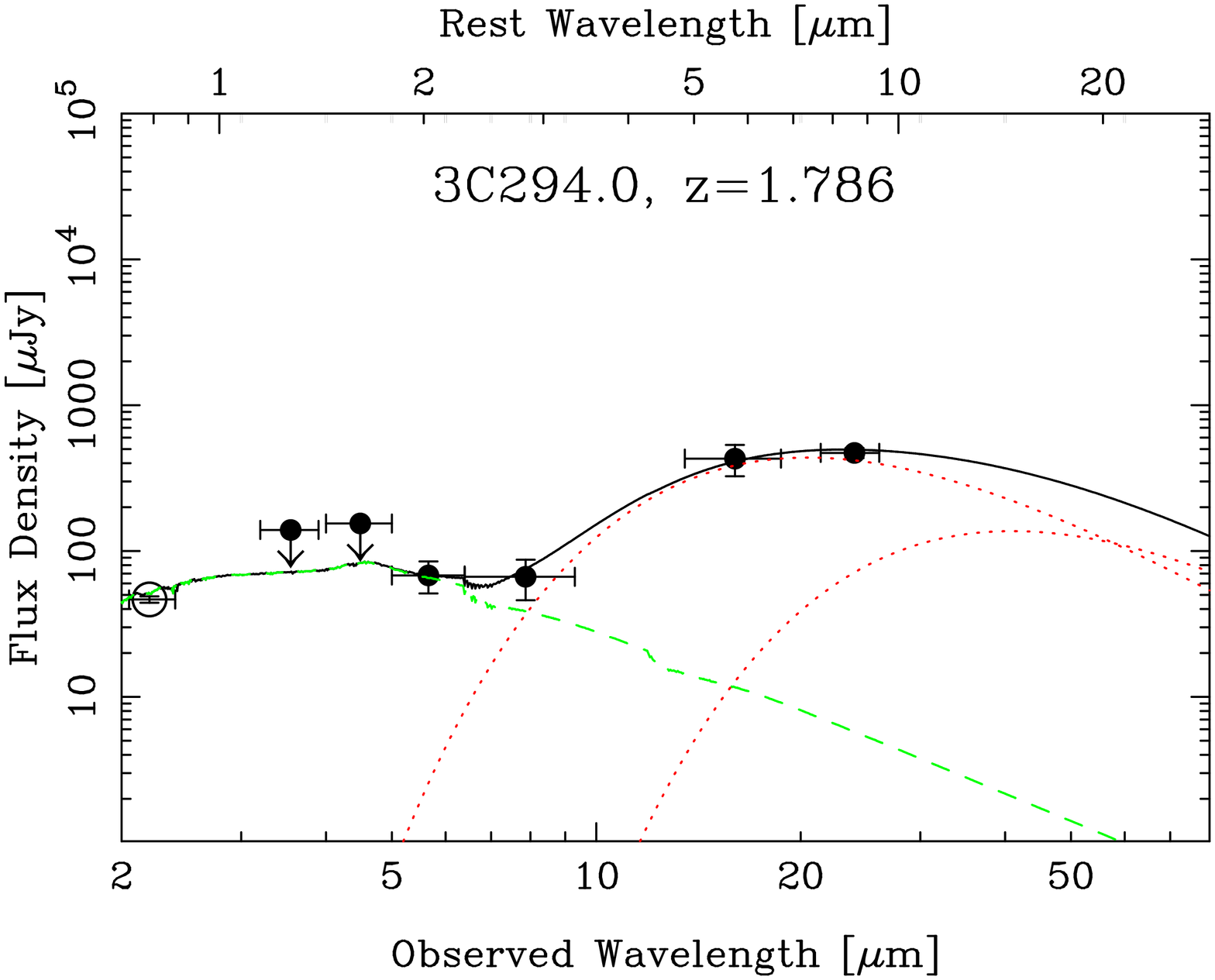} &
\includegraphics[angle=-90,width=144pt,trim=28 59  0 0,clip=true]{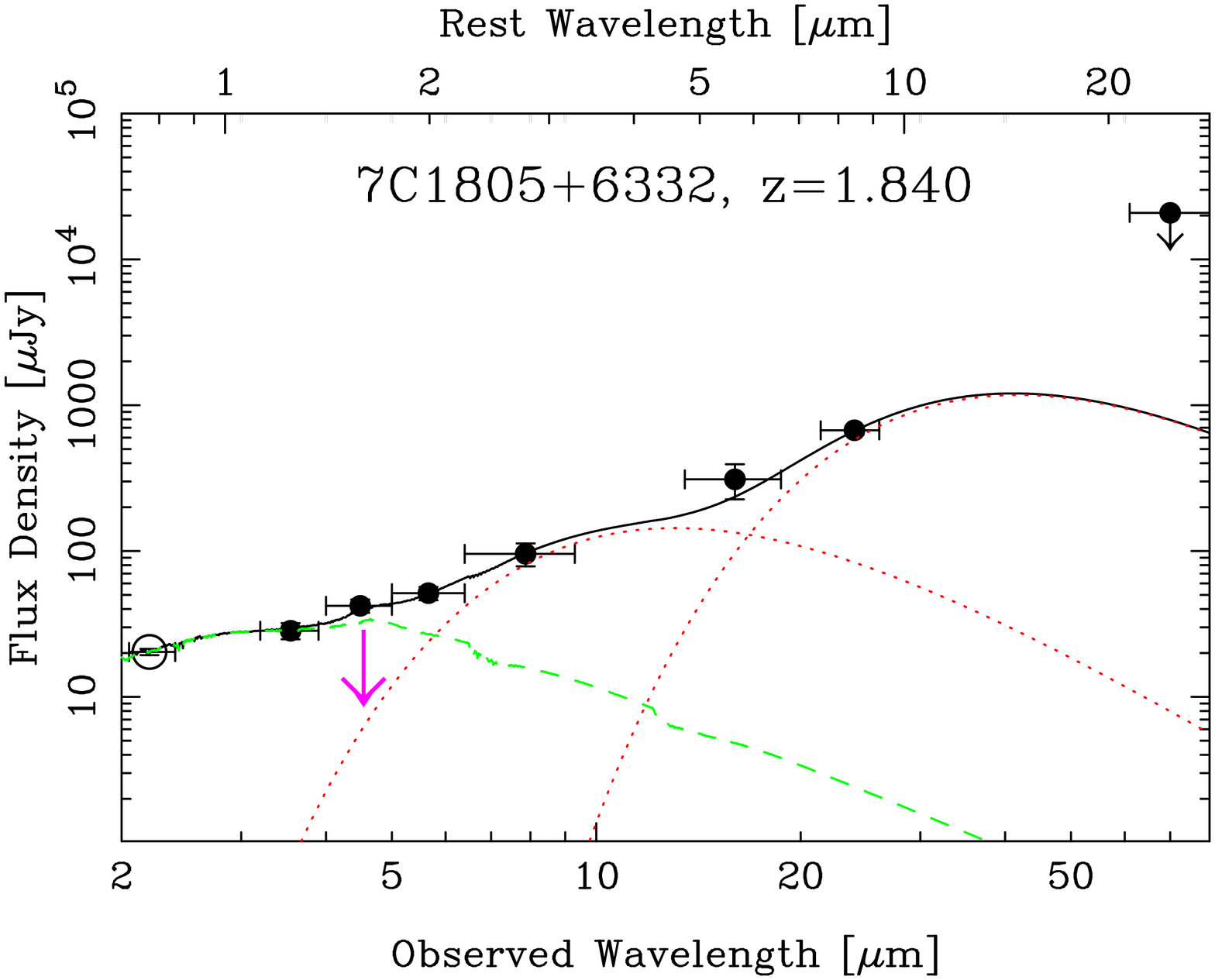} \\
\end{tabular}
\caption{\scriptsize Radio galaxy SEDs and model fits.  Lower
  abscissas are marked with observed wavelengths, and the scales are
  the same for all objects.  Upper abscissas are marked with
  restframe wavelengths, which depend on wavelength for each
  object. Filled circles with error bars denote data points used in
  the fit; open circles denote data points that could be contaminated
  by emission lines and are therefore not used in the fits.  Model
  components are a stellar population (green dashed lines) and two or
  three pure blackbodies (red dotted lines) representing dust emission.
  The sum of the model components is shown by solid black line.  The
  name and redshift of each target is marked at the top of the panel.
  A downward arrow at $\lambda_{\rm rest}$=1.6 or 5\,$\mu$m indicates
  an upper limit on the stellar or hot dust emission, respectively.}
\label{sedplots}
\end{figure}

Figure~\ref{sedplots}  presents the {\it Spitzer} SEDs for our complete HzRG sample,
including model fits to the mid-IR photometry, ordered by redshift.

\renewcommand{\baselinestretch}{1.}
\vfill\eject

\begin{figure*}
\begin{tabular}{r@{}c@{}l}
\includegraphics[angle=-90,width=160pt,trim= 0 -7 70 0,clip=true]{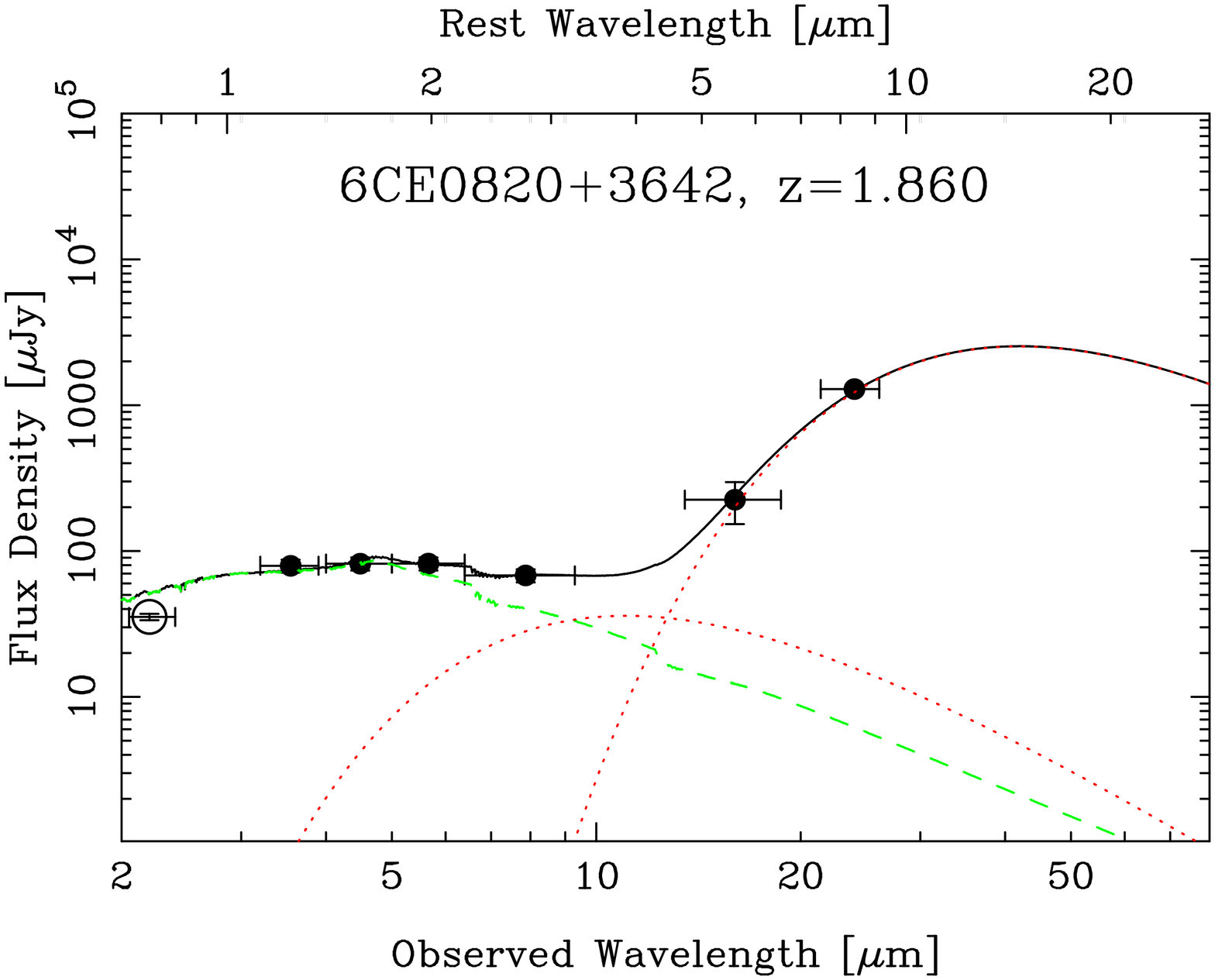} &
\includegraphics[angle=-90,width=144pt,trim= 0 59 70 0,clip=true]{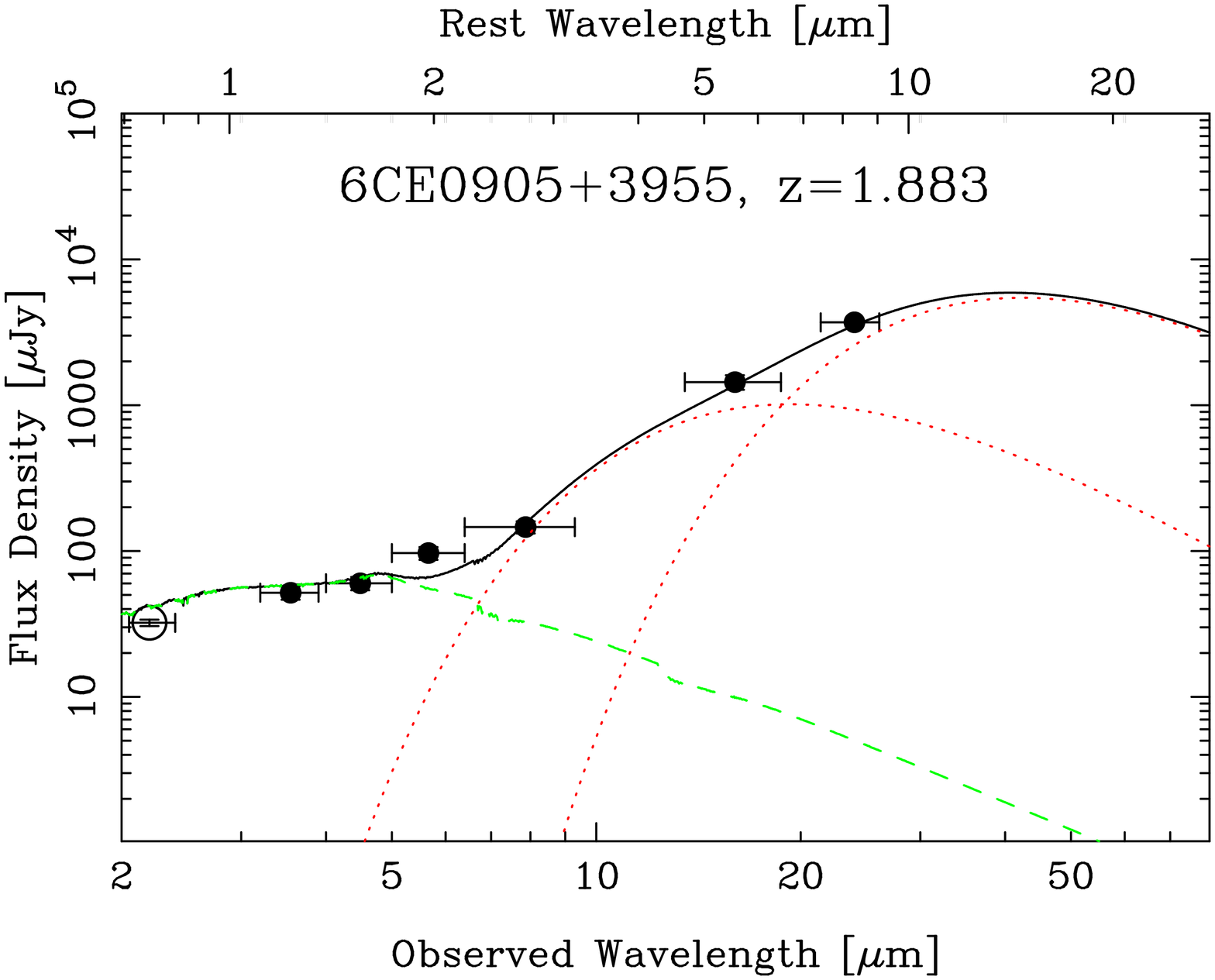} &
\includegraphics[angle=-90,width=144pt,trim= 0 59 70 0,clip=true]{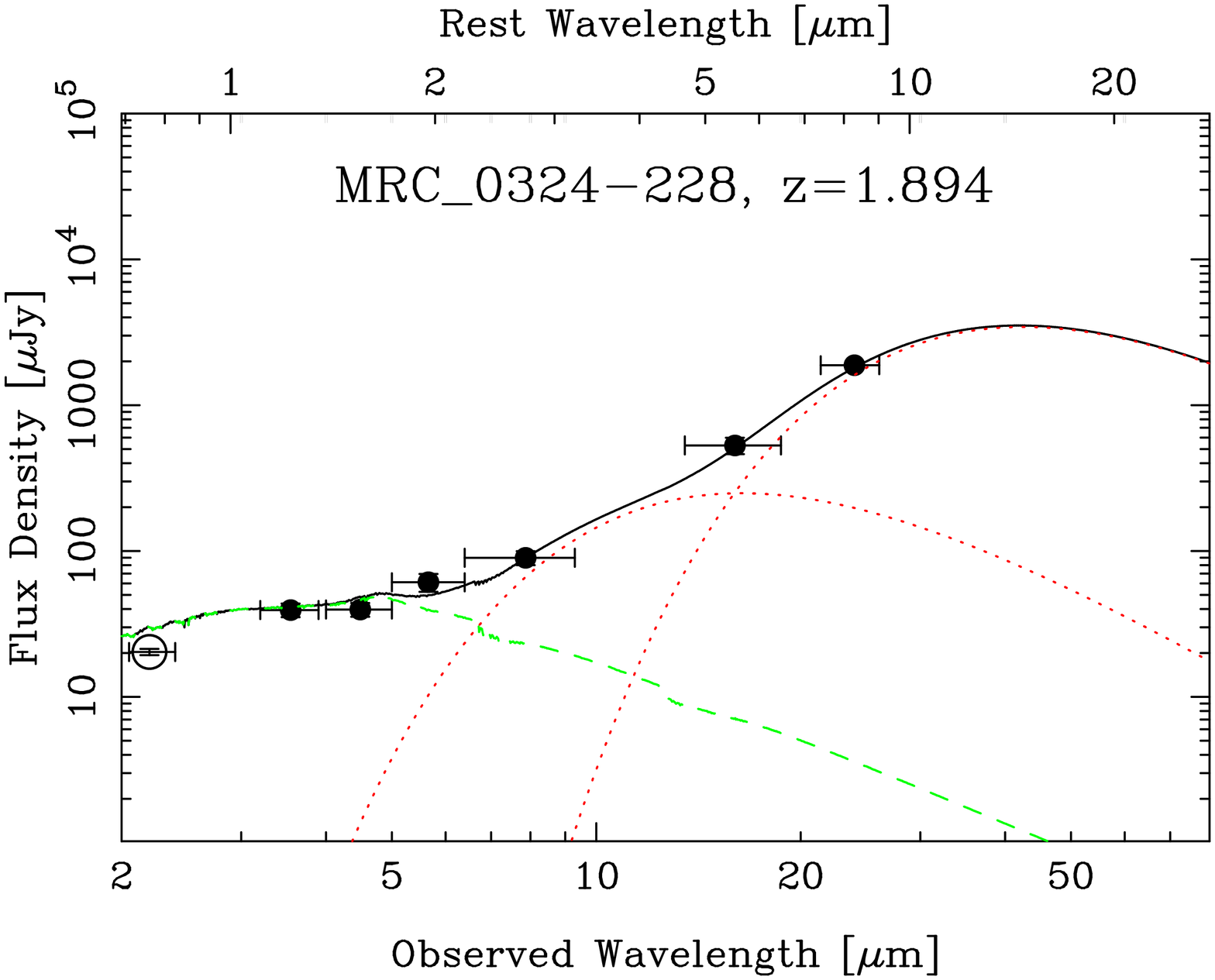} \\[-5pt]
\includegraphics[angle=-90,width=160pt,trim=28 -7 70 0,clip=true]{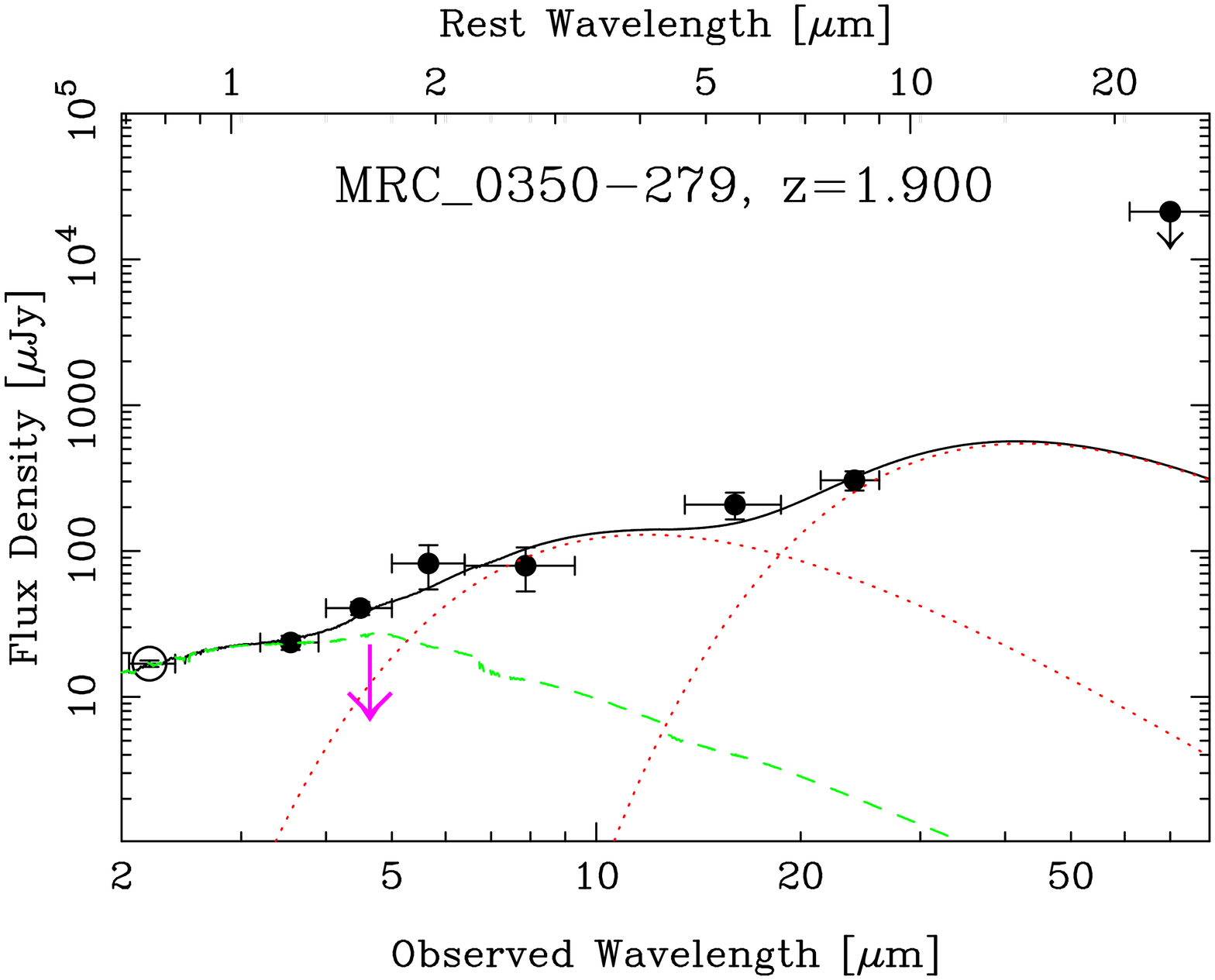} &
\includegraphics[angle=-90,width=144pt,trim=28 59 70 0,clip=true]{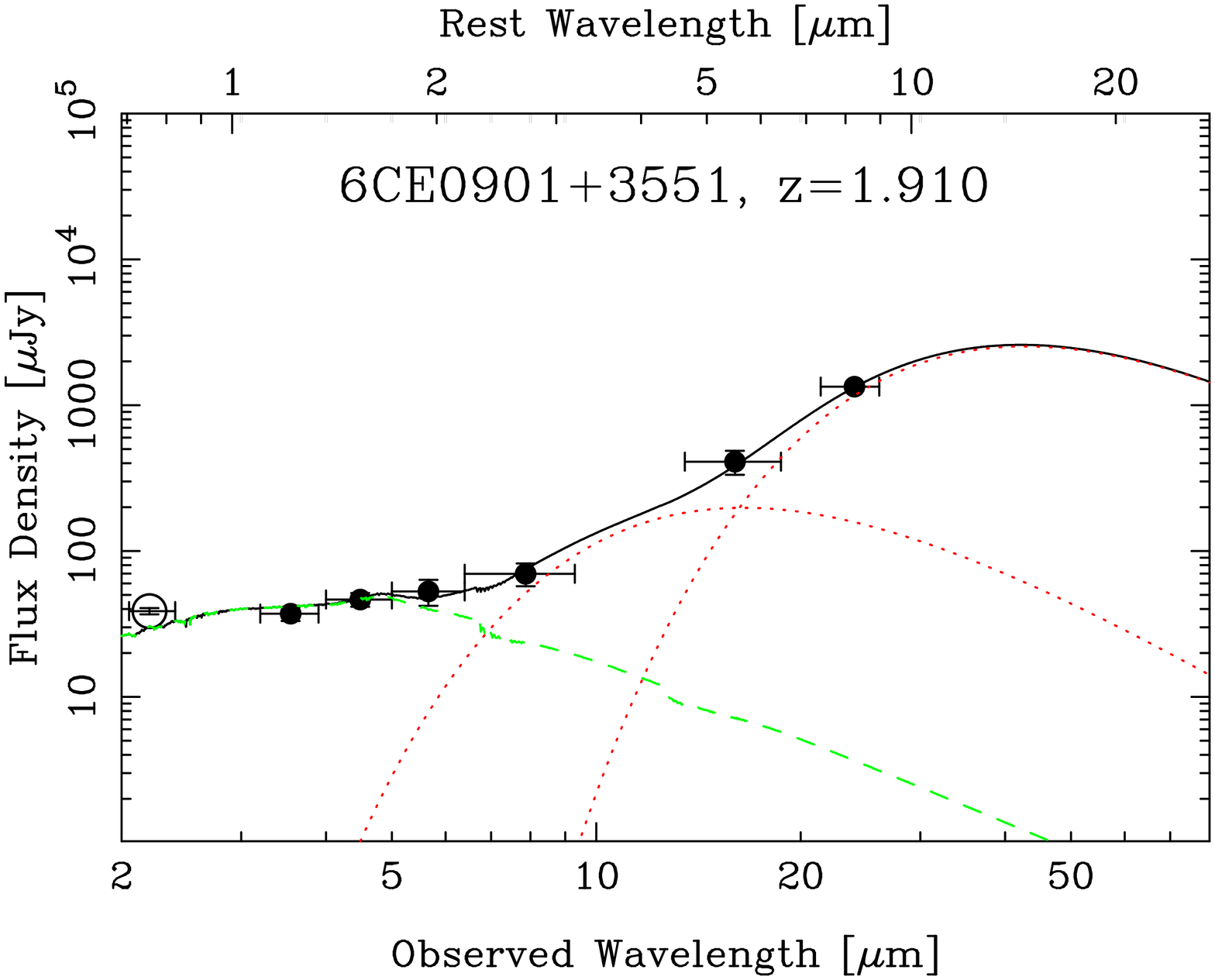} &
\includegraphics[angle=-90,width=144pt,trim=28 59 70 0,clip=true]{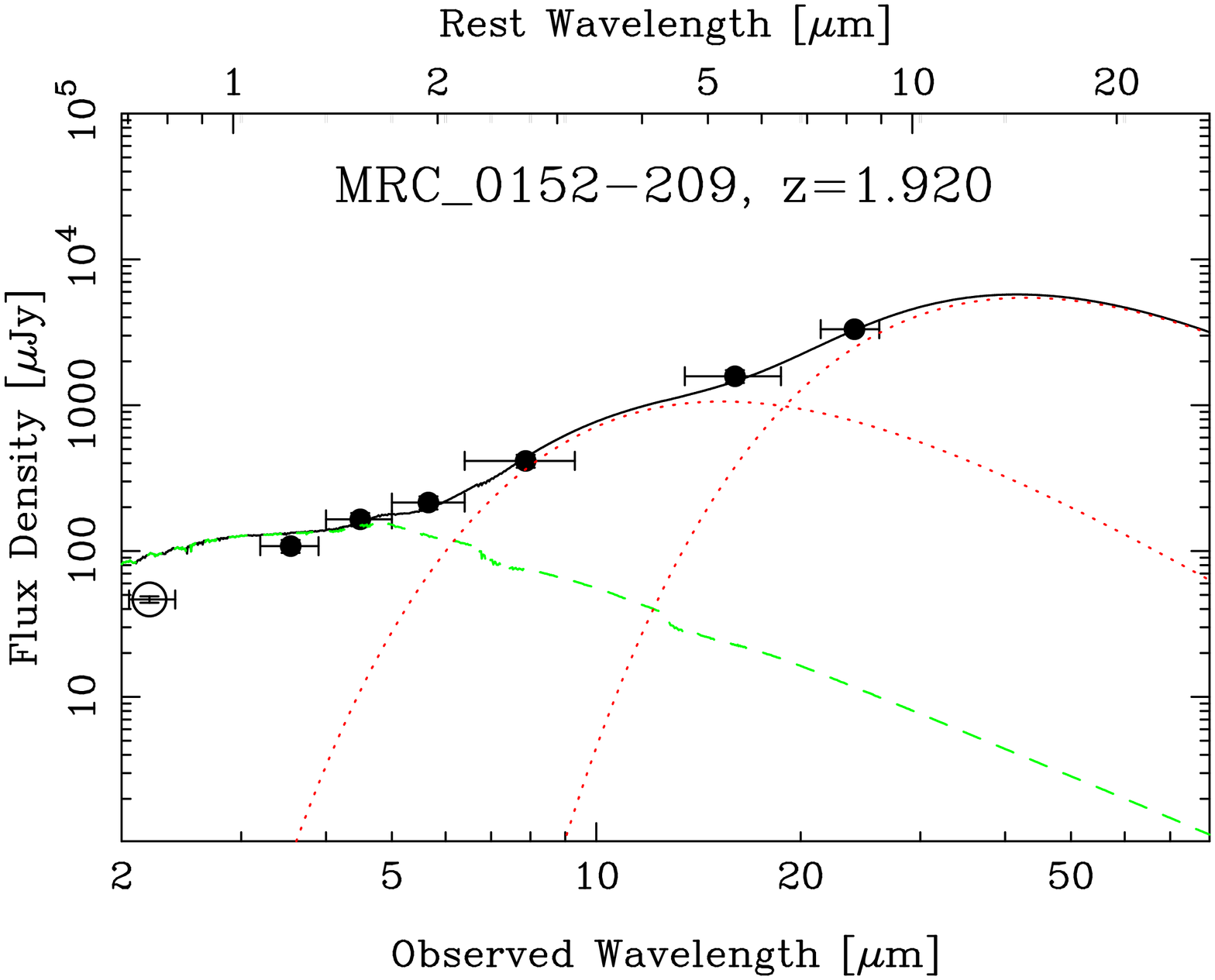} \\[-5pt]
\includegraphics[angle=-90,width=160pt,trim=28 -7 70 0,clip=true]{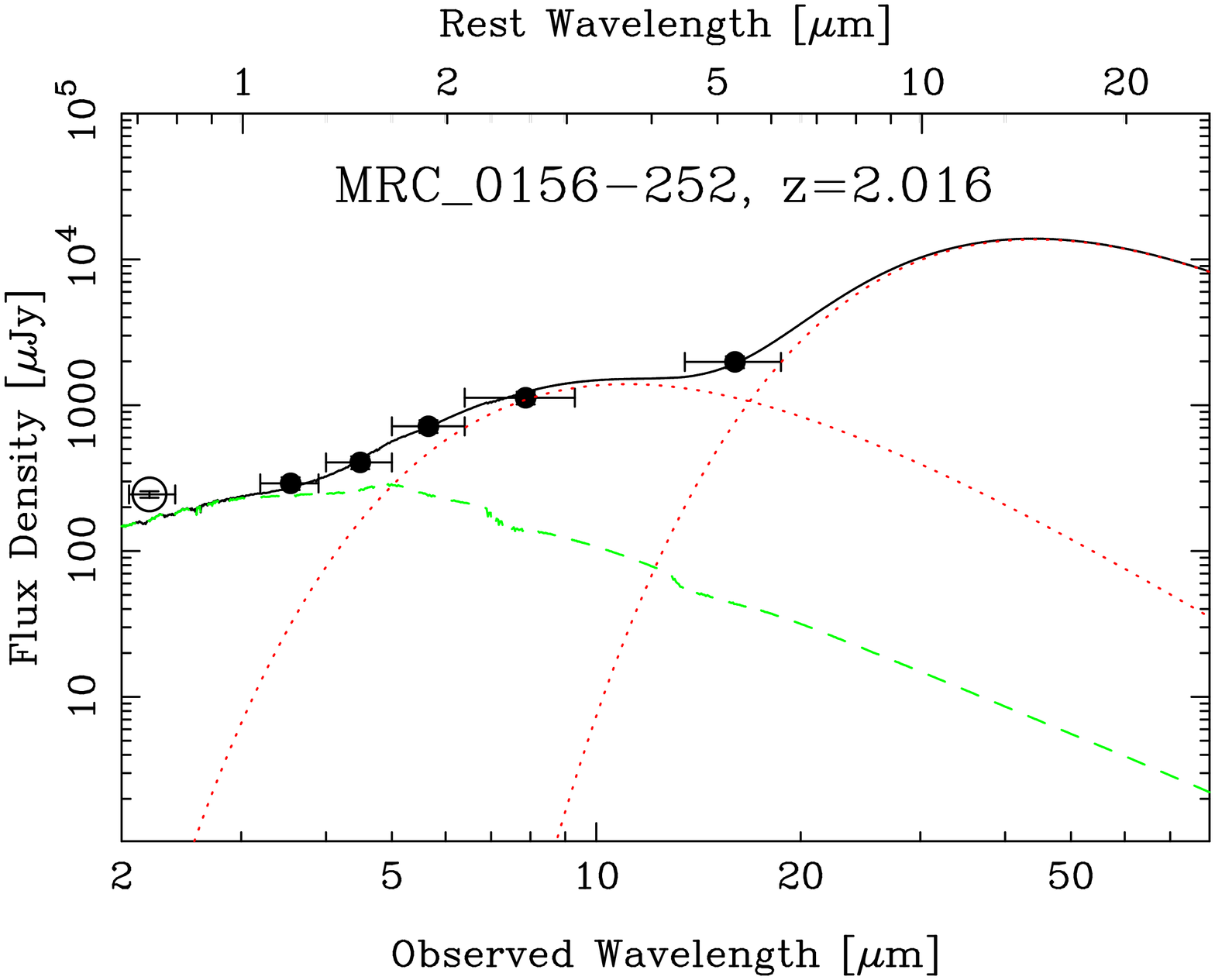} &
\includegraphics[angle=-90,width=144pt,trim=28 59 70 0,clip=true]{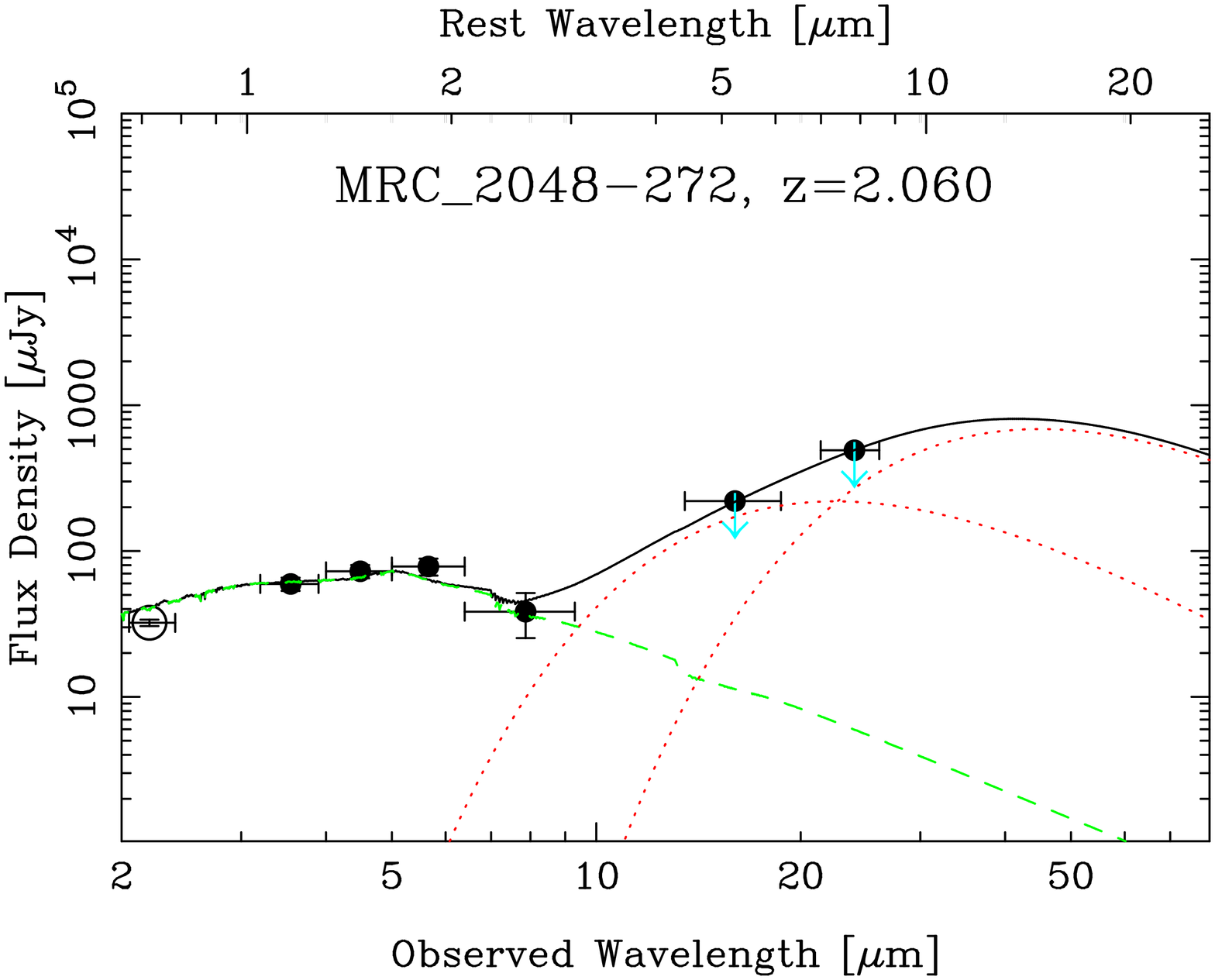} &
\includegraphics[angle=-90,width=144pt,trim=28 59 70 0,clip=true]{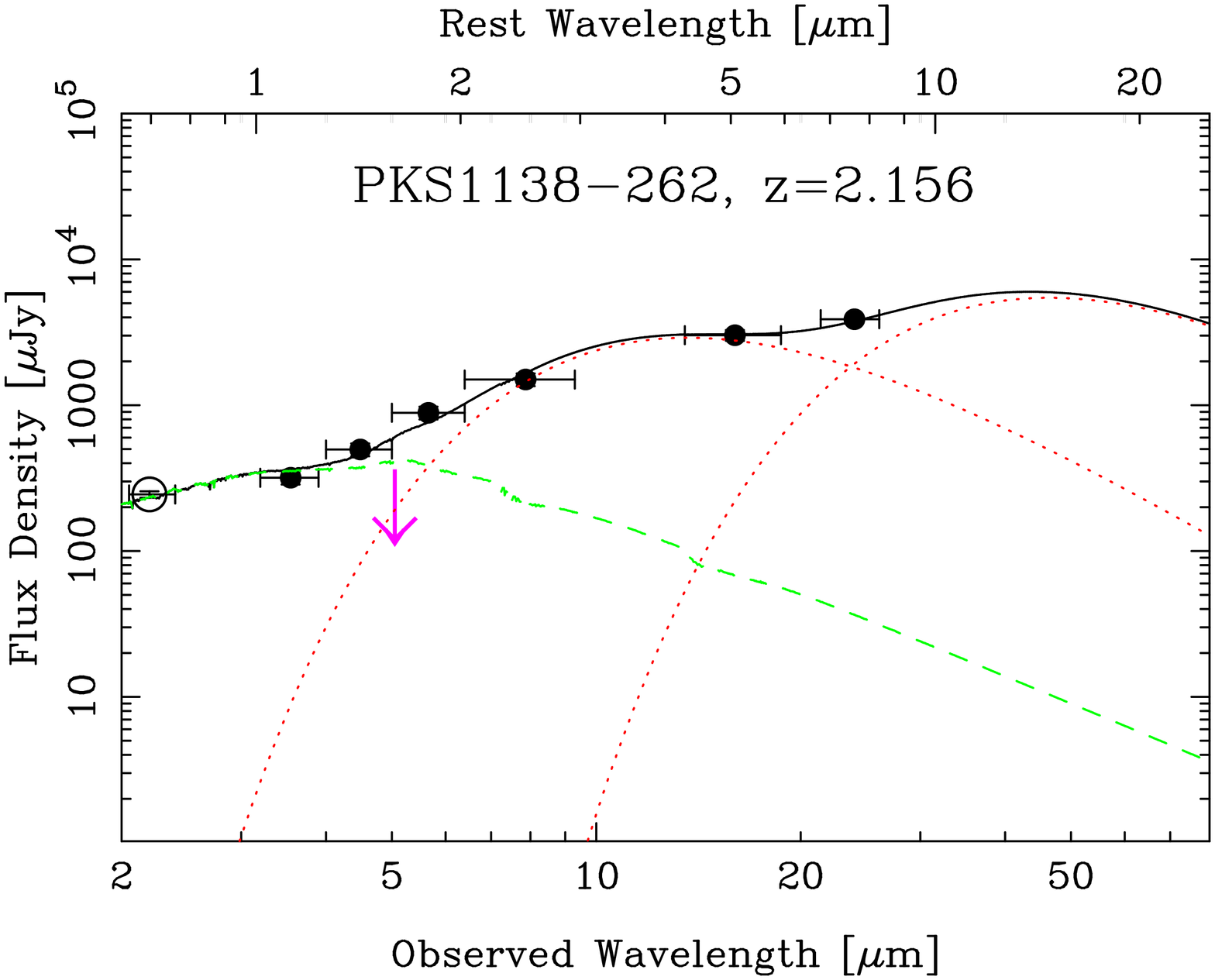} \\[-5pt]
\includegraphics[angle=-90,width=160pt,trim=28 -7 70 0,clip=true]{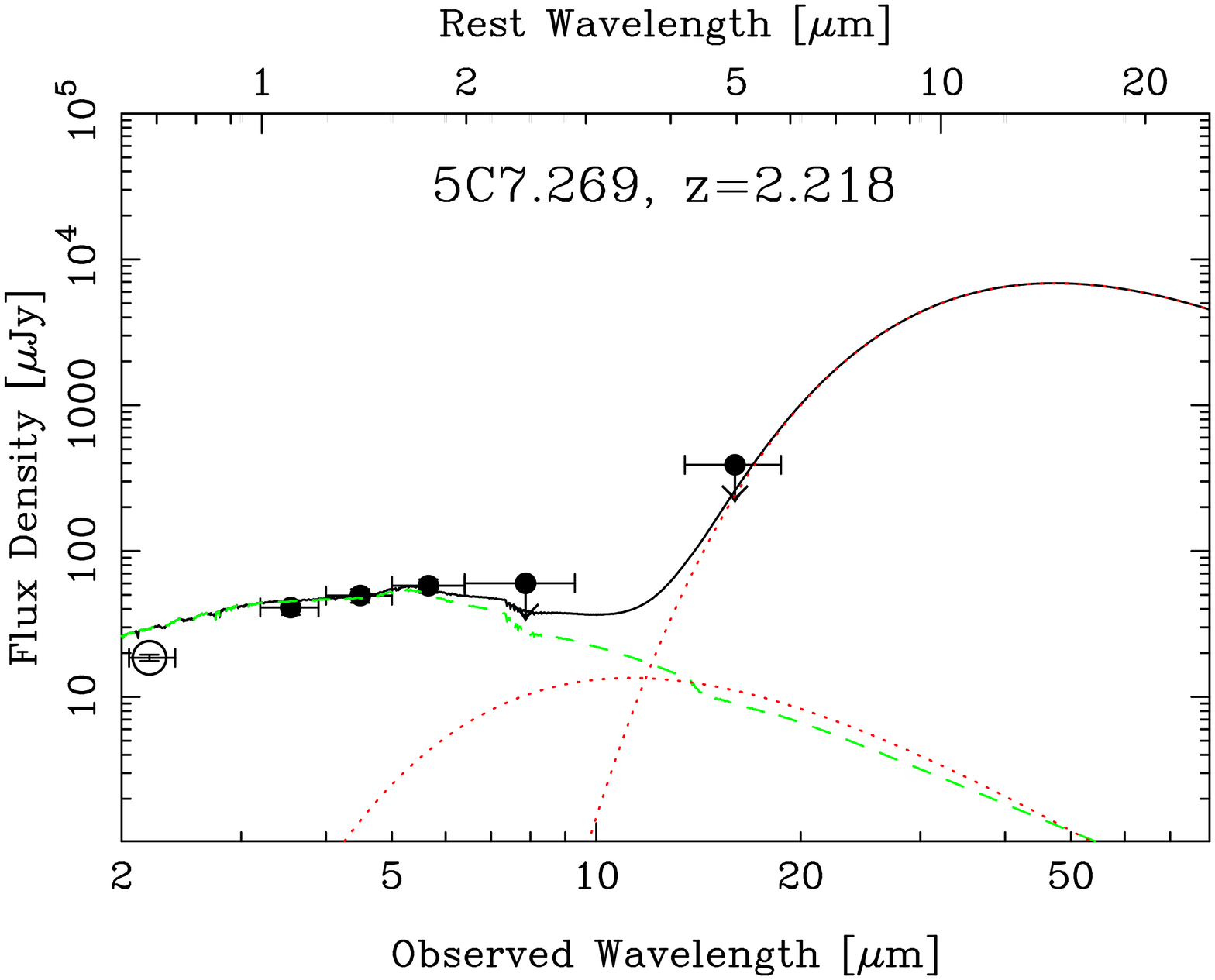} &
\includegraphics[angle=-90,width=144pt,trim=28 59 70 0,clip=true]{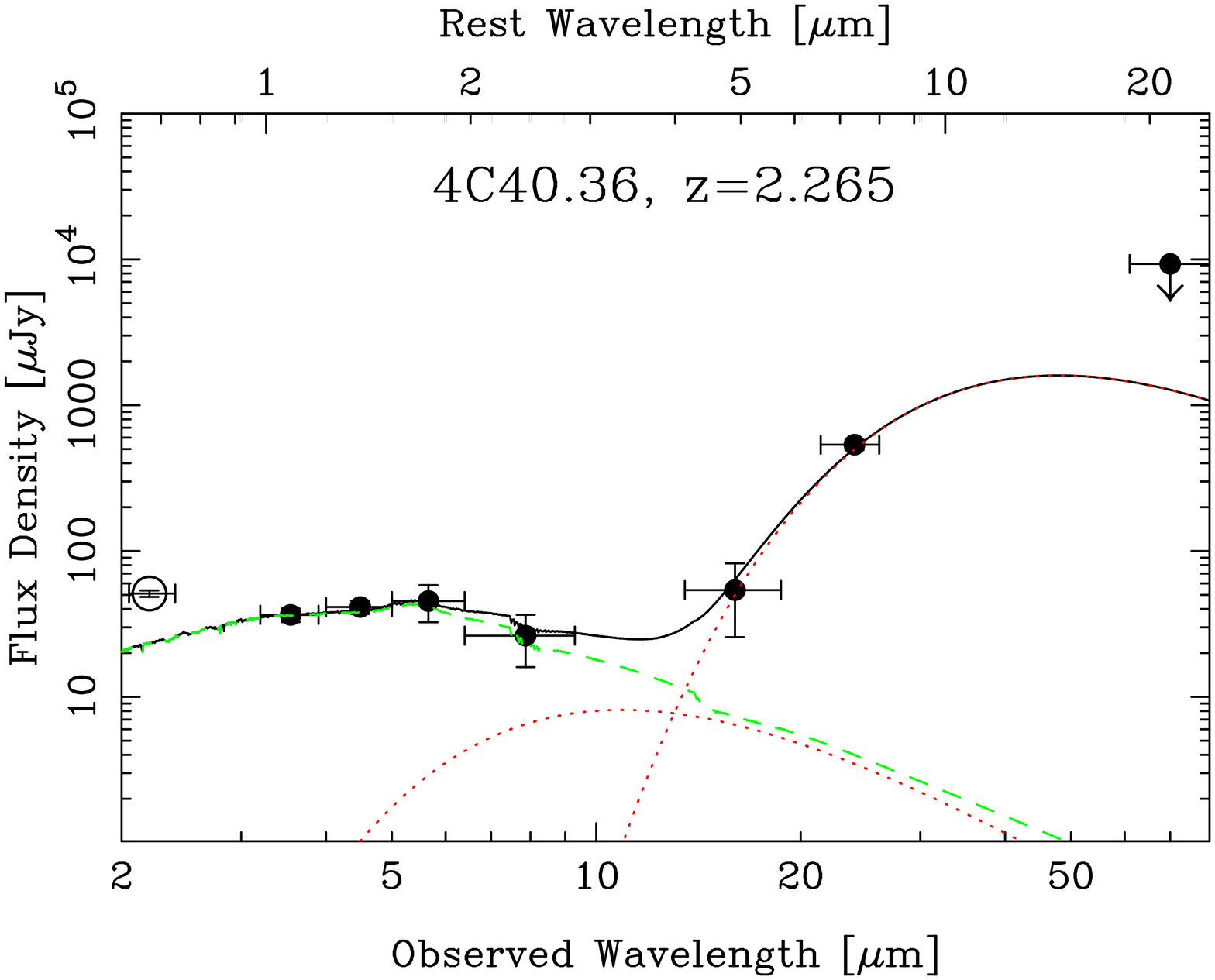} &
\includegraphics[angle=-90,width=144pt,trim=28 59 70 0,clip=true]{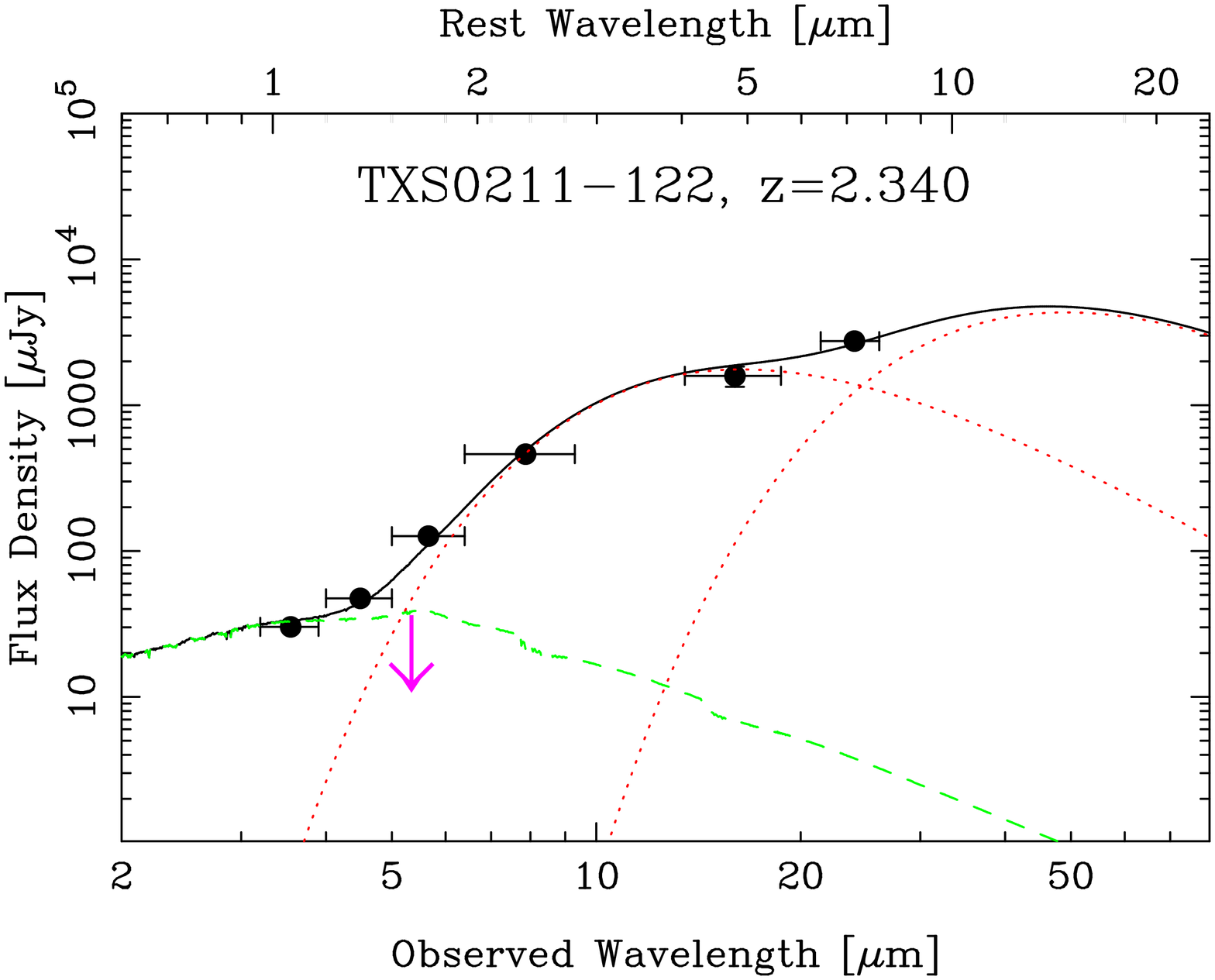} \\[-5pt]
\includegraphics[angle=-90,width=160pt,trim=28 -7 70 0,clip=true]{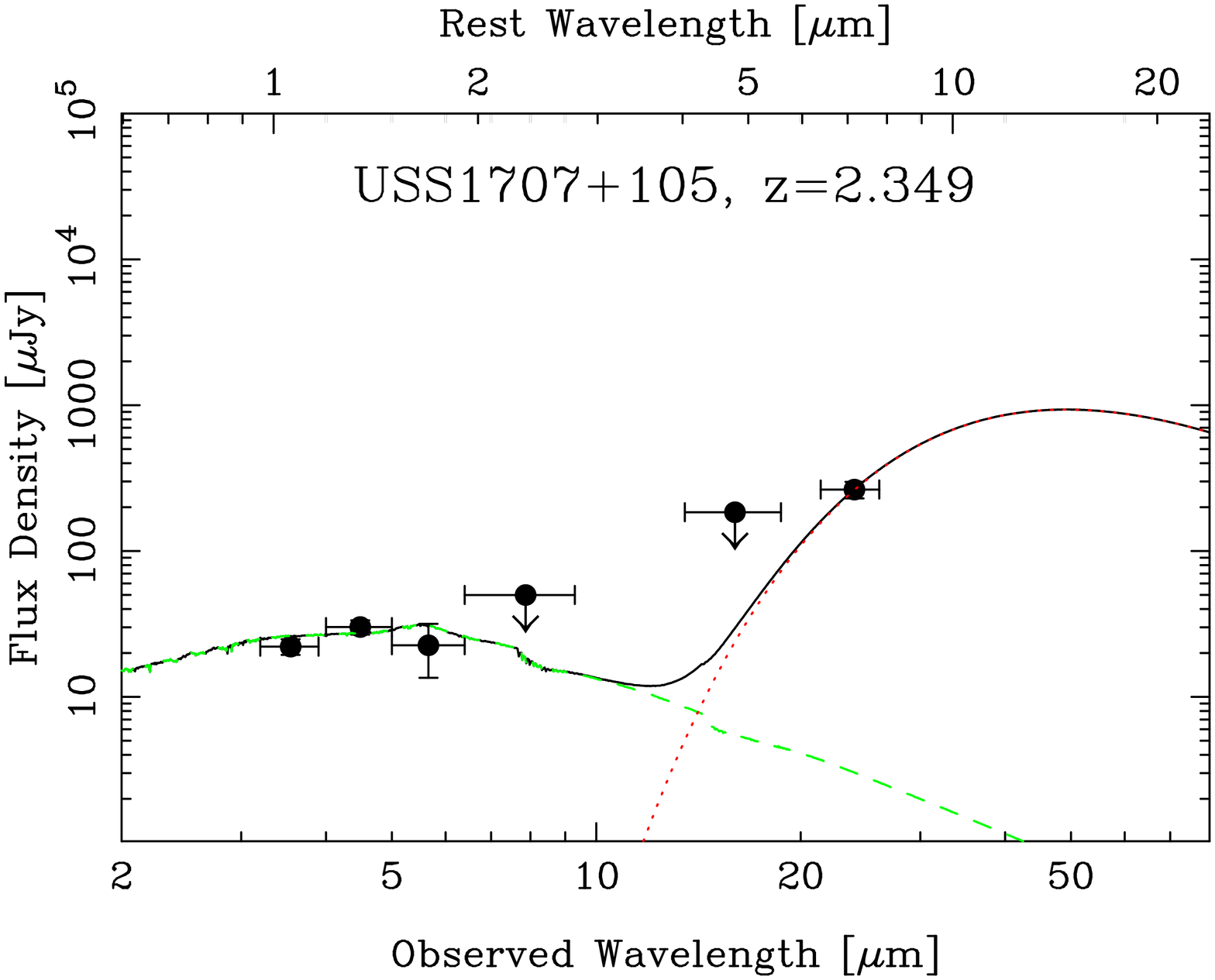} &
\includegraphics[angle=-90,width=144pt,trim=28 59 70 0,clip=true]{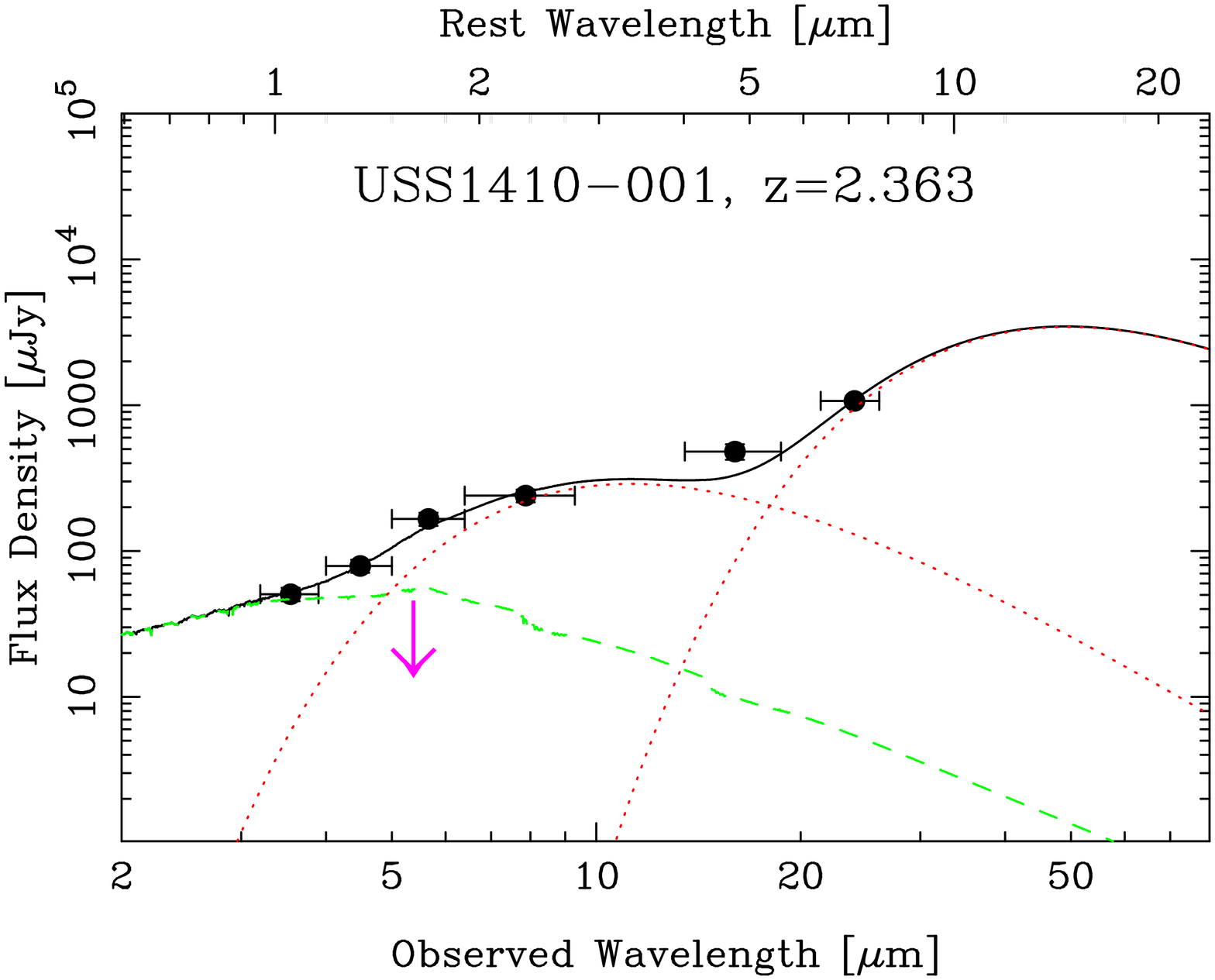} &
\includegraphics[angle=-90,width=144pt,trim=28 59 70 0,clip=true]{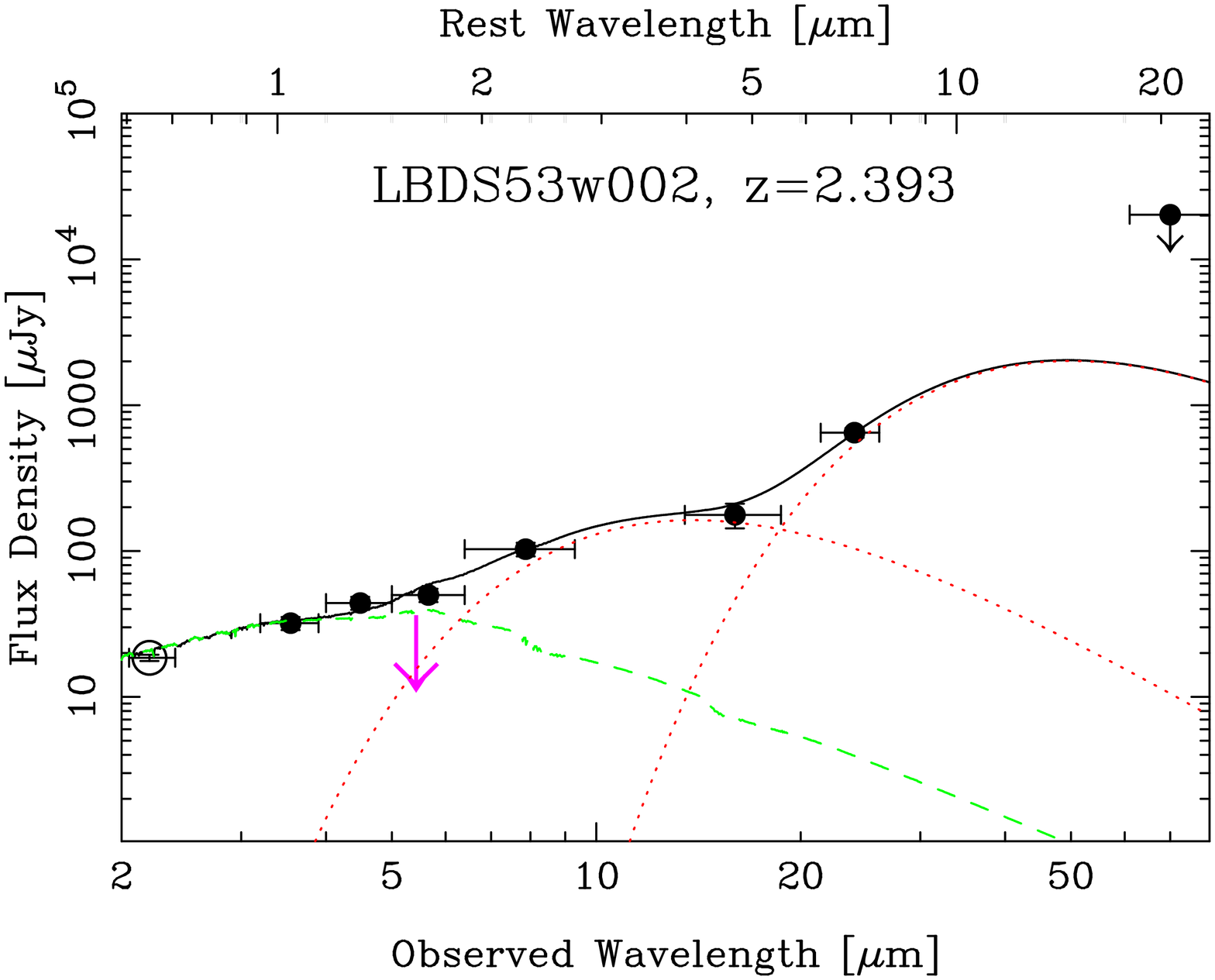} \\[-5pt]
\includegraphics[angle=-90,width=160pt,trim=28 -7  0 0,clip=true]{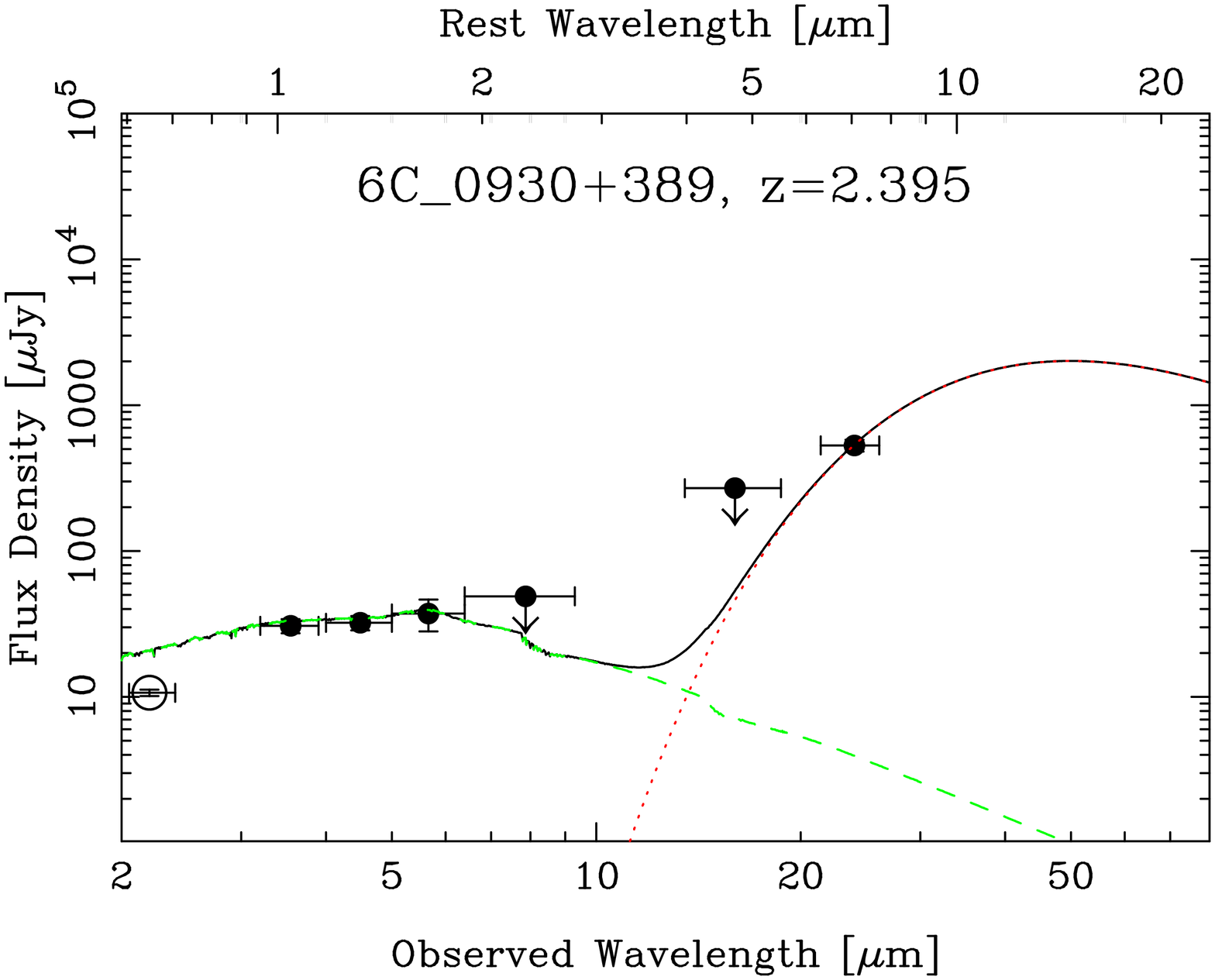} &
\includegraphics[angle=-90,width=144pt,trim=28 59  0 0,clip=true]{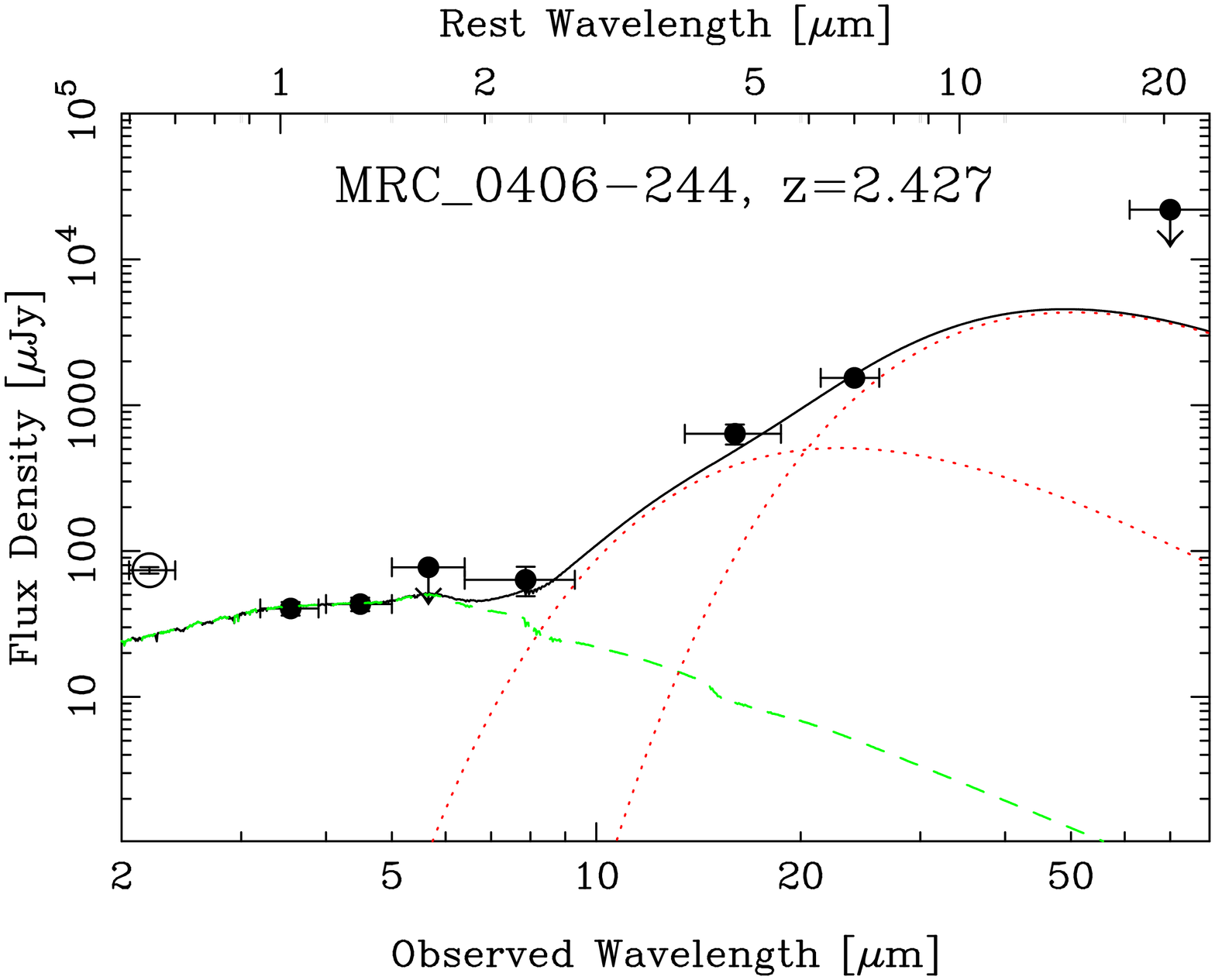} &
\includegraphics[angle=-90,width=144pt,trim=28 59  0 0,clip=true]{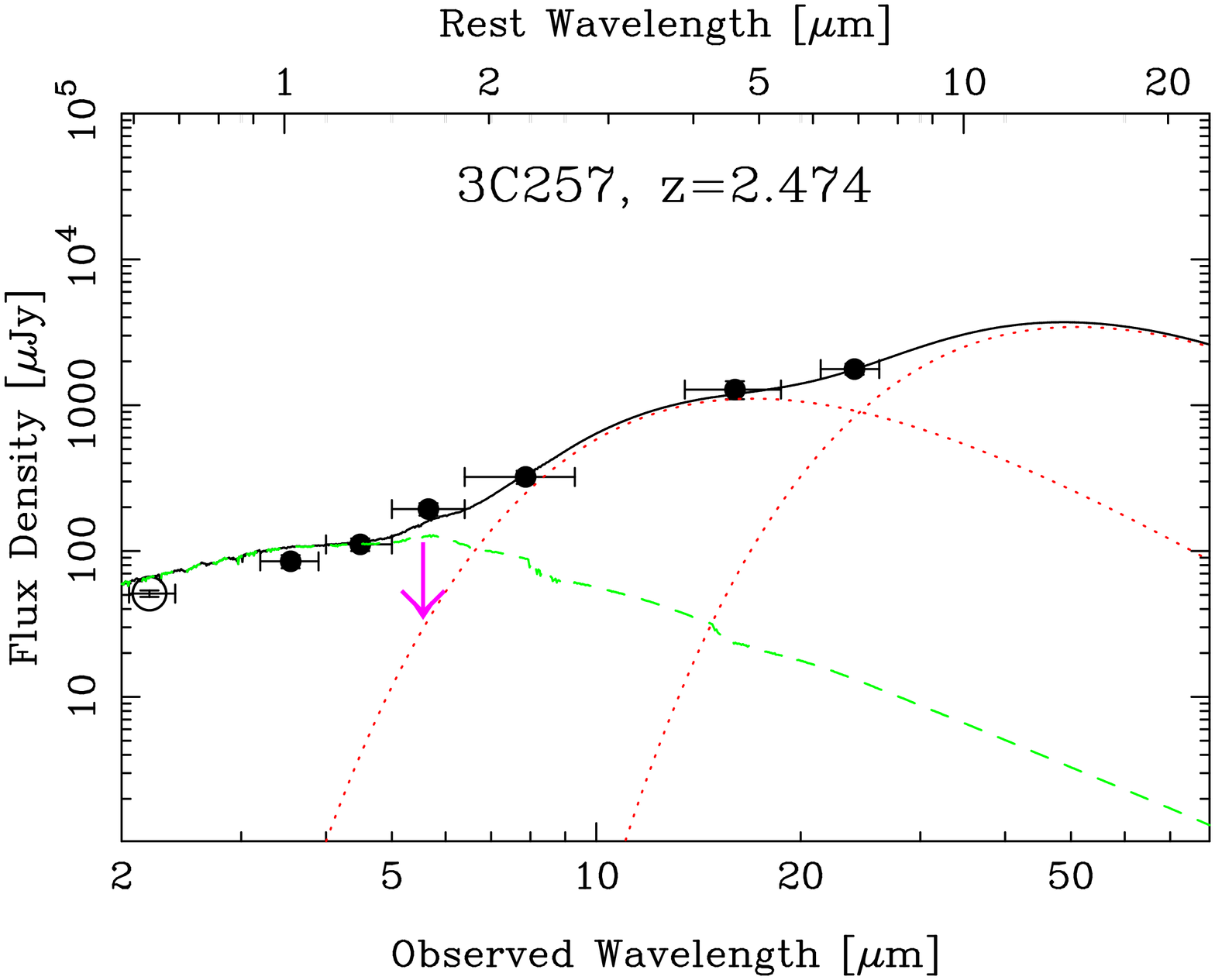} \\
\end{tabular}
\end{figure*}
\pagebreak

\begin{figure*}
\begin{tabular}{r@{}c@{}l}
\includegraphics[angle=-90,width=160pt,trim= 0 -7 70 0,clip=true]{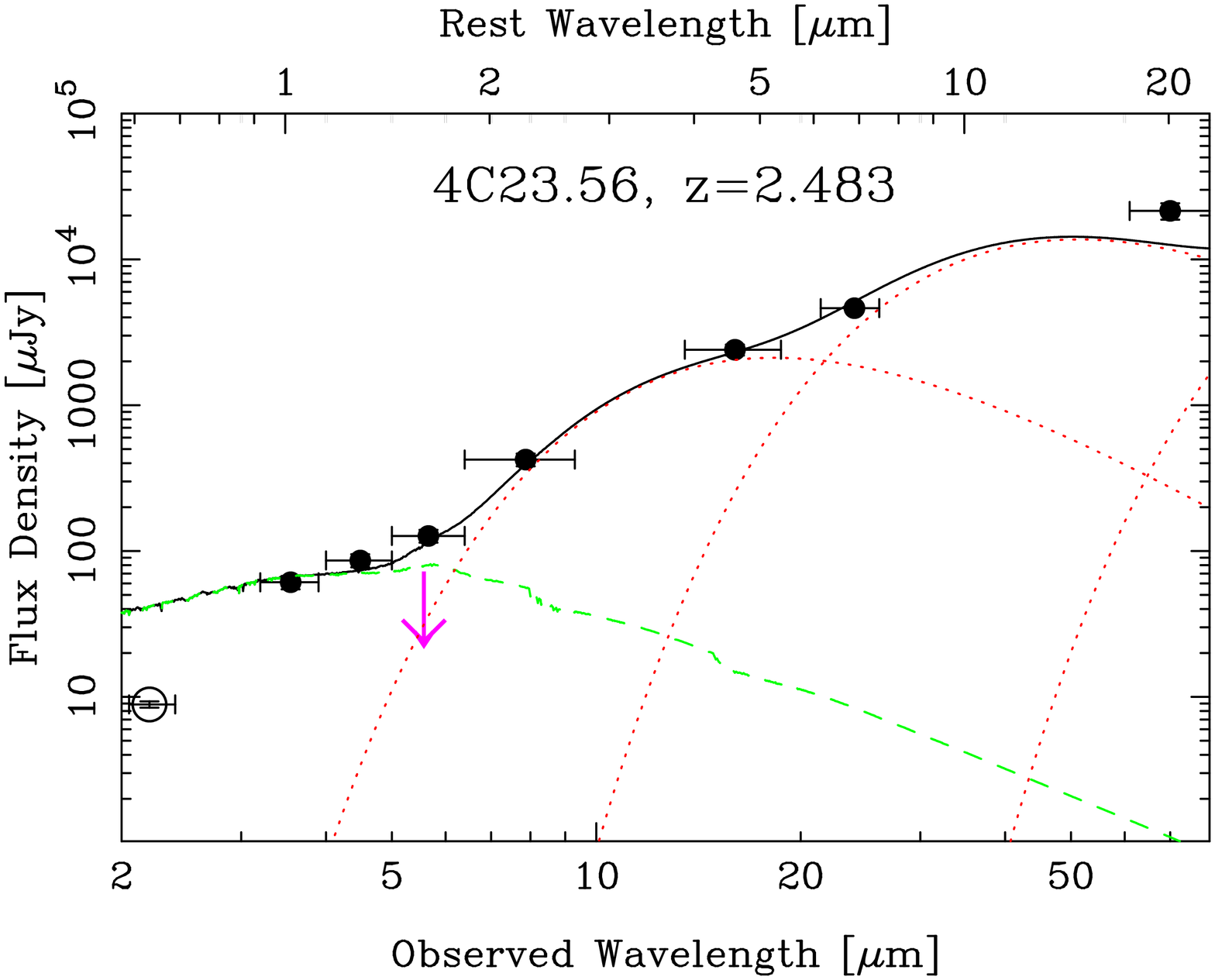} &
\includegraphics[angle=-90,width=144pt,trim= 0 59 70 0,clip=true]{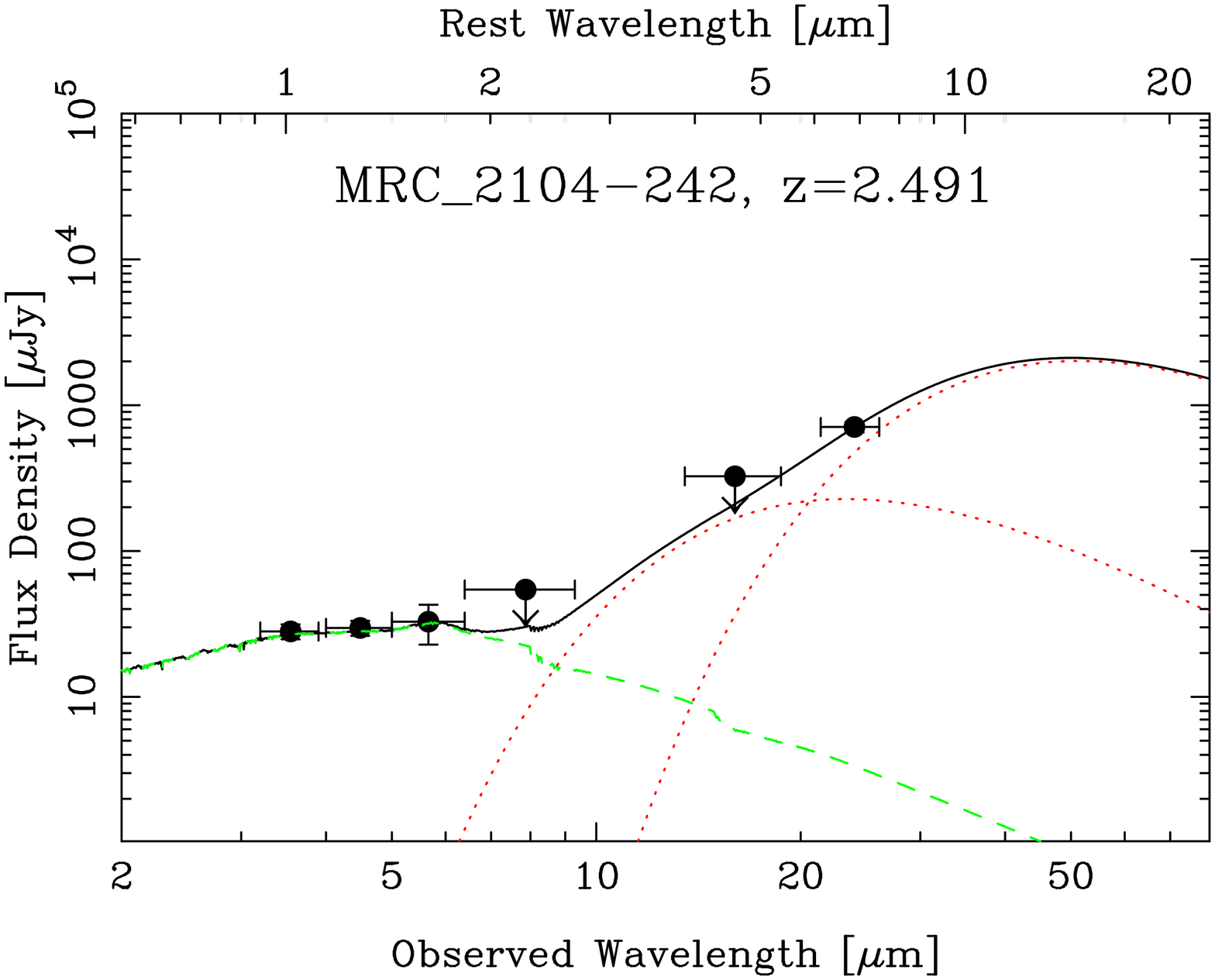} &
\includegraphics[angle=-90,width=144pt,trim= 0 59 70 0,clip=true]{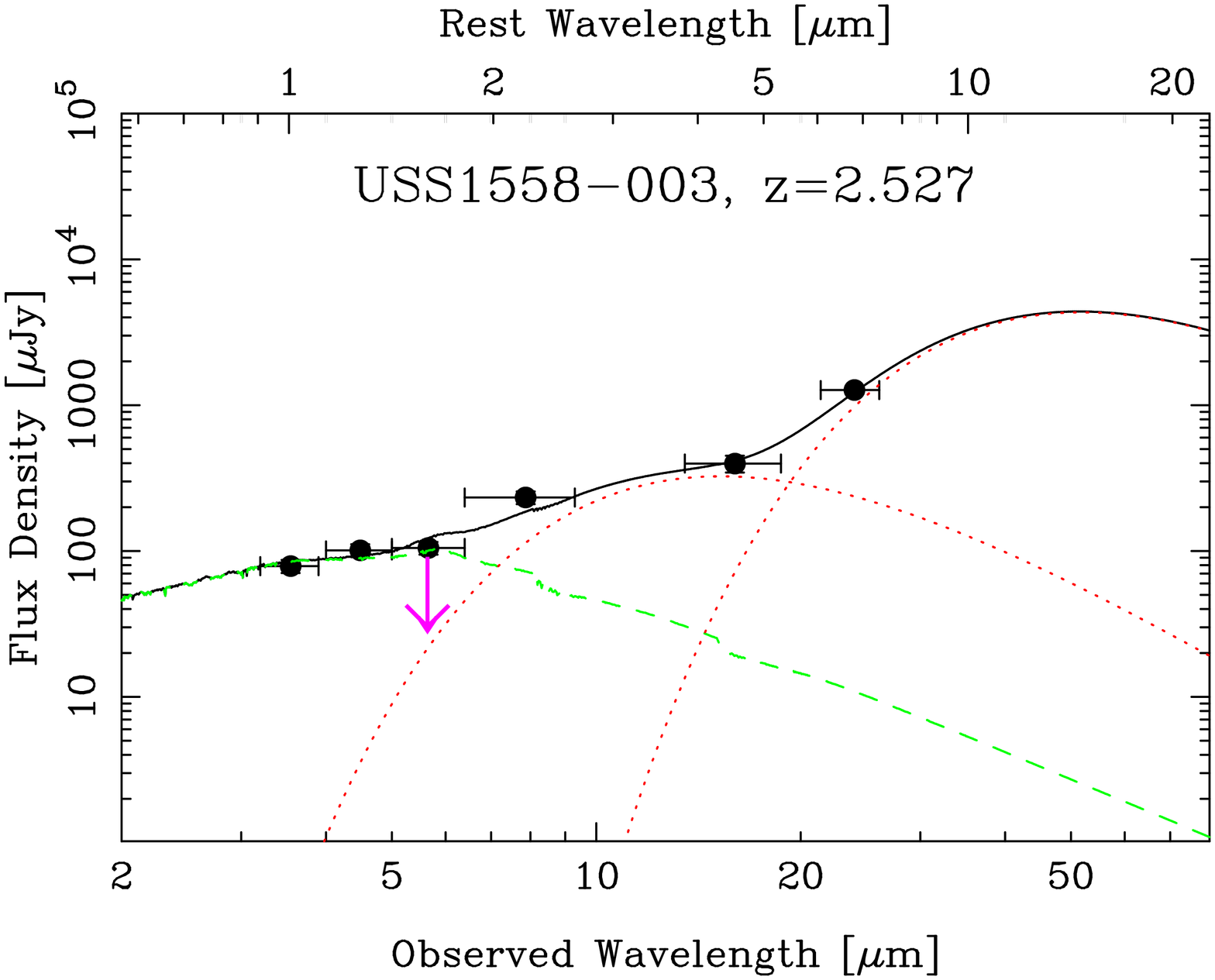} \\[-5pt]
\includegraphics[angle=-90,width=160pt,trim=28 -7 70 0,clip=true]{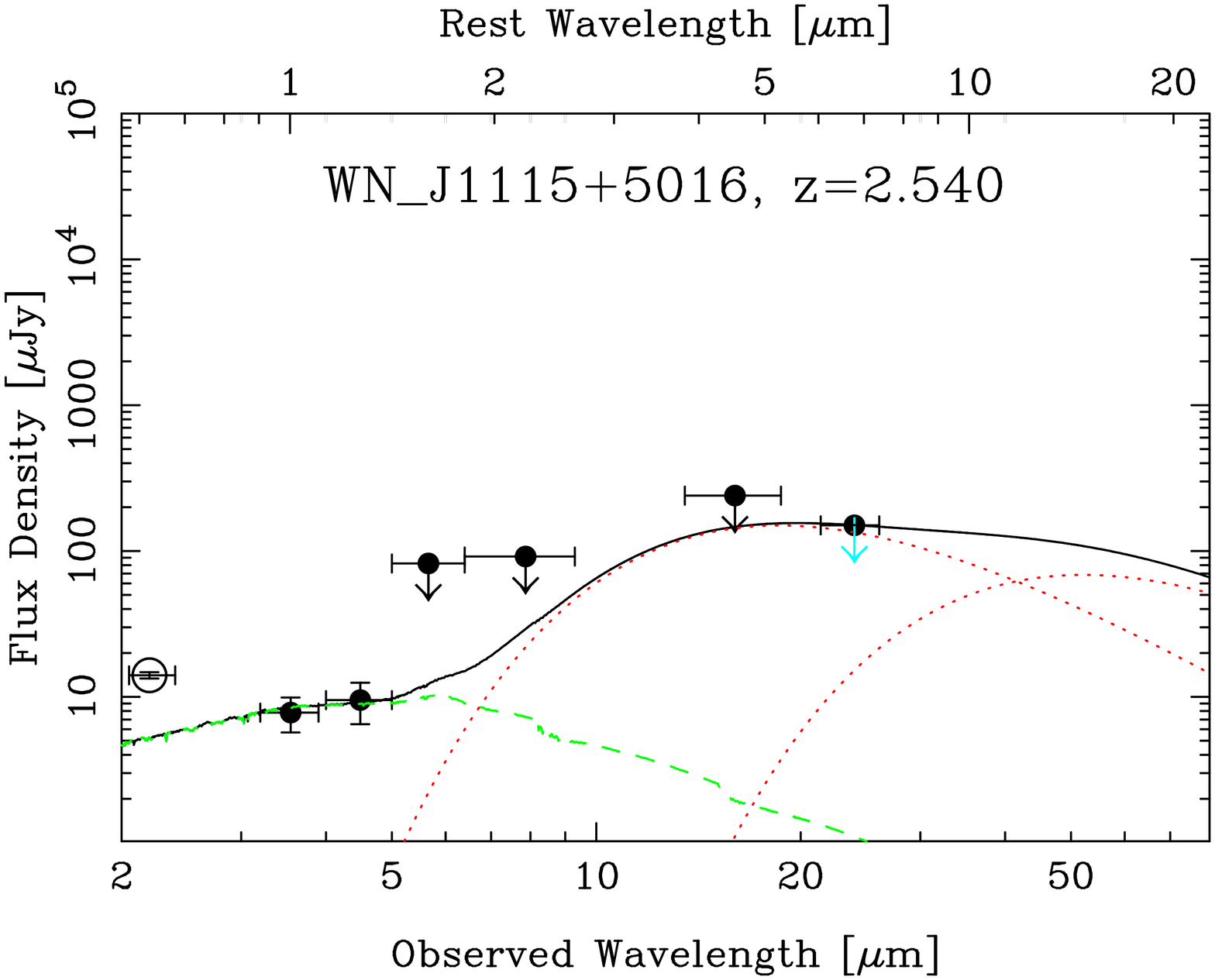} &
\includegraphics[angle=-90,width=144pt,trim=28 59 70 0,clip=true]{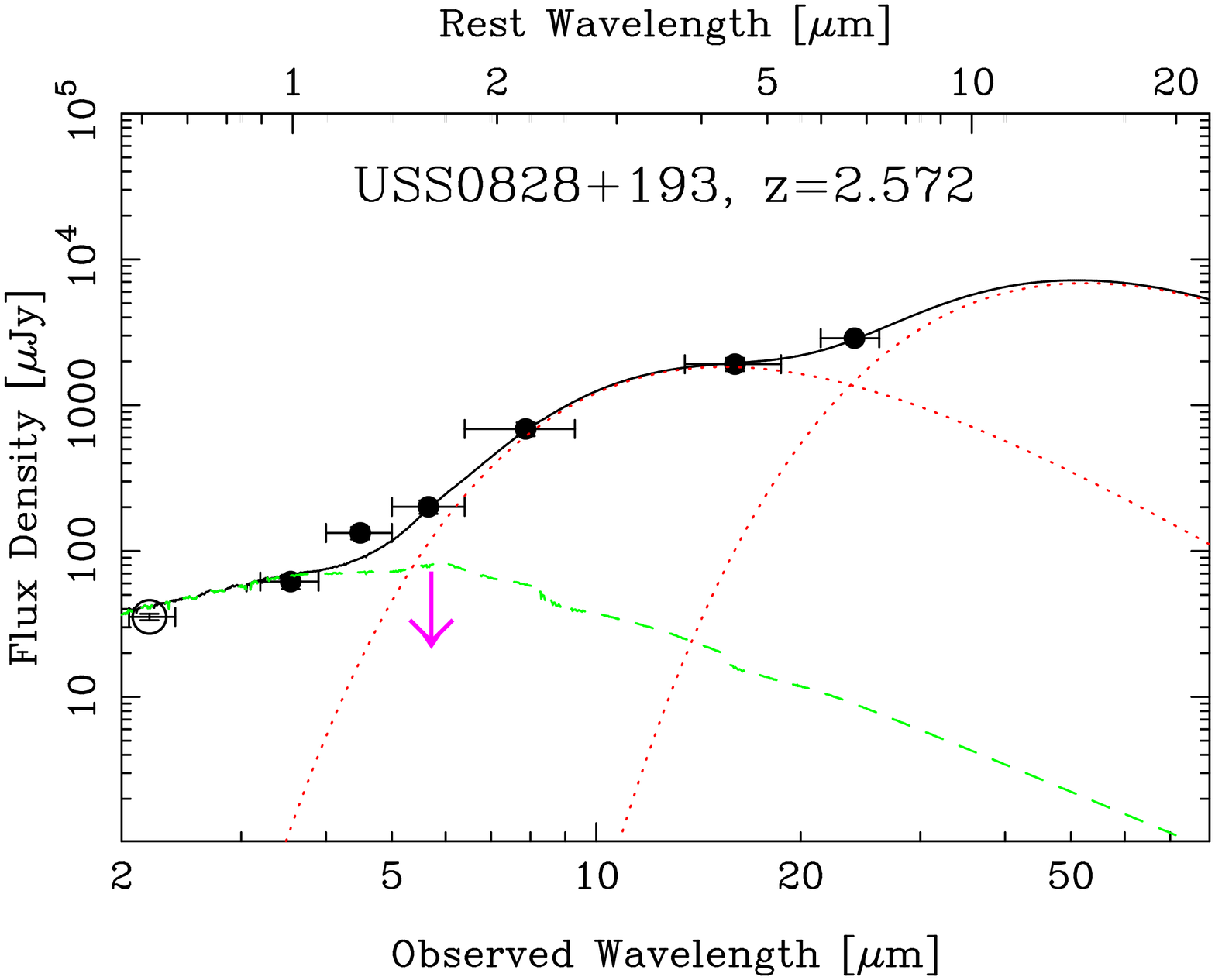} &
\includegraphics[angle=-90,width=144pt,trim=28 59 70 0,clip=true]{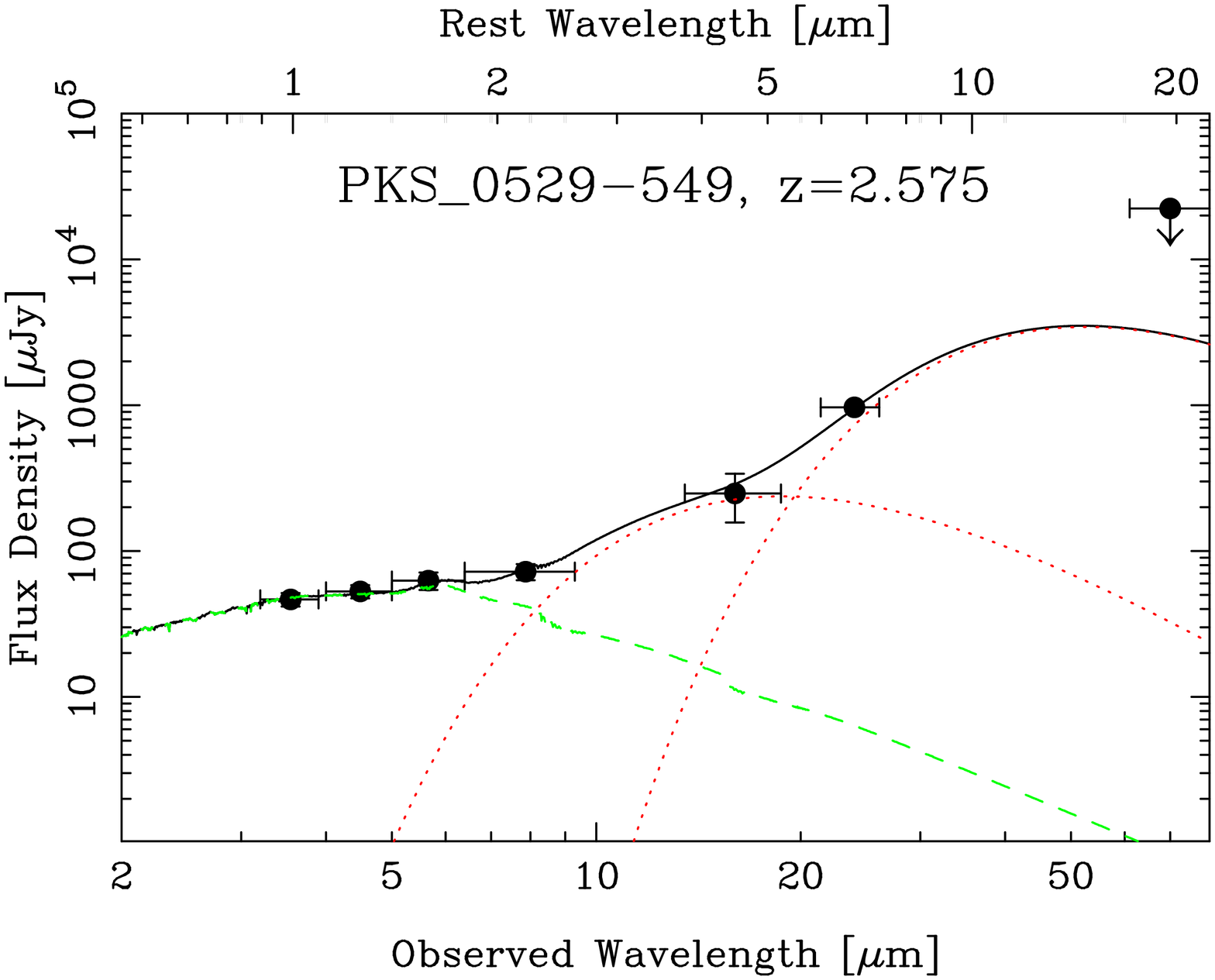} \\[-5pt]
\includegraphics[angle=-90,width=160pt,trim=28 -7 70 0,clip=true]{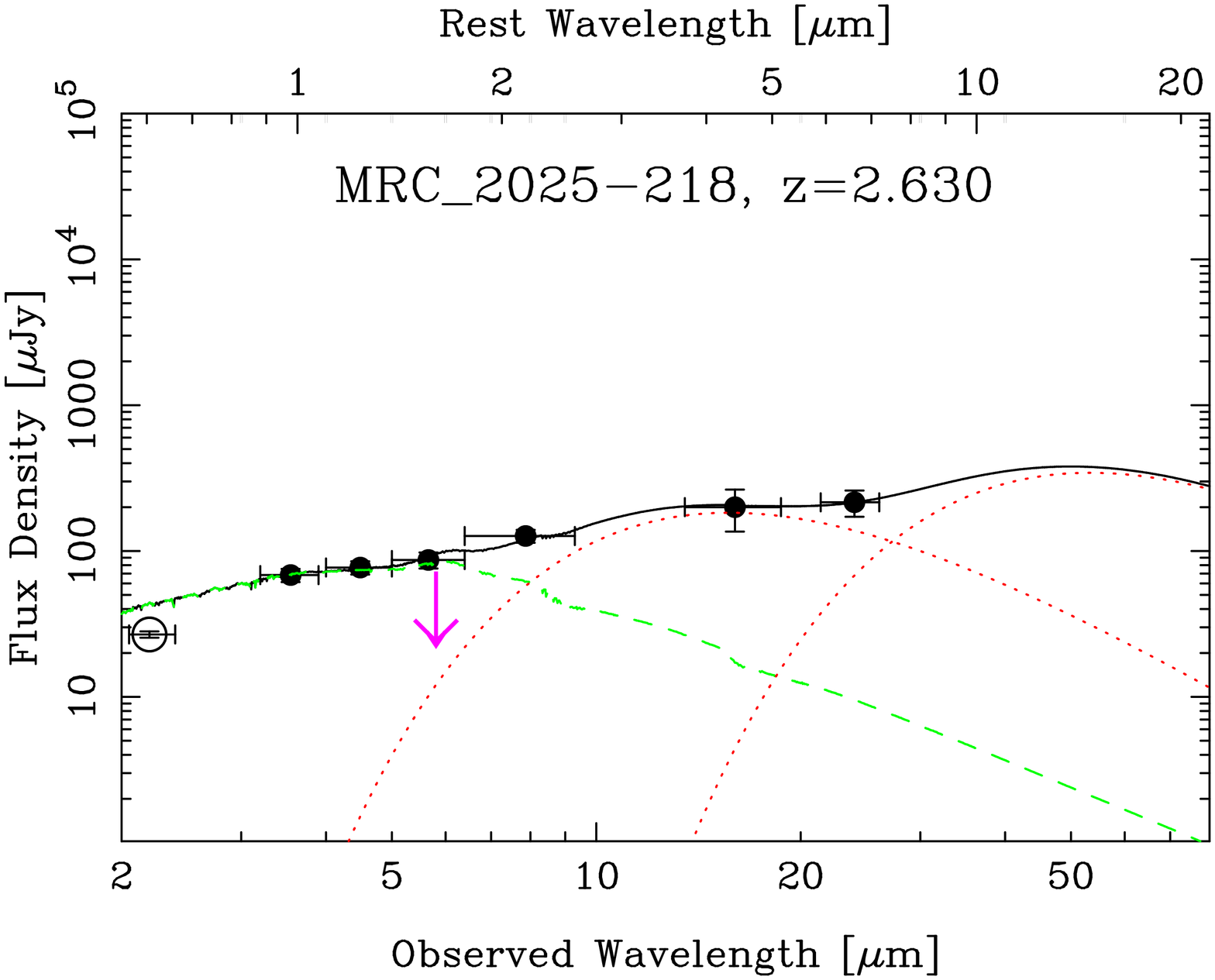} &
\includegraphics[angle=-90,width=144pt,trim=28 59 70 0,clip=true]{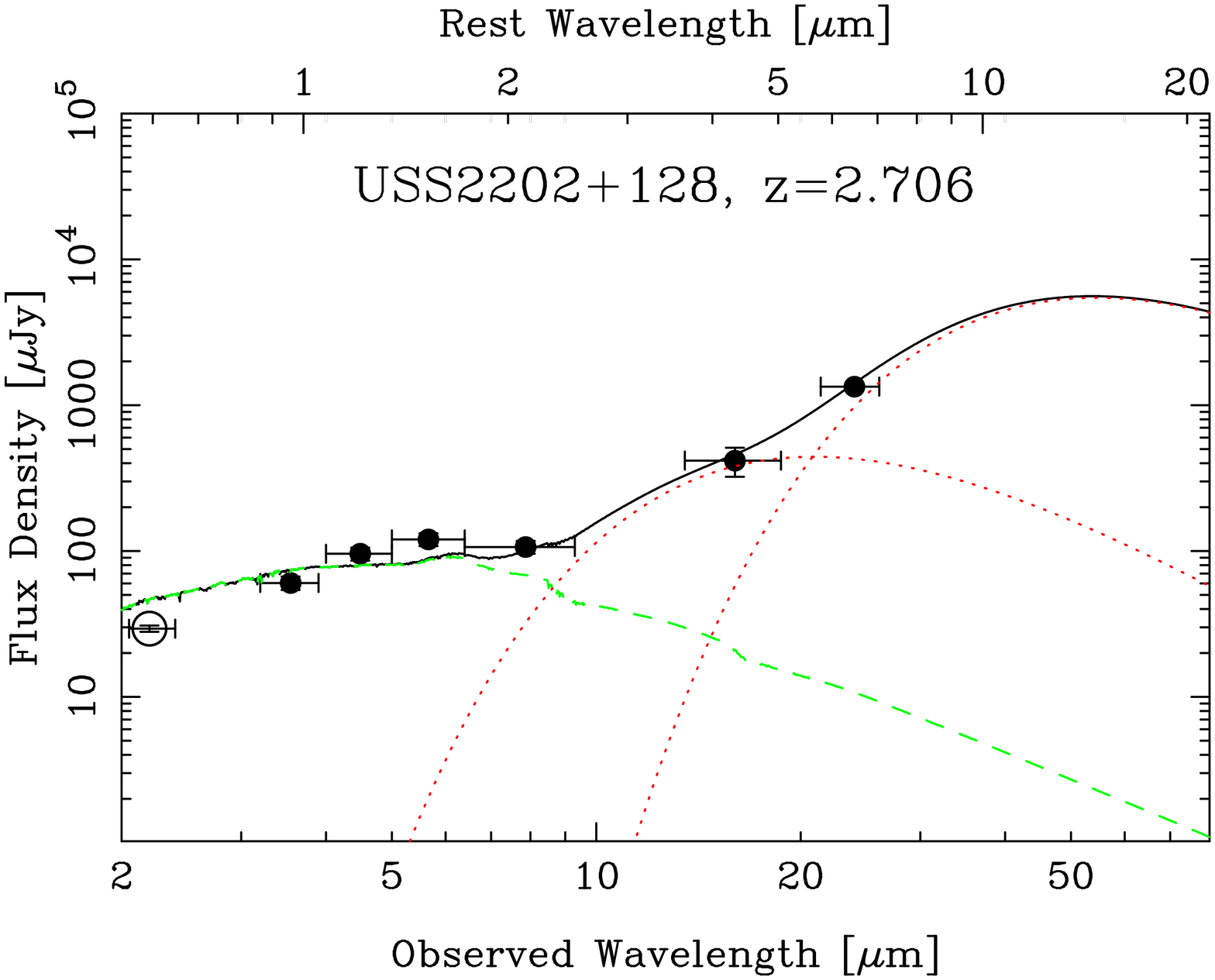} &
\includegraphics[angle=-90,width=144pt,trim=28 59 70 0,clip=true]{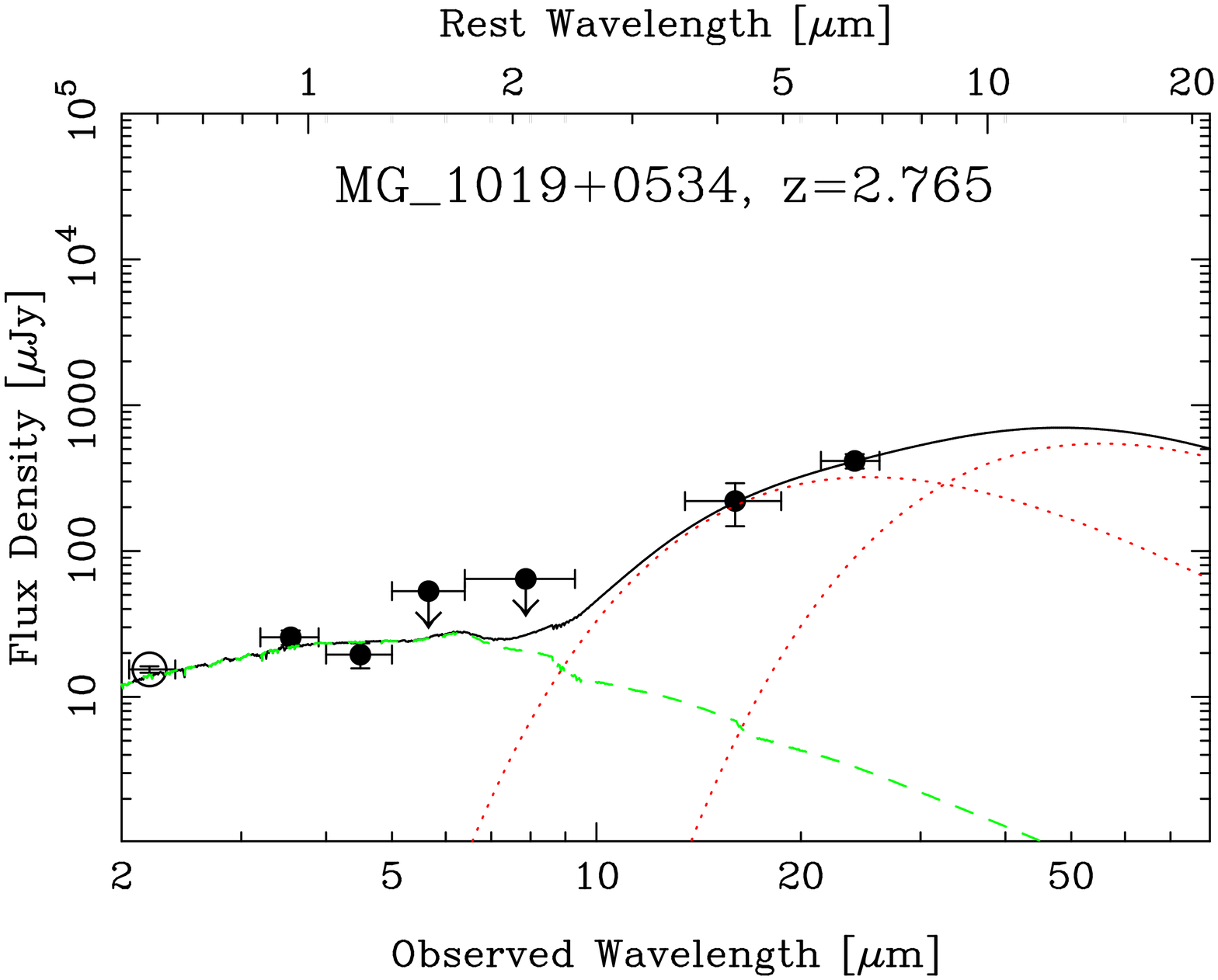} \\[-5pt]
\includegraphics[angle=-90,width=160pt,trim=28 -7 70 0,clip=true]{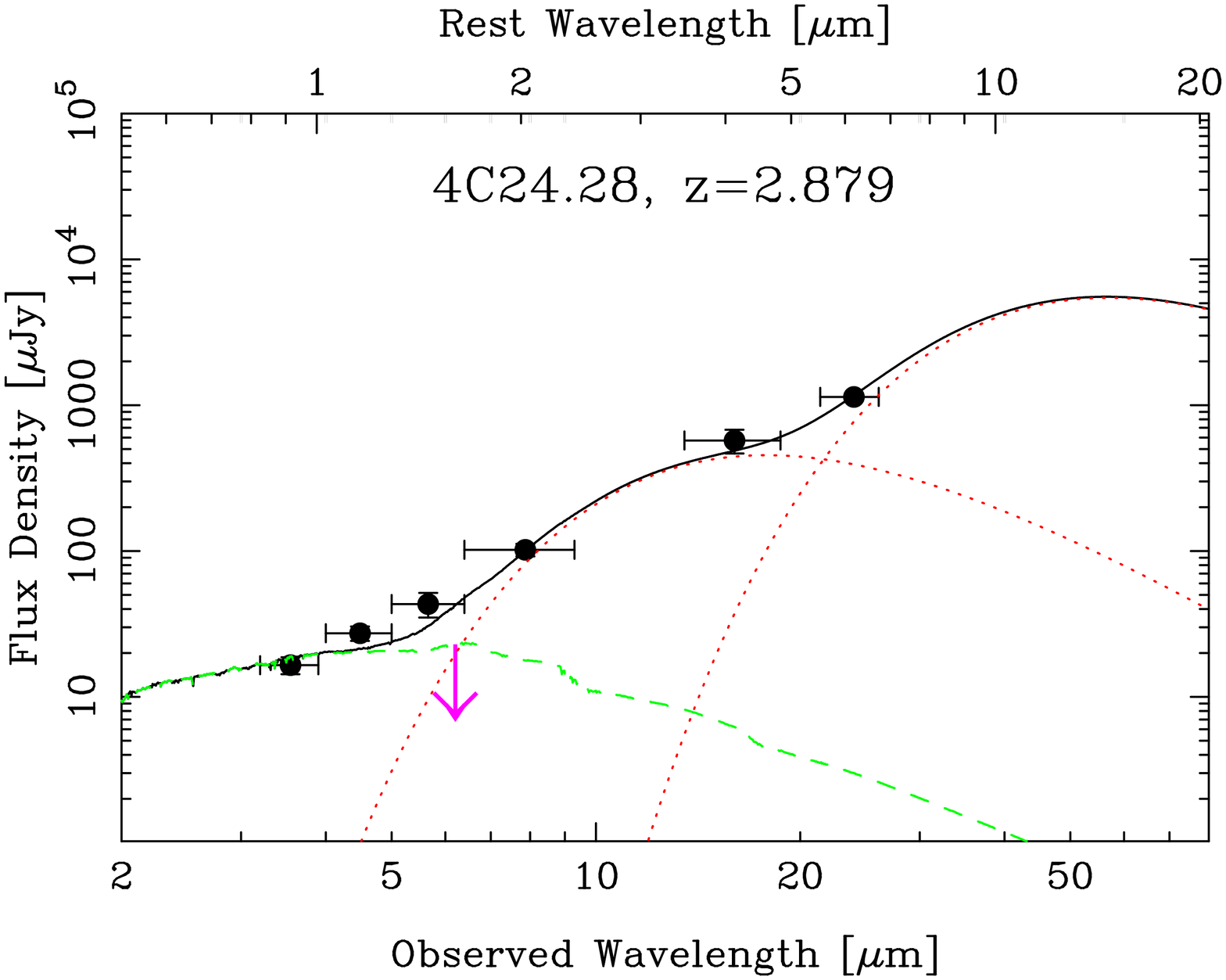} &
\includegraphics[angle=-90,width=144pt,trim=28 59 70 0,clip=true]{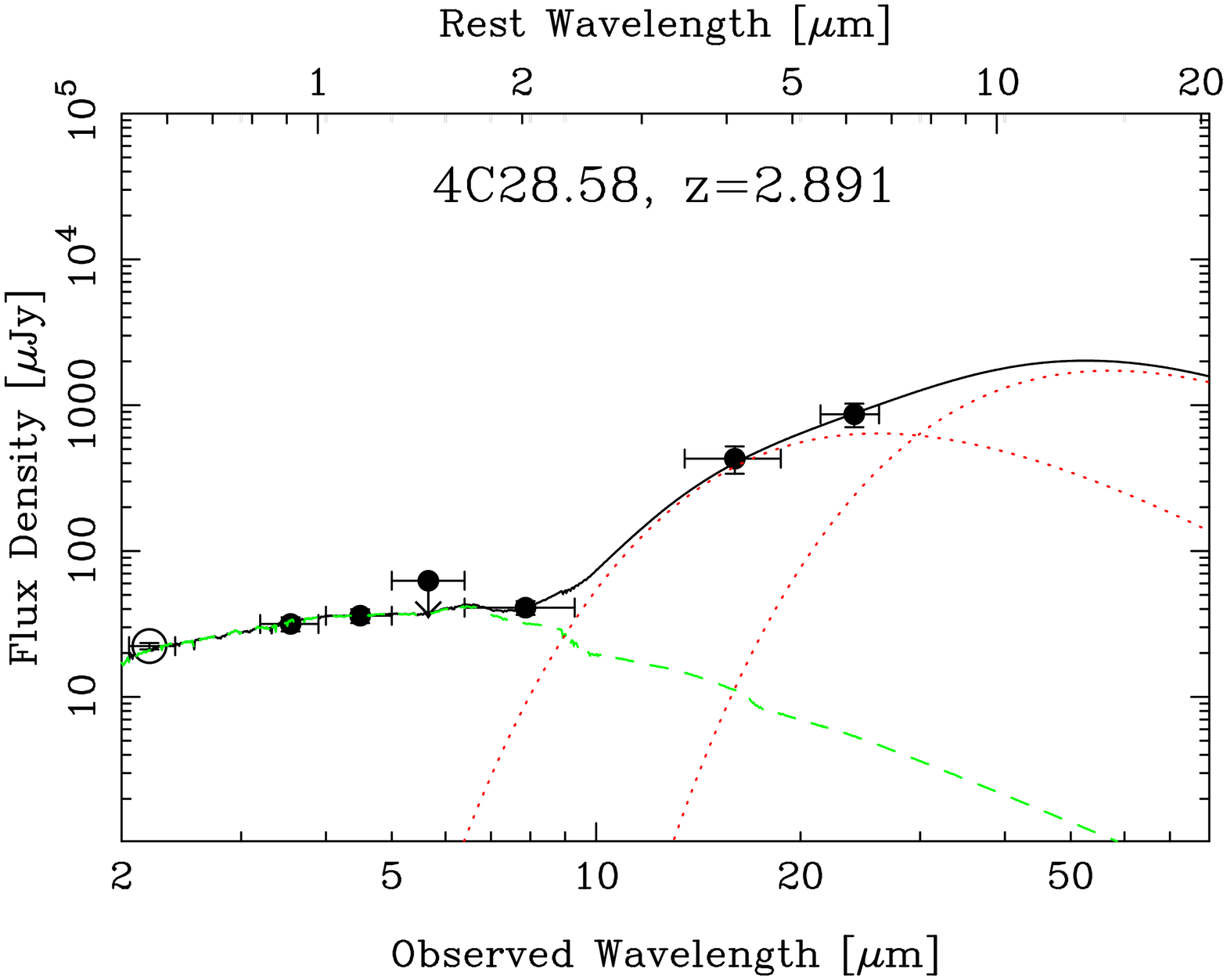} &
\includegraphics[angle=-90,width=144pt,trim=28 59 70 0,clip=true]{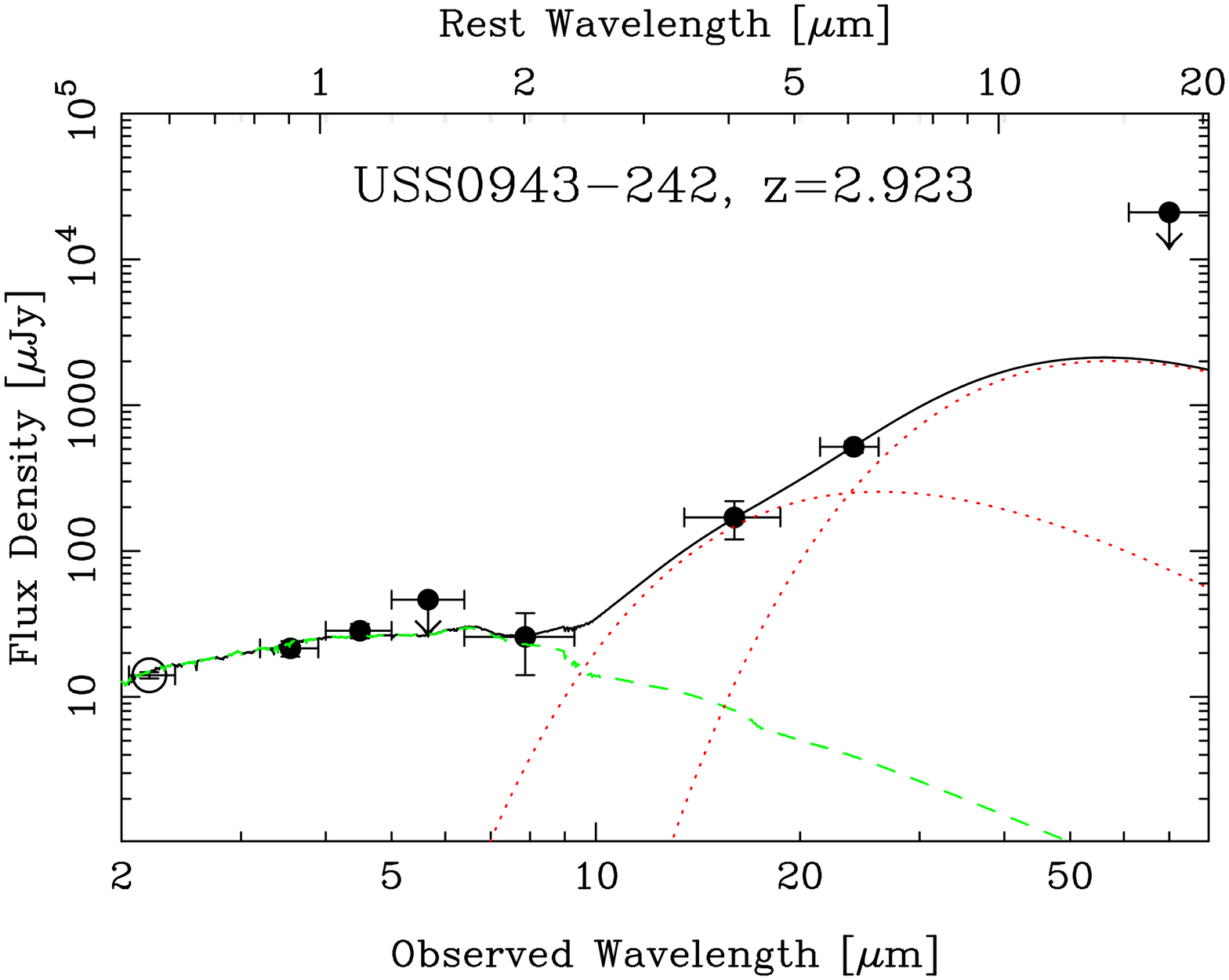} \\[-5pt]
\includegraphics[angle=-90,width=160pt,trim=28 -7 70 0,clip=true]{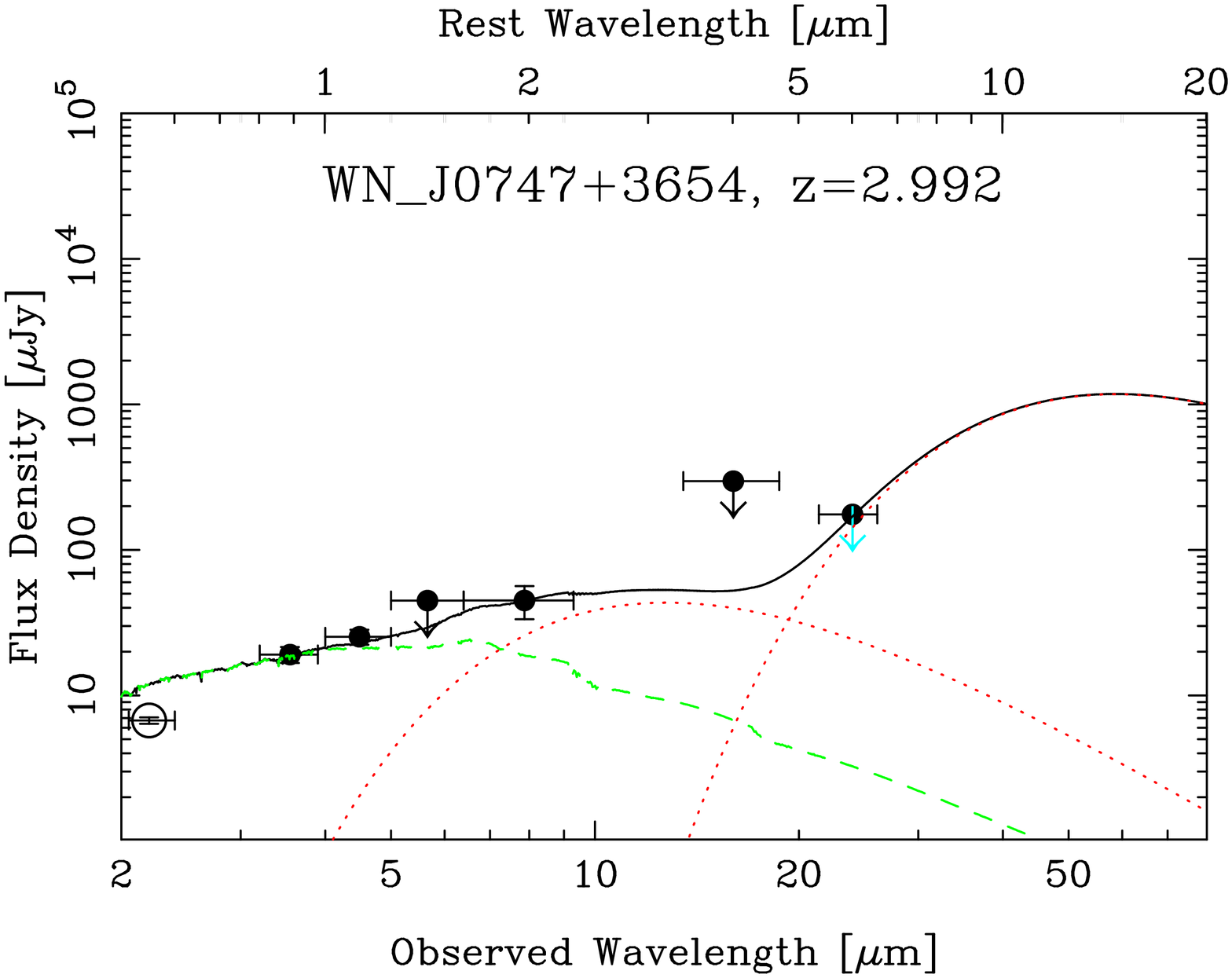} &
\includegraphics[angle=-90,width=144pt,trim=28 59 70 0,clip=true]{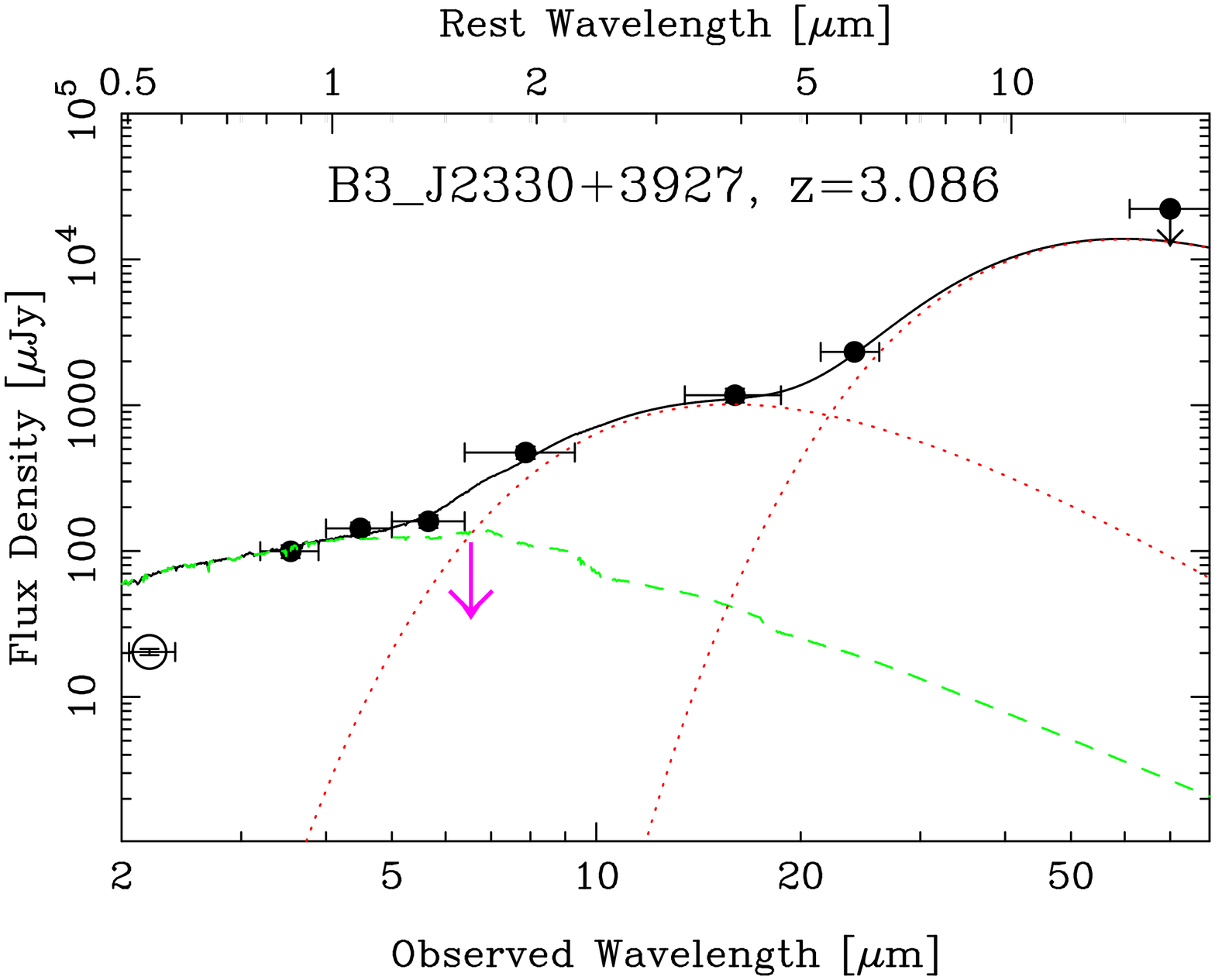} &
\includegraphics[angle=-90,width=144pt,trim=28 59 70 0,clip=true]{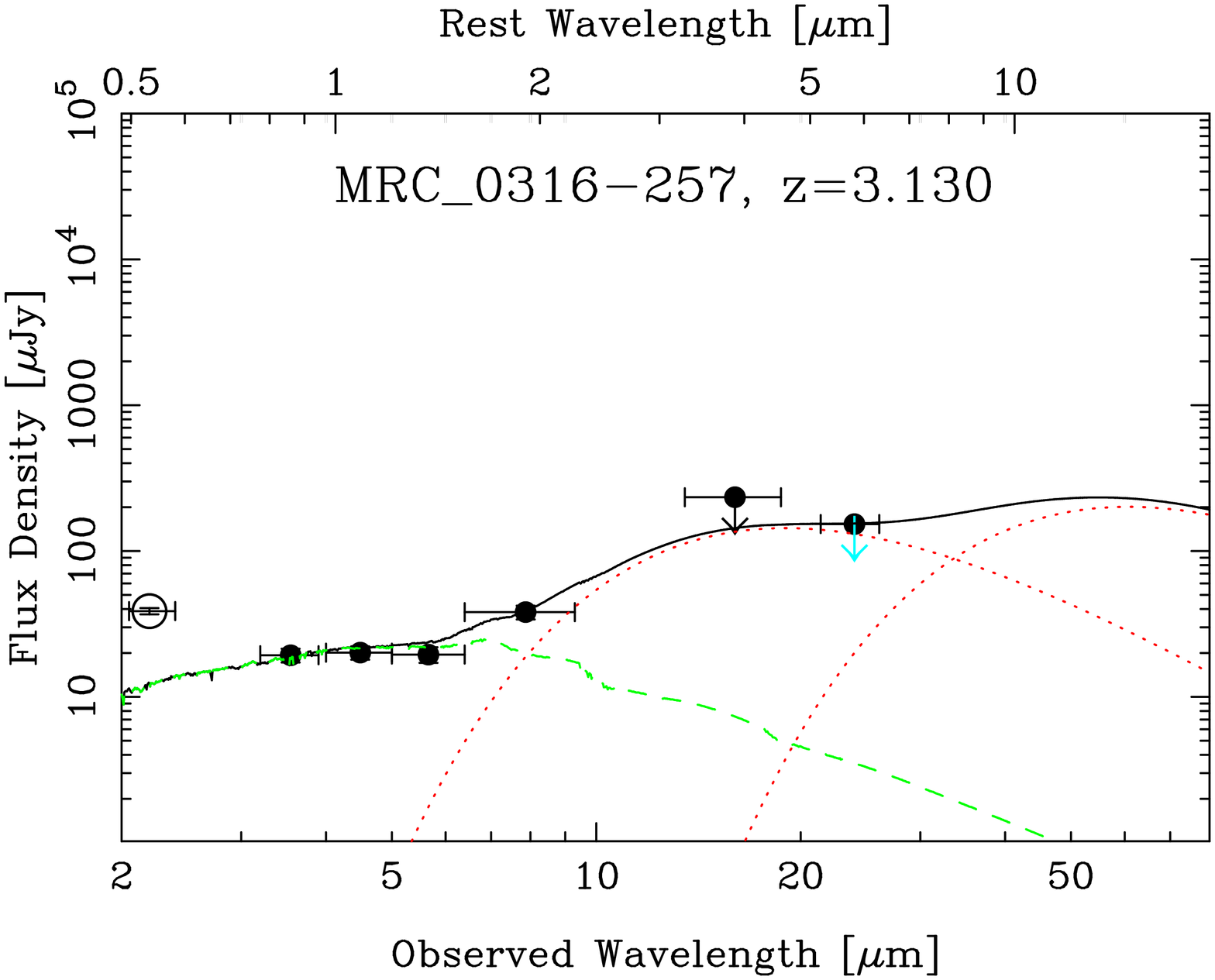} \\[-5pt]
\includegraphics[angle=-90,width=160pt,trim=28 -7  0 0,clip=true]{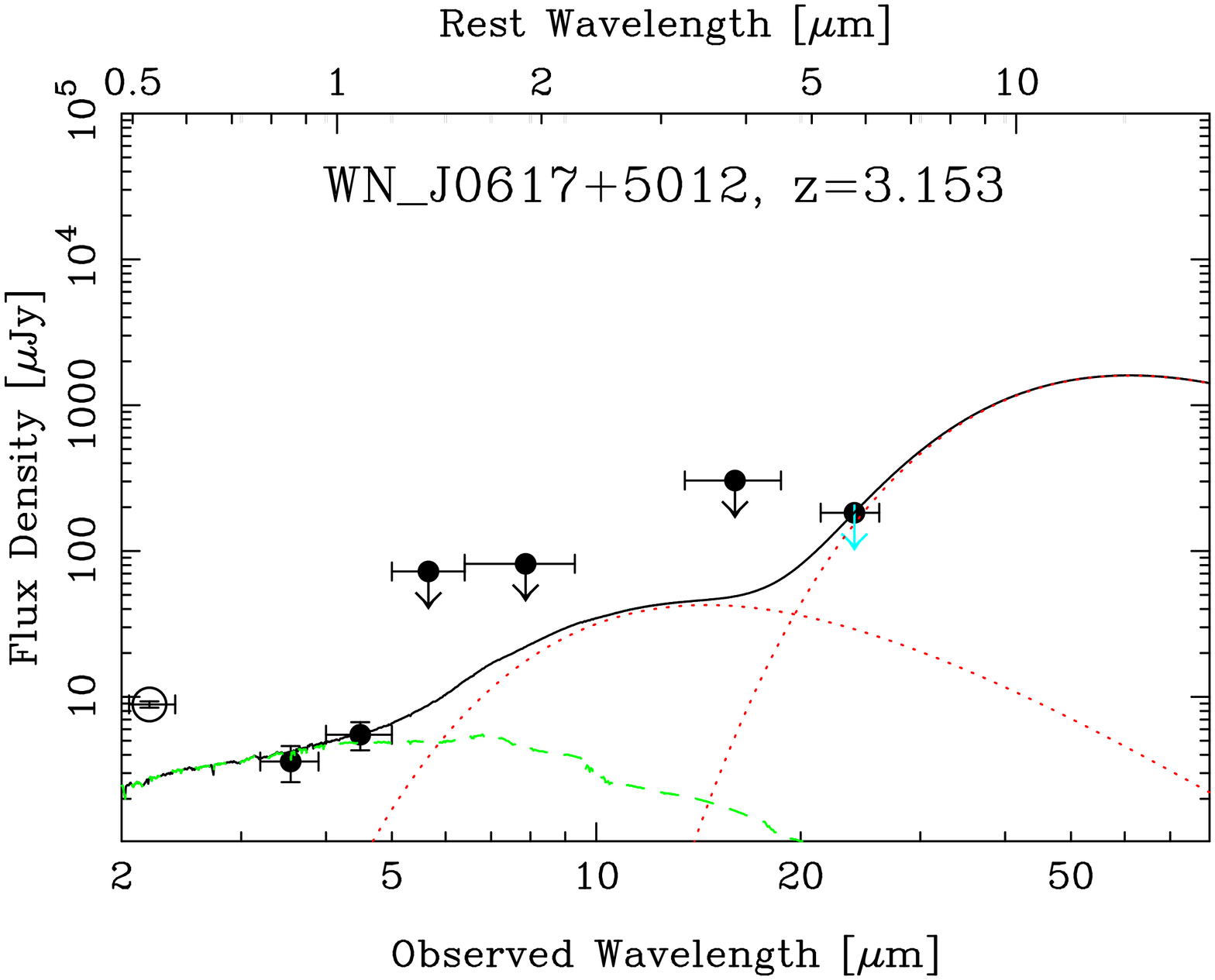} &
\includegraphics[angle=-90,width=144pt,trim=28 59  0 0,clip=true]{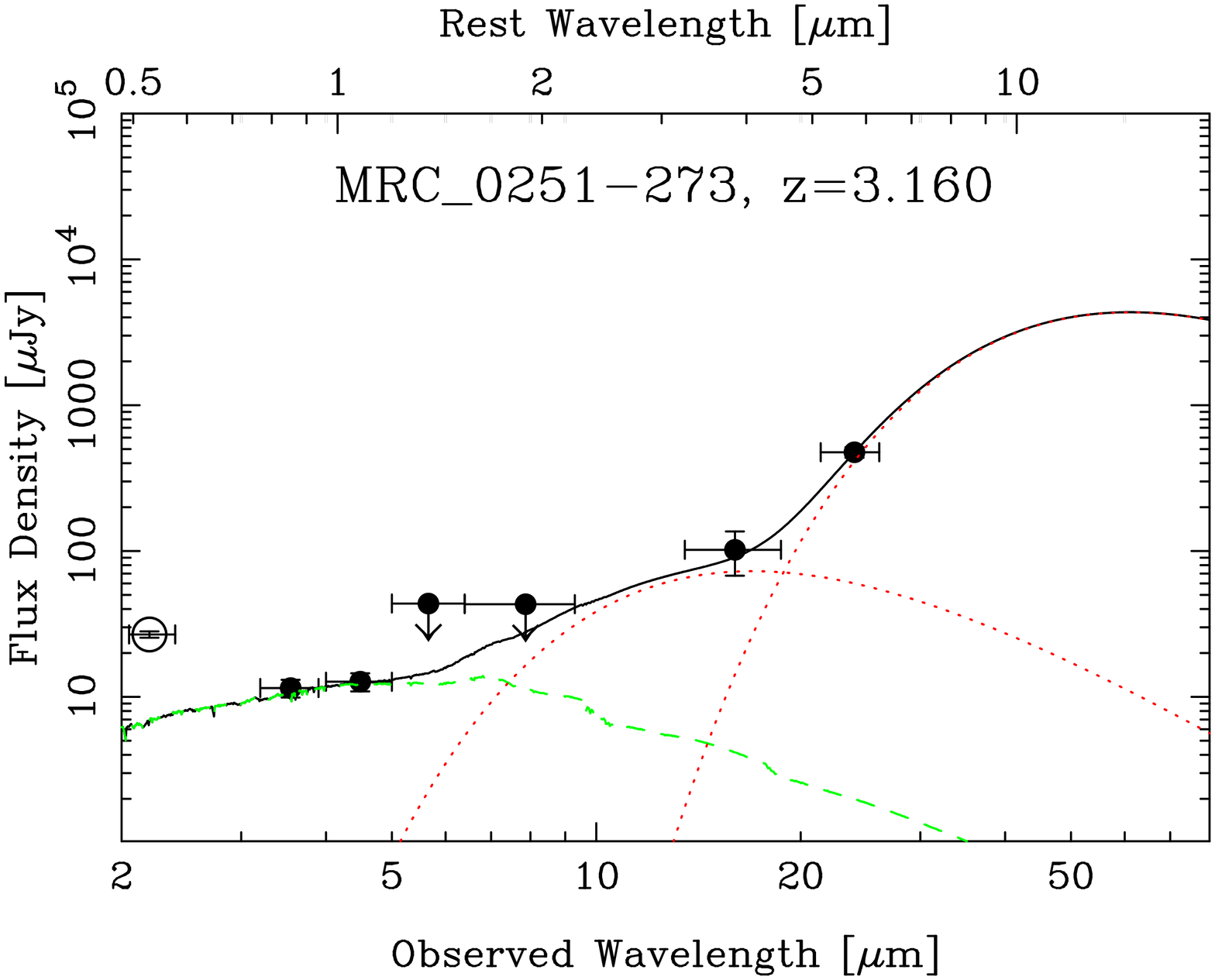} &
\includegraphics[angle=-90,width=144pt,trim=28 59  0 0,clip=true]{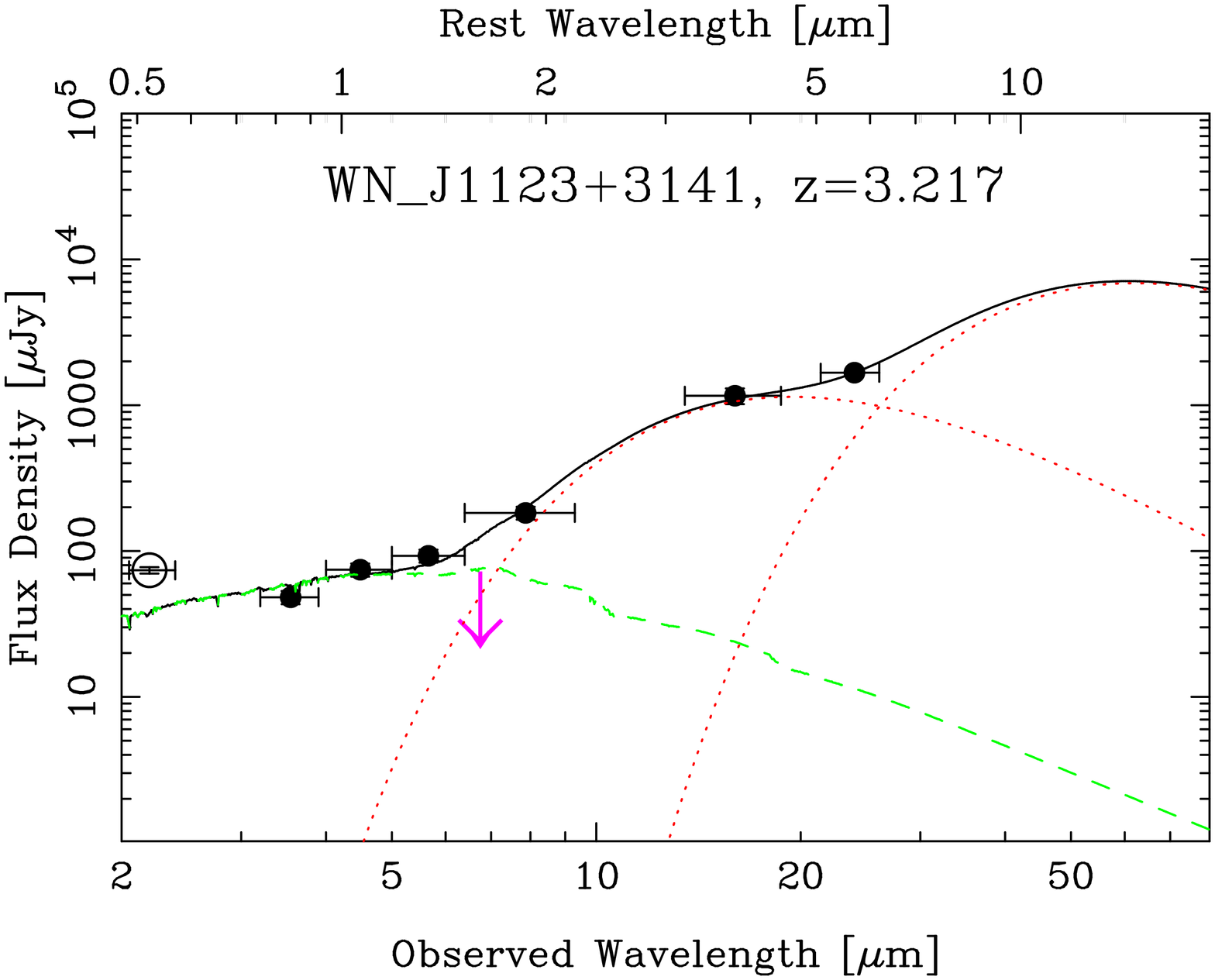} \\
\end{tabular}
\end{figure*}
\vfill\eject

\begin{figure*}
\begin{tabular}{r@{}c@{}l}
\includegraphics[angle=-90,width=160pt,trim= 0 -7 70 0,clip=true]{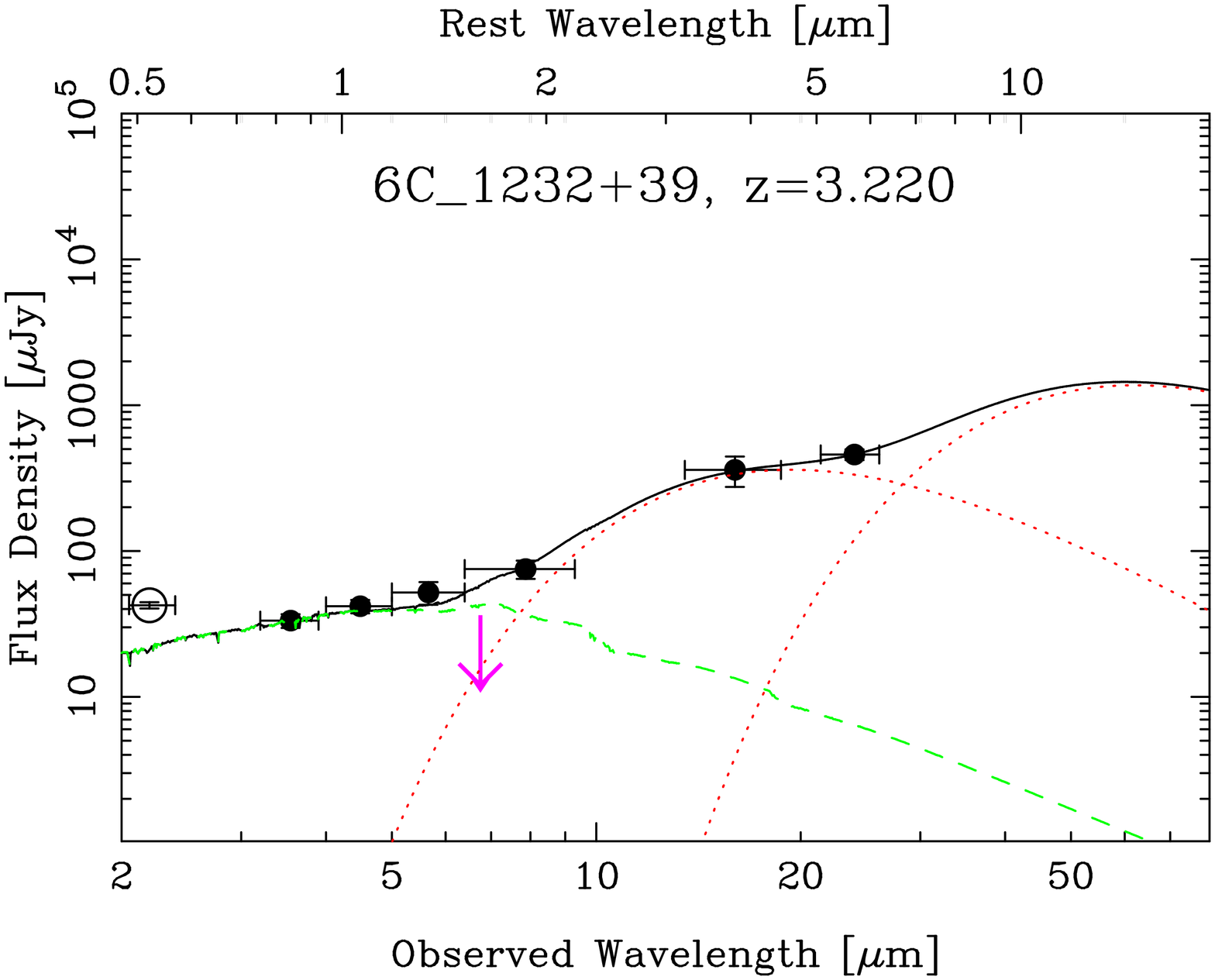} &
\includegraphics[angle=-90,width=144pt,trim= 0 59 70 0,clip=true]{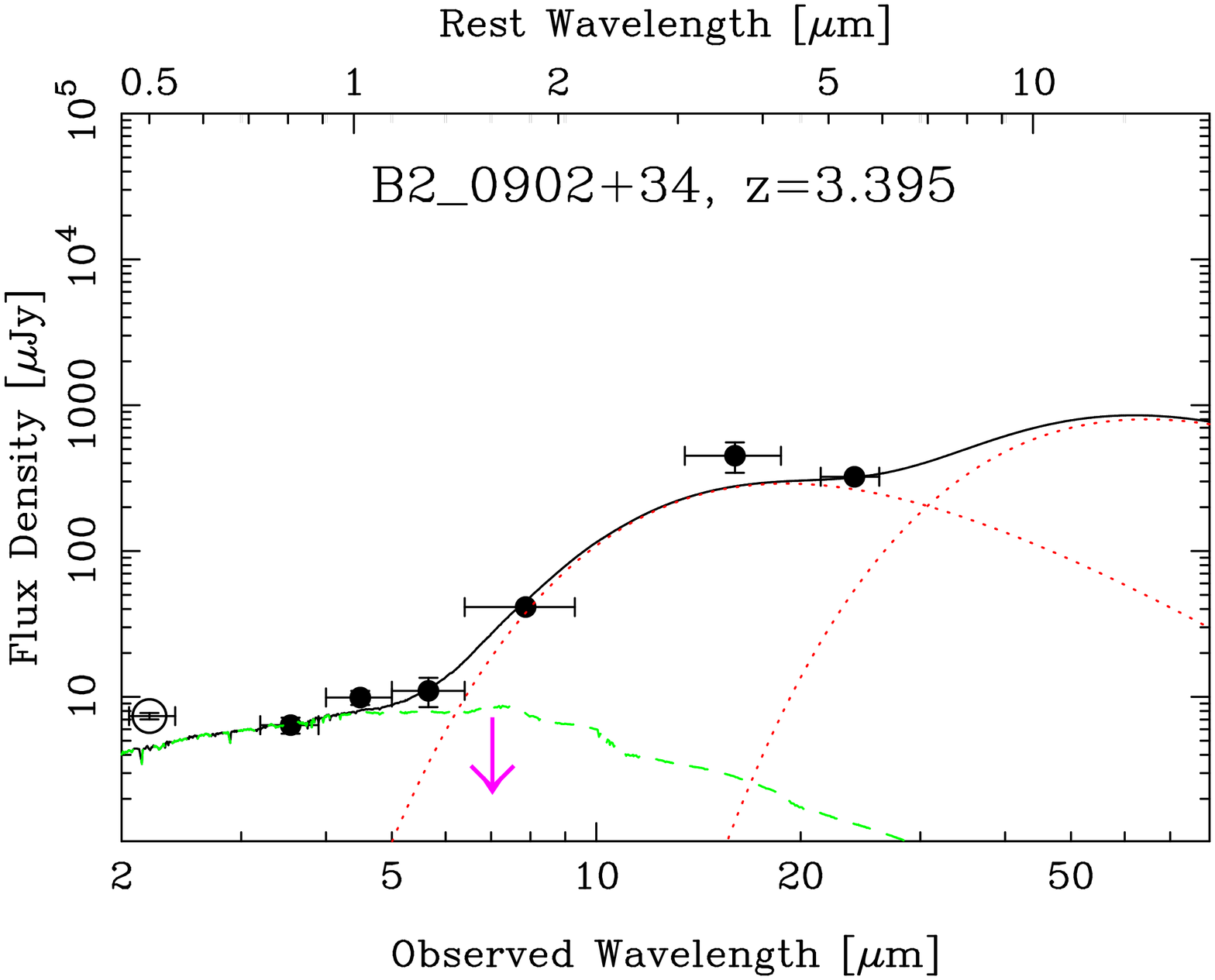} &
\includegraphics[angle=-90,width=144pt,trim= 0 59 70 0,clip=true]{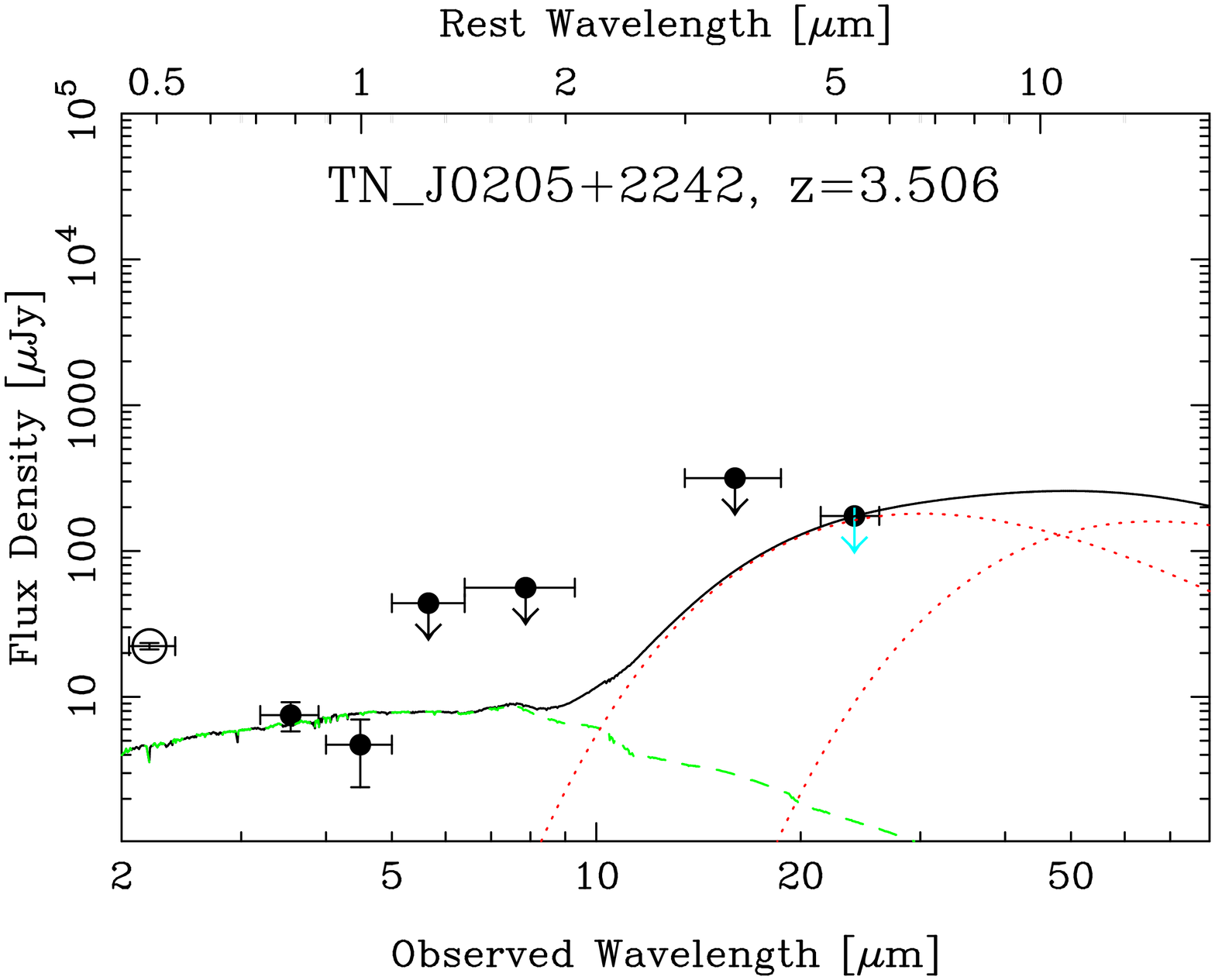} \\[-5pt]
\includegraphics[angle=-90,width=160pt,trim=28 -7 70 0,clip=true]{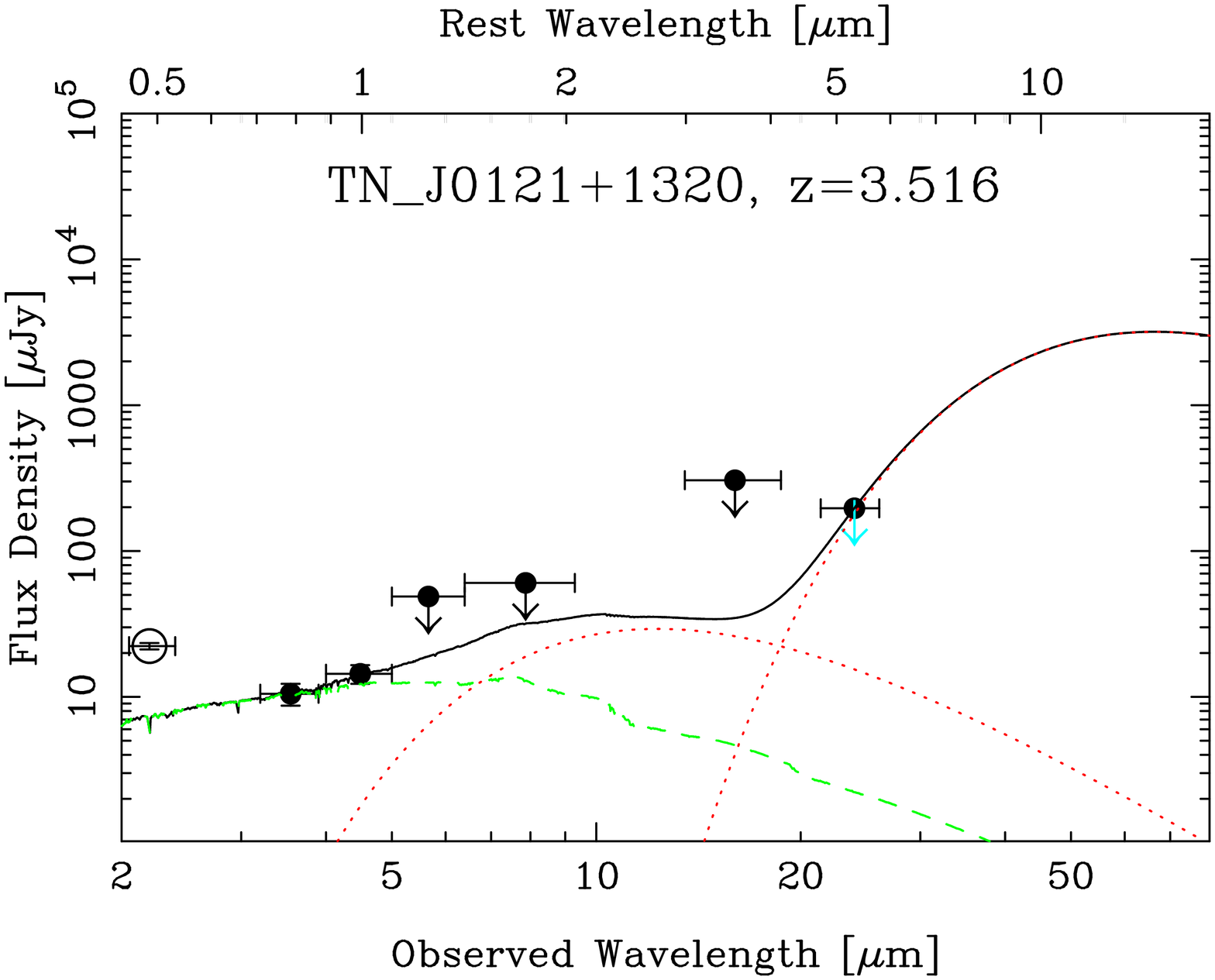} &
\includegraphics[angle=-90,width=144pt,trim=28 59 70 0,clip=true]{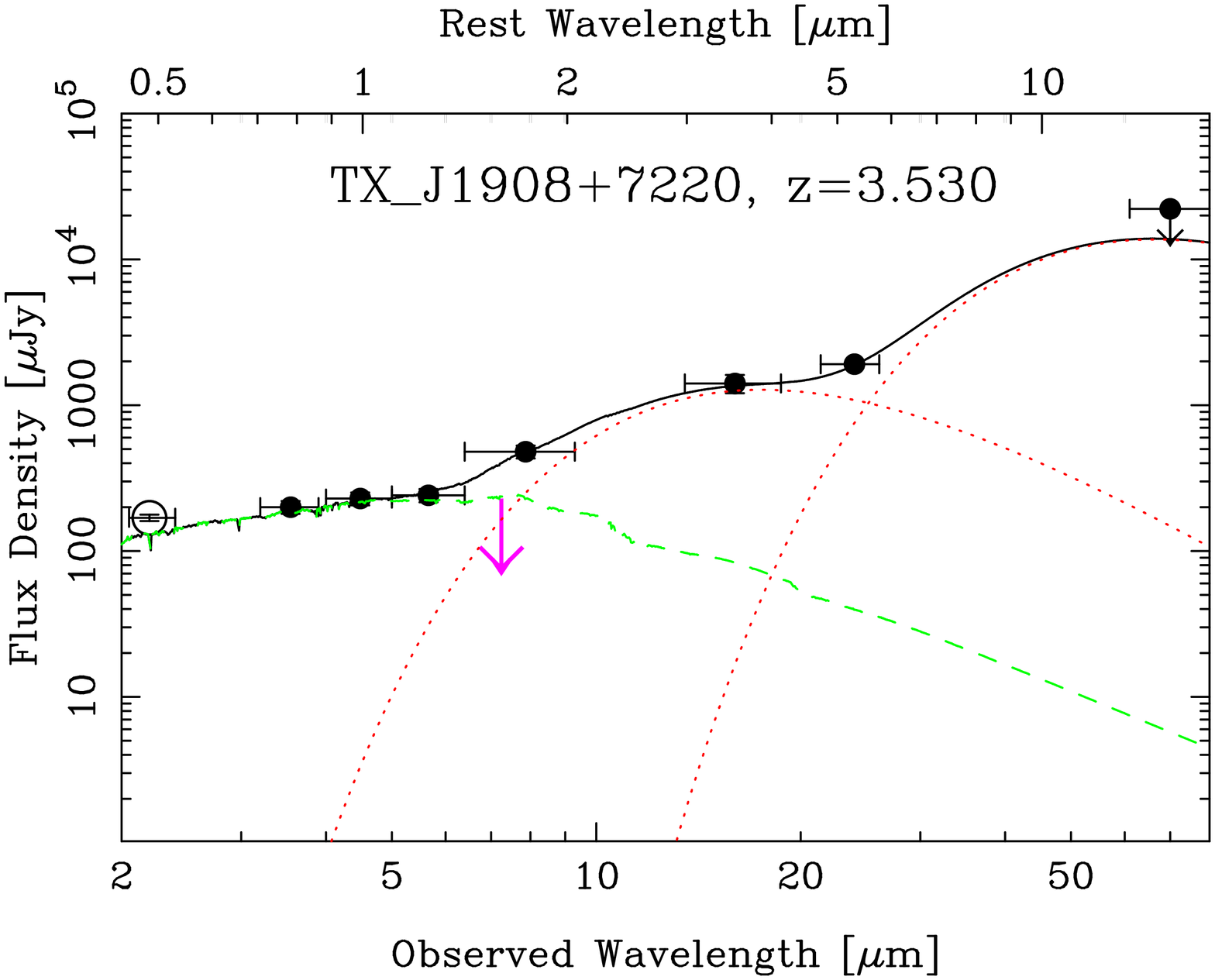} &
\includegraphics[angle=-90,width=144pt,trim=28 59 70 0,clip=true]{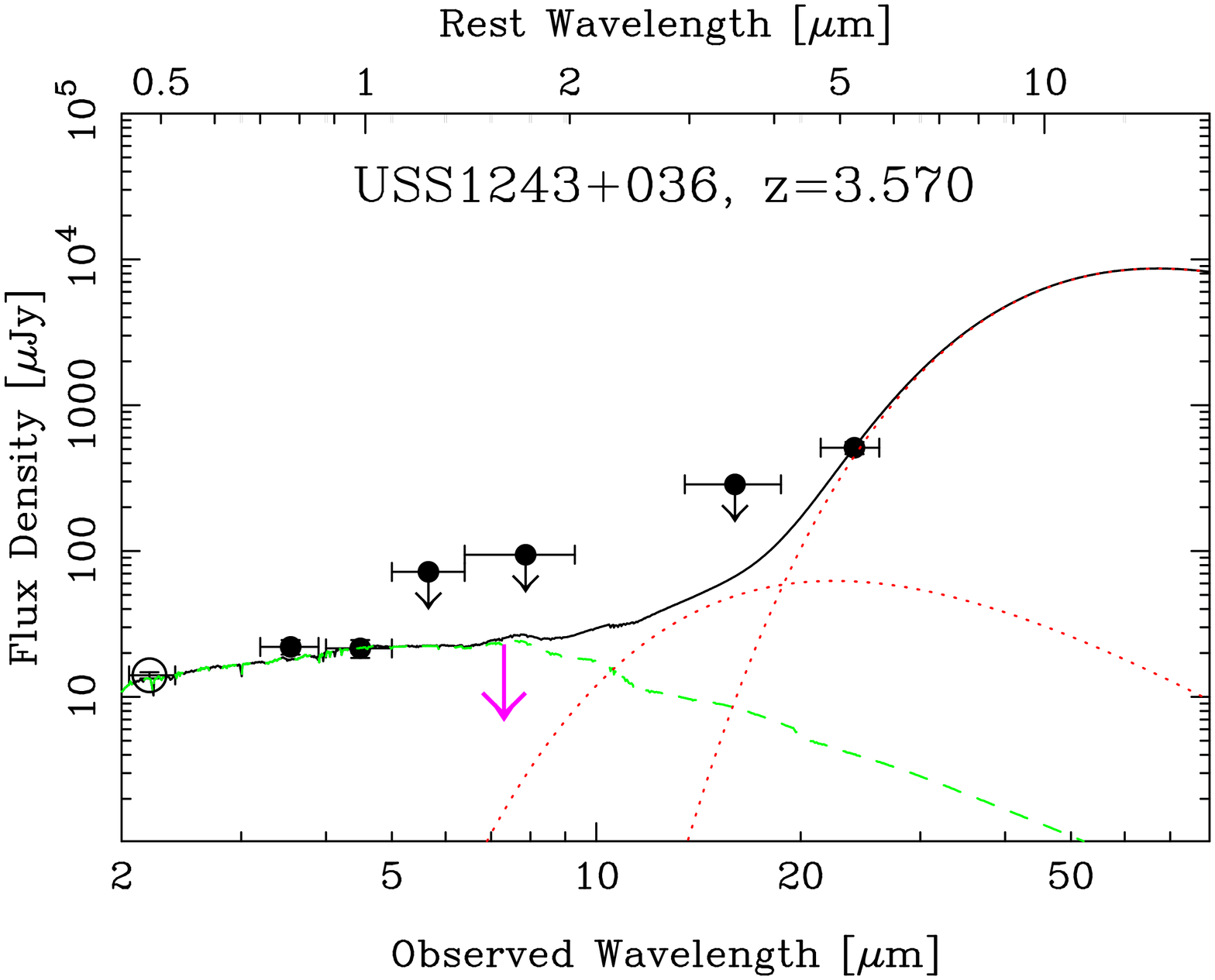} \\[-5pt]
\includegraphics[angle=-90,width=160pt,trim=28 -7 70 0,clip=true]{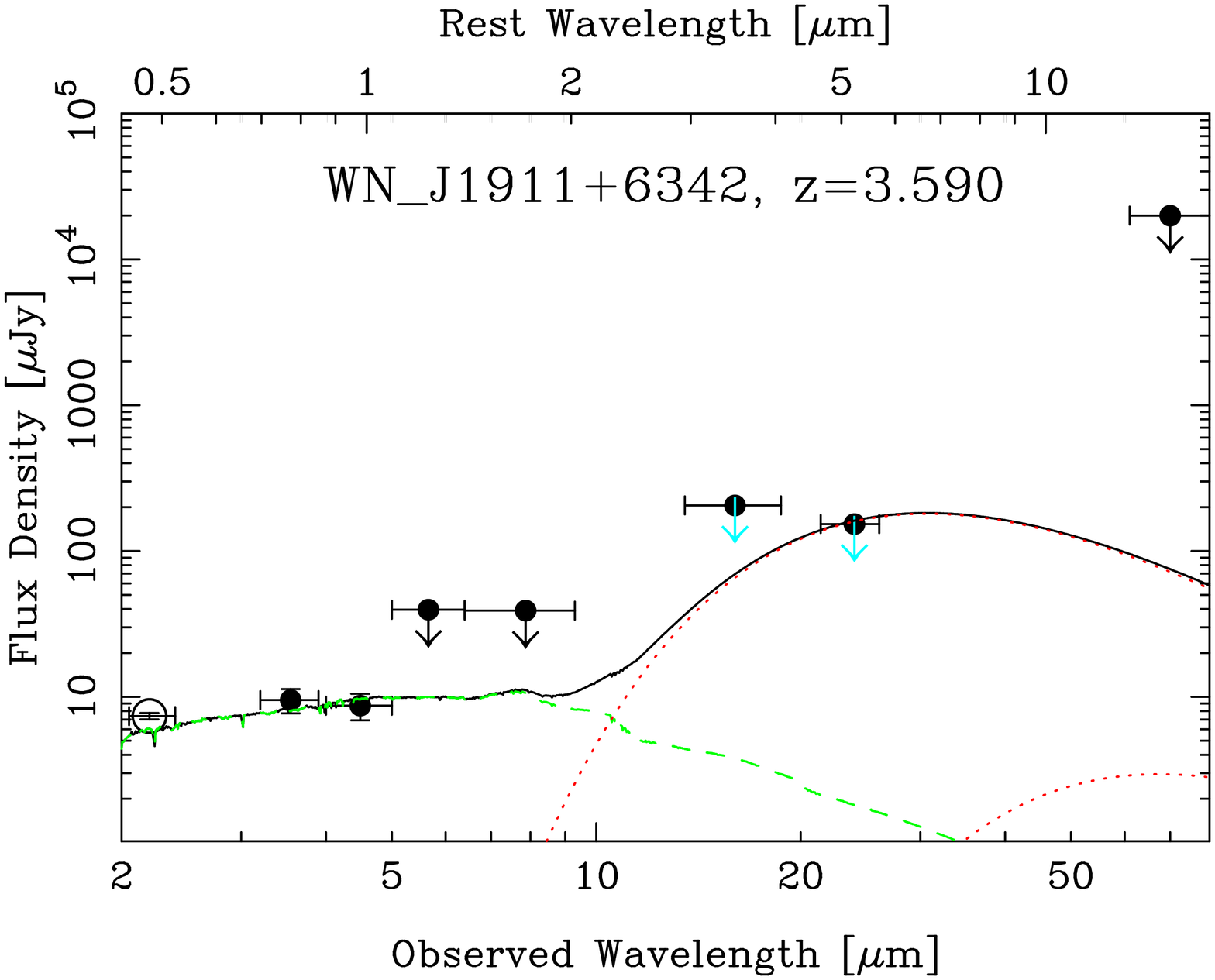} &
\includegraphics[angle=-90,width=144pt,trim=28 59 70 0,clip=true]{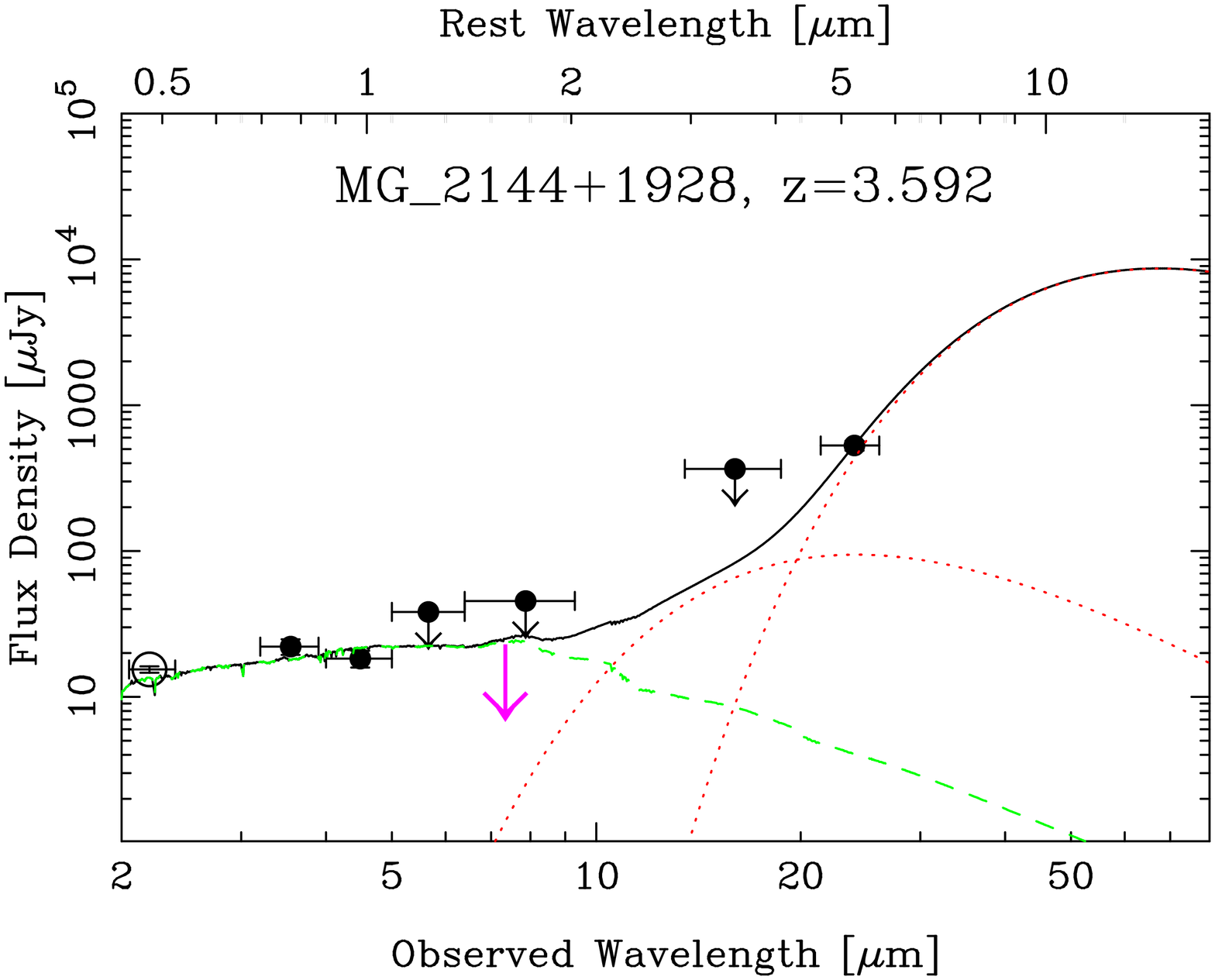} &
\includegraphics[angle=-90,width=144pt,trim=28 59 70 0,clip=true]{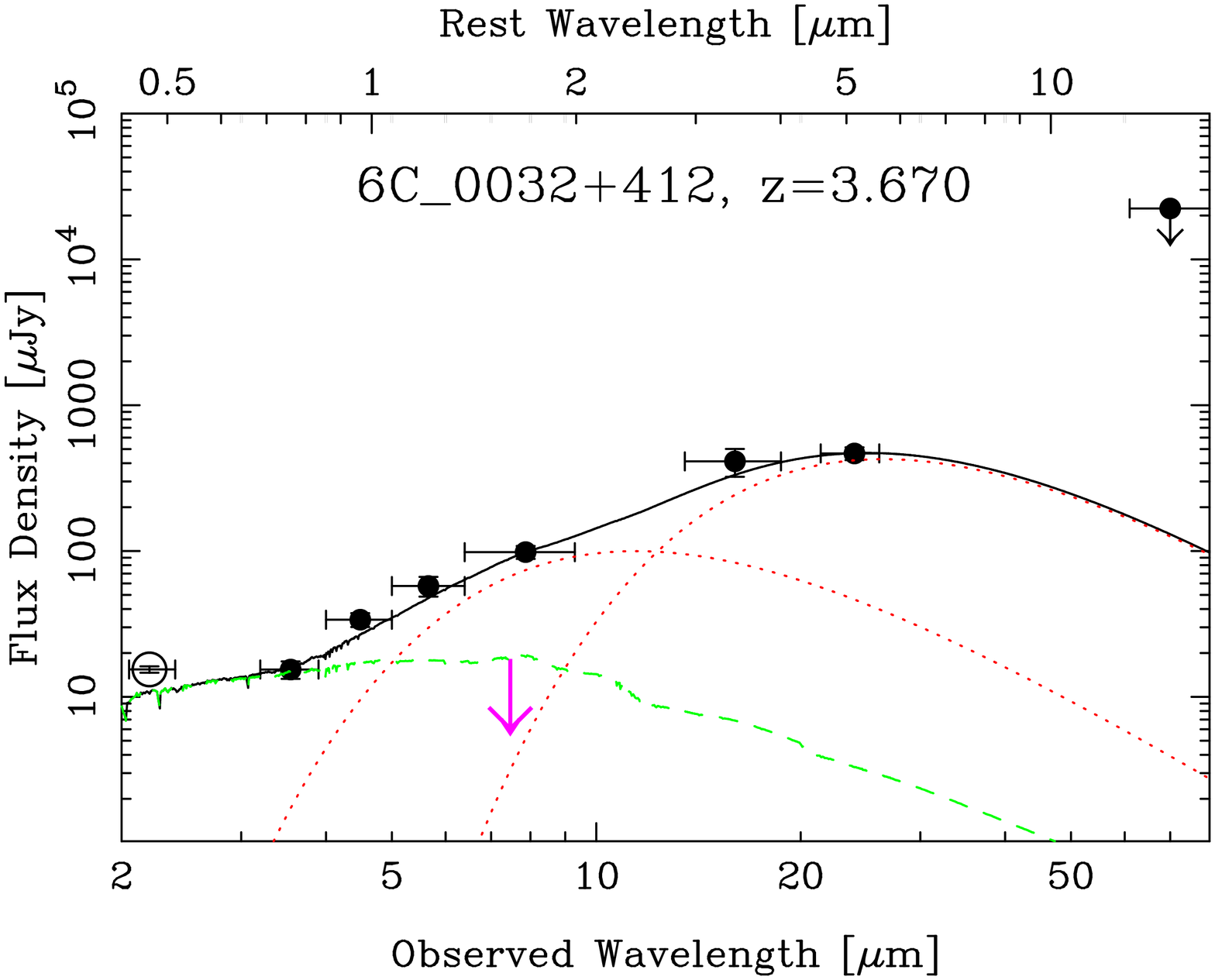} \\[-5pt]
\includegraphics[angle=-90,width=160pt,trim=28 -7 70 0,clip=true]{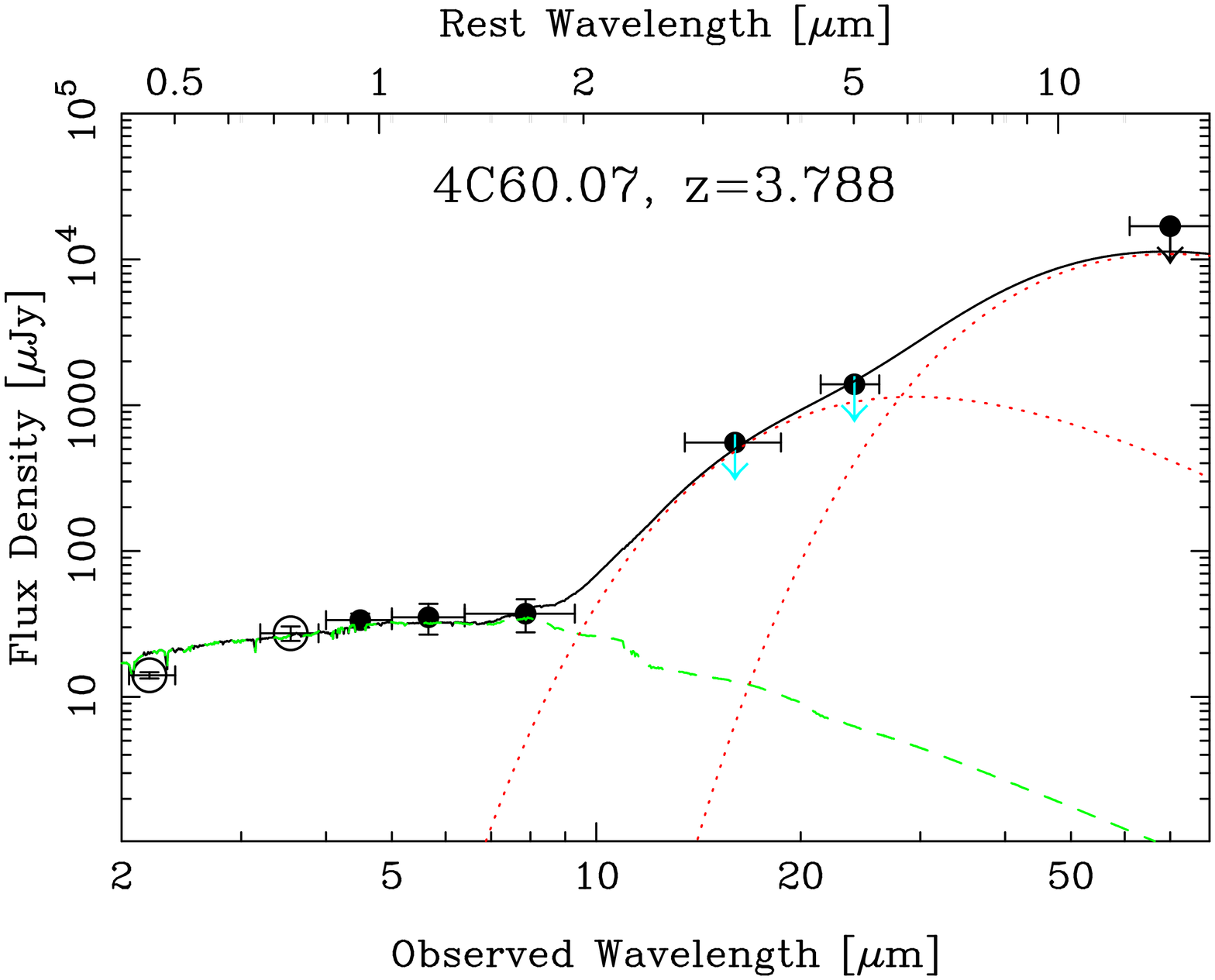} &
\includegraphics[angle=-90,width=144pt,trim=28 59 70 0,clip=true]{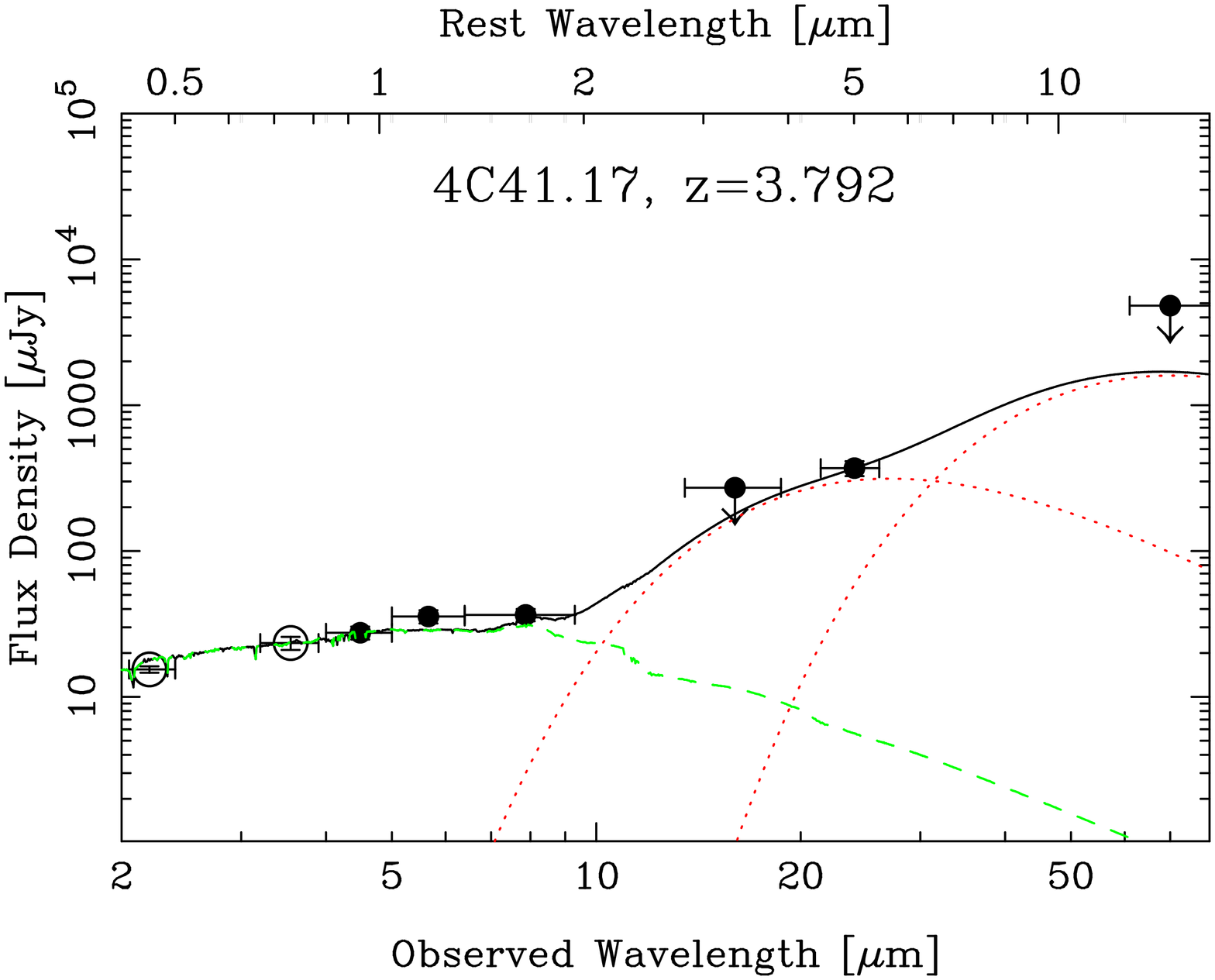} &
\includegraphics[angle=-90,width=144pt,trim=28 59 70 0,clip=true]{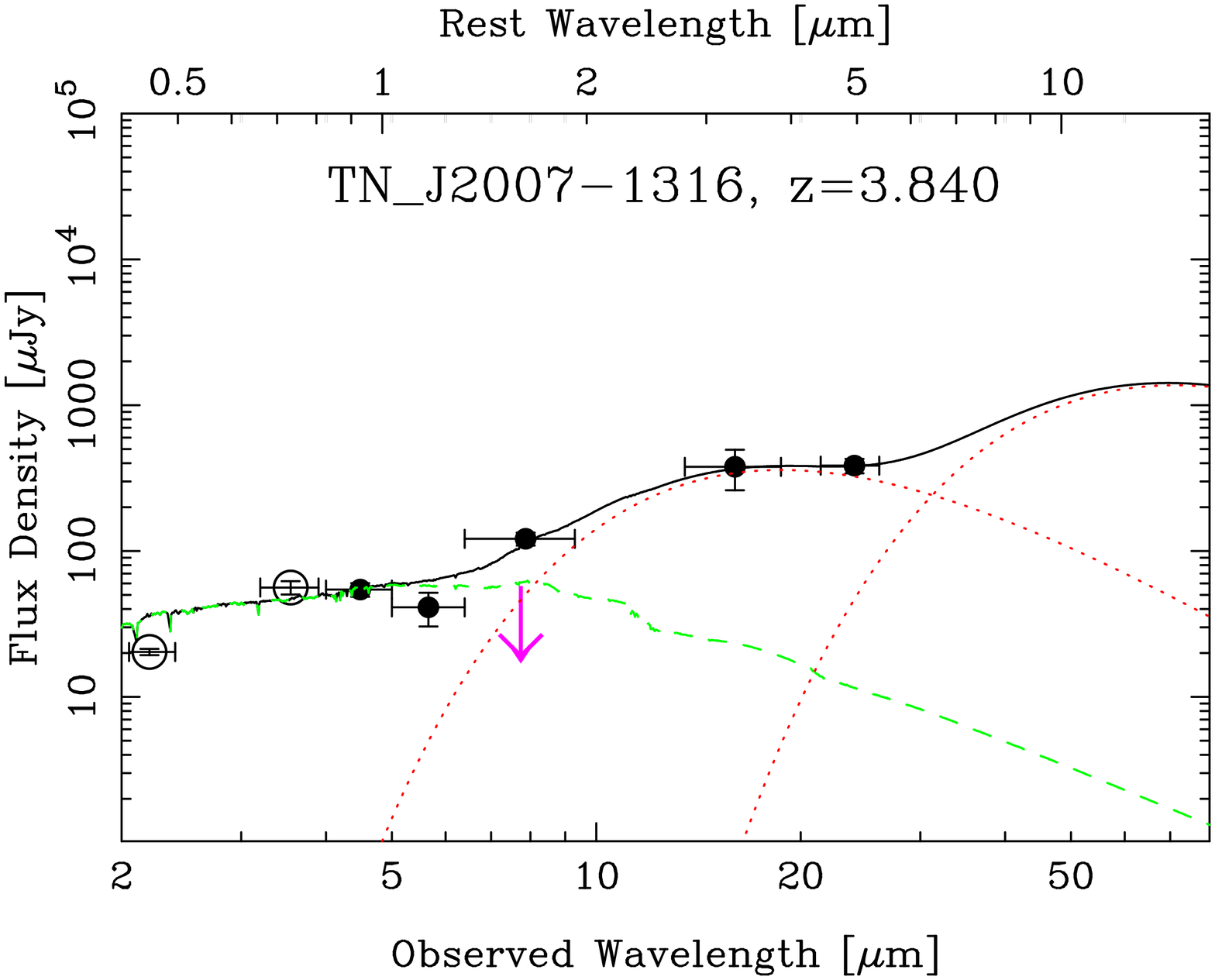} \\[-5pt]
\includegraphics[angle=-90,width=160pt,trim=28 -7 70 0,clip=true]{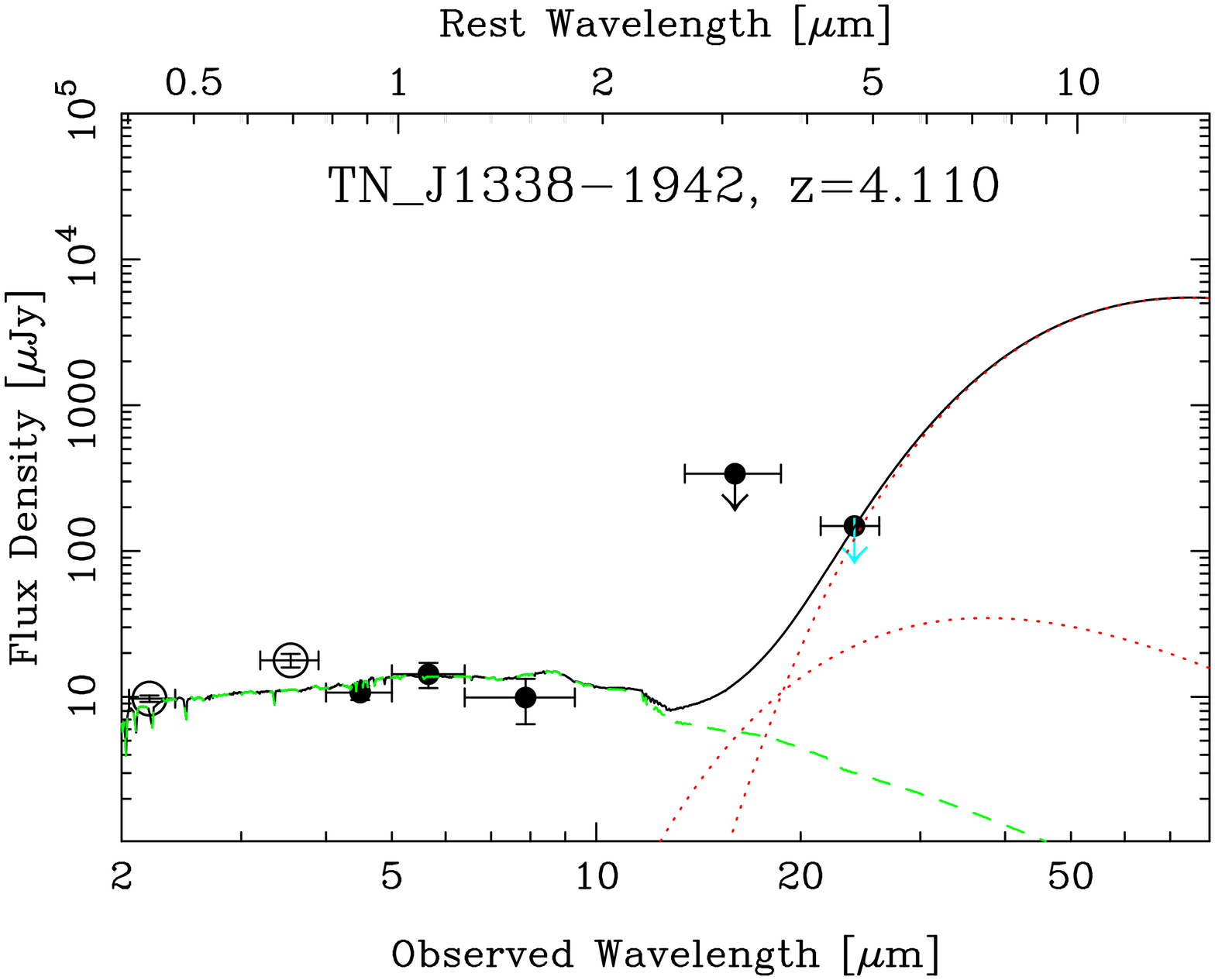} &
\includegraphics[angle=-90,width=144pt,trim=28 59 70 0,clip=true]{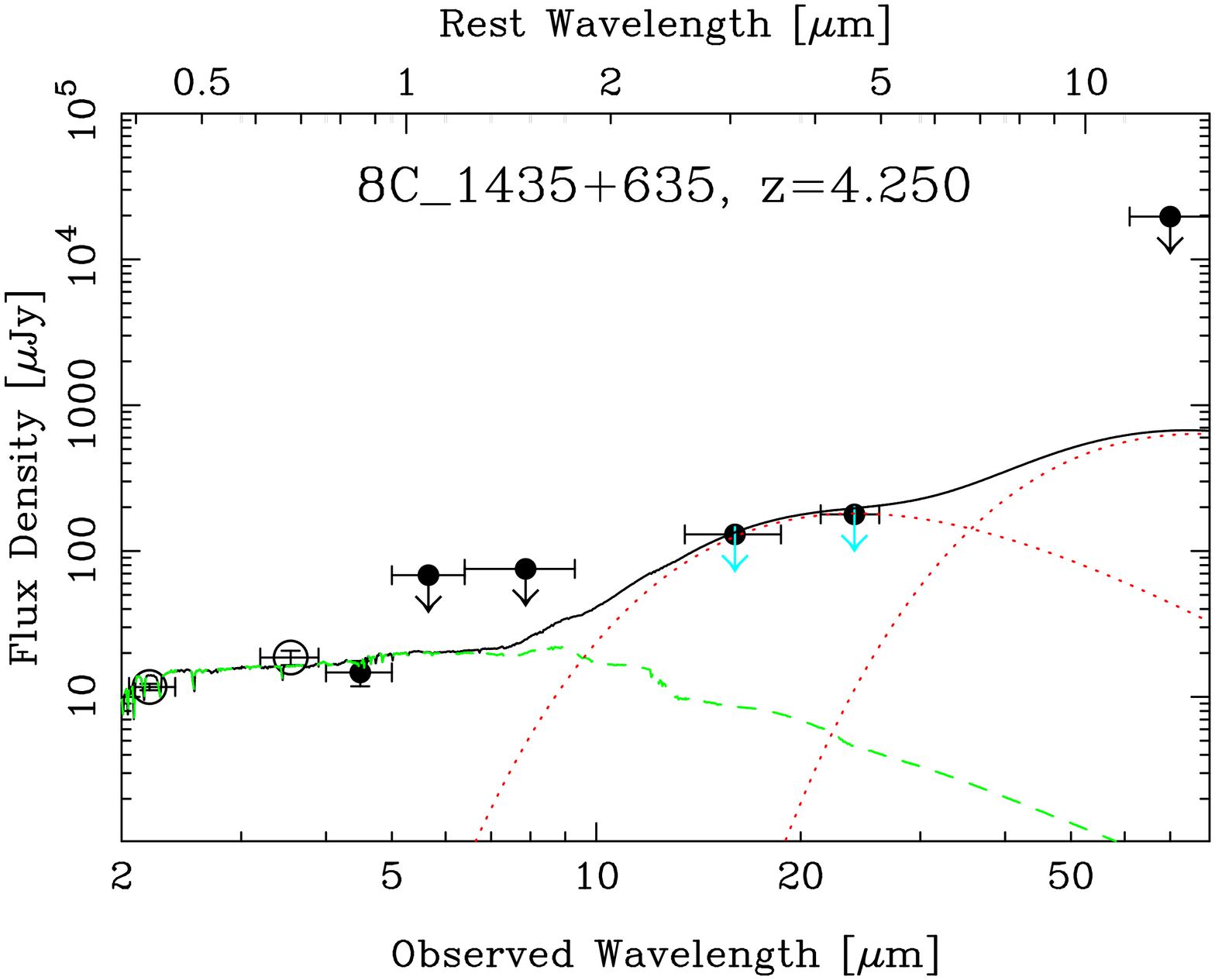} &
\includegraphics[angle=-90,width=144pt,trim=28 59 70 0,clip=true]{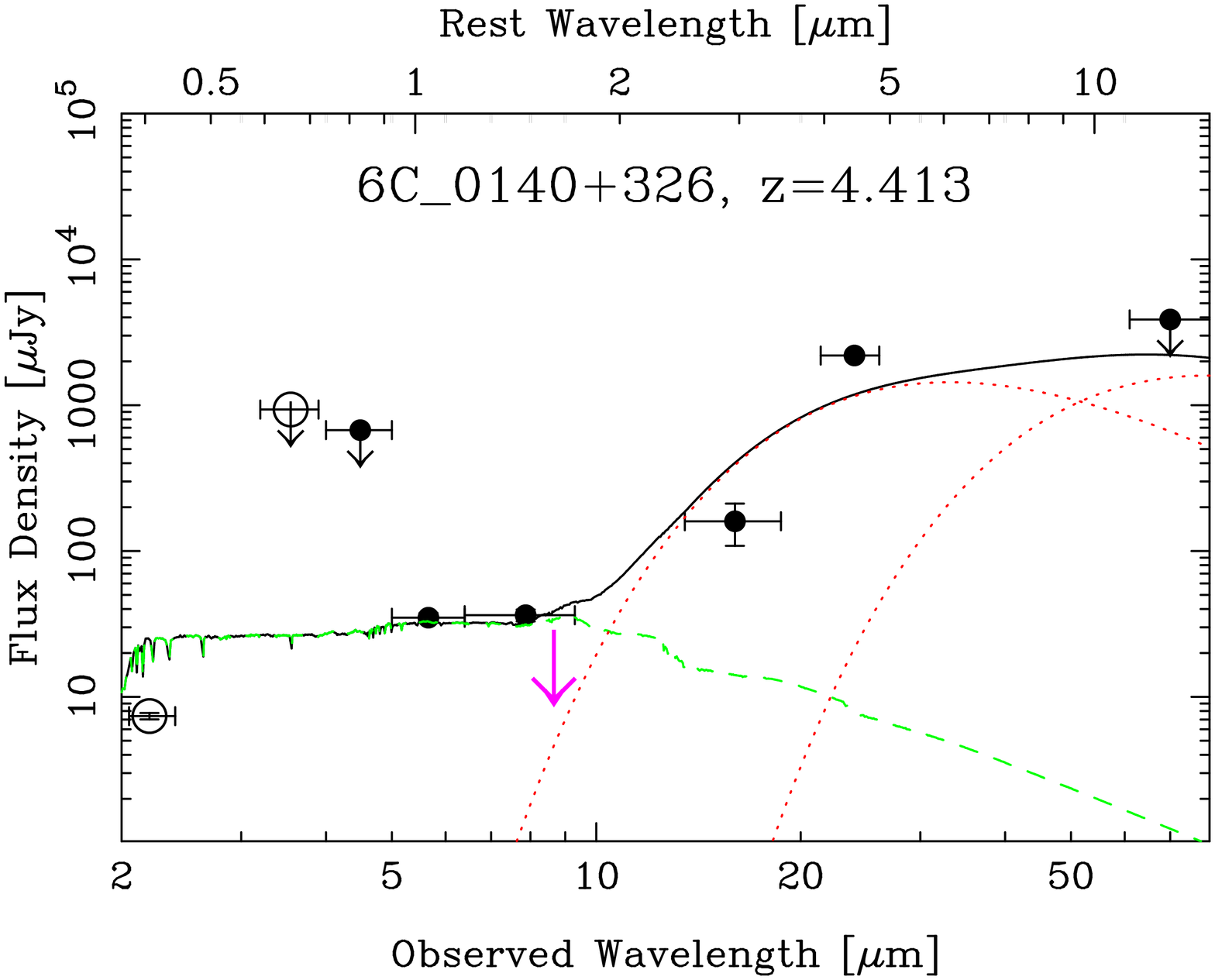} \\[-5pt]
\includegraphics[angle=-90,width=160pt,trim=28 -7  0 0,clip=true]{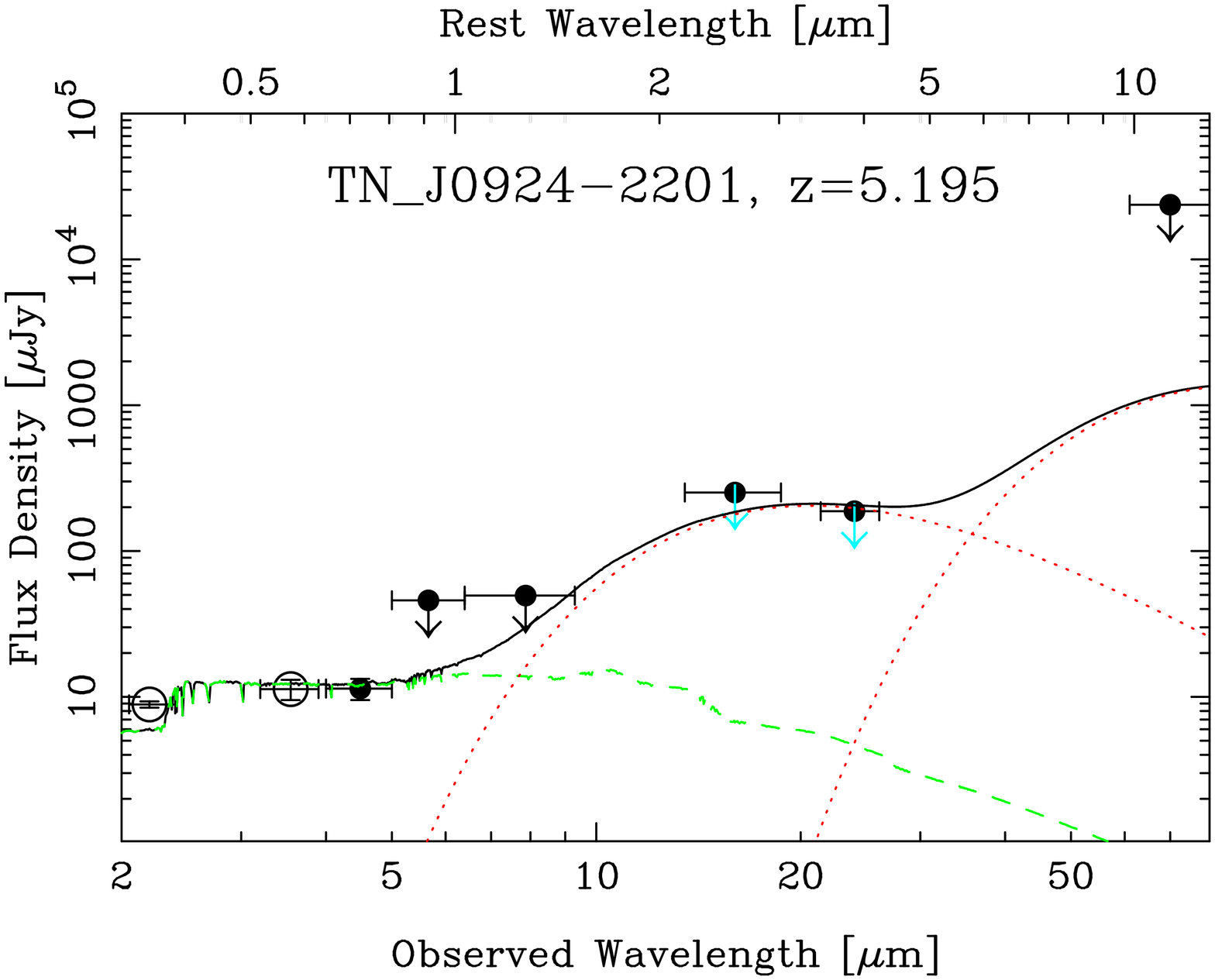} &&\\
\end{tabular}
\end{figure*}
\vfill\eject

\section{New radio maps of sources in our sample}

\begin{figure*}[ht]
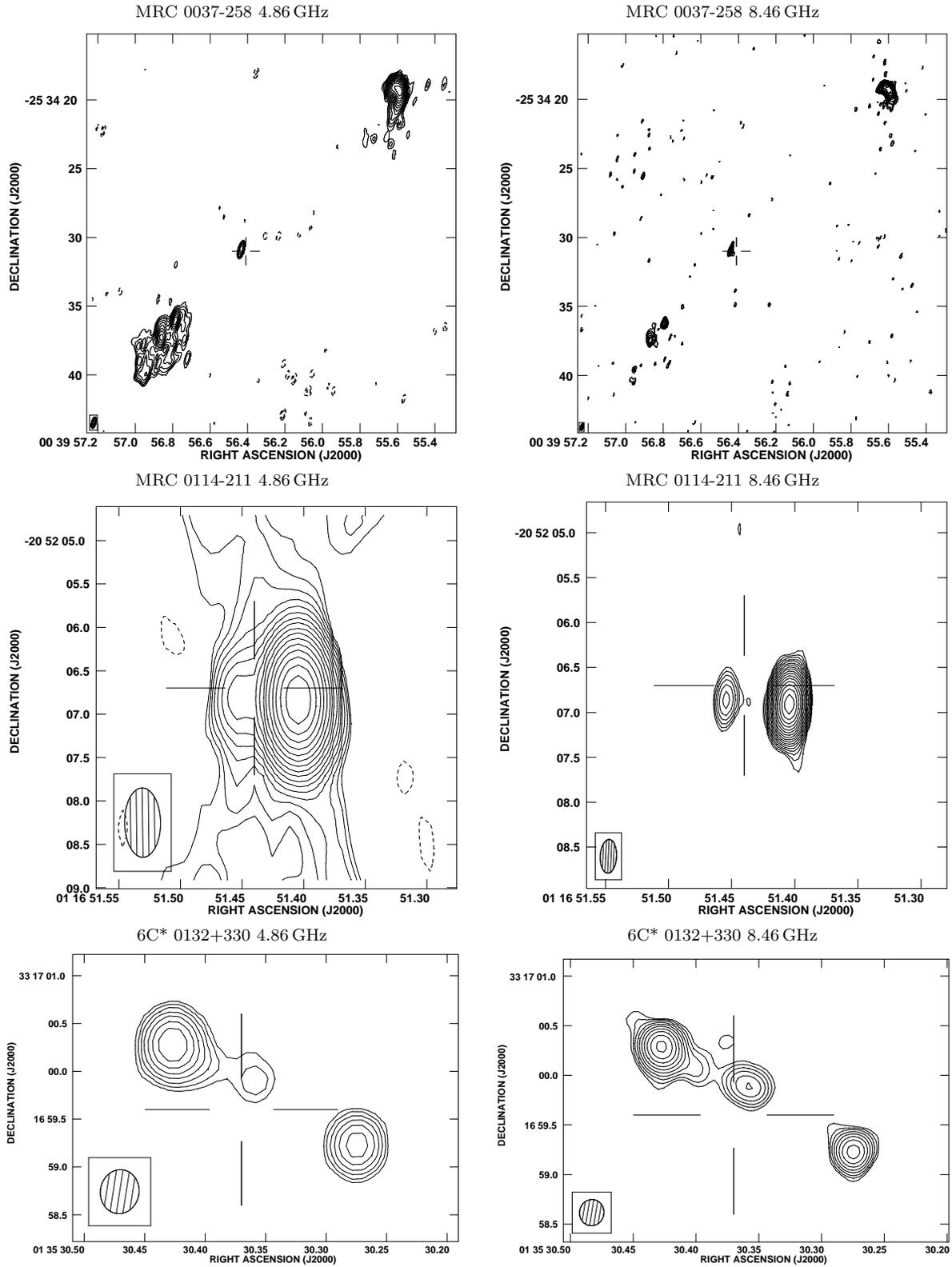

\begin{tabular}{cc}
MRC~0037-258 4.86\,GHz & MRC~0037-258 8.46\,GHz\\
\psfig{file=./MRC0037-258.C-BAND.PS,width=7.8cm,angle=-90} & \psfig{file=MRC0037-258.X-BAND.PS,width=7.8cm,angle=-90} \\
MRC~0114-211 4.86\,GHz & MRC~0114-211 8.46\,GHz\\
\psfig{file=./MRC0114-211.C-BAND.PS,width=7.8cm,angle=-90} & \psfig{file=MRC0114-211.X-BAND.PS,width=7.8cm,angle=-90} \\
6C*~0132+330 4.86\,GHz & 6C*~0132+330 8.46\,GHz\\
\psfig{file=./6CS0132+330.C-BAND.PS,width=7.8cm,angle=-90} & \psfig{file=6CS0132+330.X-BAND.PS,width=7.8cm,angle=-90} \\
\end{tabular}
\caption{VLA 4.86 and 8.46\,GHz images of 14 HzRGs in our sample. The contour scheme is a geometric progression of $\sqrt 2$, with the first contour starting at 3$\sigma$. Open crosses mark the near-IR identification position listed in Table~\ref{table.shzrg}.
\label{radiomaps}
}
\end{figure*}

Figure~\ref{radiomaps} presents the new VLA A-array radio maps.

\begin{figure*}[ht]
\begin{tabular}{cc}
MRC~0251-273 4.86\,GHz & MRC~0251-273 8.46\,GHz\\
\psfig{file=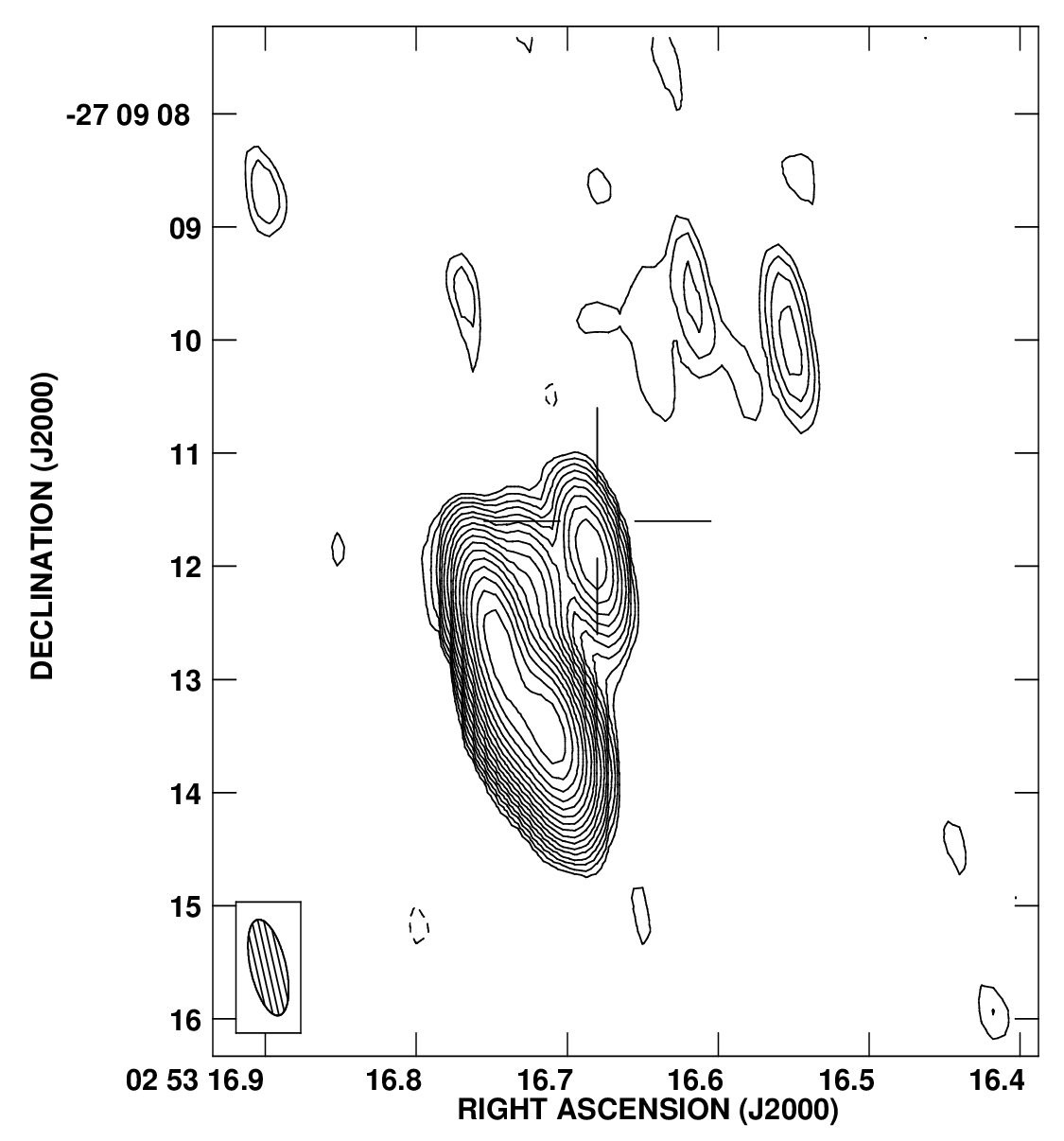,width=7.5cm} & \psfig{file=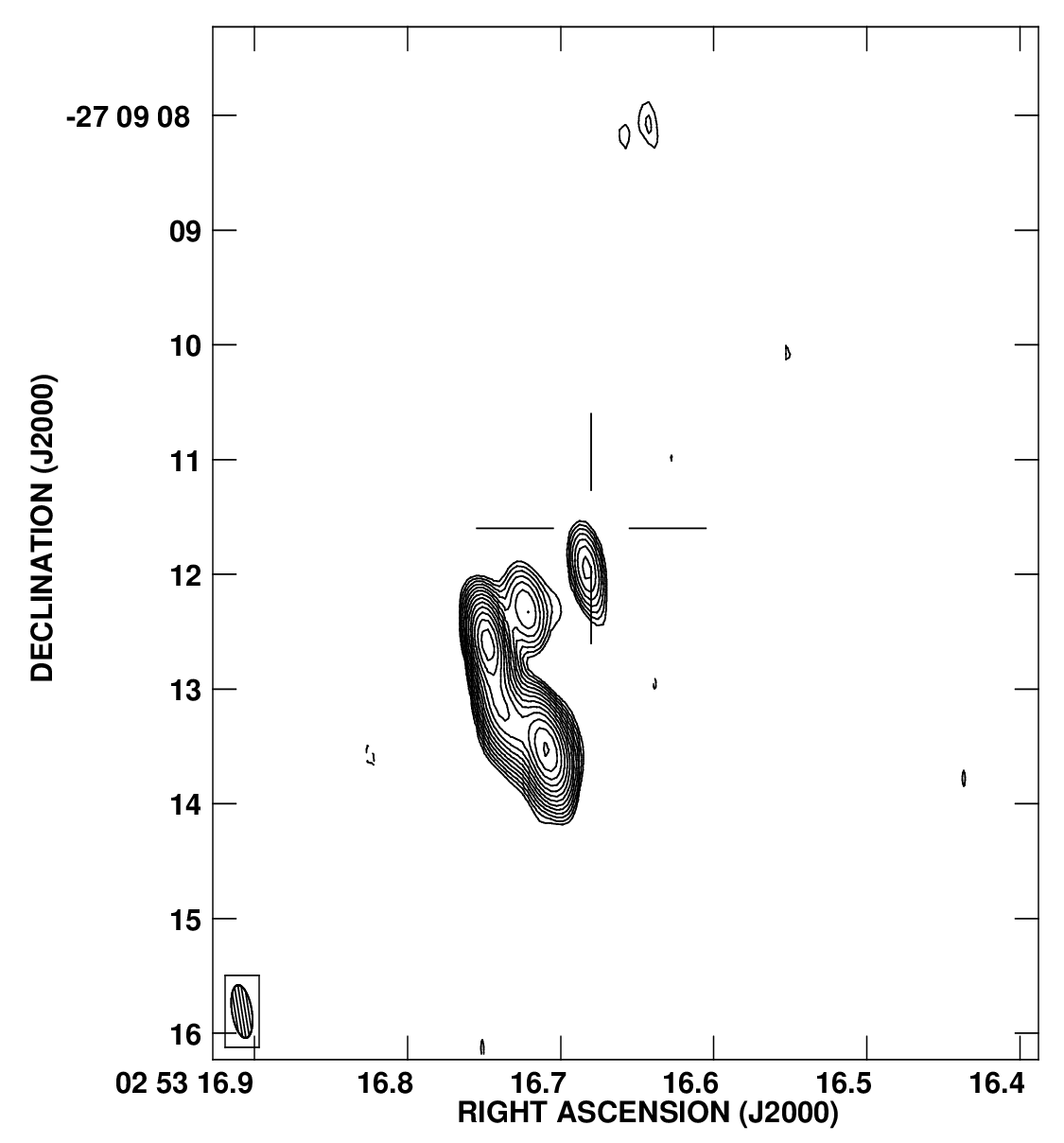,width=7.5cm} \\
MRC~0324-228 4.86\,GHz & MRC~0324-228 8.46\,GHz\\
\psfig{file=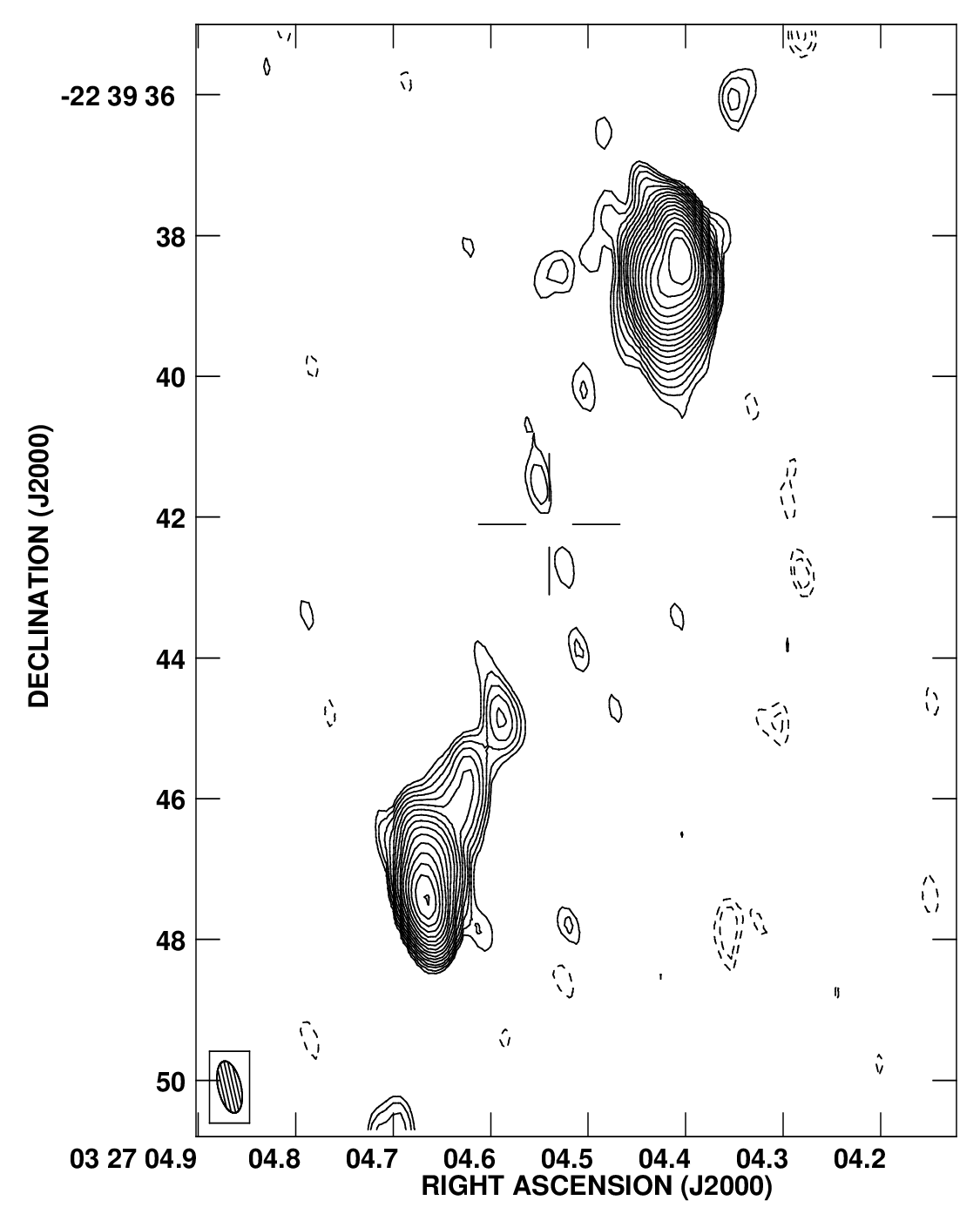,width=7.5cm} & \psfig{file=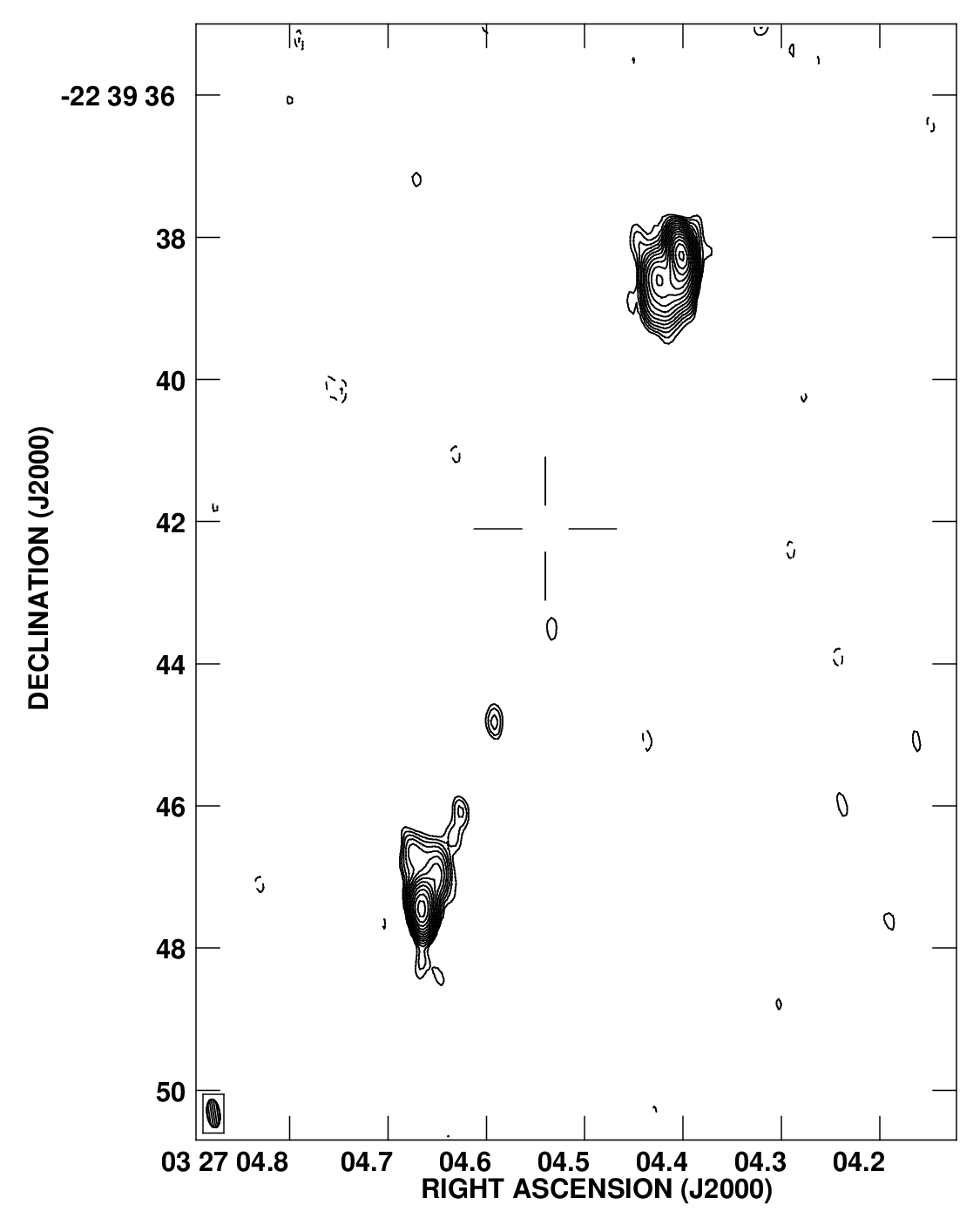,width=7.5cm} \\
MRC~0350-279 4.86\,GHz & MRC~0350-279 8.46\,GHz\\
\psfig{file=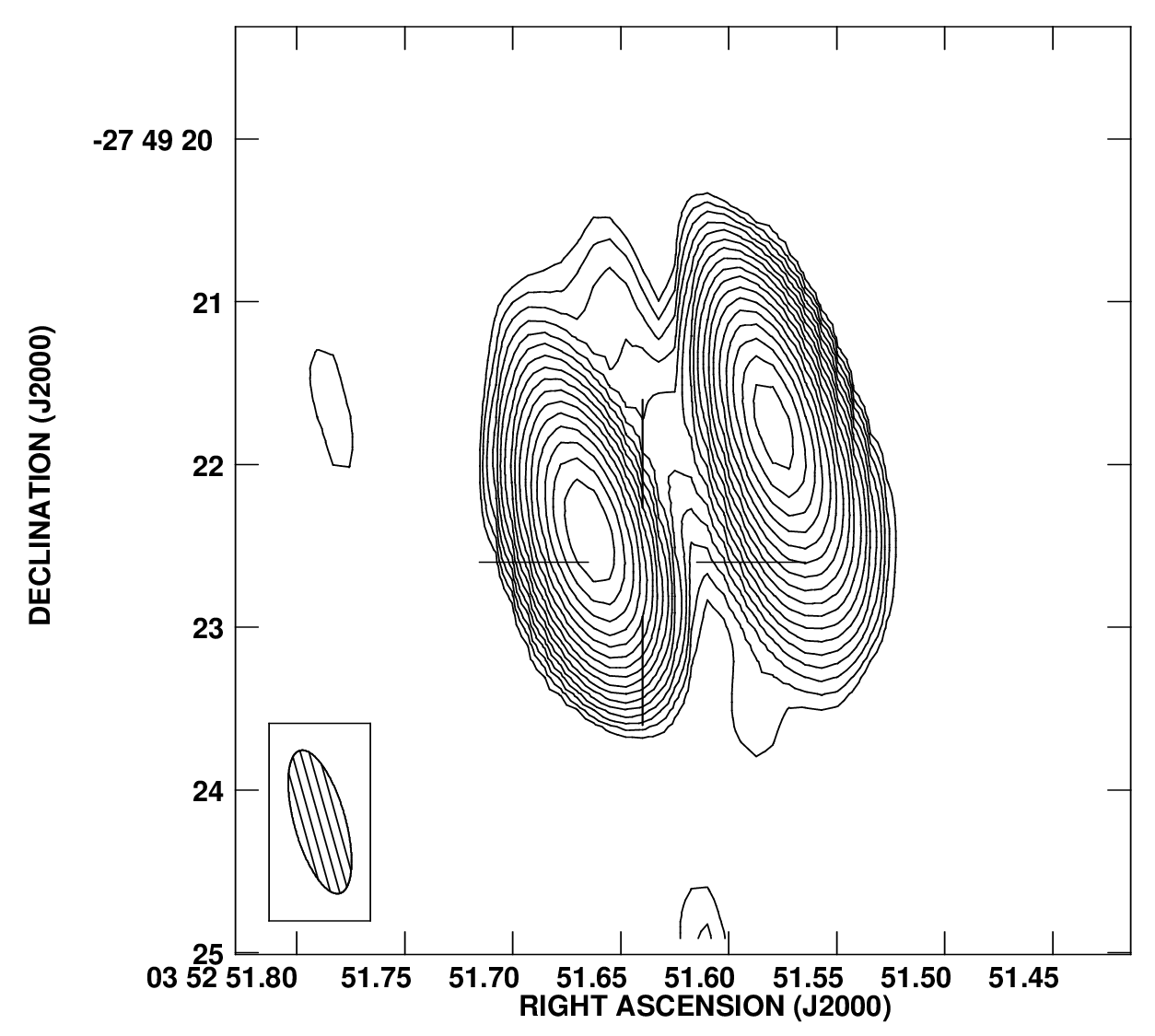,width=7.5cm,angle=-90} & \psfig{file=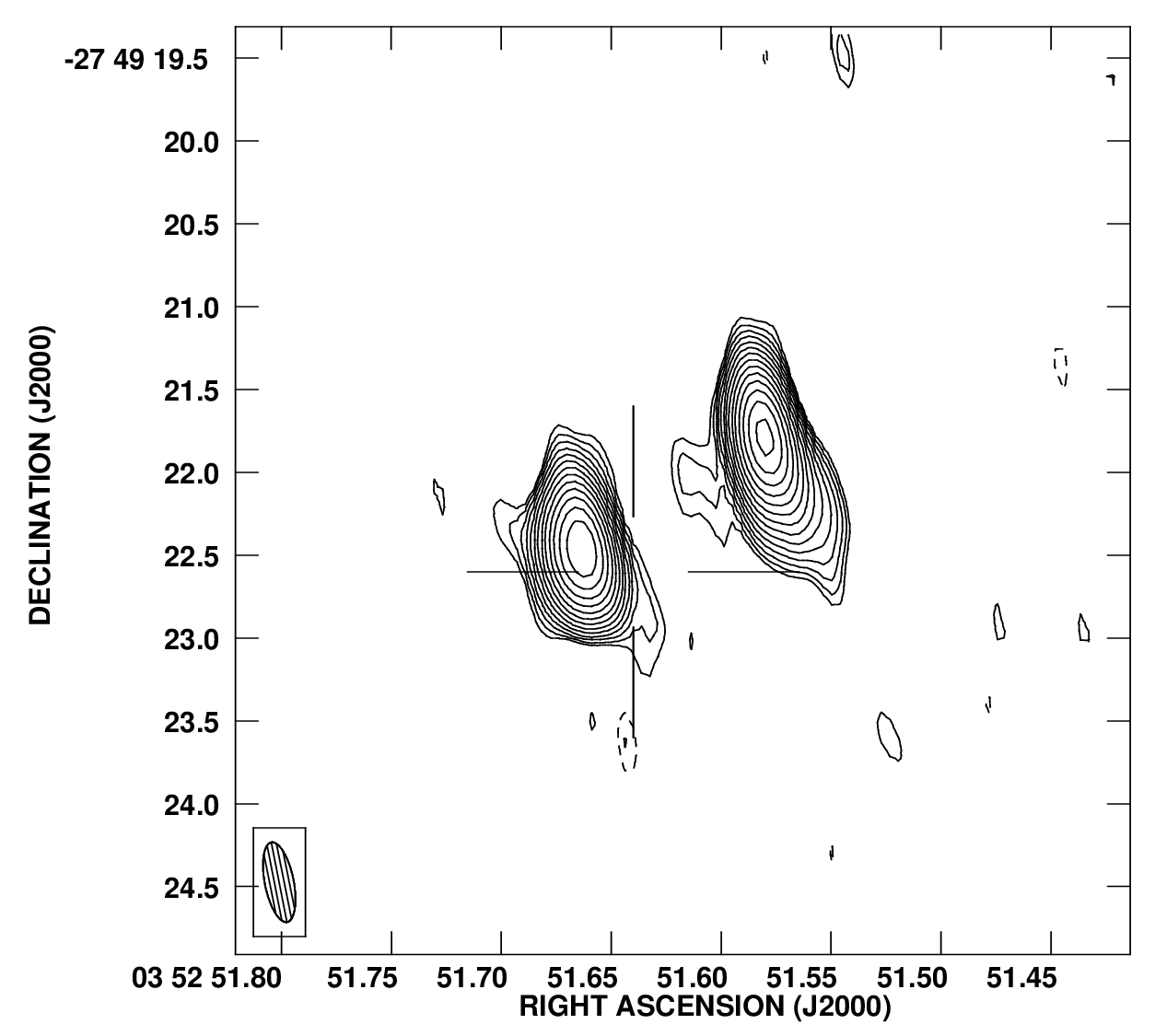,width=7.5cm,angle=-90} \\
\end{tabular}
\end{figure*}

\begin{figure*}[ht]
\begin{tabular}{cc}
5C~7.269 4.86\,GHz & 5C~7.269 8.46\,GHz\\
\psfig{file=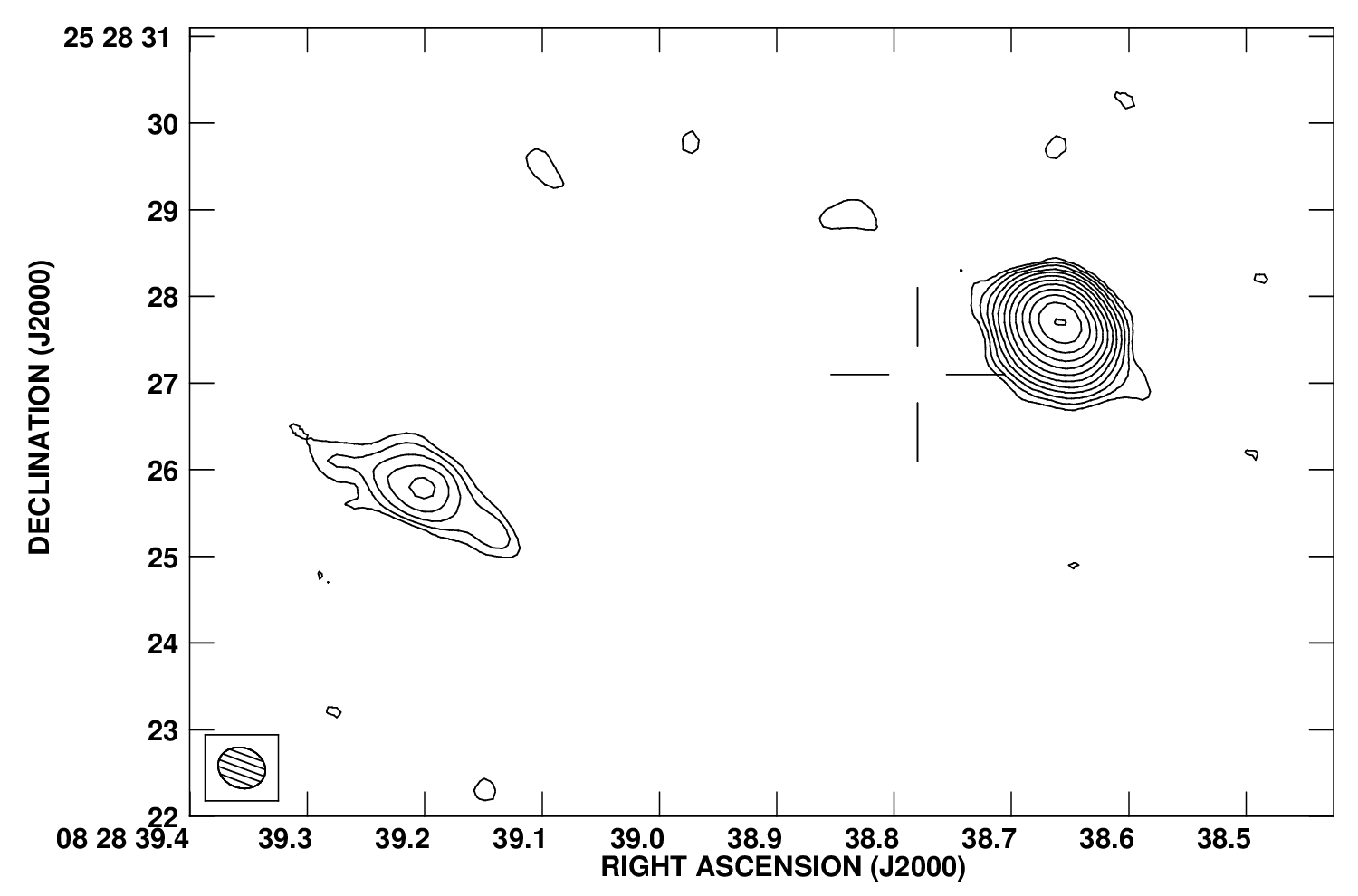,width=8.2cm,angle=-90} & \psfig{file=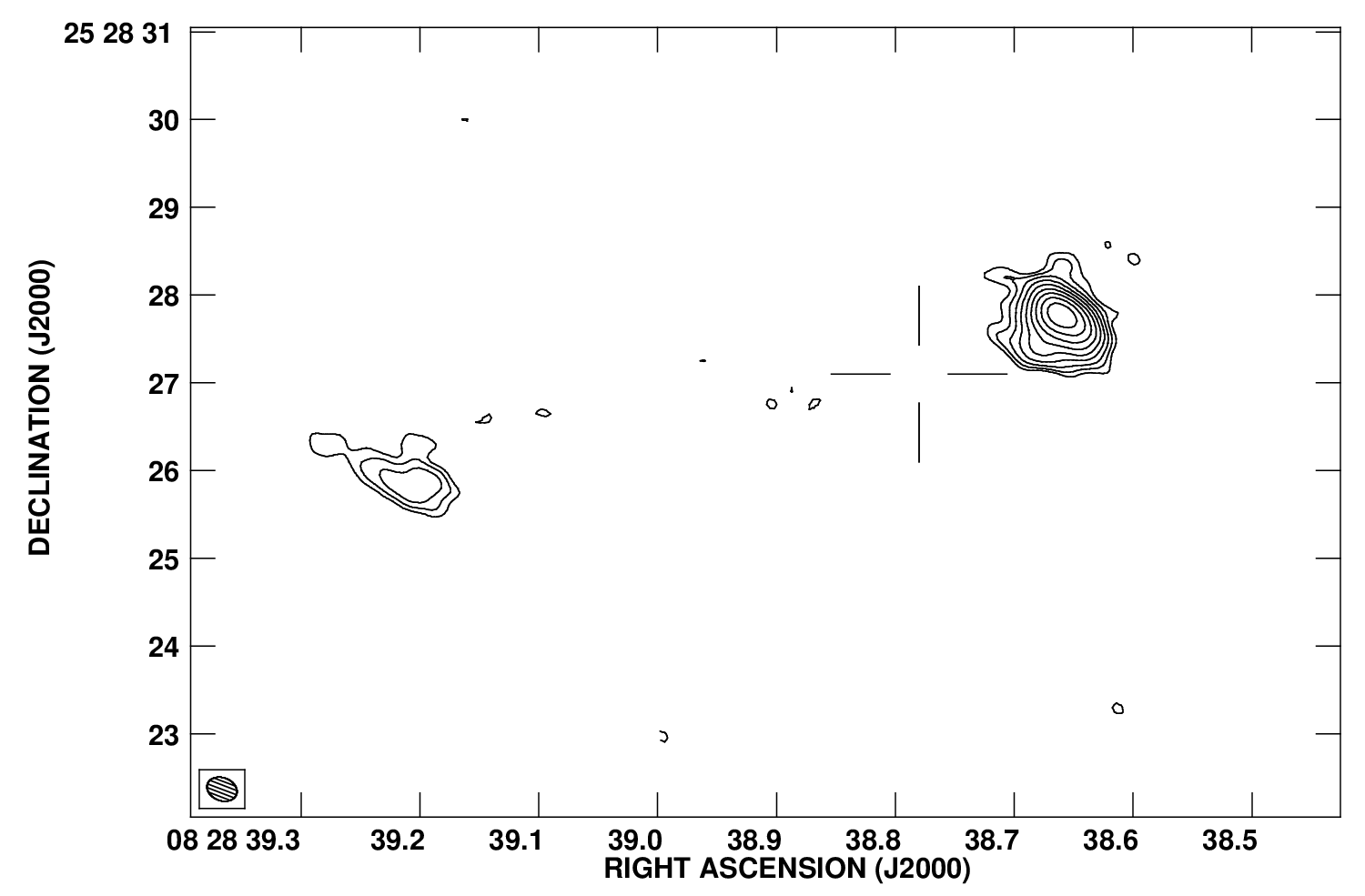,width=8.2cm,angle=-90} \\
6C~0820+3642 4.86\,GHz & 6C~0820+3642 8.46\,GHz\\
\psfig{file=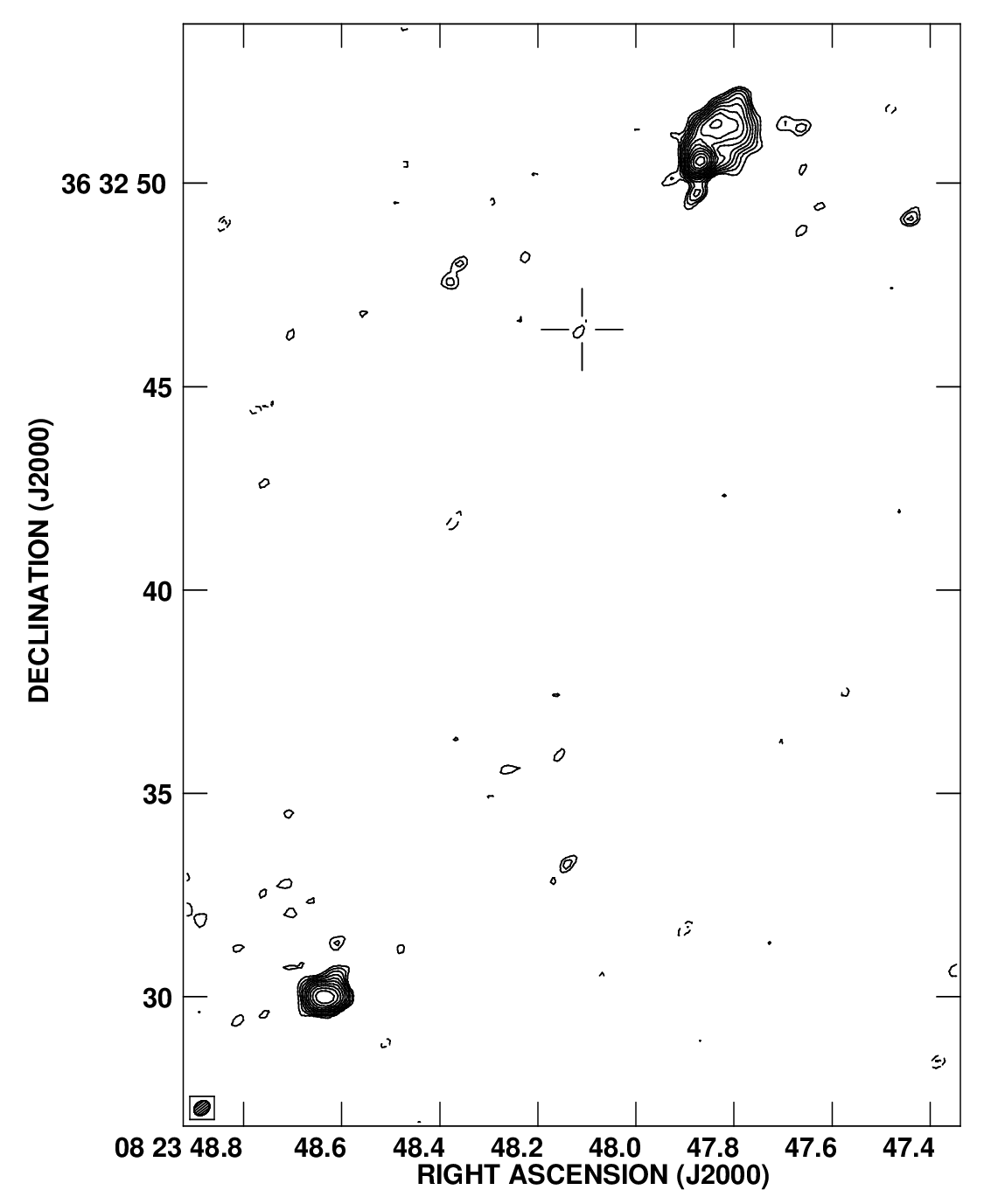,width=8.2cm} & \psfig{file=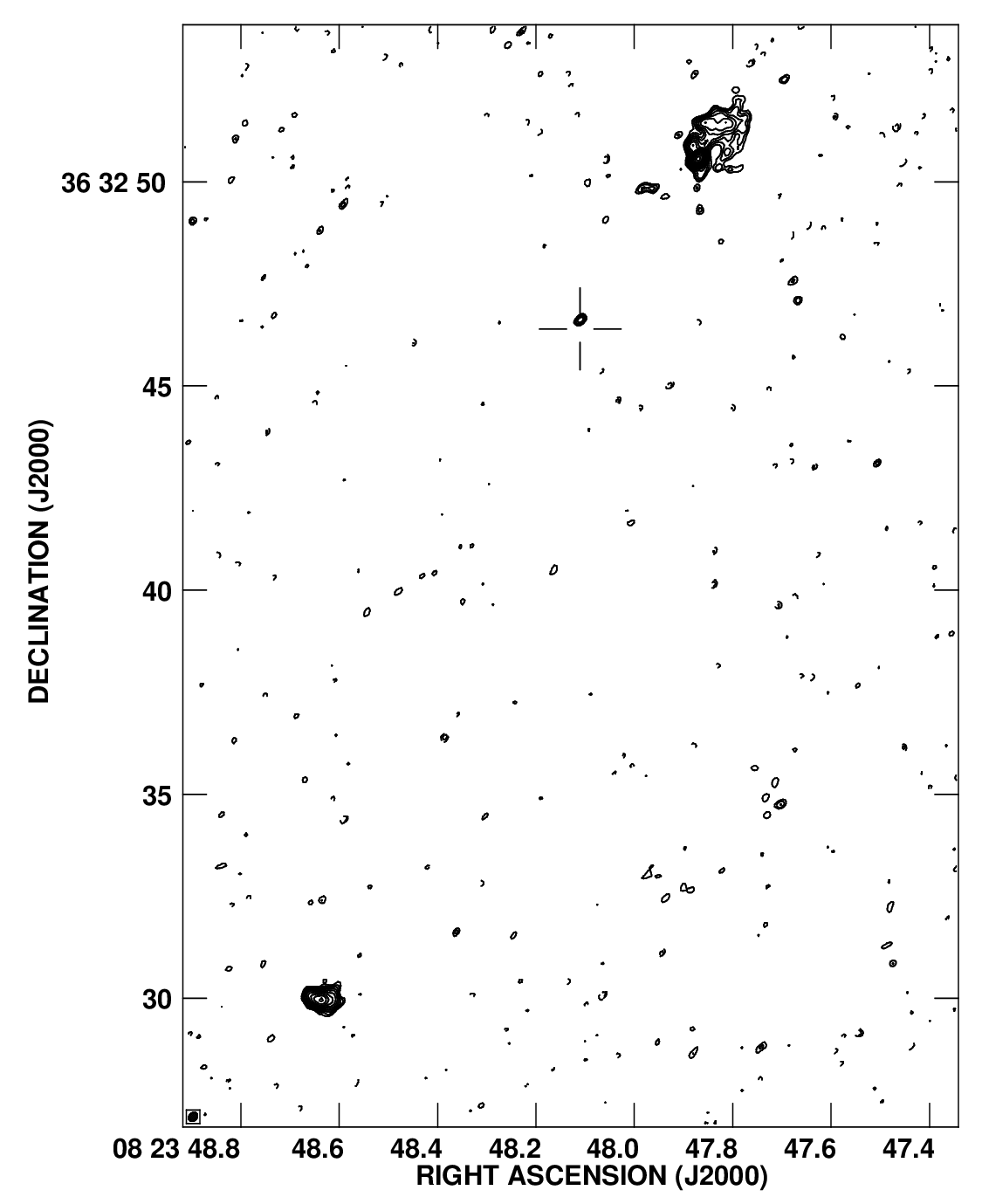,width=8.2cm} \\
6C~0901+3551 4.86\,GHz & 6C~0901+3551 8.46\,GHz\\
\psfig{file=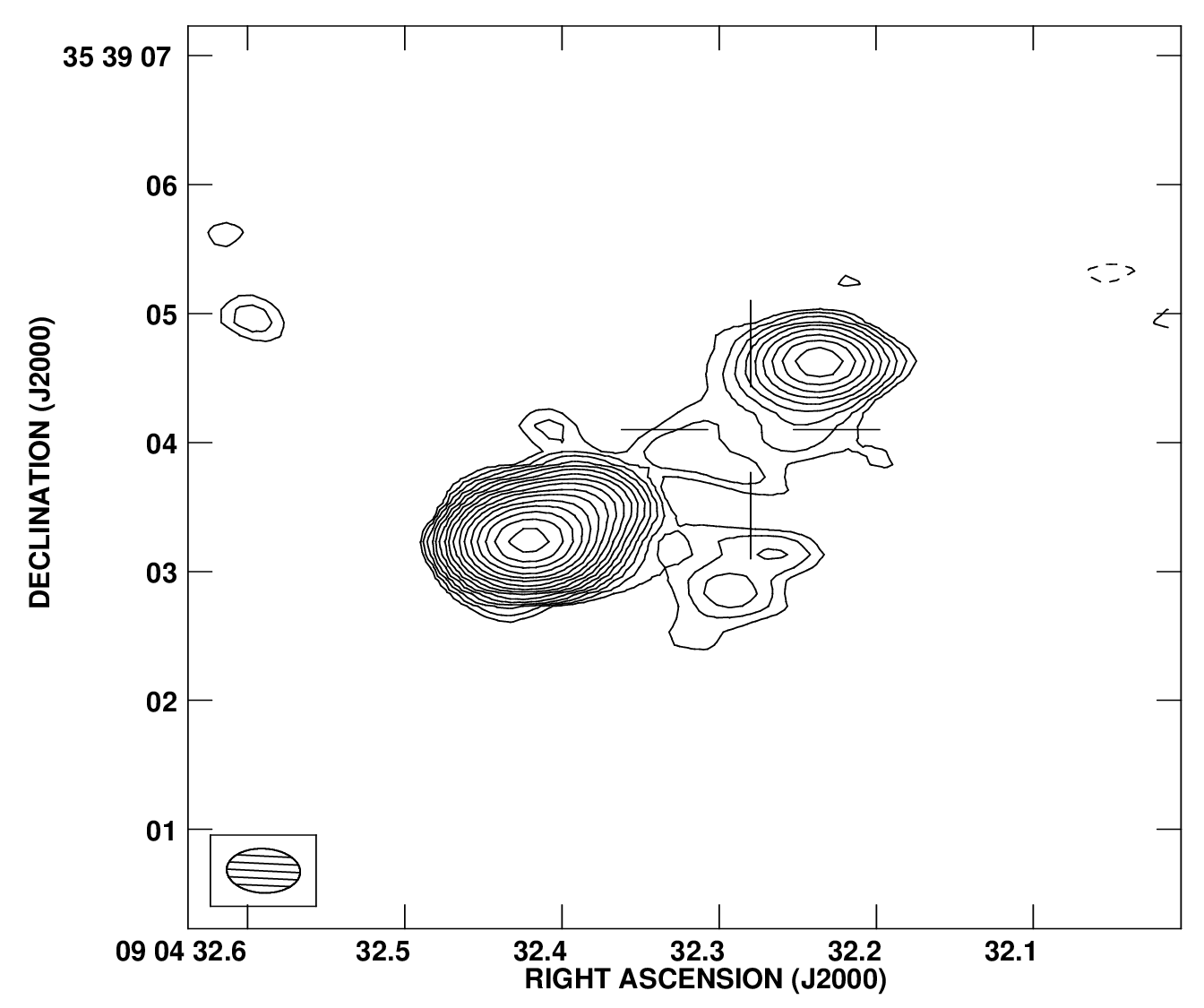,width=8.2cm,angle=-90} & \psfig{file=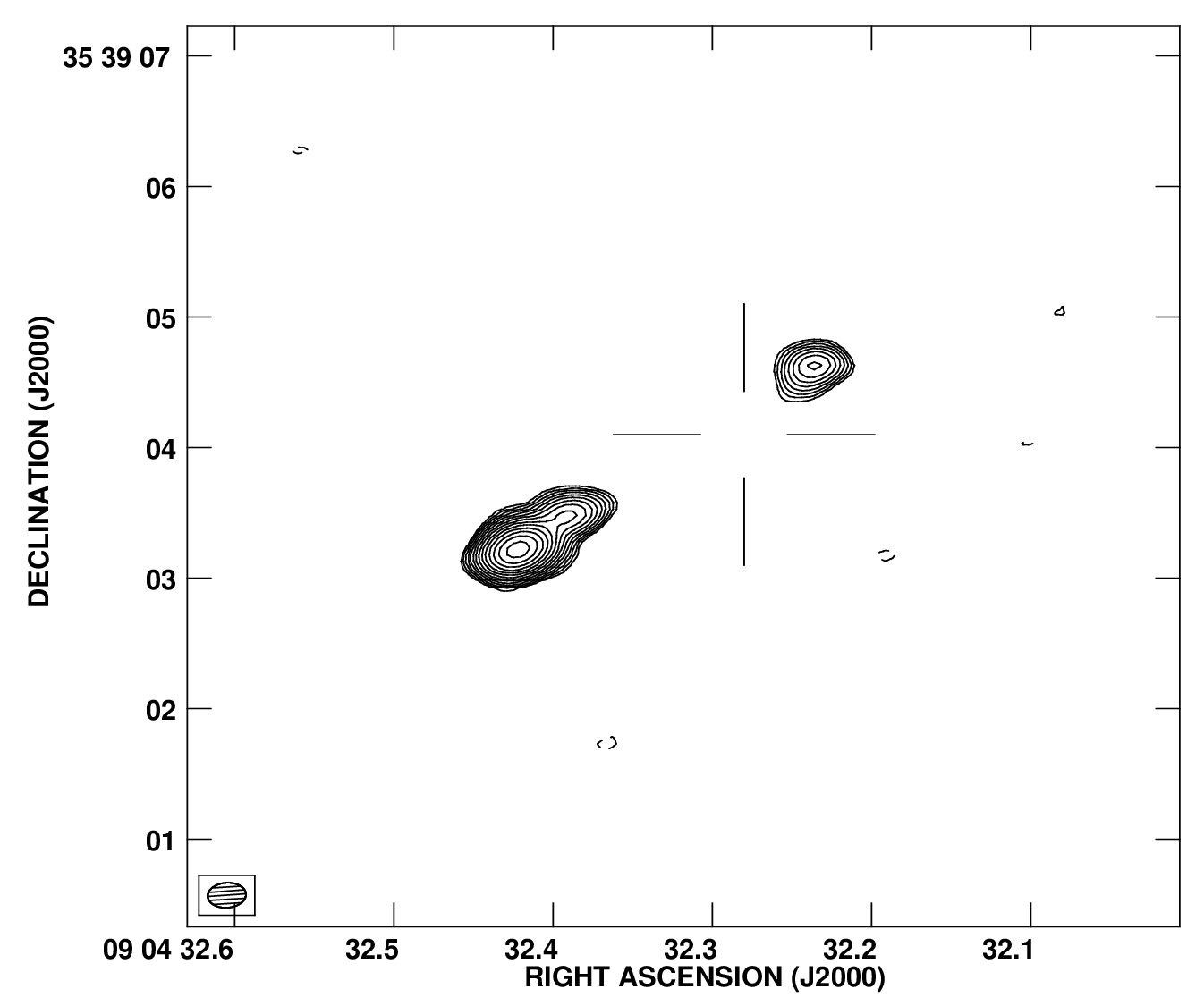,width=8.2cm,angle=-90} \\
\end{tabular}
\end{figure*}

\begin{figure*}[ht]
\begin{tabular}{cc}
USS~1707+105 4.86\,GHz & USS~1707+105 8.46\,GHz\\
\psfig{file=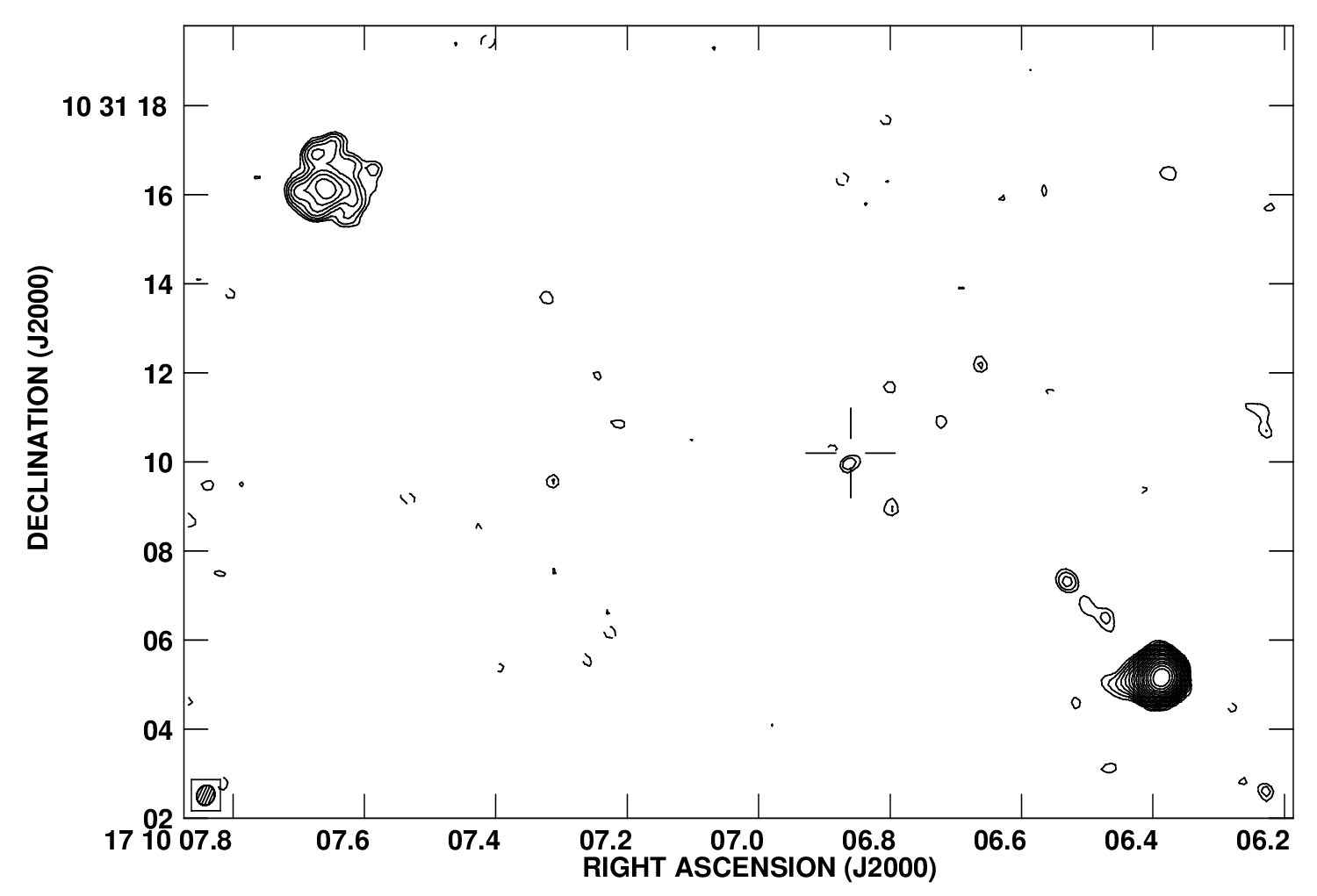,width=8.6cm,angle=-90} & \psfig{file=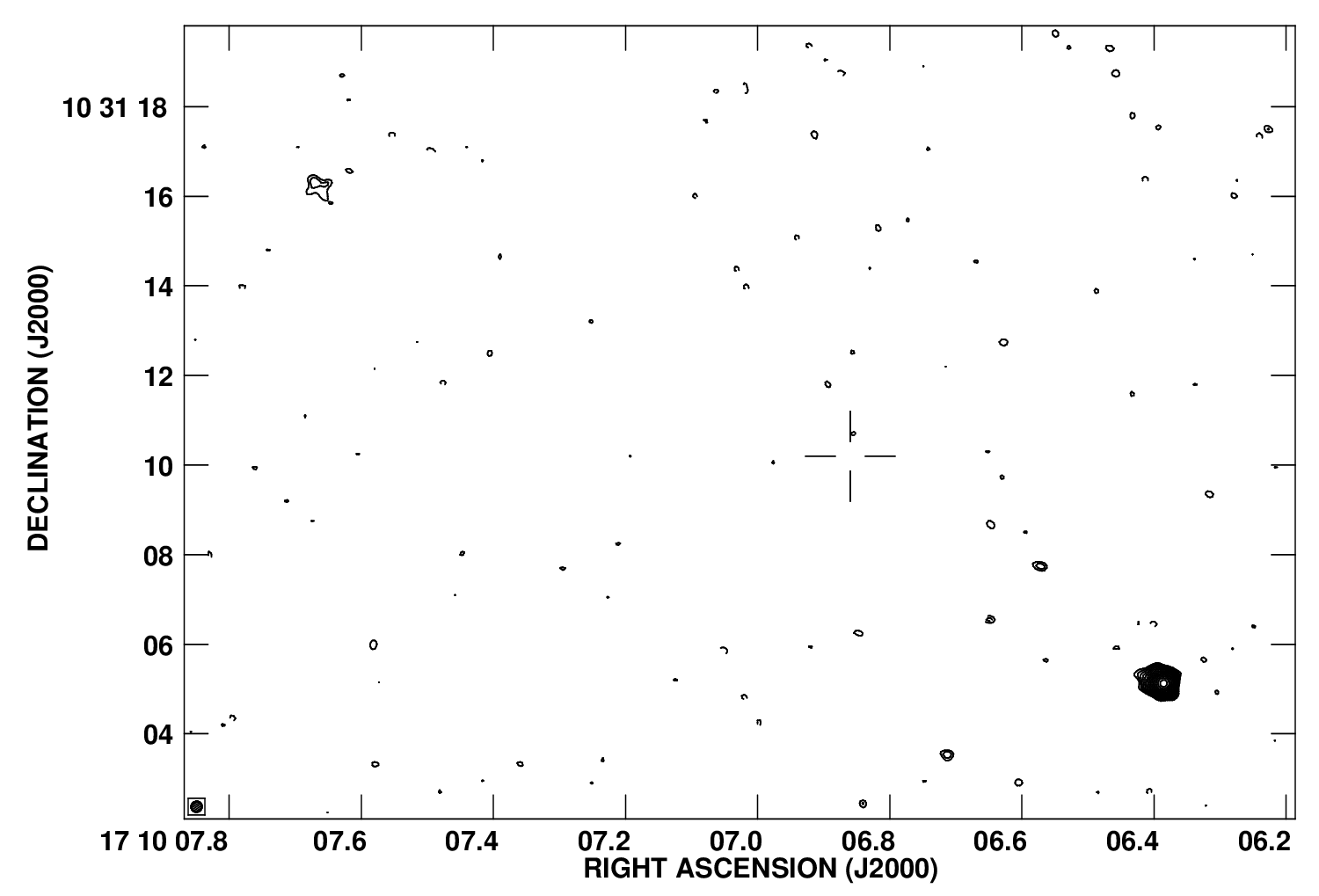,width=8.6cm,angle=-90} \\
7C~1751+6809 4.86\,GHz & 7C~1751+6809 8.46\,GHz \\
\psfig{file=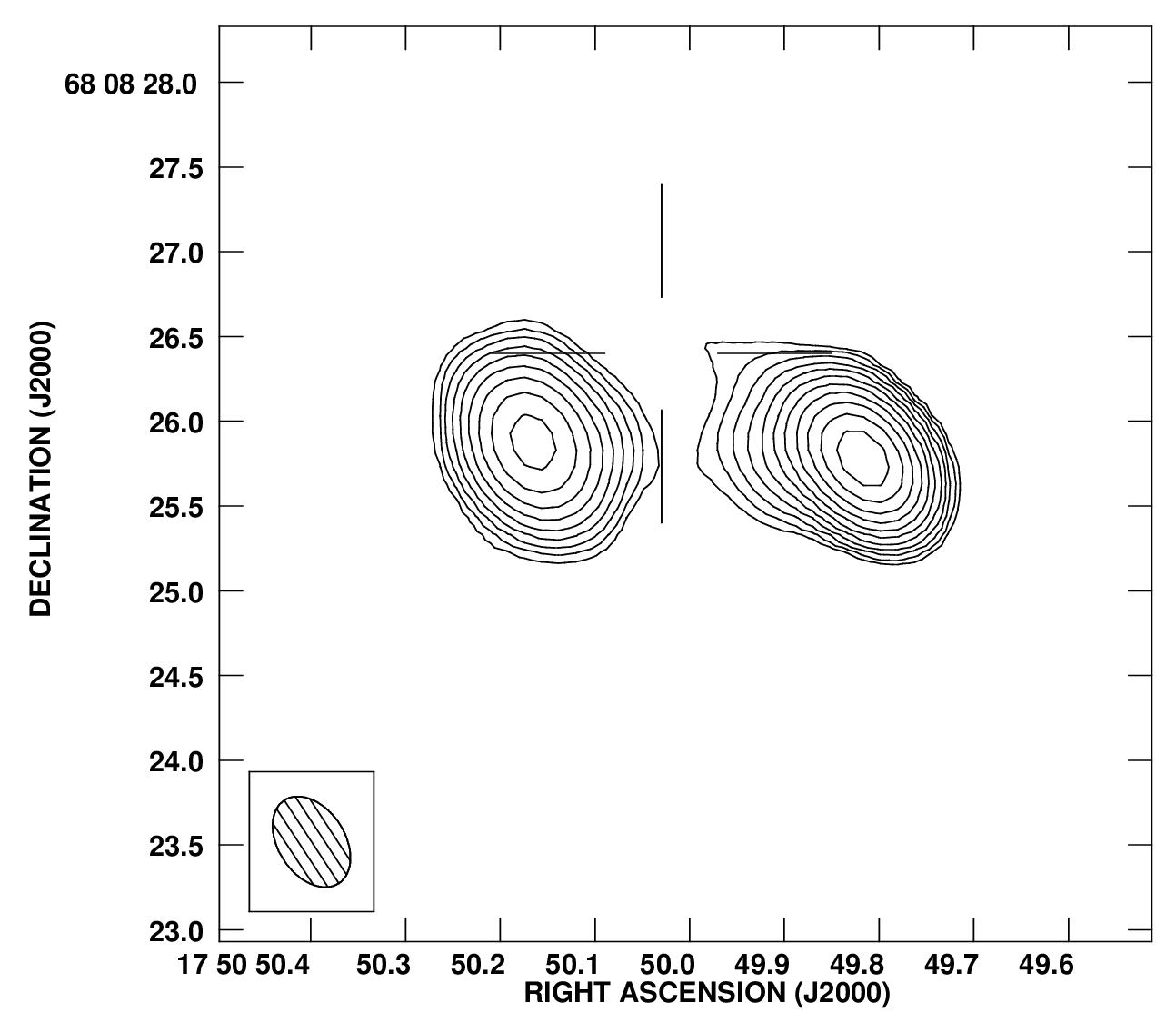,width=8.6cm,angle=-90} & \psfig{file=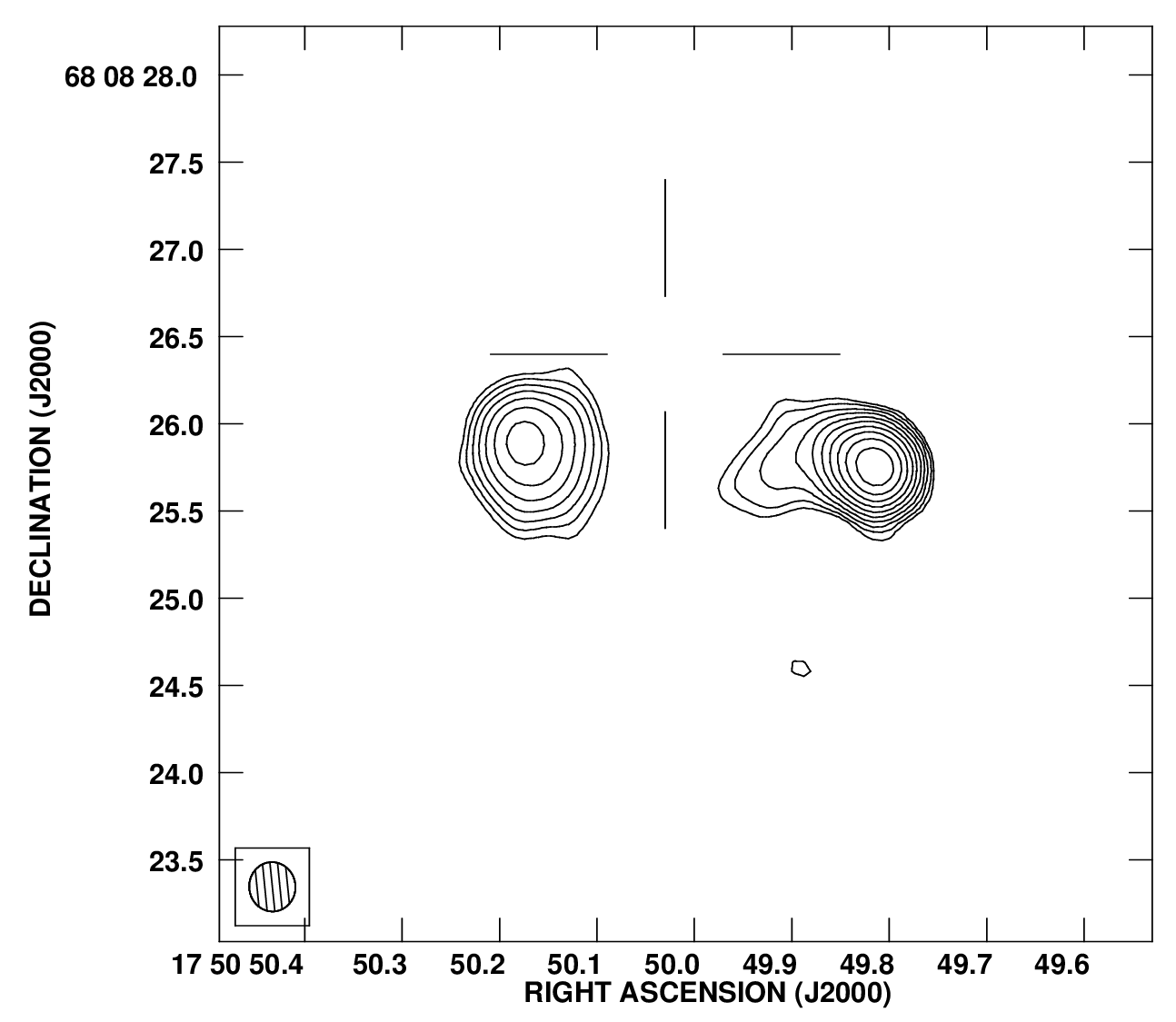,width=8.6cm,angle=-90}\\
7C~1756+6520 4.86\,GHz & 7C~1756+6520 8.46\,GHz\\
\psfig{file=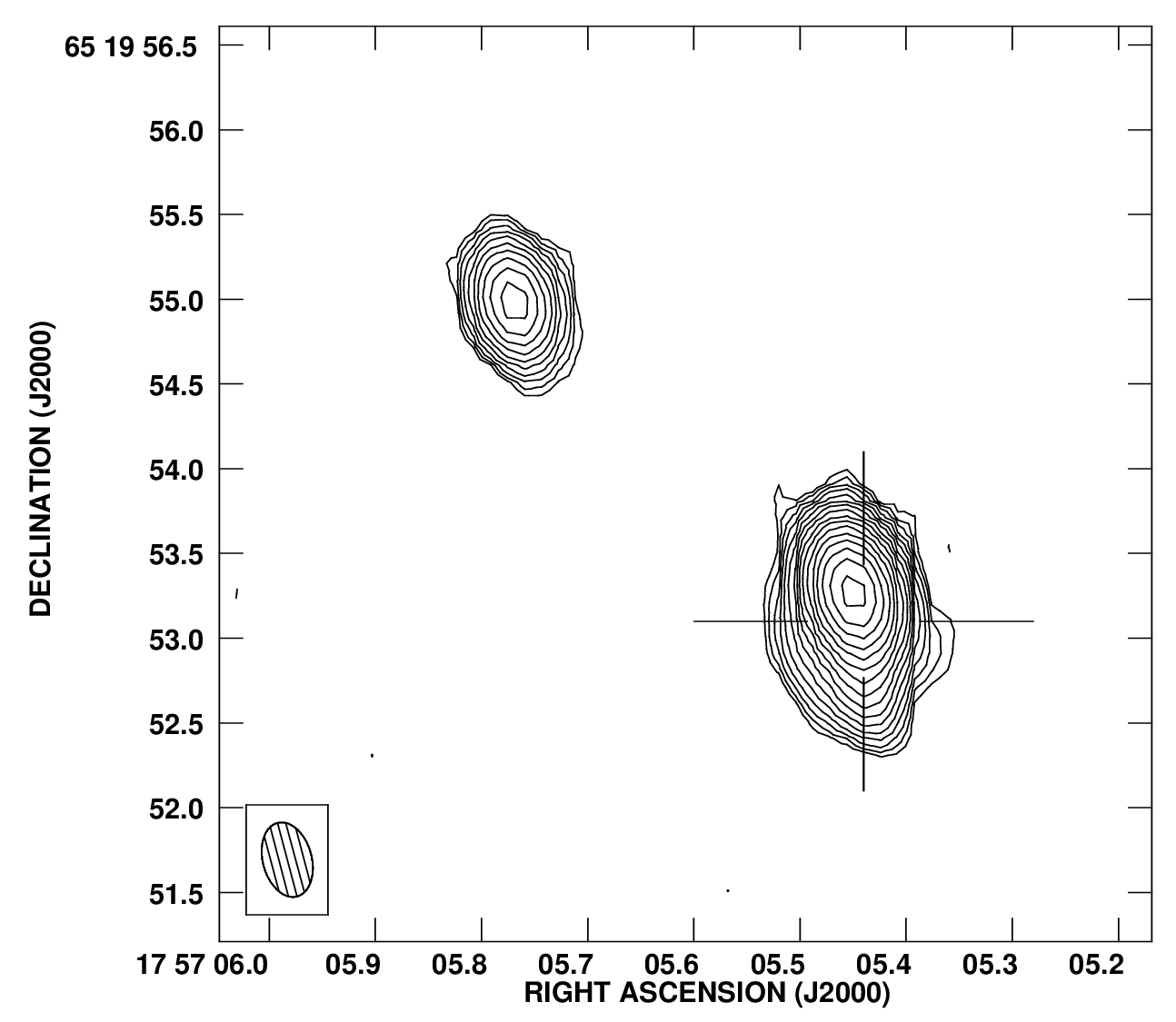,width=8.6cm,angle=-90} & \psfig{file=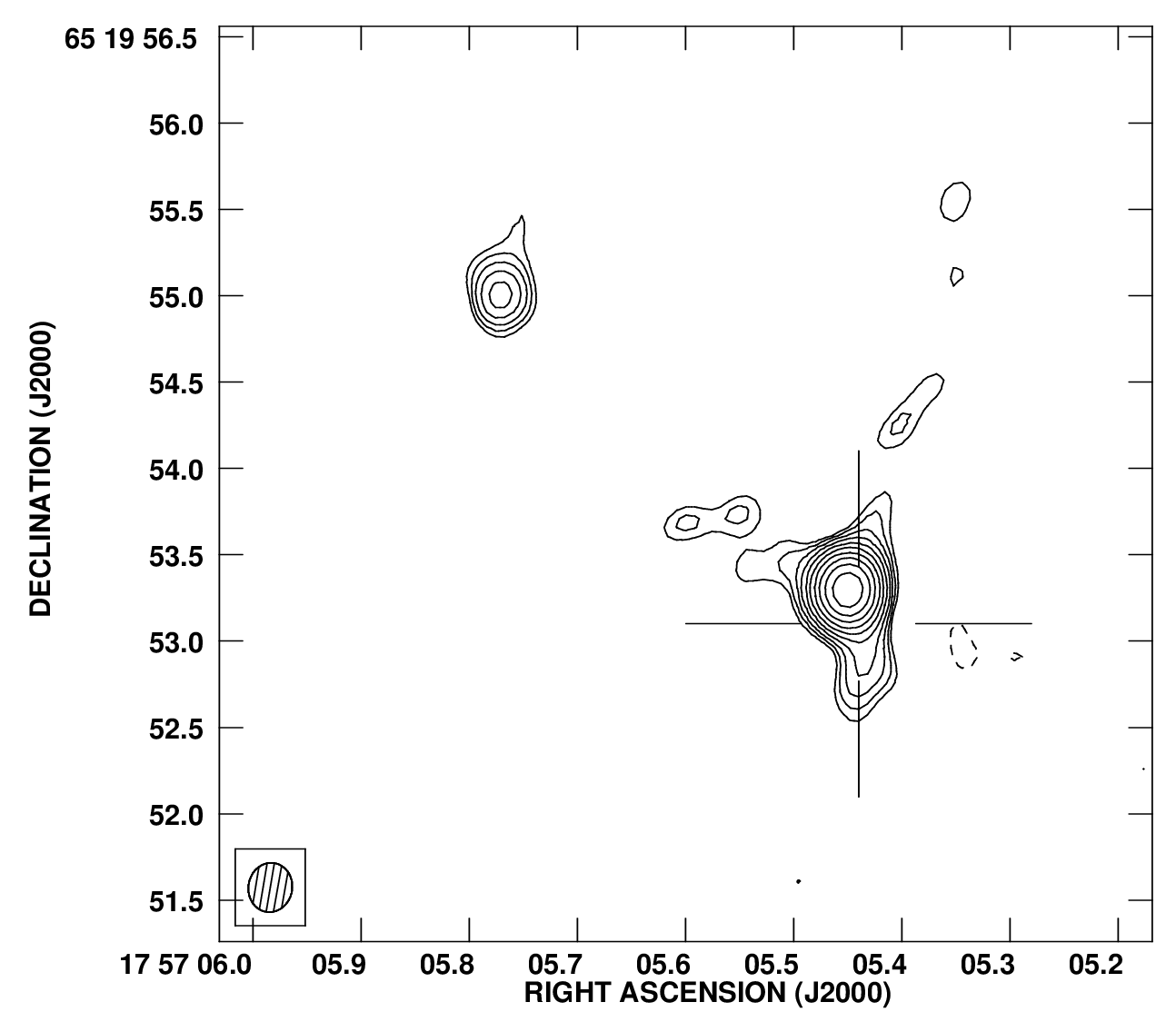,width=8.6cm,angle=-90}\\
\end{tabular}
\vspace{1.5cm}
\end{figure*}

\begin{figure*}[ht]
\begin{tabular}{cc}
7C~1805+6332 4.86\,GHz & 7C~1805+6332 8.46\,GHz \\
\psfig{file=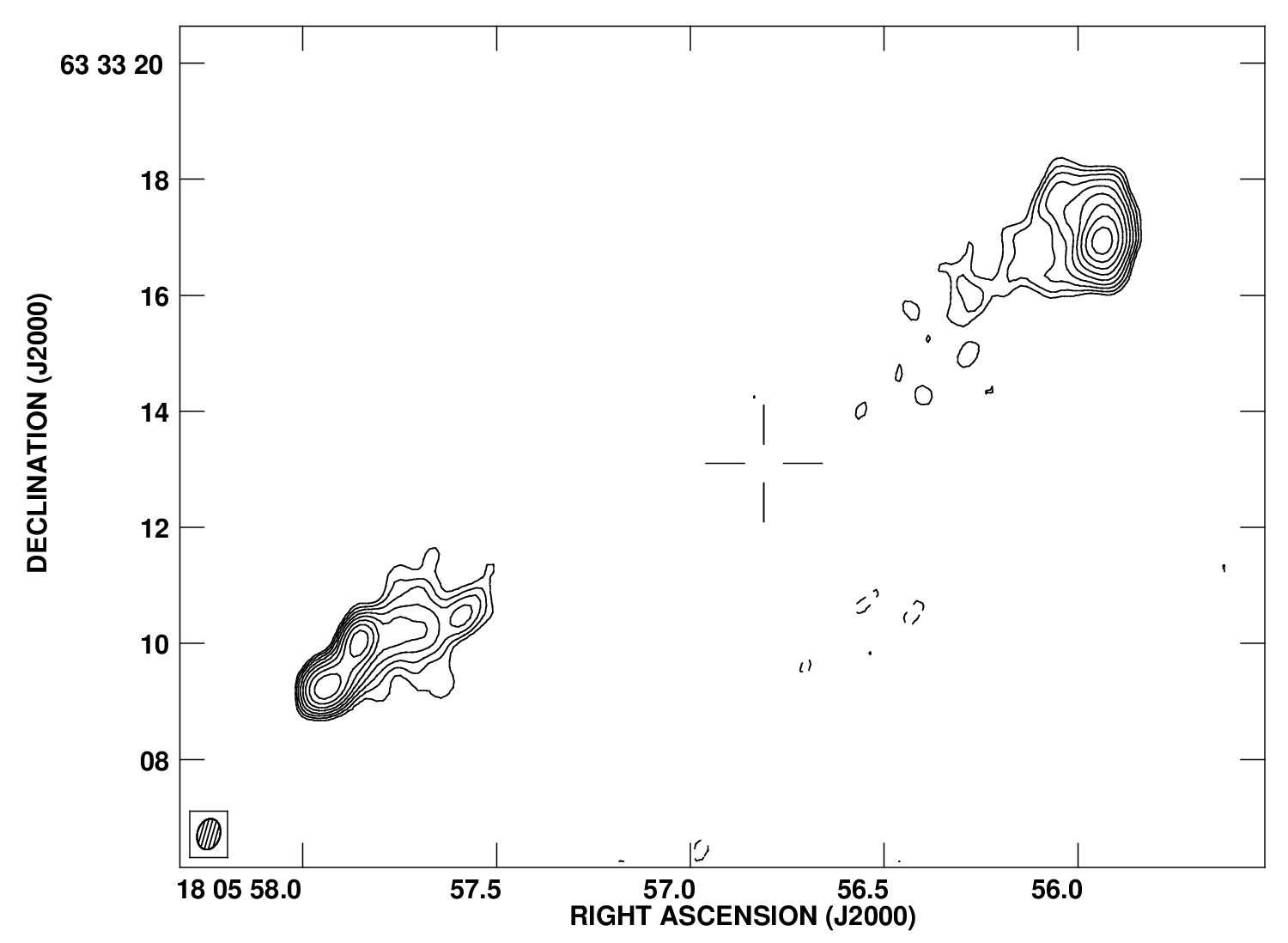,width=8cm,angle=-90} & \psfig{file=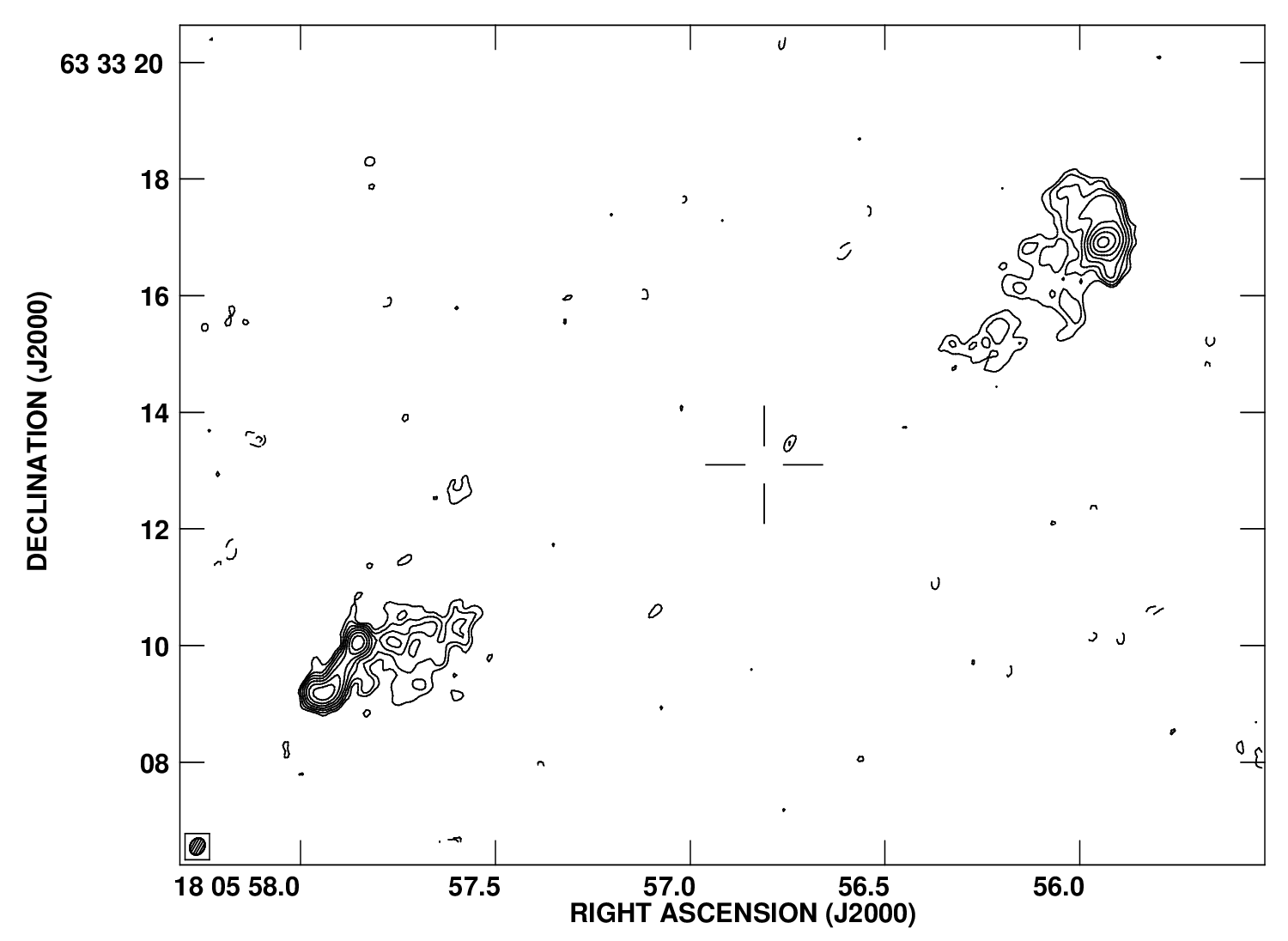,width=8cm,angle=-90}\\
TN~J2007-1316 4.86\,GHz & TN~J2007-1316 8.46\,GHz \\
\psfig{file=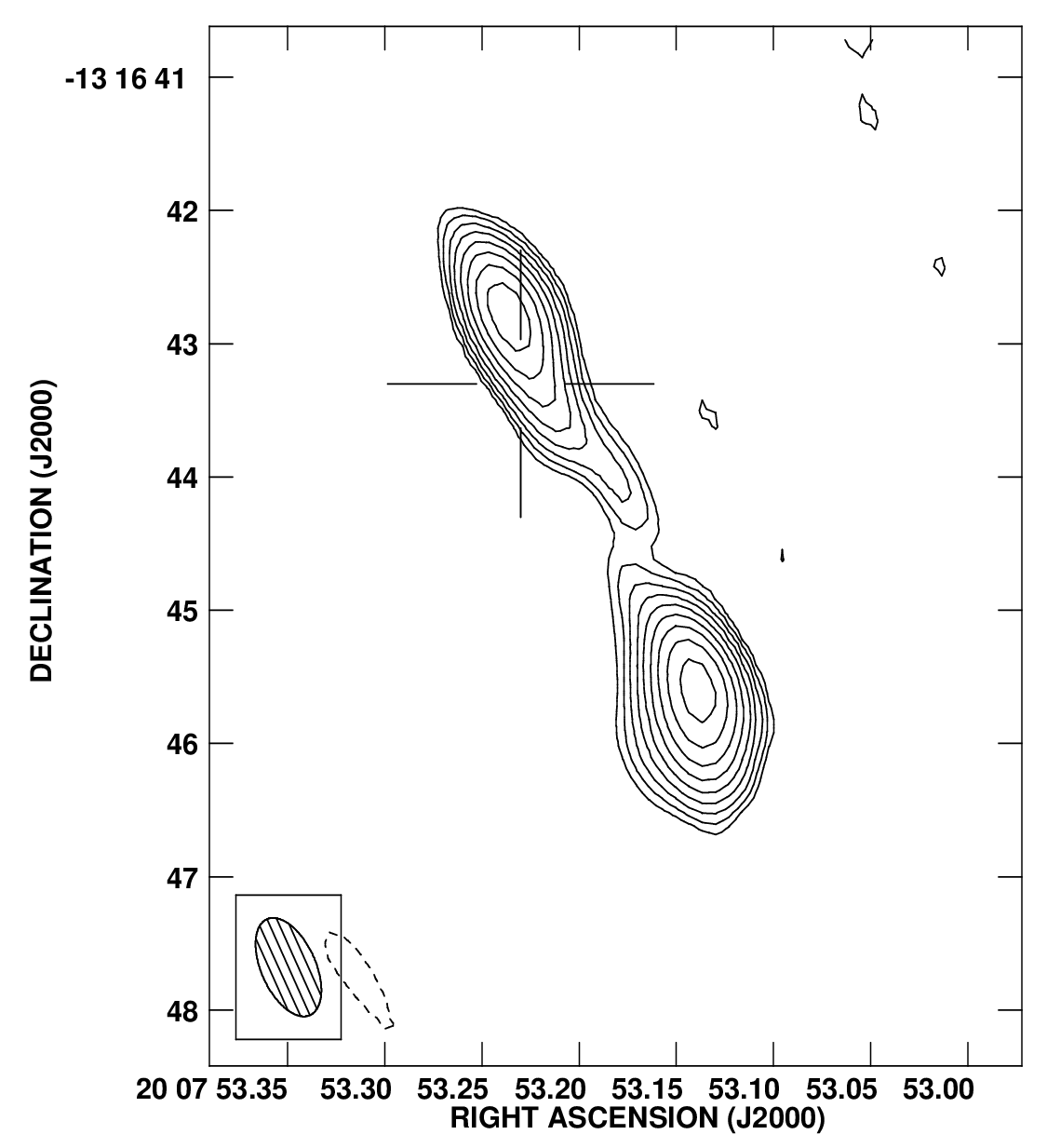,width=8cm} & \psfig{file=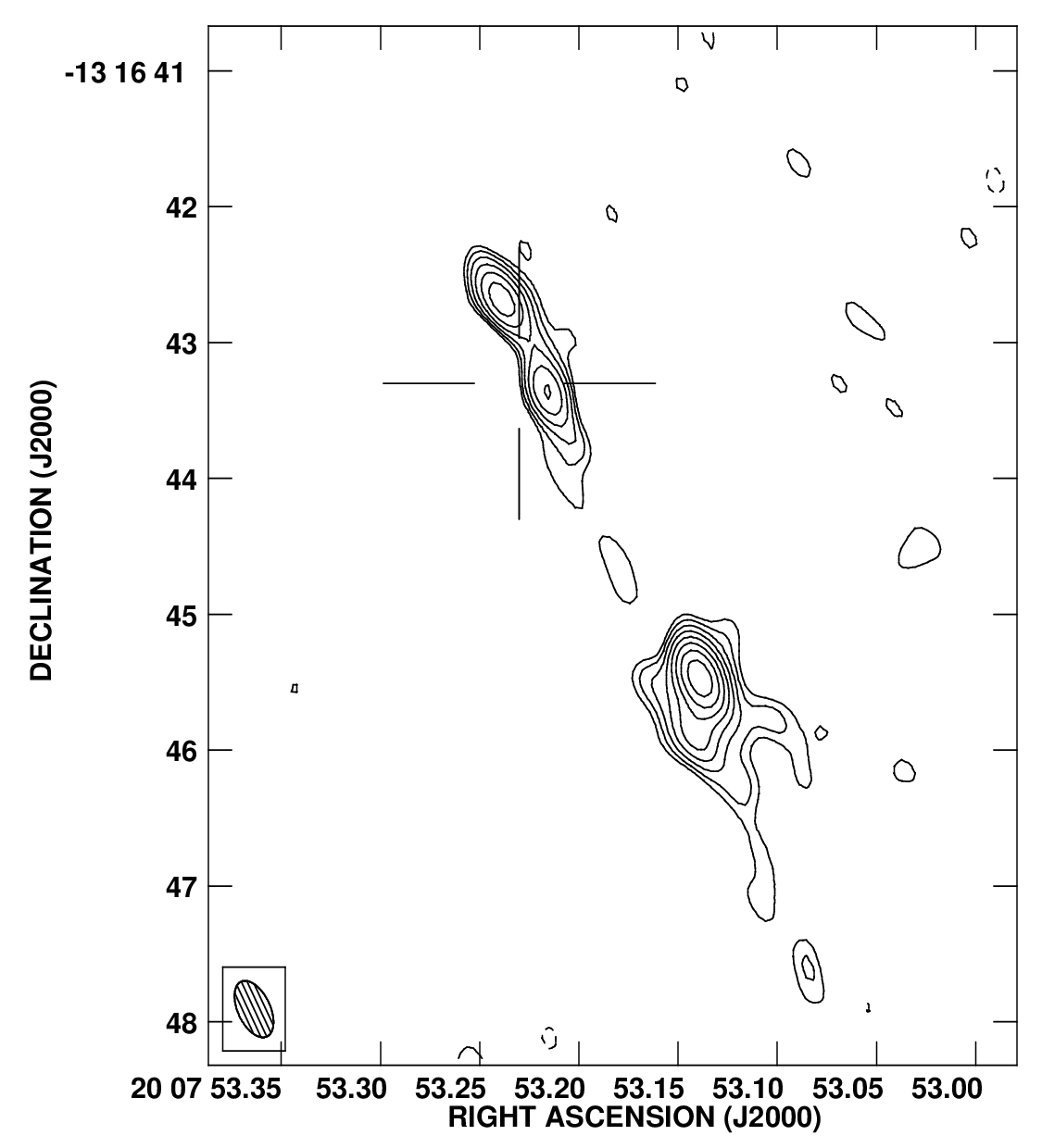,width=8cm} \\
\end{tabular}
\vspace{8cm}
\end{figure*}

\bibliographystyle{apj.bst}
\bibliography{apj-jour,bigbiblio}

\vfill\eject


\clearpage
\begin{deluxetable}{lllccrrl}
\tabletypesize{\tiny}
\tablecaption{{\it Spitzer} HzRG sample and exposure times per instrument.}
\tablewidth{0pt}
\tablehead{
\colhead{} &
\colhead{RA$^a$} & 
\colhead{Dec.$^a$} & 
\colhead{} &
\colhead{$K$-band$^b$} & 
\colhead{IRAC} &
\colhead{IRS} & 
\colhead{MIPS} \\
\colhead{HzRG} &
\colhead{(J2000)} & 
\colhead{(J2000)} & 
\colhead{Redshift} &
\colhead{(Vega)} &
\colhead{(s)} &
\colhead{(s)} & 
\colhead{24, 70, 160\,$\mu$m (s)}}
\startdata
    6C~0032+412 & 00:34:53.09 & $+41$:31:31.5 & 3.670  & 19.1 &120 & 122 & 134, 420, 881 \\ 
   MRC~0037-258 & 00:39:56.41 & $-25$:34:31.0 & 1.100  & 17.1 &120 & 305 & 267, -, - \\ 
    6C*0058+495 & 01:01:18.85 & $+49$:50:12.3 & 1.173  & 17.6 &120 & 305 & 134, 420, 881 \\ 
   MRC~0114-211 & 01:16:51.44 & $-20$:52:06.7 & 1.410  & 18.5 &120 & 305 & 267, -, - \\ 
  TN~J0121+1320 & 01:21:42.73 & $+13$:20:58.0 & 3.516  & 18.7 &120 & 122 & 267, -, - \\ 
    6C*0132+330 & 01:35:30.37 & $+33$:16:59.6 & 1.710  & 18.8 &120 & 305 & 267, -, - \\
    6C~0140+326 & 01:43:43.82 & $+32$:53:49.3 & 4.413  & 19.9 &5000& 122 & 267, 671, 2643 \\ 
   MRC~0152-209 & 01:54:55.76 & $-20$:40:26.3 & 1.920  & 17.9 &120 & 305 & 534, -, - \\ 
   MRC~0156-252 & 01:58:33.44 & $-24$:59:31.7 & 2.016  & 16.1 &120 & 122 & -, -, - \\ 
  TN~J0205+2242 & 02:05:10.69 & $+22$:42:50.4 & 3.506  & 18.7 &120 & 122 & 267, -, - \\ 
   MRC~0211-256 & 02:13:30.54 & $-25$:25:20.6 & 1.300  &\nodata&120 & 122 & 267, -, - \\ 
   TXS~0211-122 & 02:14:17.40 & $-11$:58:46.0 & 2.340  & 17.3 &120 & 305 & 267, -, - \\
           3C~65 & 02:23:43.46 & $+40$:00:52.7 & 1.176  & 16.8 &120 & 305 & 267, -, - \\ 
   MRC~0251-273 & 02:53:16.68 & $-27$:09:11.6 & 3.160  & 18.5 &120 & 122 & 267, -, - \\ 
   MRC~0316-257 & 03:18:12.06 & $-25$:35:09.7 & 3.130  & 18.1 &46000& 122 & 267, -, - \\ 
   MRC~0324-228 & 03:27:04.54 & $-22$:39:42.1 & 1.894  & 18.8 &120 & 305 & 267, -, - \\ 
   MRC~0350-279 & 03:52:51.64 & $-27$:49:22.6 & 1.900  & 19.0 &120 & 305 & 134, 420, 881 \\ 
   MRC~0406-244 & 04:08:51.46 & $-24$:18:16.4 & 2.427  & 17.4 &120 & 122 & 134, 420, 881 \\ 
        4C60.07 & 05:12:55.15 & $+60$:30:51.0 & 3.788  & 19.2 &120 & 122 & 134, 420, 881 \\ 
   PKS~0529-549 & 05:30:25.39 & $-54$:54:23.2 & 2.575  & 20.0 &120 & 122 & 134, 420, 881 \\ 
  WN~J0617+5012 & 06:17:39.35 & $+50$:12:54.2 & 3.153  & 19.7 &120 & 122 & 267, -, - \\ 
        4C41.17 & 06:50:52.23 & $+41$:30:30.1 & 3.792  & 19.1 &5000& 122 & 267, 671, 2643 \\ 
  WN~J0747+3654 & 07:47:29.38 & $+36$:54:38.1 & 2.992  & 20.0 &120 & 122 & 267, -, - \\ 
   6CE~0820+3642 & 08:23:48.11 & $+36$:32:46.4 & 1.860  & 18.2 &120 & 305 & 267, -, - \\ 
        5C~7.269 & 08:28:38.78 & $+25$:28:27.1 & 2.218  & 18.9 &120 & 122 & 267, -, - \\ 
   USS~0828+193 & 08:30:53.42 & $+19$:13:15.7 & 2.572  & 18.2 &120 & 122 & 267, -, - \\ 
   6CE~0901+3551 & 09:04:32.28 & $+35$:39:04.1 & 1.910  & 18.1 &120 & 305 & 267, -, - \\ 
     B2~0902+34 & 09:05:30.14 & $+34$:07:56.0 & 3.395  & 19.9 &1200& 122 & 2557, 2696, 3556 \\ 
   6CE~0905+3955 & 09:08:16.90 & $+39$:43:26.0 & 1.883  & 18.3 &120 & 305 & 267, -, - \\ 
  TN~J0924-2201 & 09:24:19.94 & $-22$:01:42.3 & 5.195  & 19.7 &120 & 122 & 134, 420, 881 \\ 
    6C~0930+389 & 09:33:06.91 & $+38$:41:50.1 & 2.395  & 19.5 &120 & 122 & 267, -, - \\ 
   USS~0943-242 & 09:45:32.73 & $-24$:28:49.7 & 2.923  & 19.2 &120 & 122 & 134, 420, 881 \\ 
          3C~239 & 10:11:45.42 & $+46$:28:19.8 & 1.781  & 17.8 &120 & 305 & 267, -, - \\ 
   MG~1019+0534 & 10:19:33.43 & $+05$:34:34.8 & 2.765  & 19.1 &120 & 122 & 267, -, - \\ 
   MRC~1017-220 & 10:19:49.04 & $-22$:19:59.6 & 1.768  & 17.4 &120 & 305 & 267, -, - \\ 
  WN~J1115+5016 & 11:15:06.87 & $+50$:16:23.9 & 2.540  & 19.2 &120 & 122 & 267, -, - \\ 
          3C~257 & 11:23:09.40 & $+05$:30:18.9 & 2.474  & 17.8 &120 & 122 & 267, -, - \\ 
  WN~J1123+3141 & 11:23:55.74 & $+31$:41:26.7 & 3.217  & 17.4 &120 & 122 & 267, -, - \\ 
   PKS~1138-262 & 11:40:48.38 & $-26$:29:08.8 & 2.156  & 16.1 &3000& 122 & 9000, -, - \\ 
          3C~266 & 11:45:43.37 & $+49$:46:08.2 & 1.275  & 17.1 &120 & 305 & 267, -, - \\ 
     6C~1232+39 & 12:35:04.75 & $+39$:25:38.9 & 3.220  & 18.0 &120 & 122 & 267, -, - \\ 
   USS~1243+036 & 12:45:38.36 & $+03$:23:20.7 & 3.570  & 19.2 &120 & 122 & 267, -, - \\ 
  TN~J1338-1942 & 13:38:26.09 & $-19$:42:30.7 & 4.110  & 19.6 &5000& 122 & 267, -, - \\ 
        4C~24.28 & 13:48:14.87 & $+24$:15:50.5 & 2.879  &\nodata&120 & 122 & 267, -, - \\ 
        3C~294.0 & 14:06:53.25 & $+34$:11:21.1 & 1.786  & 17.9 &120 & 305 & 267, -, - \\ 
   USS~1410-001 & 14:13:15.10 & $-00$:22:59.7 & 2.363  &\nodata&120 & 122 & 267, -, - \\ 
    8C~1435+635 & 14:36:37.21 & $+63$:19:14.4 & 4.250  & 19.4 &120 & 122 & 134, 420, 881 \\ 
   USS~1558-003 & 16:01:17.30 & $+00$:28:46.2 & 2.527  &\nodata&120 & 122 & 267, -, - \\ 
   USS~1707+105 & 17:10:06.86 & $+10$:31:10.2 & 2.349  &\nodata&120 & 122 & 267, -, -  \\ 
    LBDS~53W002 & 17:14:14.79 & $+50$:15:30.6 & 2.393  & 18.9 &3300& 122 & 134, 420, 881 \\ 
    LBDS~53W091 & 17:22:32.93 & $+50$:06:01.3 & 1.552  & 18.7 &900 & 305 & 633, 1311, 2643 \\ 
        3C~356.0 & 17:24:19.33 & $+50$:57:36.2 & 1.079  & 16.8 &120 & 305 & 134, 420, 881 \\ 
   7C~1751+6809 & 17:50:50.03 & $+68$:08:26.4 & 1.540  & 18.2 &120 & 305 & 134, 420, 881 \\ 
   7C~1756+6520 & 17:57:05.44 & $+65$:19:53.1 & 1.416  & 18.9 &120 & 305 & 134, 420, 881 \\ 
        3C~368.0 & 18:05:06.37 & $+11$:01:33.1 & 1.132  & 17.2 &120 & 305 & 134, 420, 881 \\ 
   7C~1805+6332 & 18:05:56.81 & $+63$:33:13.1 & 1.840  & 18.8 &120 & 305 & 134, 420, 881 \\ 
        4C~40.36 & 18:10:55.70 & $+40$:45:24.0 & 2.265  & 17.8 &120 & 122 & 134, 420, 881 \\ 
 TXS~J1908+7220 & 19:08:23.70 & $+72$:20:11.8 & 3.530  & 16.5 &120 & 122 & 134, 420, 881 \\ 
  WN~J1911+6342 & 19:11:49.63 & $+63$:42:09.6 & 3.590  & 19.9 &120 & 122 & 134, 420, 881 \\ 
  TN~J2007-1316 & 20:07:53.26 & $-13$:16:43.6 & 3.840  & 18.8 &120 & 122 & 267, -, - \\ 
   MRC~2025-218 & 20:27:59.48 & $-21$:40:56.9 & 2.630  & 18.5 &120 & 122 & 267, -, - \\ 
   MRC~2048-272 & 20:51:03.59 & $-27$:03:02.5 & 2.060  & 18.3 &120 & 122 & 267, -, - \\ 
   MRC~2104-242 & 21:06:58.28 & $-24$:05:09.1 & 2.491  &\nodata&120 & 122 & 267, -, - \\ 
        4C~23.56 & 21:07:14.80 & $+23$:31:45.0 & 2.483  & 19.7 &120 & 122 & 134, 420, 881 \\ 
   MG~2144+1928 & 21:44:07.56 & $+19$:29:14.6 & 3.592  & 19.1 &120 & 122 & 267, -, - \\ 
   USS~2202+128 & 22:05:14.18 & $+13$:05:33.0 & 2.706  & 18.4 &120 & 122 & 267, -, - \\ 
   MRC~2224-273 & 22:27:43.26 & $-27$:05:01.7 & 1.679  & 18.5 &120 & 305 & 267, -, - \\ 
  B3~J2330+3927 & 23:30:24.82 & $+39$:27:12.5 & 3.086  & 18.8 &120 & 122 & 134, 420, 881 \\ 
        4C~28.58 & 23:51:59.20 & $+29$:10:29.0 & 2.891  & 18.7 &120 & 122 & 267, -, - \\ 
          3C~470 & 23:58:36.00 & $+44$:04:45.0 & 1.653  & 18.5 &120 & 305 & 134, 420, 881 \\ 
\enddata
\tablenotetext{a}{J2000 coordinates of the identification in the bluest IRAC channel.}
\tablenotetext{b}{$K$-band magnitude within projected 64\,kpc radius, taken from S07.}
\normalsize
\label{table.shzrg}
\end{deluxetable}

\begin{deluxetable}{lcccrrl}
\tabletypesize{\tiny}
\tablecaption{Radio data for the SHzRG sample.}
\tablewidth{0pt}
\tablehead{
\colhead{} &
\colhead{$\log(L_{\rm 500\,MHz})$} & 
\colhead{$\log(L_{3\,GHz})$} & 
\colhead{} &
\colhead{$\theta^b$} &
\colhead{$CF_{20}^{c}$}  & 
\colhead{} \\
\colhead{HzRG} & 
\colhead{(W Hz$^{-1}$)} & 
\colhead{(W Hz$^{-1}$)} & 
\colhead{Morph$^a$} & 
\colhead{(arcsec)} &
\colhead{(\%)} &
\colhead{Reference}
}
\startdata
6C~0032+412   & 28.75 & 27.75 & T &  2.3 &  2.0  & \citet{blu98} \\
MRC~0037-258  & 27.72 & 27.07 & T & 27.6 &  1.6  & this paper; \citet{kap98} \\
6C*~0058+495  & 27.33 & 26.68 & D &  3.2 &\nodata& \citet{blu98} \\
MRC~0114-211  & 28.66 & 28.39 & T &  0.7 &  0.3  & this paper; \citet{kap98} \\
TN~J0121+1320 & 28.49 & 27.41 & D &  0.3 &\nodata& \citet{deb00} \\
6C*~0132+330  & 27.64 & 26.60 & T &  2.2 &  9    & this paper \\
6C~0140+326   & 28.73 & 27.89 & D &  2.5 &\nodata& \citet{blu98} \\
MRC~0152-209  & 28.20 & 27.77 & D &  2.2 &\nodata& \citet{pen00} \\
MRC~0156-252  & 28.46 & 27.79 & T &  8.3 & 10.0  & \citet{car97} \\
TN~J0205+2242 & 28.46 & 27.43 & D &  2.7 &\nodata& \citet{deb00} \\
MRC~0211-256  & 27.78 & 26.22 & S &$<$1.5&\nodata& \citet{kap98} \\
TXS~0211-122  & 28.48 & 27.81 & T & 17.0 &  3.8  & \citet{car97} \\
3C~65         & 28.63 & 28.06 & T & 17.5 &  0.1  & \citet{best97,cor98} \\
MRC~0251-273  & 28.54 & 28.09 & T &  3.9 &  3.5  & this paper; \citet{kap98} \\
MRC~0316-257  & 28.95 & 28.26 & D &  6.7 &\nodata& \citet{mcc91,ath97} \\
MRC~0324-228  & 28.49 & 27.81 & T &  9.5 &  0.2  & this paper; \citet{mcc91} \\
MRC~0350-279  & 28.25 & 27.61 & D &  1.2 &\nodata& this paper; \citet{kap98} \\
MRC~0406-244  & 29.03 & 28.11 & T & 10.0 &  2.6  & \citet{car97} \\
4C~60.07      & 29.20 & 27.91 & T &  9.0 &  2.1  & \citet{car97} \\
PKS~0529-549  & 29.16 & 28.27 & D &  1.2 &$<$10  & \citet{bro07} \\
WN~J0617+5012 & 28.02 & 26.97 & D &  3.4 &\nodata& \citet{deb00} \\
4C~41.17      & 29.18 & 28.17 & T & 20.0 &  2.4  & \citet{car94} \\
WN~J0747+3654 & 28.14 & 27.02 & S &  2.1 &\nodata& \citet{deb00} \\
6CE~0820+3642 & 28.28 & 27.49 & T & 23.4 &  0.9  & this paper; \citet{law95a} \\
5C~7.269      & 27.82 & 27.08 & D &  7.6 &\nodata& this paper \\
USS~0828+193  & 28.44 & 27.47 & T & 12.8 & 21.0  & \citet{car97} \\
6CE~0901+3551 & 28.19 & 27.47 & T &  2.9 &  0.4  & this paper \\
B2~0902+34    & 28.78 & 28.27 & T &  4.2 & 18.0  & \citet{car94} \\
6CE~0905+3955 & 28.17 & 27.49 & T & 111  &  0.6  & \citet{law95b} \\
TN~J0924-2201 & 29.51 & 27.83 & D &  1.2 &\nodata& \citet{deb00} \\
6C~0930+389   & 28.41 & 27.79 & T &  4.2 &  0.8  & \citet{pen00} \\
USS~0943-242  & 28.62 & 27.95 & D &  3.9 &\nodata& \citet{car97} \\
3C~239        & 29.00 & 28.19 & T & 11.9 &  0.3  & \citet{best97} \\
MG~1019+0534  & 28.57 & 28.13 & T &  2.2 &  4.7  & \citet{pen00} \\
MRC~1017-220  & 27.94 & 28.11 & S &$<$0.2&\nodata& \citet{pen00} \\
WN~J1115+5016 & 27.82 & 26.87 & D &  0.2 &\nodata& \citet{deb00} \\
3C~257        & 29.16 & 28.62 & D & 12   &\nodata& \citet{wvb98} \\
WN~J1123+3141 & 28.51 & 27.42 & T & 25.8 &  5.5  & \citet{whi97} \\
PKS~1138-262  & 29.07 & 28.14 & T & 15.8 &  3.1  & \citet{car97} \\
3C~266        & 28.54 & 27.80 & D &  4.6 &$<$0.1 & \citet{best97} \\
6C~1232+39    & 28.93 & 28.01 & T &  8.0 &  0.7  & \citet{car97} \\
USS1243+036   & 29.23 & 28.25 & T &  6.0 &  1.7  & \citet{ojik96} \\
TN~J1338-1942 & 28.71 & 27.90 & T &  5.2 &  1.9  & \citet{pen00} \\
4C~24.28      & 29.05 & 28.25 & T &  2.3 &  0.7  & \citet{car97} \\
3C~294.0      & 28.96 & 28.12 & T & 14.5 &  0.3  & \citet{mcc90} \\
USS~1410-001  & 28.41 & 27.69 & T & 24.0 &  6.7  & \citet{car97} \\
8C~1435+635   & 29.40 & 28.55 & T &  4.3 &  3.3  & \citet{car97} \\
USS~1558-003  & 28.82 & 28.00 & T &  9.2 &  1.6  & \citet{pen00} \\
USS~1707+105  & 28.63 & 27.78 & D & 22.5 &\nodata& \citet{pen01} \\
LBDS~53W002   & 27.78 & 27.02 & S &$<$1.5&\nodata& \citet{fom02} \\
LBDS~53W091   & 27.04 & 26.29 & D &  4.3 &\nodata& \citet{kap90,spi97} \\
3C~356.0      & 28.35 & 27.65 & T & 76.2 &  0.1  & \citet{best97} \\
7C~1751+6809  & 27.46 & 27.01 & D &  2.0 &\nodata& this paper; \citet{lacy92} \\
7C~1756+6520  & 27.40 & 27.00 & D &  2.7 &\nodata& this paper; \citet{lacy92} \\
3C~368.0      & 28.52 & 27.63 & T &  8.8 &  0.2  & \citet{best97} \\
7C~1805+6332  & 27.78 & 27.12 & T & 15.6 &  0.6  & this paper; \citet{lacy92} \\
4C~40.36      & 28.79 & 27.94 & D &  4.0 &\nodata& \citet{car97} \\
TXS~J1908+7220& 29.12 & 28.15 & T & 15.4 &  4.8  & \citet{pen00} \\
WN~J1911+6342 & 28.14 & 27.03 & S &  1.8 &\nodata& \citet{deb00} \\
TN~J2007-1316 & 29.13 & 27.79 & T &  3.5 & 15    & this paper \\
MRC~2025-218  & 28.74 & 27.96 & T &  5.1 &  0.7  & \citet{car97} \\
MRC~2048-272  & 28.72 & 27.85 & D &  6.7 &$<$0.2 & \citet{pen00} \\
MRC~2104-242  & 28.84 & 27.88 & T & 23.7 &  0.7  & \citet{pen00} \\
4C~23.56      & 28.93 & 28.26 & T & 53.0 & 14.6  & \citet{car97} \\
MG~2144+1928  & 29.08 & 28.27 & D &  8.9 &\nodata& \citet{car97} \\
USS~2202+128  & 28.54 & 27.75 & T &  4.2 &  0.9  & \citet{car97} \\
MRC~2224-273  & 27.52 & 27.39 & S &$<$0.2&\nodata& \citet{pen00} \\
B3~J2330+3927 & 28.33 & 27.59 & T &  3.0 & 50    & \citet{per05} \\
4C~28.58      & 28.91 & 27.89 & T & 14.5 &  2.0  & \citet{cai02,cha96} \\
3C~470        & 28.79 & 28.24 & T & 24.6 &  0.4  & \citet{best97} \\
\enddata
\tablenotetext{a}{Radio source morphology: S=Single component,
D=Double component, T=Three or more components, including a radio
core.}
\tablenotetext{b}{{Largest angular size of the radio source; see \citet{car97}.}}
\tablenotetext{c}{Core fraction at restframe 20\,GHz; see \citet{car97}.}
\normalsize
\label{table.radiodata}
\end{deluxetable}

\clearpage
\begin{deluxetable}{ccccccccc}
\tabletypesize{\tiny}
\tablecaption{{\it Spitzer} photometry for HzRGs ($\uJy$).}
\tablewidth{0pt}
\tablehead{
\colhead{HzRG} &
\colhead{$f_{3.6\um}$} &
\colhead{$f_{4.5\um}$} &
\colhead{$f_{5.8\um}$} &
\colhead{$f_{8.0\um}$} &
\colhead{$f_{16\um}$} &
\colhead{$f_{24\um}$} &
\colhead{$f_{70\um}$} &
\colhead{$f_{160\um}$}}
\startdata
   6C~0032+412 &  15.4$\pm$ 2.1 &  33.8$\pm$ 3.7 &   57.6$\pm$ 8.9 &   98.3$\pm$ 10.0 &   412$\pm$ 83 &  467$\pm$ 42 & $<  3950$      & $< 96000$ \\ 
  MR~C0037-258 & 221.0$\pm$22.0 & 248.0$\pm$25.0 &  286.0$\pm$29.0 &  518.0$\pm$ 52.0 &   877$\pm$100 & 1740$\pm$ 39 &    \nodata     & \nodata   \\ 
   6C*0058+495 &  82.0$\pm$ 8.3 &  86.7$\pm$ 8.8 &   93.2$\pm$ 9.5 &  309.0$\pm$ 31.0 &  1280$\pm$100 & 1490$\pm$ 50 & 18900$\pm$1962 & $< 94600$ \\ 
  MRC~0114-211 &  87.3$\pm$ 8.9 & 117.0$\pm$12.0 &  157.0$\pm$16.0 &  398.0$\pm$ 40.0 &  1690$\pm$100 & 2090$\pm$ 40 &    \nodata     & \nodata   \\ 
 TN~J0121+1320 &  10.5$\pm$ 1.8 &  14.4$\pm$ 2.1 &  $<  32.5$      & $<  40.3$        &  $<$204       & $<$131       &    \nodata     & \nodata   \\ 
  6C*~0132+330 &  31.6$\pm$ 3.6 &  41.3$\pm$ 4.5 &   51.9$\pm$12.4 &  109.0$\pm$ 11.0 &  $<$280       &  109$\pm$ 25 &    \nodata     & \nodata   \\ 
   6C~0140+326 & $< 623.0$      & $< 450.0$      &  $<$ 34.9       & $<$36.3          &  $<$160       & $<$2190      & $<  2580$      & $< 20700$ \\ 
  MRC~0152-209 & 108.0$\pm$11.0 & 165.0$\pm$17.0 &  215.0$\pm$22.0 &  415.0$\pm$ 42.0 &  1580$\pm$100 & 3320$\pm$133 &    \nodata     & \nodata   \\ 
  MRC~0156-252 & 291.0$\pm$29.0 & 405.0$\pm$41.0 &  717.0$\pm$72.0 & 1125.0$\pm$113.0 &  1980$\pm$100 & \nodata      &    \nodata     & \nodata   \\ 
 TN~J0205+2242 &   7.5$\pm$ 1.7 &   4.7$\pm$ 2.3 &  $<  29.3$      & $<  37.4$        &  $<$211       & $<$116       &    \nodata     & \nodata   \\ 
  MRC~0211-256 & 166.0$\pm$17.0 & 197.0$\pm$20.0 &  222.0$\pm$22.0 &  278.0$\pm$ 28.0 &   378$\pm$ 86 &  710$\pm$ 35 &    \nodata     & \nodata   \\ 
  TXS~0211-122 &  30.2$\pm$0.7  &  47.4$\pm$1.1  &  126.5$\pm$6.3  &  462.2$\pm$7.9   &  1590$\pm$220 & 2750$\pm$ 40 &    \nodata     & \nodata   \\
         3C~65 & 161.0$\pm$32.0 &   \nodata      &  320.0$\pm$66.0 &    \nodata       &  1220$\pm$330 & 2072$\pm$ 30 &    \nodata     & \nodata   \\ 
  MRC~0251-273 &  11.5$\pm$ 1.6 &  12.7$\pm$ 1.8 &  $<  29.1$      & $<  28.8$        &   102$\pm$ 33 &  476$\pm$ 33 &    \nodata     & \nodata   \\ 
  MRC~0316-257 &  19.3$\pm$ 2.1 &  20.1$\pm$ 2.1 &   19.5$\pm$ 2.4 &   38.1$\pm$  4.1 &  $<$156       & $<$102       &    \nodata     & \nodata   \\ 
  MRC~0324-228 &  39.4$\pm$ 4.2 &  39.7$\pm$ 4.3 &   61.1$\pm$ 8.6 &   89.9$\pm$  9.9 &   530$\pm$ 54 & 1880$\pm$ 35 &    \nodata     & \nodata   \\ 
  MRC~0350-279 &  23.6$\pm$ 2.6 &  40.6$\pm$ 4.2 &   82.2$\pm$27.6 &   79.3$\pm$ 26.5 &   208$\pm$ 40 &  306$\pm$ 44 & $<  1040$      & $< 88800$ \\ 
  MRC~0406-244 &  40.4$\pm$ 4.3 &  43.3$\pm$ 4.6 &  $<  51.6$      &   63.5$\pm$ 14.5 &   637$\pm$ 86 & 1540$\pm$ 40 & 24700$\pm$2306 & $< 47700$ \\ 
       4C60.07 &  20.6$\pm$ 3.1 &  27.0$\pm$ 2.0 &  $<  26.0$      & $<  39.0$        &  \nodata$^a$  & \nodata$^a$  & $<  3750$      & $< 64600$ \\ 
  PKS~0529-549 &  46.6$\pm$ 4.9 &  52.9$\pm$ 5.5 &   62.7$\pm$ 8.6 &   72.2$\pm$  9.1 &   248$\pm$ 89 &  966$\pm$ 40 & $<  4110$      & $< 74100$ \\ 
 WN~J0617+5012 &   3.6$\pm$ 1.0 &   5.5$\pm$ 1.2 &  $<  48.3$      & $<  54.5$        &  $<$203       & $<$122       &    \nodata     & \nodata   \\ 
       4C41.17 &  23.4$\pm$ 2.4 &  27.5$\pm$ 2.8 &   35.6$\pm$ 3.7 &   36.5$\pm$  3.5 &  $<$181       &  370$\pm$ 40 & $<  3210$      & $< 26300$ \\ 
 WN~J0747+3654 &  19.1$\pm$ 2.4 &  25.3$\pm$ 3.0 &  $<  29.9$      &   44.9$\pm$ 11.5 &  $<$198       & $<$117       &    \nodata     & \nodata   \\ 
 6CE~0820+3642 &  79.2$\pm$ 8.1 &  81.9$\pm$ 8.4 &   82.0$\pm$ 8.4 &   68.0$\pm$  7.0 &   225$\pm$ 70 & 1290$\pm$ 40 &    \nodata     & \nodata   \\ 
       5C7.269 &  41.0$\pm$ 4.5 &  49.5$\pm$ 5.3 &   57.8$\pm$ 6.1 & $<  40.1$        &  $<$260       & \nodata      &    \nodata     & \nodata   \\ 
  USS~0828+193 &  61.7$\pm$ 6.9 & 133.0$\pm$13.0 &  201.0$\pm$21.0 &  687.0$\pm$ 74.0 &  1910$\pm$130 & 2880$\pm$ 40 &    \nodata     & \nodata   \\ 
 6CE~0901+3551 &  37.2$\pm$ 4.1 &  46.5$\pm$ 5.0 &   52.8$\pm$10.7 &   69.8$\pm$ 12.4 &   410$\pm$ 70 & 1340$\pm$ 40 &    \nodata     & \nodata   \\ 
    B2~0902+34 &   6.4$\pm$ 0.8 &   9.9$\pm$ 1.1 &   11.0$\pm$ 2.5 &   41.3$\pm$  2.3 &   450$\pm$100 &  323$\pm$ 12 &    \nodata     & $< 18500$ \\ 
 6CE~0905+3955 &  51.8$\pm$ 5.4 &  60.1$\pm$ 6.2 &   96.8$\pm$ 9.8 &  146.0$\pm$ 14.0 &  1440$\pm$120 & 3700$\pm$ 40 &    \nodata     & \nodata   \\ 
 TN~J0924-2201 &  11.3$\pm$ 1.8 &  11.4$\pm$ 1.9 &  $<  30.6$      & $<  33.0$        &  $<$168       & $<$125       & $<  3330$      & $< 52900$ \\ 
   6C~0930+389 &  30.7$\pm$ 3.4 &  32.2$\pm$ 3.6 &   37.3$\pm$ 9.2 & $<  32.6$        &  $<$180       &  530$\pm$ 43 &    \nodata     & \nodata   \\ 
  USS~0943-242 &  21.5$\pm$ 2.6 &  28.4$\pm$ 3.2 &  $<  30.9$      &   25.8$\pm$ 11.7 &   170$\pm$ 48 &  518$\pm$ 40 & $<  3390$      & $< 50900$ \\ 
        3C~239 &  96.4$\pm$ 9.8 & 111.0$\pm$11.0 &  130.0$\pm$12.0 &  142.0$\pm$ 14.0 &   848$\pm$ 90 & 1890$\pm$ 60 &    \nodata     & \nodata   \\ 
  MG~1019+0534 &  25.6$\pm$ 2.9 &  19.5$\pm$ 3.8 &  $<  35.4$      & $<  42.9$        &   220$\pm$ 70 &  415$\pm$ 43 &    \nodata     & \nodata   \\ 
  MRC~1017-220 & 119.0$\pm$12.0 & 179.0$\pm$18.0 &  273.0$\pm$27.0 &  360.0$\pm$ 36.0 &   740$\pm$120 & 1140$\pm$ 30 &    \nodata     & \nodata   \\ 
 WN~J1115+5016 &   7.8$\pm$ 2.1 &   9.5$\pm$ 3.0 &  $<  54.8$      & $<  61.1$        &  $<$160       & $<$100       &    \nodata     & \nodata   \\ 
        3C~257 &  85.0$\pm$ 8.7 & 111.0$\pm$11.0 &  194.0$\pm$19.0 &  322.0$\pm$ 33.0 &  1280$\pm$150 & 1770$\pm$130 &    \nodata     & \nodata   \\ 
 WN~J1123+3141 &  48.2$\pm$ 5.0 &  74.4$\pm$ 7.6 &   92.7$\pm$ 9.4 &  182.6$\pm$ 18.4 &  1160$\pm$110 & 1670$\pm$ 37 &    \nodata     & \nodata   \\ 
  PKS~1138-262 & 318.0$\pm$32.0 & 497.0$\pm$50.0 &  887.0$\pm$89.0 & 1500.0$\pm$150.0 &  3020$\pm$100 & 3890$\pm$ 20 &    \nodata     & \nodata   \\ 
        3C~266 &  67.9$\pm$ 7.0 &  73.1$\pm$ 7.5 &   45.1$\pm$ 4.7 &  102.6$\pm$ 10.4 &   660$\pm$120 &  985$\pm$ 22 &    \nodata     & \nodata   \\ 
    6C~1232+39 &  33.3$\pm$ 3.6 &  41.8$\pm$ 4.4 &   52.0$\pm$ 9.2 &   75.2$\pm$ 10.8 &   360$\pm$ 80 &  459$\pm$ 33 &    \nodata     & \nodata   \\ 
  USS~1243+036 &  22.0$\pm$ 2.5 &  21.5$\pm$ 3.0 &  $<  48.0$      & $<  62.9$        &  $<$260       &  511$\pm$ 43 &    \nodata     & \nodata   \\ 
 TN~J1338-1942 &  17.8$\pm$ 1.9 &  10.7$\pm$ 1.2 &   14.3$\pm$ 2.8 &    9.9$\pm$  3.4 &  $<$226       & $<$99        &    \nodata     & \nodata   \\ 
      4C~24.28 &  16.5$\pm$ 2.2 &  27.3$\pm$ 3.1 &   43.3$\pm$ 8.3 &  102.0$\pm$ 10.0 &   573$\pm$ 96 & 1140$\pm$ 40 &    \nodata     & \nodata   \\ 
      3C~294.0 & $<  93.0$      & $< 103.0$      &   68.0$\pm$16.8 &   66.6$\pm$ 20.6 &   430$\pm$ 99 &  471$\pm$ 31 &    \nodata     & \nodata   \\ 
  USS~1410-001 &  50.6$\pm$ 5.3 &  79.0$\pm$ 8.1 &  166.0$\pm$17.0 &  240.0$\pm$ 24.0 &   660$\pm$120 & 1070$\pm$ 40 &    \nodata     & \nodata   \\ 
   8C~1435+635 &  18.6$\pm$ 2.1 &  14.7$\pm$ 2.9 &  $<  45.5$      & $<  50.3$        &   $<$86       & $<$119       & $<  4130$      & $< 33900$ \\ 
  USS~1558-003 &  78.8$\pm$ 8.1 & 101.0$\pm$10.3 &  105.0$\pm$10.5 &  233.0$\pm$ 23.4 &   993$\pm$100 & 1270$\pm$ 37 &    \nodata     & \nodata   \\ 
  USS~1707+105 &  22.1$\pm$ 2.7 &  30.1$\pm$ 3.4 &   22.6$\pm$ 9.1 & $<  33.3$        &   $<$123      &  264$\pm$ 32 &    \nodata     & \nodata   \\ 
   LBDS~53W002 &  32.0$\pm$ 3.3 &  44.0$\pm$ 4.5 &   49.9$\pm$ 5.2 &  103.0$\pm$ 11.0 &   593$\pm$114 &  648$\pm$ 40 & $<  4300$      & $< 65100$ \\ 
   LBDS~53W091 &  43.3$\pm$ 6.3 &  51.4$\pm$ 5.3 &   23.9$\pm$ 6.1 &   26.6$\pm$  6.4 &   $<$166      & $<$45        & $<  1610$      & $< 23900$ \\ 
 3C~356{\em a} & 108.0$\pm$11.0 & 110.0$\pm$11.0 &  122.0$\pm$14.0 &  434.0$\pm$ 47.0 &  2280$\pm$220 & 4170$\pm$ 40 & $<  4400$      & $< 70200$ \\ 
  7C~1751+6809 &  46.6$\pm$ 4.9 &  50.8$\pm$ 5.3 &  $<  40.9$      &   36.5$\pm$ 16.0 &   147$\pm$ 57 &  342$\pm$ 34 & $<  3600$      & $< 51600$ \\ 
  7C~1756+6520 &  39.6$\pm$ 4.2 &  46.9$\pm$ 5.0 &   34.7$\pm$ 7.7 &   42.6$\pm$  8.6 & 220$^a\pm$114 &  95$^a\pm$40 & $<  5900$      & $<151000$ \\ 
      3C~368.0 & 126.0$\pm$13.0 & 112.0$\pm$11.0 &  112.0$\pm$11.0 &  210.0$\pm$ 21.0 &  1620$\pm$180 & 3350$\pm$ 50 & 28800$\pm$2710 & $< 39000$ \\ 
  7C~1805+6332 &  28.4$\pm$ 3.6 &  42.1$\pm$ 4.4 &   51.4$\pm$ 5.4 &   95.6$\pm$ 17.1 &   310$\pm$ 80 &  673$\pm$ 35 & $<  4330$      & $<111000$ \\ 
      4C~40.36 &  36.5$\pm$ 3.9 &  41.3$\pm$ 4.3 &   45.4$\pm$12.9 &   26.3$\pm$ 10.3 &    54$\pm$ 28 &  536$\pm$ 37 & $<  4750$      & $< 64500$ \\ 
TXS~J1908+7220 & 200.0$\pm$20.0 & 229.0$\pm$23.0 &  241.0$\pm$25.0 &  480.0$\pm$ 48.0 &  1410$\pm$170 & 1910$\pm$ 49 & 16200$\pm$1905 & $< 63300$ \\ 
 WN~J1911+6342 &   9.5$\pm$ 1.8 &   8.7$\pm$ 1.8 &  $<  26.4$      & $<  26.0$        &  $<$137       & $<$102       & $<  3710$      & $< 57100$ \\ 
 TN~J2007-1316 &  56.2$\pm$ 5.9 &  54.4$\pm$ 5.8 &   41.1$\pm$10.7 &  121.3$\pm$ 12.3 &   378$\pm$113 &  385$\pm$ 40 &    \nodata     & \nodata   \\ 
  MRC~2025-218 &  68.4$\pm$ 7.1 &  77.1$\pm$ 8.0 &   86.8$\pm$10.9 &  126.8$\pm$ 12.9 &   200$\pm$ 62 &  216$\pm$ 43 &    \nodata     & \nodata   \\ 
  MRC~2048-272 &  59.5$\pm$ 6.2 &  72.6$\pm$ 7.5 &   78.3$\pm$10.4 &   38.4$\pm$ 13.1 &  $<$220       & \nodata$^a$  &    \nodata     & \nodata   \\ 
  MRC~2104-242 &  28.1$\pm$ 3.3 &  29.7$\pm$ 3.5 &   32.8$\pm$10.0 & $<  36.3$        &  $<$217       &  709$\pm$ 48 &    \nodata     & \nodata   \\ 
      4C~23.56 &  61.1$\pm$ 6.4 &  86.2$\pm$ 8.8 &  126.9$\pm$12.8 &  423.7$\pm$ 42.5 &  2400$\pm$ 90 & 4630$\pm$ 40 & 30300$\pm$2958 & $< 70500$ \\ 
  MG~2144+1928 &  22.1$\pm$ 2.7 &  18.3$\pm$ 2.4 &  $<  25.5$      & $<  30.3$        &  $<$244       &  529$\pm$ 33 &    \nodata     & \nodata   \\ 
  USS~2202+128 &  60.4$\pm$ 6.3 &  95.8$\pm$ 9.8 &  120.1$\pm$12.1 &  106.6$\pm$ 10.8 &   417$\pm$ 88 & 1341$\pm$ 35 &    \nodata     & \nodata   \\ 
  MRC~2224-273 &  61.6$\pm$ 6.4 &  86.1$\pm$ 8.8 &   98.4$\pm$10.0 &  203.9$\pm$ 20.5 &   625$\pm$117 & 1060$\pm$ 40 &    \nodata     & \nodata   \\ 
 B3~J2330+3927 &  99.6$\pm$10.1 & 143.0$\pm$14.0 &  160.0$\pm$16.0 &  474.0$\pm$ 47.0 &  1170$\pm$ 90 & 2320$\pm$ 40 & $<  4670$      & $< 64300$ \\ 
      4C~28.58 &  31.6$\pm$ 3.5 &  36.0$\pm$ 3.9 &  $<  41.7$      &   40.9$\pm$  4.4 &   430$\pm$ 85 &  866$\pm$155 &    \nodata     & \nodata   \\ 
        3C~470 &  49.5$\pm$10.4 &  75.2$\pm$11.8 &   70.9$\pm$10.4 &  266.0$\pm$ 30.0 &  1540$\pm$180 & 2750$\pm$ 40 & $<  5570$      & $<102000$ \\ 
\enddata
\tablenotetext{a}{Contaminated by nearby object.}
\label{table.photometry}
\end{deluxetable}

\begin{deluxetable}{lccccrr}
\tabletypesize{\scriptsize}
\tablecaption{Results of SED fitting.}
\tablewidth{0pt}
\tablehead{
\colhead{Source} &
\colhead{$\log(L_{1.6\,\micron}/\Lsun)$} & 
\colhead{$\log(L_{3\,\micron}/\Lsun)$} & 
\colhead{$\log(L_{5\,\micron}/\Lsun)$} & 
\colhead{$f_{\rm stel}^a$} &
\colhead{$\log(L_{\rm H,*}/\Lsun)$} &
\colhead{$\log(M_{\rm *}/\Msun)$}
}
\startdata
  6C~0032+412 & 12.07 & 12.28 & 12.29 & 0.21 & $<$11.37 & $<$11.18 \\
 MRC~0037-258 & 11.49 & 11.50 & 11.53 & 0.95 &    11.44 &    11.56 \\
  6C*~0058+495 & 11.19 & 11.20 & 11.53 & 0.99 &    11.16 &    11.26 \\
 MRC~0114-211 & 11.40 & 11.70 & 11.96 & 0.96 &    11.36 &    11.39 \\
TN~J0121+1320 & 11.55 & 11.36 & 11.73 & 0.46 &    11.19 &    11.02 \\
  6C*~0132+330 & 11.14 & 11.33 & 11.23 & 0.84 & $<$11.04 & $<$11.03 \\
  6C~0140+326 & 11.84 & 12.61 & 12.89 & 0.88 & $<$11.76 & $<$11.42 \\
 MRC~0152-209 & 11.88 & 12.15 & 12.26 & 0.91 &    11.80 &    11.76 \\
 MRC~0156-252 & 12.40 & 12.57 & 12.43 & 0.54 &    12.11 &    12.05 \\
TN~J0205+2242 & 11.03 & 11.42 & 11.79 & 0.97 &    10.99 &    10.82 \\
 MRC~0211-256 & 11.58 & 11.49 & 11.24 & 0.83 & $<$11.47 & $<$11.54 \\
  TXS~0211-122 & 11.80 & 12.43 & 12.40 & 0.31 & $<$11.26 & $<$11.16 \\
         3C~65 & 11.43 & 11.66 & 11.73 & 0.92 & $<$11.37 & $<$11.47 \\
 MRC~0251-273 & 11.33 & 11.57 & 11.88 & 0.66 &    11.12 &    10.96 \\
 MRC~0316-257 & 11.49 & 11.77 & 11.70 & 0.79 &    11.37 &    11.20 \\
 MRC~0324-228 & 11.33 & 11.44 & 11.71 & 0.96 &    11.29 &    11.25 \\
 MRC~0350-279 & 11.24 & 11.44 & 11.30 & 0.69 & $<$11.04 & $<$11.00 \\
 MRC~0406-244 & 11.53 & 11.65 & 12.10 & 0.99 &    11.49 &    11.38 \\
      4C~60.07 & 11.73 & 12.51 & 12.91 & 0.88 &    11.65 &    11.44 \\
 PKS~0529-549 & 11.66 & 11.75 & 11.95 & 0.94 &    11.60 &    11.46 \\
WN~J0617+5012 & 11.18 & 11.38 & 11.49 & 0.37 &    10.72 &    10.55 \\
      4C~41.17 & 11.66 & 12.00 & 12.21 & 0.93 &    11.60 &    11.39 \\
WN~J0747+3654 & 11.55 & 11.43 & 11.38 & 0.63 &    11.33 &    11.15 \\
 6CE~0820+3642 & 11.57 & 11.19 & 11.27 & 0.95 &    11.52 &    11.48 \\
      5C~7.269 & 11.51 & 11.06 & 11.89 & 0.95 &    11.46 &    11.38 \\
  USS~0828+193 & 12.18 & 12.75 & 12.68 & 0.40 & $<$11.75 & $<$11.60 \\
 6CE~0901+3551 & 11.34 & 11.37 & 11.61 & 0.97 &    11.30 &    11.25 \\
   B2~0902+34 & 11.52 & 12.15 & 12.06 & 0.30 & $<$10.97 & $<$10.81 \\
 6CE~0905+3955 & 11.47 & 11.73 & 12.16 & 0.98 &    11.43 &    11.39 \\
TN~J0924-2201 & 12.19 & 12.40 & 12.18 & 0.21 &    11.50 &    11.10 \\
  6C~0930+389 & 11.32 & 11.44 & 11.80 & 0.99 &    11.28 &    11.17 \\
  USS~0943-242 & 11.43 & 11.49 & 11.93 & 0.99 &    11.40 &    11.22 \\
        3C~239 & 11.66 & 11.57 & 11.90 & 0.99 &    11.63 &    11.60 \\
 MG~1019+0534 & 11.36 & 11.53 & 11.89 & 0.99 &    11.32 &    11.15 \\
 MRC~1017-220 & 11.84 & 11.97 & 11.85 & 0.81 & $<$11.72 & $<$11.70 \\
WN~J1115+5016 & 10.95 & 11.48 & 11.55 & 0.82 &    10.84 &    10.70 \\
        3C~257 & 12.03 & 12.42 & 12.44 & 0.81 & $<$11.91 & $<$11.78 \\
WN~J1123+3141 & 12.13 & 12.67 & 12.68 & 0.60 & $<$11.89 & $<$11.72 \\
  PKS~1138-262 & 12.52 & 12.84 & 12.73 & 0.68 & $<$12.33 & $<$12.26 \\
        3C~266 & 11.13 & 10.90 & 11.10 & 0.99 &    11.10 &    11.18 \\
   6C~1232+39 & 11.80 & 12.18 & 12.15 & 0.73 & $<$11.64 & $<$11.47 \\
  USS~1243+036 & 11.62 & 12.17 & 12.28 & 0.71 & $<$11.45 & $<$11.27 \\
TN~J1338-1942 & 11.37 & 11.55 & 11.92 & 0.99 &    11.34 &    11.04 \\
      4C~24.28 & 11.58 & 12.18 & 12.27 & 0.53 & $<$11.28 & $<$11.11 \\
      3C~294.0 & 11.41 & 11.29 & 11.56 & 0.99 &    11.38 &    11.36 \\
  USS~1410-001 & 11.92 & 12.02 & 11.86 & 0.42 & $<$11.52 & $<$11.41 \\
  8C~1435+635 & 11.80 & 12.04 & 12.08 & 0.56 &    11.53 &    11.21 \\
  USS~1558-003 & 11.95 & 12.05 & 12.05 & 0.82 & $<$11.83 & $<$11.70 \\
  USS~1707+105 & 11.29 & 10.86 & 10.96 & 0.89 &    11.21 &    11.11 \\
   LBDS~53w002 & 11.55 & 11.72 & 11.68 & 0.71 & $<$11.38 & $<$11.27 \\
   LBDS~53w091 & 11.20 & 10.59 & 10.20 & 1.00 &    11.18 &    11.19 \\
      3C~356.0 & 11.21 & 11.23 & 11.62 & 0.99 &    11.18 &    11.29 \\
  7C~1751+6809 & 11.20 & 10.78 & 10.91 & 1.00 &    11.17 &    11.19 \\
  7C~1756+6520 & 11.15 & 10.92 & 11.32 & 1.00 &    11.12 &    11.15 \\
      3C~368.0 & 11.35 & 11.10 & 11.41 & 1.00 &    11.32 &    11.43 \\
  7C~1805+6332 & 11.21 & 11.39 & 11.40 & 0.84 & $<$11.11 & $<$11.07 \\
      4C~40.36 & 11.43 & 10.93 & 11.14 & 0.95 &    11.38 &    11.29 \\
TX~J1908+7220 & 12.70 & 12.91 & 12.82 & 0.59 & $<$12.44 & $<$12.27 \\
WN~J1911+6342 & 11.38 & 11.94 & 11.87 & 0.57 &    11.11 &    10.93 \\
TN~J2007-1316 & 12.19 & 12.42 & 12.24 & 0.56 & $<$11.92 & $<$11.69 \\
 MRC~2025-218 & 11.87 & 11.86 & 11.69 & 0.87 & $<$11.78 & $<$11.62 \\
 MRC~2048-272 & 11.56 & 11.30 & 11.68 & 1.00 &    11.53 &    11.47 \\
 MRC~2104-242 & 11.35 & 11.55 & 11.94 & 0.98 &    11.32 &    11.19 \\
      4C~23.56 & 11.89 & 12.60 & 12.74 & 0.71 & $<$11.71 & $<$11.59 \\
 MG~2144+1928 & 11.49 & 12.22 & 12.32 & 0.70 & $<$11.31 & $<$11.13 \\
  USS~2202+128 & 11.97 & 11.61 & 11.89 & 0.70 &    11.78 &    11.62 \\
 MRC~2224-273 & 11.47 & 11.63 & 11.78 & 0.95 &    11.42 &    11.41 \\
B3~J2330+3927 & 12.43 & 12.71 & 12.66 & 0.51 & $<$12.11 & $<$11.94 \\
      4C~28.58 & 11.57 & 11.84 & 12.24 & 0.98 &    11.54 &    11.36 \\
        3C~470 & 11.35 & 11.72 & 12.02 & 0.95 &    11.30 &    11.30 \\
\enddata
\tablenotetext{a}{Stellar fraction at 1.6\,$\mu$m in our SED model (see \S 3.2).}
\label{table.parameters}
\end{deluxetable}

\begin{deluxetable}{ccccccccc}
\tabletypesize{\scriptsize}
\tablecaption{Binary AGNs}
\tablewidth{0pt}
\tablehead{
\colhead{} &
\colhead{RA} &
\colhead{Dec.} &
\colhead{$f_{3.6\,\um}$} &
\colhead{$f_{4.5\,\um}$} &
\colhead{$f_{5.8\,\um}$} &
\colhead{$f_{8.0\,\um}$} &
\colhead{$f_{16\,\um}$} &
\colhead{$f_{24\,\um}$} \\
\colhead{Source} &
\colhead{(J2000)} &
\colhead{(J2000)} &
\colhead{($\mu$Jy)} &
\colhead{($\mu$Jy)} &
\colhead{($\mu$Jy)} &
\colhead{($\mu$Jy)} &
\colhead{($\mu$Jy)} &
\colhead{($\mu$Jy)} 
}
\startdata
4C~60.07B      & 05 12 54.76 & $+$60 30 48.9 & $<$3.3       &  7.8$\pm$1.0 & 23.5$\pm$3.3 &  43.3$\pm$ 5.1 &  555$\pm$ 83 & 1390$\pm$50 \\ 
3C~356{\em b}  & 17 24 19.02 & $+$50 57 40.3 & 72.6$\pm$7.3 & 71.1$\pm$7.2 & 69.4$\pm$7.5 & 231.9$\pm$23.4 & 2280$\pm$220 & 4170$\pm$40 \\ 
Cl~1756.7     & 17 57 04.98 & $+$65 19 51.0 & 23.0$\pm$2.3 & 33.7$\pm$3.4 & 28.1$\pm$3.8 &  37.6$\pm$ 4.7 & 220$\pm$150  & 370$\pm$40  \\ 
MRC~2048$-$272B & 20 51 03.30 & $-$27 03 03.2 & 24.6$\pm$2.5 & 32.9$\pm$3.4 & 35.2$\pm$4.7 &  34.3$\pm$ 5.0 &  $<$220      &  491$\pm$42 \\ 
\enddata
\label{table.binaryAGN}
\end{deluxetable}



\end{document}